\documentclass[12pt,a4paper, twoside]{report}

\newcommand{\eol}{\notag \\}

\def \bi{\bibitem}

\usepackage{import}
\usepackage{Preamble}
\newcommand{\N}{{\cN}}

\newcommand{\OI}{{\overline{I}}}
\newcommand{\OJ}{{\overline{J}}}
\newcommand{\OK}{{\overline{K}}}
\newcommand{\OL}{{\overline{L}}}
\newcommand{\OM}{{\overline{M}}}

\newcommand{\UI}{{\underline{I}}}
\newcommand{\UJ}{{\underline{J}}}
\newcommand{\UK}{{\underline{K}}}
\newcommand{\UL}{{\underline{L}}}
\newcommand{\UM}{{\underline{M}}}

\newcommand{\1}{{\underline{1}}}
\newcommand{\2}{{\underline{2}}}

\newcommand{\Iu}{\underline{I}}

\newcommand{\Ju}{\underline{J}}

\newcommand{\Io}{{\overline{I}}}

\newcommand{\Jo}{\overline{J}}

\begin{document}

\pagestyle{fancy}
\pagenumbering{Roman}

\begin{titlepage}
\begin{center}
	
	\vspace*{1cm}
	
	{\LARGE \bf Aspects of superconformal symmetry} 
	
	\vspace{1.7cm}
	
	{\Large{\textbf{Emmanouil Sergios Nektarios Raptakis}}}\\
	
	\large  
	\vspace{1cm}
	Supervisor: ~~~~~~~~~~~~~~~~Prof. Sergei M. Kuzenko\\
	Co-supervisor:   ~~~~~~~~~~~~A/Prof. Evgeny I. Buchbinder

	\vspace{2cm}

	\includegraphics[width=0.3\textwidth]{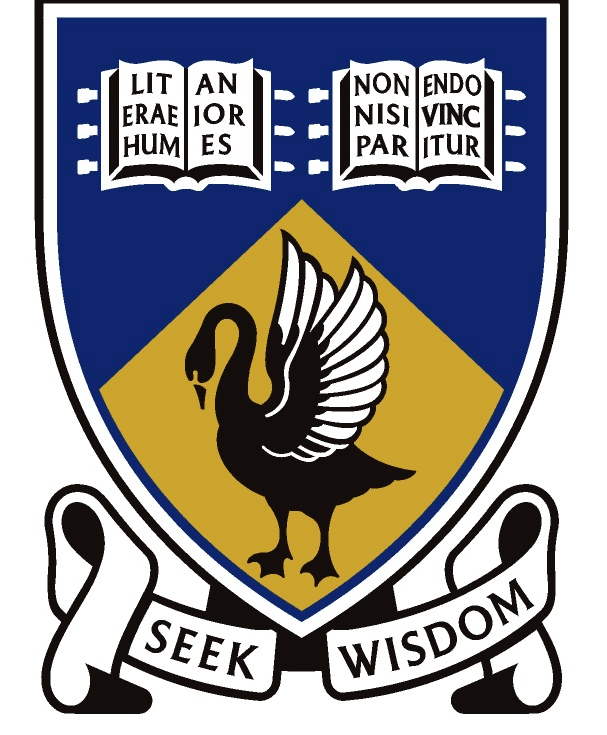}

	\vspace{2.5cm}

	This thesis is presented for the degree of Doctor of Philosophy\\
	The University of Western Australia\\
	Department of Physics\\
	July 2023
	
\end{center} 

\flushleft{
	\underline{Examiners:}\\
	
	Prof. Ivo Sachs ~ \hfill (University of Munich, Germany) \\
	Prof. Paul Townsend ~\hfill ~(University of Cambridge, England)}

\centering
\end{titlepage}
%
%
%
%
%
%
%
%
%
%
%

\chapter*{Abstract}%
\newenvironment{changemargin}[2]{%
	\begin{list}{}{%
			\setlength{\topsep}{0pt}%
			\setlength{\leftmargin}{#1}%
			\setlength{\rightmargin}{#2}%
			\setlength{\listparindent}{\parindent}%
			\setlength{\itemindent}{\parindent}%
			\setlength{\parsep}{\parskip}%
		}%
		\item[]}{\end{list}}

\vspace{-0.6cm}

\begin{changemargin}{-0.1cm}{-0.1cm}
	
	In this thesis we study classical aspects of superconformal field theory via symmetry principles. Specifically, by employing the powerful setup of conformal superspace, we obtain a plethora of new results in the fields of geometric and higher symmetries, (super)conformal higher-spin theory and conformal supergravity. These findings open up numerous novel research pathways.

	After reviewing the necessary background material, we begin by studying the symmetries of curved (super)space. In particular, we show that the conformal isometries of a given background are characterised by a single vector (super)field obeying a closed form equation. 
	By coupling the background to some conformal compensator(s), we also describe its isometries. Associated with each (conformal) isometry is a first-order differential operator preserving the equations of motion of each field theory propagating on the background. As a natural generalisation, we compute the higher-derivative symmetries for the kinetic operators of several fundamental matter multiplets. The latter are shown to provide geometric realisations of (super)conformal higher-spin ((S)CHS) algebras.
	
	Following this, we study various aspects of (S)CHS theories. We begin by identifying all possible higher-spin conformal supercurrents in general curved backgrounds. Then, via the method of supercurrent multiplets, we deduce the structure of the (S)CHS gauge prepotentials and derive their free actions on general conformally flat backgrounds. The latter are then shown to possess $\sU(1)$ duality-invariance at the level of the equations of motion. As a natural extension, we formulate the theory of $\sU(1)$ duality rotations for (S)CHS gauge multiplets and derive several novel models, including a higher-spin extension of ModMax electrodynamics. Finally, we construct a model for conformal matter interacting with an infinite tower of CHS fields of integer spin and discuss its $\cN=1$ superconformal extension.
	
	In the final part, we formulate two-dimensional conformal $(p,q)$ supergravity as a gauge theory of the superconformal group $\sOSp_0(p|2;\mathbb{R}) \times \sOSp_0(q|2;\mathbb{R})$ in a superspace setting. By fixing certain gauge symmetries, we show that this geometry reduces to a conformally flat superspace with structure group $\sSO_0(1,1) \times \sSO(p) \times \sSO(q)$. To conclude, we study several applications and generalisations of our formalism. Specifically, we describe: (i) $\cN$-extended AdS superspace as a maximally supersymmetric geometry in the $p=q=\cN$ case; (ii) supersymmetric extensions of the Gauss-Bonnet invariant and Fradkin-Tseytlin term; and (iii) alternative superconformal groups for the case that $p=2n$.
	
\end{changemargin}
\chapter*{Authorship Declaration}
\label{AuthorDec}

This thesis is based on twelve published papers \cite{KR19,KPR,KPR2,KLRTM,KR21,KR21-2,KPR22,KRTM1,KRTM2,KR22,KR23,HKR}. Their details are as follows:
\begin{enumerate}
	
	\item S.~M.~Kuzenko and E.~S.~N.~Raptakis, \\
	{\it Symmetries of supergravity backgrounds and supersymmetric field theory,}\\
	JHEP \textbf{04}, 133 (2020)
	\href{https://arxiv.org/abs/1912.08552}{[arXiv:1912.08552 [hep-th]]}. \\
	{\bf Location in thesis:} Chapter \ref{Chapter3}.
	\vspace{15pt}
	
	\item S.~M.~Kuzenko, M.~Ponds and E.~S.~N.~Raptakis,  \\
	{\it New locally (super)conformal gauge models in Bach-flat backgrounds,}\\
	JHEP \textbf{08}, 068 (2020)
	\href{https://arxiv.org/abs/2005.08657}{[arXiv:2005.08657 [hep-th]]}. \\
	{\bf Location in thesis:} Chapter \ref{Chapter4}.
	\vspace{15pt}
	
	\item S.~M.~Kuzenko, M.~Ponds and E.~S.~N.~Raptakis,  \\
	{\it Generalised superconformal higher-spin multiplets,}\\
	JHEP \textbf{03}, 183 (2021)
	\href{https://arxiv.org/abs/2011.11300}{[arXiv:2011.11300 [hep-th]]}.\\
	{\bf Location in thesis:} Chapter \ref{Chapter4}.
	\vspace{15pt}
	
	\item S.~M.~Kuzenko, U.~Lindstr\"om, E.~S.~N.~Raptakis and G.~Tartaglino-Mazzucchelli,  \\
	{\it Symmetries of $ \mathcal{N} $ = (1,0) supergravity backgrounds in six dimensions,}\\
	JHEP \textbf{03}, 157 (2021)
	\href{https://arxiv.org/abs/2012.08159}{[arXiv:2012.08159 [hep-th]]}. \\
	{\bf Location in thesis:} Chapters \ref{Chapter2} and \ref{Chapter3}.
	\vspace{15pt}
	
	\item S.~M.~Kuzenko and E.~S.~N.~Raptakis, \\
	{\it Extended superconformal higher-spin gauge theories in four dimensions,}\\
	JHEP \textbf{12}, 210 (2021)
	\href{https://arxiv.org/abs/2104.10416}{[arXiv:2104.10416 [hep-th]]}. \\
	{\bf Location in thesis:} Chapter \ref{Chapter4}.
	\vspace{15pt}
	
	\newpage
	
	\item S.~M.~Kuzenko and E.~S.~N.~Raptakis, \\
	{\it Duality-invariant superconformal higher-spin models,}\\
	Phys. Rev. D \textbf{104}, no.12, 125003 (2021)
	\href{https://arxiv.org/abs/2107.02001}{[arXiv:2107.02001 [hep-th]]}. \\
	{\bf Location in thesis:} Chapter \ref{Chapter4}.
	\vspace{15pt}
	
	\item S.~M.~Kuzenko, M.~Ponds and E.~S.~N.~Raptakis,  \\
	{\it Conformal interactions between matter and higher-spin (super)fields,}\\
	Fortsch. Phys. \textbf{71}, no.1, 1 (2023)
	\href{https://arxiv.org/abs/2208.07783}{[arXiv:2208.07783 [hep-th]]}.\\
	{\bf Location in thesis:} Chapter \ref{Chapter4}.
	\vspace{15pt}
	
	\item S.~M.~Kuzenko, E.~S.~N.~Raptakis and G.~Tartaglino-Mazzucchelli,  \\
	{\it Superspace approaches to $\mathcal{N}=1$ supergravity,} \\
	Invited chapter for the ``Handbook of Quantum Gravity" (Eds. C. Bambi, L. Modesto and I.L. Shapiro, Springer, expected in 2023)
	\href{https://arxiv.org/abs/2210.17088}{[arXiv:2210.17088 [hep-th]]}. \\
	{\bf Location in thesis:} Chapter \ref{Chapter2}.
	\vspace{15pt}
	
	\item S.~M.~Kuzenko, E.~S.~N.~Raptakis and G.~Tartaglino-Mazzucchelli,  \\
	{\it Covariant superspace approaches to ${\cal N}=2$ supergravity,}\\
	Invited chapter for the ``Handbook of Quantum Gravity" (Eds. C. Bambi, L. Modesto and I.L. Shapiro, Springer, expected in 2023)
	\href{https://arxiv.org/abs/2211.11162}{[arXiv:2211.11162 [hep-th]]}.\\
	{\bf Location in thesis:} Chapter \ref{Chapter2}.
	\vspace{15pt}
	
	\item S.~M.~Kuzenko and E.~S.~N.~Raptakis, \\
	{\it Conformal $(p, q)$ supergeometries in two dimensions,}\\
	JHEP \textbf{02}, 166 (2023)
	\href{https://arxiv.org/abs/2211.16169}{[arXiv:2211.16169 [hep-th]]}.\\
	{\bf Location in thesis:} Chapter \ref{Chapter5}.
	\vspace{15pt}
	
	\item S.~M.~Kuzenko and E.~S.~N.~Raptakis, \\
	{\it On higher-spin ${\mathcal{N}=2}$ supercurrent multiplets,}\\
	JHEP \textbf{05}, 056 (2023)
	\href{https://arxiv.org/abs/2301.09386}{[arXiv:2301.09386 [hep-th]]}.\\
	{\bf Location in thesis:} Chapter \ref{Chapter3}.
	\vspace{15pt}
	
	\item D.~Hutchings, S.~M.~Kuzenko and E.~S.~N.~Raptakis, \\
	{\it The $\mathcal{N}=2$ superconformal gravitino multiplet,} \\
	Phys. Lett. B \textbf{845}, 138132 (2023)
	\href{https://arxiv.org/abs/2305.16029}{[arXiv:2305.16029 [hep-th]]}.\\
	{\bf Location in thesis:} Chapter \ref{Chapter4}.
	\vspace{15pt}
	
	Permission has been granted to use the above work. \\
	
	Daniel Hutchings \\ \\ \\
	Ulf Lindstr\"om \\ \\ \\
	Sergei Kuzenko \\ \\ \\
	Michael Ponds \\ \\ \\
	Gabriele Tartaglino-Mazzucchelli
	
\end{enumerate}
\chapter*{Acknowledgements}
While only my name appears on the authorship section on the title page of this thesis, the writing of this work has hardly been a solo effort. In particular, if I have achieved anything at all, it is only because of the knowledge and unwavering support of those I have leaned on over the past four and a half years.

Firstly, I would like to express my gratitude to my supervisor, Sergei Kuzenko, for his guidance and (endless) patience throughout my studies. Additionally, I thank my co-supervisor, Evgeny Buchbinder, for his various comments on this thesis. Further, I would also like to extend my gratitude to Ulf Lindstr\"om, Daniel Hutchings, Michael Ponds and Gabriele Tartaglino-Mazzucchelli for very enjoyable collaborations on our joint works.

I am also grateful to all past and present members of the field theory and quantum gravity group for their support. In particular, I would like to single out Daniel Hutchings, for the enjoyable academic journey we have shared since 2015, Darren Grasso, for always being a figure of support and a source of enjoyable martial arts discussions and Benjamin Stone, for the many interesting and hilarious conversations we have shared. Additionally, I extend my thanks to James La Fontaine, Jessica Hutomo, Nowar Koning, Conway Li, Joshua Pinelli, Jake Stirling and Kai Turner. Also, I would like to thank Darren Grasso, Daniel Hutchings, Benjamin Stone and Gabriele Tartaglino-Mazzucchelli for their invaluable comments on my thesis.

I also wish to thank my thesis examiners, Prof. Ivo Sachs and Prof. Paul Townsend, for their valuable comments and suggestions on this work.

Although this thesis is a purely academic work, its completion would have been impossible without the assistance of my friends and family. First, I would like to thank my parents, Ioannis and Polyxeni, stepmother Georgina, grandparents Theodore and Rose, and sister Theodora for everything they have done in order for me to reach this moment, especially the things I have yet to realise. I am also grateful to my close friends Andrew, Elouise, Lara, Marcella, Monica and Reuben for always being there for me in difficult periods and for our shared laughs.

The completion of my PhD would hardly have been possible without my wonderful partner Lila (and her amazing family), who has been a constant source of support, always looked out for my best interests through this challenging period and, of course, opened my eyes to the cuteness of cats. It would be impossible to mention Lila without Fish, the cutest tuxedo cat there ever was. I will always cherish the time we spent together and not a day goes by where I don't miss you.

During the course of my PhD, the ``Cat Cafe Purrth'' was a constant source of mental relief during the many strenuous periods I faced. While its recent and sudden closure has been a travesty, I am eternally grateful to Pepi, Chris and all of their staff for the many memories I made there. To all the veteran cats of the cafe, I miss you every day and thank you for your service.

I would also like to extend my thanks to all the coaches and friends I have made over the past three and a half years at the UWA Judo Club. Your mentorship has allowed me to grow immensely and I am confident that this thesis would hardly take the form it does today without your guidance. 

This research was supported in part by by the Hackett Postgraduate Scholarship UWA,
under the Australian Government Research Training Program and by the Australian 
Research Council, project No. DP230101629.

{\hypersetup{hidelinks}
\tableofcontents
}
\clearpage\pagenumbering{arabic}

\chapter{Introduction} \label{Chapter1}

\noindent
\textit{
	``Man had always assumed that he was more intelligent than dolphins because he had achieved so much—the wheel, New York, wars and so on—whilst all the dolphins had ever done was muck about in the water having a good time. But conversely, the dolphins had always believed that they were far more intelligent than man—for precisely the same reasons.''} \\
\hfill --- Douglas Adams, The Hitchhiker's Guide to the Galaxy
\\

Symmetry principles have played a pivotal role in the study of fundamental physics since the dawn of the 20th century. This is perhaps best exemplified by Einstein's breakthrough discovery of special relativity in 1905. Specifically, he suggested that the laws of nature should be covariant with respect to the Poincar\'e group (also known as the inhomogeneous Lorentz group), which encodes the global geometric symmetries of spacetime, severely constraining the possible physical laws. Since then, such symmetry principles have proven to be an invaluable tool to probe the structure of fundamental physics.

Within the modern framework of theoretical physics, the Poincar\'e group $\mathsf{IO}(d-1,1)$ is understood as the isometry group of $d$-dimensional Minkowski space; Poincar\'e transformations are those which preserve the spacetime interval
\begin{align}
	\label{1.1}
	\rd s^2 = \eta_{a b} \rd x^a \rd x^b ~, \qquad \eta_{a b} = \text{diag}(-1,1,\dots,1)~, \qquad a,b = 0\,, \dots \,, d-1~.
\end{align}
Of particular interest is its connected component, $\mathsf{ISO}_0(d-1,1)$, which is the symmetry group of every closed relativistic system.\footnote{This means that we exclude from our study the discrete $\mathsf{P}$ and $\mathsf{T}$ transformations, which are not symmetries of all physical systems. While it was originally believed that this was the case, it was discovered in the 1950s that weak interactions do not obey parity invariance. Additionally, in the 1960s, it was discovered that time-reversal symmetry is only upheld approximately. We refer the reader to \cite{WeinbergVol1} and references therein for further details. } 

Classically, Poincar\'e invariance is mathematically realised by the requirement that action functionals must take the form $\cS[\mathcal{L}] = \int \rd^dx \, \cL(x)$, where the Lagrangian $\cL$ is scalar with respect to Lorentz transformations, which places significant restrictions on the possible physical laws. Further, owing to Noether's theorem, Poincar\'e symmetry implies the existence of $d(d+1)/2$ conserved quantities -  one for each generator of $\mathsf{ISO}_0(d-1,1)$. At the quantum level, the latter generate symmetry transformations in the Hilbert space of physical states and obey the commutation relations:
\begin{align}
	\label{1.2}
	[M_{ab},M_{cd}]=2\eta_{c[a}M_{b]d}-2\eta_{d[a}M_{b]c}~, \qquad
	[M_{ab},P_c]=2\eta_{c[a}P_{b]}~,
\end{align}
where $M_{ab} = - M_{ba}$ and $P_a$ are the Lorentz and translation generators, respectively. 
Further, according to the Wigner classification \cite{Wigner}, see e.g. \cite{WeinbergVol1} for a review, the irreducible unitary representations of the Poincar\'e group are identified with free elementary particles. It would not be an exaggeration to say that this classification forms the foundation of our understanding of quantum field theory.

\subsubsection{Conformal symmetry}

In the present day, our quest for new fundamental laws of physics is largely guided by considering new symmetries of nature. Of particlar importance are the so-called conformal symmetries, which, alongside their supersymmetric extensions, will play a central role in this thesis. Historically, such symmetries have been of great interest since it was shown in 1909 by Cunningham \cite{Cunningham} and Bateman \cite{Bateman} that Maxwell's equations are conformally-invariant. Such symmetries also play a pivotal role within the landscape of modern theoretical physics, in particular through the AdS/CFT correspondence \cite{Maldacena,GKP,Witten} and string theory (see e.g. \cite{AdSCFTReview} and \cite{BLT,GSW1,GSW2} for reviews of the former and latter, respectively).

We recall that the conformal group of (compactified) Minkowski space, $\mathsf{O}(d,2)/\mathbb{Z}_2$, is spanned by those transformations preserving the spacetime interval \eqref{1.1} up to a local scaling
\begin{align}
	\rd s^2 \xrightarrow{\mathsf{O}(d,2)/\mathbb{Z}_2} \re^{2 \s(x)} \rd s^2~.
\end{align}
Its associated Lie algebra is a natural extension of the Poincar\'e algebra \eqref{1.2}, obtained by incorporating the dilatation $\mathbb{D}$ and special conformal $K_a$ generators:
\begin{subequations}
	\label{1.4}
	\begin{align}
		[\mathbb{D},P_a]&=P_a~, \qquad \qquad [\mathbb{D},K_a]=-K_a~,\\
		[M_{ab},K_c]&=2\eta_{c[a}K_{b]}~, \qquad [K_a,P_b]=2\eta_{ab}\mathbb{D}+2M_{ab}~.
	\end{align}
\end{subequations}
For a field theory to be conformal, it suffices for its action to be Poincar\'e-invariant and for its Lagrangian to be a primary field of dimension $d$ \cite{WessConf,MackSalam}, see also \cite{FGG,FP,Todorov} for a review. Hence, by employing the principles of conformal symmetry, the space of physical theories has become significantly constrained. Such simplifications are, philosophically, at the heart of this work.

\subsubsection{Supersymmetry and superspace}

Within the realm of theoretical physics, few mathematical discoveries since the end of the Second World War rival that of supersymmetry, a concept of fundamental importance which has profoundly reshaped our understanding of fundamental physics. We recall that supersymmetry is a symmetry relating bosons and fermions, discovered during the early 1970s by Golfand and Likhtman \cite{GL}, Volkov and Akulov \cite{VA1,VA2} and Wess and Zumino \cite{WZ74,WZ74-2}.\footnote{It should be emphasised that our discussion here is specific to the $d=4$ case. 
} Remarkably, supersymmetry offers a solution to the infamous hierarchy problem plaguing the standard model, see \cite{Martin,WeinbergVol3} for a review. Further, the marriage of local supersymmetry and Poincar\'e invariance yields the supersymmetric extension of general relativity, namely supergravity \cite{FvNF,DZ}. The latter possesses many remarkable features, see e.g. \cite{WB,BK,GGRS,FVP} for a review, and thus, almost 50 years since its discovery,\footnote{We refer the reader to \cite{SG40} for a historical account of supergravity.} it remains a prominent area of research.

The specific feature of supersymmetric field theories is that they possess fermionic operators (supercharges) which generate the supersymmetries; such operators take bosonic fields to fermionic fields and vice versa. As the structure of spinor representations depends on the dimension of spacetime, we will now fix $d=4$ for convenience. The generators of supersymmetries are then denoted $Q_\a^i$, $\a = 1 ,2$ and $i = 1 \,,\dots \,, \cN$, where $\a$ is a Weyl spinor index, while $i$ is an isospinor index and $4 \cN$ indicates the total number of real supercharges. 
Along with their conjugates, they obey the anti-commutation relations
\begin{align}
	\label{1.5}
	\{ Q_\a^i , \bar{Q}_{\ad j} \} = - 2 \ri \d_j^i (\s^a)_\aa P_a~.
\end{align}
Here $\s^a=(\mathds{1} , \vec{\s})$, where $\vec{\s}$ denotes the usual Pauli spin matrices. Importantly, if one appends this relation to the Poincar\'e algebra \eqref{1.2}, the resulting structure is closed\footnote{Strictly speaking, one also needs to include the commutation relation $[M_{ab}, Q_\g^i \big] = (\s_{ab})_\g{}^\d Q_\d^i$ (and its conjugate), where $\s_{ab}$ is defined in \eqref{2.182}, see e.g \cite{WeinbergVol3} for more details.} and forms a superalgebra.

While supersymmetric field theories may be formulated on ordinary spacetime as those described by Lagrangians invariant under supersymmetries, it is more natural to work within the manifestly supersymmetric framework of superspace. Such a formalism was introduced independently by Volkov and Akulov \cite{AV} and Salam and Strathdee \cite{SS}.\footnote{It should be noted that Volkov and Akulov restricted their use of superspace to nonlinear realisations of supersymmetry while Salam and Strathdee proposed it for the purpose of field theory.} This framework is obtained by appending spacetime with additional anti-commuting (fermionic) coordinates. Supersymmetric field theories are then described in terms of fields on superspace, known as superfields, which naturally define representations of the supersymmetry algebra,\footnote{In practice, one must impose additional covariant constraints on the superfields such that they define irreducible representations of this algebra.} hence the supersymmetry is kept manifest. Thus, this approach will be extensively employed in this thesis. For thorough introductions to superspace, we refer the reader to \cite{GGRS,BK,WB}.

\subsubsection{Superconformal symmetry}

Given the interesting properties exhibited by field theories exhibiting both supersymmetry and conformal symmetry, it is natural to study them in unison. Such theories are said to be superconformal and will play a key role throughout this thesis. In four dimensions, the superconformal algebra was introduced in the seminal work \cite{HLS} and studied further by Sohnius \cite{Sohnius}.\footnote{We point out that $\cN=1$ superconformal symmetry had been discovered earlier by Wess and Zumino \cite{WZ74}.} In particular, it was shown in the latter that superconformal transformations preserve the `metric' of flat superspace up to a local scaling.

The superconformal algebra is obtained from its non-supersymmetric cousin \eqref{1.4} by appending to it the supercharges $Q_\a^i$ described above. For the resulting algebra to be closed, it is then necessary to introduce generators for $S$-supersymmetry and $R$-symmetry transformations. The former are fermionic analogues of the special conformal tranformations, while the latter constitute outer automorphisms of the supercharges. This superalgebra will be described in further detail in chapter \ref{Chapter2.1}.

\subsubsection{Conformal supergravity and the superconformal tensor calculus}

When studying superconformal field theories, it is useful to work within a framework which allows one to keep the superconformal symmetry manifest. Such a formalism, known as the superconformal tensor calculus (STC), was pioneered by Kaku, Townsend and van Nieuwenhuizen \cite{KTvN1,KTvN2,FKTvN,KakuTownsend} as the gauge theory of the superconformal group in spacetime, see \cite{FVP} for a review.\footnote{In the original work \cite{KTvN1}, the authors also formulated conformal gravity as the gauge theory of the conformal group.} Specifically, one introduces a gauge field for each superconformal generator.\footnote{It should be emphasised that the connection associated with the translational generators $P_a$ is the spacetime vielbein.} Then, to obtain the usual Weyl multiplet of conformal supergravity, one imposes certain covariant constraints on the gauge fields, much like when one imposes the torsion-free constraint in (pseudo-)Riemannian geometry to determine the Levi-Civita connection.

The STC has lead to a plethora of new results in the context of supergravity theories. In particular, as follows from the construction of Kaku and Townsend \cite{KakuTownsend}, Poincar\'e supergravity may be described by coupling the superconformal gauge multiplet to certain compensating multiplets. Further, this framework was employed by Ferrara {\it et al.} \cite{FGKvP} to elucidate the relationship between different off-shell\footnote{An action is are said to `off-shell' if the algebra of supersymmetry transformations closes without needing to impose the dynamical equations.} formulations of $\cN=1$ supergravity-matter systems. In the $\cN=2$ case, the STC was utilised to construct one of the first off-shell supergravity formulations \cite{deWvHVP} and then further developed in \cite{BdeRdeW,deWvHVP2,deWPV} to derive many important results for supergravity-matter systems \cite{deWit:1983xhu,deWit:1984rvr}. More recently, it has been utilised to obtain off-shell constructions of $\cN=3$ conformal supergravity \cite{HMS,MP,HS}.

While the STC is powerful in formulating supergravity-matter systems, it has some limitations. For instance, in the $\cN=2$ case, it is not equipped to describe off-shell charged hypermultiplets.\footnote{These issues are also present in the $d=5,~ \cN=1$ \cite{Ohashi,Bergshoeff5D} and $d=6,~ \cN=(1,0)$ \cite{BSvP} superconformal tensor calculi.} Specifically, the latter are either on-shell or involve a gauged central charge. Such realisations for the hypermultiplet are unable to provide an off-shell description for the most general locally supersymmetric nonlinear $\s$ model. Further, if such a formulation were to exist, it would necessarily be described by off-shell hypermultiplets possessing an infinite tail of auxiliary fields. Consequently, it is exceptionally difficult to work with off-shell hypermultiplets at the component level. These issues can be mitigated via superspace methods. The appropriate techniques were first developed using harmonic \cite{GIKOS} superspace and later introduced into projective superspace \cite{LR1}.\footnote{See \cite{GIOS} and \cite{LR2008,K2010} for reviews of harmonic and projective superspace, respectively.}  Additionally, the STC does not seem to offer powerful insights into the construction of higher-derivative supergravity actions. We refer the reader to \cite{VPHD} for further details.

\subsubsection{Conformal superspace}

As follows from the discussion above, it is natural to consider a superspace analogue of the STC, that is, a gauge theory of the superconformal algebra in superspace. Owing to the manifest off-shell supersymmetry inherent to superspace formulations, such a construction would be ideal for describing superconformal field theories. Such a formulation was developed for the $\cN=1$ and $\cN=2$ cases by Butter \cite{ButterN=1,ButterN=2} and is known as conformal superspace. Shortly thereafter, extensions of this construction to $3 \leq d \leq 6$ dimensions were described in the works \cite{BKNT-M1,BKNT-M3,BKNT}. These formulations have lead to numerous applications, including: (i) off-shell actions for $3 \leq \cN \leq 6$ conformal supergravity \cite{BKNT-M2,KNTM3D} and on-shell constructions of $4 \leq \cN \leq 8$ topologically massive supergravity theories \cite{LS1,LS2} in three dimensions; (ii) the construction of all $\cN = 4$ conformal supergravity actions in four dimensions  \cite{ButterN=4}; (iii) new tools to generate higher-derivative off-shell supergravity invariants and to study general supergravity-matter systems in five dimensions \cite{BKNT-M3}; and (iv) the construction of new invariants for $\cN=(1,0)$ conformal supergravity in six dimensions \cite{BKNT}. 

By definition, conformal superspace is a supermanifold $\cM^{4|4\cN}$ having the superconformal group as its local structure group. Thus, the geometry of $\cM^{4|4\cN}$ is encoded in the conformally covariant derivatives
\bea
\label{1.6}
\nabla_A = (\nabla_a, \nabla_{\a}^i, \bar{\nabla}_i^\ad) = E_A{}^M \pa_M - \O_A{}^{\underline{B}} X_{\underline{B}} ~.
\eea
Here $E_A{}^M$ denotes the inverse superspace vielbein and $\O_A{}^{\underline{B}}$ collectively denotes the connections associated with non-supertranslational generators of the superconformal group $X_{\underline{B}}$. The algebra of covariant derivatives $[\nabla_A, \nabla_B\}$ is expressed solely in terms of the conformal curvatures of superspace. In what follows, we will identify this geometry with the pair $(\cM^{4|4\cN},\nabla)$.

The power of conformal superspace lies in the manifest covariance of $\nabla_A$ with respect to the superconformal group; it varies under structure group transformations as follows:
\bea
\d_\mathscr{K} \nabla_A &=& [\mathscr{K} , \nabla_A] ~, \qquad 
\qquad \mathscr{K}:= \x^B  \nabla_B + \L^{\underline{B}}  X_{\underline{B}} ~, 
\label{1.7}
\eea
where the gauge parameters $\x^B$ and $\L^{\underline{B}}$ obey standard reality conditions but are otherwise arbitrary. Further, given a tensor superfield $\Phi$ (with suppressed indices), its superconformal transformation law is simply
\begin{align}
	\d_\mathscr{K} \Phi = \mathscr{K} \Phi~.
\end{align}

As will be shown in the following chapter, the gauge transformations \eqref{1.7} may be utilised to impose a gauge where the supervielbein is the only independent geometric object. This process is known as degauging and reduces the structure group to $\sSL(2,\mathbb{C}) \times \sU(\cN)_R$, where $\sU(\cN)_R$ is the $R$-symmetry group. The resulting supergeometry coincides with the $\sU(\cN)$ superspace due to Howe \cite{Howe}.
In addition to structure group transformations, the gauge group of such supergeometries also includes super-Weyl transformations.\footnote{These transformations naturally originate within conformal superspace. This connection will be elucidated in the next chapter.} By employing the latter, one may further reduce the structure group to $\sSL(2,\mathbb{C}) \times \sSU(\cN)_R$. In the $\cN=1$ and $\cN=2$ cases, this leads to the Grimm-Wess-Zumino (GWZ) and $\sSU(2)$ superspace formulations of \cite{GWZ} and \cite{Grimm}, respectively.

While the $\sU(\cN)$, GWZ and $\sSU(2)$ formulations have a longer tenure in the literature than conformal superspace, when studying superconformal field theories, they possess some limitations as compared with the latter. Specifically: (i) the resulting geometry is described by several non-conformally covariant torsions; (ii) the algebra of covariant derivatives is significantly more complex (especially in the case of a conformally-flat background); and (iii) it becomes, especially in the case of higher-derivative models, quite non-trivial to show super-Weyl invariance.

While conformal superspace is formulated such that it ensures manifest superconformal invariance, it is also convenient for describing off-shell Poincar\'e supergravity. This may be achieved by coupling the background to certain compensating multiplets $\Xi$, in accordance with the universal approach of Kaku and Townsend \cite{KakuTownsend}. Conventional superspace descriptions of Poincar\'e supergravity may be obtained from this description by utilising the superconformal symmetry to fix the gauge $\Xi = 1$.
Further, any field theory in curved superspace may be made superconformal by coupling it to such a conformal compensator. 

\subsubsection{Geometric symmetries of curved backgrounds}

The approach to curved supergeometries sketched above is particularly powerful in the description of geometric symmetries of curved backgrounds. The latter are particularly important in the study of supersymmetric field theories propagating on such backgrounds as they imply highly non-trivial restrictions on its dynamics. A general framework to study (conformal) isometries of curved superspace in $3 \leq d \leq 6$ dimensions was developed by Kuzenko in \cite{K15}, following the description of (conformal) isometries of curved backgrounds in old minimal supergravity \cite{BK}. Within four-dimensional conformal superspace the scheme of \cite{K15} can be summarised as follows. A real supervector field $\x= \x^B E_B$ on $(\cM^{4 |4 \cN}, \nabla)$ is said to be conformal Killing if the transformation \eqref{1.7} preserves the covariant derivatives
\bea
\d_{\mathscr{K}[\xi]} \nabla_A = [\xi^B \nabla_B + \L^{\underline{B}}  X_{\underline{B}} , \nabla_A] = 0~.
\label{1.9}
\eea
It is expected that \eqref{1.9} implies the following properties:
\begin{itemize}
	\item The parameters $\L^{\underline{B}}$ are uniquely determined 
	in terms of $\x^B$; $\L^{\underline{B}} = \L^{\underline{B}}[\xi]$.
	\item The spinor parameters $\x^{\b}_j$ and $\bar{\x}_\bd^j$ are uniquely determined in terms of $\x^b$, which is known as a conformal Killing vector superfield
	\item The vector parameter $\x^b$ obeys a closed-form equation containing complete information regarding the  conformal Killing supervector field.
	\item The set of conformal Killing supervector fields forms a finite-dimensional Lie superalgebra with respect to the standard Lie bracket. This is the superconformal algebra of $(\cM^{4 |4 \cN}, \nabla)$.
\end{itemize}
Analogous properties to those above have been established for diverse dimensions in several publications \cite{BK,KLRST-M,BIL,KNT-M,KLRTM,LO}, however, the $\N\geq2$ case in four dimensions remains largely unexplored. One of our goals herein is to bridge this gap.

In addition, the scheme sketched above may be generalised to describe isometries of Poincar\'e supergravity backgrounds. Recalling that the latter may be described within the framework of conformal superspace by coupling the background to some compensator $\Xi$. Then, the isometry transformations are those conformal isometries, eq. \eqref{1.9}, which also preserve $\Xi$:
\begin{align}
	\label{1.10}
	\d_{\mathscr{K}[\xi]} \Xi = (\x^B  \nabla_B + \L^{\underline{B}}[\xi]  X_{\underline{B}}) \Xi = 0~.
\end{align}
It should be emphasised that when the gauge $\Xi = 1$ is fixed, the isometry conditions \eqref{1.9} and \eqref{1.10} yield a supersymmetric extension of the usual Killing equation.

\subsubsection{Higher symmetries}

Remarkably, conformal isometries \eqref{1.9} define symmetries of every superconformal field theory defined on the associated background.
Specifically, they leave invariant every superconformal action $\cS[\F]$ describing the propagation of some supermultiplet $\F$ on $(\cM^{4|4\cN},\nabla)$.
Consequently, such symmetries preserve all off-shell constraints obeyed by $\F$ (e.g. its superconformal properties) and its dynamical equations.
For simplicity, we group these constraints together into a single equation, $\mathfrak{L} \Phi = 0$, for some appropriate operator $\mathfrak{L}$. We emphasise that the operator $\mathfrak{L}$ defines a superconformal wave equation.

Then, given a conformal Killing supervector field $\xi$ on $(\cM^{4 |4 \cN}, \nabla)$, the first order operator $\mathfrak{D}^{(1)} = \mathscr{K}[\xi]$ preserves the superconformal wave equation $\mathfrak{L} \Phi = 0$
\begin{align}
	\mathfrak{L} \Phi = 0 \quad \implies \quad \mathfrak{L} \mathfrak{D}^{(1)} \Phi = 0~.
\end{align}
Hence, the operator $\mathfrak{D}^{(1)}$ is said to be a symmetry of $\mathfrak{L}$. It should also be emphasised that, as follows from the discussion of conformal isometries given above, $\mathfrak{D}^{(1)}$ is determined solely in terms of a single parameter $\xi^a$, which is a conformal Killing vector superfield.

One of the main objectives of this work is to study higher-derivative symmetries of superconformal kinetic operators $\mathfrak{L}$. These are $n^{\text{th}}$-order ($n \geq 2$) linear differential operators $\mathfrak{D}^{(n)}$ preserving the wave equation of $\mathfrak{L} \F = 0$. The operator $\mathfrak{D}^{(n)}$ is said to be a `higher symmetry' of $\mathfrak{L}$. As will be shown, higher symmetries are described in terms of a single tensorial parameter $\xi^{a(n)}$ obeying a closed-form constraint which generalises the usual conformal Killing equation. Thus, such parameters are said to be conformal Killing tensor superfields. It should also be mentioned that, in the case that $\mathfrak{L}$ is dependent on some conformal compensator(s), it is necessary to impose some supplementary constraints on $\xi^{a(n)}$ for $\mathfrak{D}^{(n)}$ to constitute a higher symmetry of $\mathfrak{L}$. Such constraints generalise the familiar Killing equation and so $\xi^{a(n)}$ is said to be a Killing tensor superfield.

The study of higher symmetries has a rich history in the mathematical literature, see e.g. \cite{KM,ShSh,Eastwood}, \cite{MSWMath} and \cite{KMW}, for the higher symmetries of the conformal d'Alembertian, Dirac and Maxwell operators respectively. This interest is largely due to their relationship with that of separable coordinate systems for partial differential equations, see e.g. \cite{KM,WF}. More recently, the higher symmetries of supersymmetric kinetic operators were initiated by Howe and Lindstr\"om in flat superspace \cite{HL2}. Specifically, they considered higher symmetries of so-called super-Laplacians.\footnote{In the original work \cite{HL2}, super-Laplacians were defined as ``a set of differential operators in superspace whose highest-dimensional component is given by the spacetime Laplacian."} Such symmetries are particularly important in the context of supersymmetric higher-spin theory as they form super(conformal) higher-spin algebras \cite{FL-algebras,Bekaert}.

\subsubsection{Superconformal higher-spin theory}

The gauge theory of the superconformal higher-spin algebras described above yields what is known as superconformal higher-spin (SCHS) theory \cite{FT, FL-algebras}. The latter describes interactions between conformal supermultiplets of all superspins, consistent with superconformal invariance.\footnote{One may also consider conformal higher-spin (CHS) theory, which is the gauge theory of fields of all integer spins, consistent with conformal invariance \cite{FT}.} By employing a component field approach, gauge-invariant actions for SCHS multiplets were constructed at the cubic level in \cite{FL}. To probe the structure of SCHS theories further, it would be advantageous to employ a manifestly supersymmetric construction via superspace. Such studies were initiated by Howe, Stelle and Townsend \cite{HST}, who postulated an infinite family of $\cN$-extended SCHS gauge prepotentials $H_{\a(s) \ad(s)}$, $s\geq1$,\footnote{It should be emphasised that, for $\cN=1$, $H_\aa$ describes the conformal supergravity multiplet \cite{FZ2}, which does not contain higher-spin fields.} in Minkowski superspace. More recently, Kuzenko, Manvelyan and Theisen \cite{KMT} postulated the $\cN=1$ SCHS gauge prepotentials $H_{\a(m) \ad(n)}$, $m,n\geq1$, and derived their corresponding gauge-invariant actions in flat superspace. Their construction was subsequently extended to general conformally-flat backgrounds by Kuzenko and Ponds in \cite{KP}. This thesis is, in part, aimed at the further development of these constructions. 

The off-shell\footnote{Supermultiplets are said to be `off-shell' if the algebra of supersymmetry transformations closes without needing to impose their dynamical equations.} superconformal higher-spin multiplets were derived in \cite{HST}, several years before the works \cite{FT,FL-algebras,FL}, though the corresponding kinetic actions were not considered. Their structure was deduced by employing the method of supercurrent multiplets \cite{HST,OS,BdeRdeW}; once the structure of a conformal higher-spin supercurrent $\mathfrak{J}^\cA$ is known, its associated gauge prepotential $\U_\cA$ is determined by requiring that the Noether coupling
\begin{align}
	\label{1.12}
	\mathcal{S}_{\rm N.C.} = \int \rd \mu \, \U_\cA  \mathfrak{J}^\cA ~,
\end{align}
is locally superconformal and gauge-invariant, where $\rd \mu$ is an appropriately defined superspace measure. In this thesis we will employ superconformal symmetry principles to first derive the most general higher-spin conformal supercurrents $\mathfrak{J}^\cA$, which in turn will allow us to deduce all possible SCHS gauge prepotentials $\U_\cA$, including some exotic families of multiplets not considered in \cite{HST}. The free actions for the conformal gauge supermultiplets $\U_\cA$ are, in general, not yet known.\footnote{Free actions for the CHS gauge fields were first given in \cite{FT}.} Specifically, in the $\cN=1$, case such actions were recently derived in Minkowski superspace \cite{KMT} and subsequently extended to conformally-flat backgrounds \cite{KP}. The $\cN\geq2$ story is yet to be studied, and its completion is one of our goals herein.

As discussed above, the ultimate goal of this superspace construction is to obtain a manifestly off-shell supersymmetric description of the complete nonlinear SCHS theory. The natural approach to such a formulation, as advocated for in \cite{KMT}, is a superspace analogue of the effective action approach to CHS theory of \cite{Tseytlin,Segal}. Such a formulation was given, in the $\cN=1$ case, in \cite{KP23} and is as follows: given a gauge-invariant action $\cS[\S,\U]$ describing the interactions between some matter supermultiplets $\S$ and an infinite tower of background SCHS gauge multiplets $\U_\cA$, the action for SCHS theory is formally defined as the logarithmically divergent part of the 1-loop determinant of the operator determining matter couplings.

\subsubsection{Conformal supergeometries in two dimensions}

As we have seen in the four-dimensional context above, symmetry principles are a powerful tool by which to study superconformal field theories. We now extend the scope of our consideration to the two-dimensional case. Historically, local superconformal symmetry in two dimensions played a major role in the study of string theory and supergravity. In particular, the $\cN=1$ superstring can be formulated as a linear sigma model coupled either to $\cN=1$ Poincar\'e supergravity \cite{DeserZumino,BDVH} with super-Weyl invariance \cite{Howe1979} or  $\cN=1$ conformal supergravity \cite{vanNieuwenhuizen:1985an}. Similar analyses for the $\cN=2$ superstring modelled as a Weyl invariant matter-coupled $\cN=2$ supergravity theory and as a linear sigma model coupled to $\cN=2$ conformal supergravity were conducted in \cite{BrinkSchwarz} and \cite{vanNieuwenhuizen:1985an}, respectively.

Unlike the four-dimensional case, the superconformal group in two dimensions proves to be non-simple; it takes the form $G = G_L \times G_R$, where the simple supergroups  $G_L$ and $G_R$ were classified by G\"unaydin, Sierra and Townsend \cite{GST} (see also \cite{McCabe:1986vb}). Within the framework of the superconformal tensor calculus, conformal $(p,q)$ supergravity in two dimensions was described as the gauge theory of the superconformal group $\sOSp_0 (p|2; {\mathbb R} ) \times  \sOSp_0 (q|2; {\mathbb R} )$, for $p,q \leq 2$, in the mid 1980s \cite{vanNieuwenhuizen:1985an, Uematsu:1984zy, Uematsu:1986de, Hayashi:1986ev, Uematsu:1986aa, Bergshoeff:1985qr, Bergshoeff:1985gc, McCabe:1986jg}. Additionally, in 1985, an off-shell formulation for $\cN=4 \equiv (4,4)$ conformal supergravity as a gauge theory of the superconformal group
$\sPSU (1,1|2 ) \times  \sPSU (1,1|2 )$ was described by Pernici and van Nieuwenhuizen \cite{Pernici:1985dq}. One of the main goals of this thesis is to construct, as a generalisation of the component results, superspace formulations for conformal $(p,q)$ supergravity, $p,q\geq0$, as a gauge theory of the superconformal group $\sOSp_0 (p|2; {\mathbb R} ) \times  \sOSp_0 (q|2; {\mathbb R} )$.

\subsubsection{Thesis structure}
This thesis is organised as follows. We begin in chapter \ref{Chapter2} by reviewing the salient aspects of (primarily four-dimensional) conformal (super)gravity and (super)conformal field theory fundamental to our studies in the subsequent chapters. This chapter is partially based on the publication \cite{KLRTM} and the review articles \cite{KRTM1,KRTM2}. In particular, we provide a derivation of the $\cN$-extended (super)conformal algebra in section \ref{Chapter2.1}. In sections \ref{Chapter2.2} and \ref{Chapter2.3}, we describe in detail conformal (super)space as a gauge theory of the (super)conformal group. The `degauging' procedure necessary to relate such conformal geometries to their conventional counterparts is then reviewed in section \ref{Chapter2.4}. The main body of chapter \ref{Chapter2} is accompanied by three technical appendices. In appendix \ref{Appendix2A}, we review the four-dimensional spinor conventions employed throughout this thesis. Appendix \ref{Appendix2B} briefly outlines the method of compensators in conformal (super)gravity. Finally, appendix \ref{Appendix2C} is devoted to a review of $\cN=(1,0)$ conformal supergravity in six dimensions.

The material original to this thesis is presented in chapters \ref{Chapter3} to \ref{Chapter5}, which are based on the publications \cite{KR19,KPR,KPR2,KLRTM,KR21,KR21-2,KPR22,KR22,KR23,HKR}. Further, the precise locations of each publication's results is described in the authorship declaration section, located on page \pageref{AuthorDec}. This information is reiterated at the beginning of each relevant chapter, along with an outline of its content and structure.

The first of these chapters, namely chapter \ref{Chapter3}, is devoted to the study of (conformal) isometries of curved backgrounds and higher symmetries of kinetic operators for matter theories defined on the latter. As will be discussed, these higher symmetries are intimately related to (super)conformal higher-spin theory, which is the subject of chapter \ref{Chapter4}. Finally, in chapter \ref{Chapter5}, we study conformal supergeometries in two spacetime dimensions. Concluding comments and a future outlook is given in chapter \ref{Chapter6}. 

We emphasise that, as this thesis is intended to be an original work of scholarship, the results of \cite{KR19,KPR,KPR2,KLRTM,KR21,KR21-2,KPR22,KR22,KR23,HKR} appearing in this thesis are presented as if they have appeared for the first time. As mentioned above, we refer the interested reader to the authorship declaration section for the appropriate citations.

\chapter{Background material} \label{Chapter2}

This chapter is a devoted to a review of aspects of conformal (super)gravity and (super)conformal field theory pertinent to this thesis. It is not intended to be a complete description of these topics and we refer the interested reader to \cite{GGRS,WB,BK,FVP,KRTM1,KRTM2} for further details. This chapter is based on the publications \cite{KLRTM,KRTM1,KRTM2}.

\section{Rigid superconformal transformations of $\mathbb{M}^{4|4\cN}$} \label{Chapter2.1}

In the case of a $d$-dimensional superconformal field theory in Minkowski superspace ${\mathbb M}^{d|\d}$, where $\d$ denotes the number of fermionic coordinates,
its symmetries are formulated in terms of the conformal Killing supervector fields of  ${\mathbb M}^{d|\d}$.
In this section we describe the conformal Killing supervector fields of 
$\mathbb{M}^{4|4\cN}$, the $\mathcal{N}$-extended Minkowski superspace in four dimensions. Their initial description in the literature was provided by Park \cite{Park}, see also \cite{KT}. The four-dimensional spinor conventions used in this thesis are reviewed in appendix \ref{Appendix2A}.

\subsection{The conformal Killing supervector fields of $\mathbb{M}^{4|4\mathcal{N}}$}
\label{Section2.1}

Minkowski superspace $\mathbb{M}^{4|4 \mathcal{N}}$
is  parametrised by the real coordinates $z^{A} = (x^{a},\q^\a_i, \bar{\q}_\ad^i)$, where $a=0,1,2,3$, $\a = 1,2$, $\ad = \dot{1}, \dot{2}$ and $i = \underline{1}, \dots , \underline{\cN}$. Its covariant derivatives $D_A = (\partial_{a}, D_\a^i , \bar{D}^{\ad}_i)$ take the form
\bea
\pa_{a}:=\frac{\pa}{\pa x^{a}}~, \quad
D_{\a}^{i}
:=\frac{\pa}{\pa \q^{\a}_i} + \ri \bar{\q}^{\ad i} (\s^a)_\aa \pa_a
~, \quad
\bar{D}^{\ad}_{i}
:= -\frac{\pa}{\pa \bar{\q}_{\ad}^i} - \ri {\q}_i^\a (\s^a)_\a{}^\ad \pa_a~,
\eea
and satisfy the algebra:
\bea
\big \{ D_\a^i , \bar{D}^\ad_j \big\} = - 2 \ri \d^i_j (\s^a)_\a{}^{\ad} \partial_a = - 2 \ri \d^i_j \partial_\a{}^{\ad}~,
\eea
where all other graded commutators vanish.

The conformal Killing supervector fields of $\mathbb{M}^{4|4 \mathcal{N}}$,
\be 
\xi = \xi^{a} \partial_{a} + \xi^\a_i D_\a^i + \bar{\xi}_\ad^i \bar{D}^\ad_i = \bar{\xi} ~,
\ee
may be defined to satisfy the constraints
\be 
\label{24}
[\xi , D_\a^i ] = - (D_\a^i \xi_j^\b) D_\b^j ~, \qquad
[\xi , \bar{D}^\ad_i ] = - (\bar{D}^\ad_i \bar{\xi}_\bd^j) \bar{D}^\bd_j ~.
\ee
These imply the conformal Killing vector equation
\bea
\label{25}
D_{(\a}^i \xi_{\b) \bd} = 0 ~, \qquad \bar{D}_{(\ad i} \xi_{\b \bd)} = 0 \quad \implies \quad \partial_{(\a (\ad} \xi_{\b) \bd)} = 0~,
\eea
and yield expressions for the spinor parameters
\begin{align}
	\label{26}
	\xi^\a_i = - \frac{\ri}{8} \bar{D}_{\ad i} \xi^{\aa} ~, \qquad \bar{\xi}_\ad^i = - \frac{\ri}{8} D^{\a i} \xi_\aa~.
\end{align}
The general solution to these constraints was given in \cite{Park}, see also \cite{K06}.

Taking \eqref{25} into account, equations \eqref{24} 
can be rewritten in the form
\begin{subequations}
	\begin{align}
		[\xi , D_\a^i] &= - \hf \s[\xi] D_\a^i - \ri \r[\xi] D_\a^i - K[\xi]_\a{}^\b D_\b^i + \chi[\xi]^i{}_j D_\a^j  ~, \\
		[\xi , \bar{D}^\ad_i] &= - \hf \s[\xi]\bar{D}^\ad_i + \ri \r[\xi]\bar{D}^\ad_i - \bar{K}[\xi]^\ad{}_\bd \bar{D}^\bd_i - \chi[\xi]^j{}_i \bar{D}^\ad_j ~,
	\end{align}
\end{subequations}
where we have defined the following parameters:
\begin{subequations}
	\label{28}
	\begin{align}
		\s[\xi] &:= - \frac 1 8 \partial_\aa \xi^\aa ~, \\
		\r [\xi] &:= - \frac {1}{32 \mathcal{N}} \big[D_\a^i , \bar{D}_{\ad i}\big] \xi^{\aa} ~, \\
		K[\xi]_{\a \b} &:= \frac 1 4 \partial_{(\a}{}^\ad \xi_{\b) \ad}~, \\
		\chi[\xi]^i{}_j &:= \frac{\ri}{32} \Big(\big[D_\a^i , \bar{D}_{\ad j}\big] - \frac{1}{\mathcal{N}} \d^i_j \big[D_\a^k , \bar{D}_{\ad k}\big] \Big) \xi^{\aa}~.
	\end{align}
\end{subequations}
Their $z$-independent components generate scale, $\sU(1)_R$, Lorentz and $\sSU(\mathcal{N})_R$ transformations, respectively. Employing \eqref{25}, it is possible to show that the parameters 
\eqref{28} satisfy 
\begin{subequations}
	\label{29}
	\begin{align}
		D_\a^i K[\xi]_{\b \g} &= - 2 \ve_{\a(\b} D_{\g)}^i \s[\xi] ~, \qquad D_\a^i \bar{K}[\xi]_{\bd \gd} = 0 ~,\\
		D_\a^i \r[\xi] &= \frac{-\ri(\mathcal{N} - 4)}{2\mathcal{N}} D_\a^i \s[\xi] ~, \qquad D_\a^i \chi[\xi]^{j}{}_k = 2 \d^i_k D_\a^j \s[\xi] - \frac{2}{\mathcal{N}} \d_k^j D_\a^i \s [\xi] ~, \\
		D_\a^i D_\b^j \s [\xi] &= 0 ~, \qquad D_\a^i \bar{D}_{\bd j} \s [\xi] = - \ri \d_j^i \pa_{\a \bd} \s [\xi] ~, \qquad D_\a^i \pa_{\b \bd} \s[\xi] = 0~.
	\end{align}
\end{subequations}
In particular, we see that when $\cN=4$, the $\sU(1)_R$ parameter $\r[\xi]$ is constant.

Given two conformal Killing supervectors $\xi_1$ and $\xi_2$, their commutator is another conformal Killing supervector $\xi_3$
\begin{subequations}
	\begin{align} 
		[\xi_1 , \xi_2] = \xi_3^A D_A = \xi_3~,
	\end{align}
	where we have made the definition
	\begin{align}
		\label{2.10b}
		\xi_3^{\aa} := - \hf \xi^{\bb}_1 \pa_{\bb} \xi_2^\aa - \frac{\ri}{16} \bar{D}_{\bd i} \xi^{\a \bd}_1 D_{\b}^i \xi_2^{\b \ad} - (1 \leftrightarrow 2)~,
	\end{align}
\end{subequations}
and the spinor parameters are determined by eq. \eqref{26}. This result, in conjunction with \eqref{29}, implies that the superconformal algebra of $\mathbb{M}^{4|4\mathcal{N}}$ is finite dimensional.

\subsection{The superconformal algebra}

The results of the previous subsection mean that we may parametrise conformal Killing supervector fields as
\be 
\xi \equiv \xi(\l(P)^{a}, \l(Q)^\a_i, \l(\bar{Q})_\ad^i ,  \l(M)^{ab} , \l(\mathbb{D}) , \l(\mathbb{Y}), \l(\mathbb{J})^i{}_j, \l(K)_a , \l(S)_\a^i , \l(\bar{S})^\ad_i ) ~,
\ee
where we have defined the constant parameters
\bsubeq
\begin{align} 
	\quad \l(P)^a &:= \xi^a|_{z=0} ~,\quad \l(Q)^\a_i := \xi^\a_i|_{z=0} ~, \quad \l(\bar{Q})_\ad^i := \bar{\xi}_\ad^i|_{z=0} ~, \\
	\l(M)^{ab} &:= K[\xi]^{ab}|_{z=0} ~, \qquad \l(\mathbb{D}) := \s[\xi]|_{z=0} ~, \\
	\l(\mathbb{Y}) &:= \r[\xi]|_{z=0} ~, \qquad \l(\mathbb{J})^i{}_j := \chi[\xi]^i{}_j|_{z=0} ~, \\
	\quad \l(K)_a &:= \hf \pa_a \s[\xi] |_{z=0} ~, \quad \l(S)_\a^i := \hf D_\a^i \s[\xi] |_{z=0} ~, \quad \l(\bar{S})^\ad_i := \hf \bar{D}^\ad_i \s[\xi] |_{z=0} ~.
\end{align}
\esubeq
In particular, $\xi$ may be represented as
\be
\xi = \l(X)^{\tilde{A}} X_{\tilde{A}} ~,
\ee
where we make use of the following notation for the superconformal parameters
\begin{subequations}
	\begin{align}
		\l(X)^{\tilde{A}} &= (\l(P)^{A}, \l(M)^{ab} , \l(\mathbb{D}), \l(\mathbb{Y}), \l(\mathbb{J})^i{}_j , \l(K)_{A})~, \\
		\l(P)^A &= (\l(P)^{a}, \l(Q)^\a_i, \l(\bar{Q})_\ad^i) ~, \qquad \l(K)_A = (\l(K)_{a},  \l(S)_\a^i , \l(\bar{S})^\ad_i ) ~,
	\end{align}
\end{subequations}
and for generators of the superconformal algebra
\begin{subequations}
	\label{2.15}
	\begin{align}
		X_{\tilde{A}} &= (P_A, M_{ab} , \mathbb{D} , \mathbb{Y}, \mathbb{J}^i{}_j , K^A)~, \\
		P_A &= (P_{a}, Q_\a^i, \bar{Q}^\ad_i) ~, \qquad K^A = (K^{a},  S^\a_i , \bar{S}_\ad^i )~.
	\end{align}
\end{subequations}

The above results allow us to derive the graded commutation relations for the superconformal algebra via the important relation
\bea
[\xi_1,\xi_2] = - \l(X)_2^{\tilde{B}} \l(X)_1^{\tilde{A}} \big[ X_{\tilde{A}}, X_{\tilde{B}} \big \}~.
\eea  
The commutation relations for the conformal algebra are as follows:\footnote{It should be emphasised that \eqref{2.17} is the conformal algebra for all dimensions $d \geq 2$.}
\begin{subequations} 
	\label{2.17}
	\begin{align}
		&[M_{ab},M_{cd}]=2\eta_{c[a}M_{b]d}-2\eta_{d[a}M_{b]c}~, \phantom{inserting blank space inserting} \\
		&[M_{ab},P_c]=2\eta_{c[a}P_{b]}~, \qquad \qquad \qquad \qquad ~ [\mathbb{D},P_a]=P_a~,\\
		&[M_{ab},K_c]=2\eta_{c[a}K_{b]}~, \qquad \qquad \qquad \qquad [\mathbb{D},K_a]=-K_a~,\\
		&[K_a,P_b]=2\eta_{ab}\mathbb{D}+2M_{ab}~.
	\end{align}
\end{subequations}

The $R$-symmetry generators $\mathbb{Y}$ and $\mathbb{J}^i{}_j$ commute with all the generators of the conformal group. Amongst themselves, they obey the algebra
\begin{align}
	[\mathbb{J}^{i}{}_j,\mathbb{J}^{k}{}_l] = \d^i_l \mathbb{J}^k{}_j - \d^k_j \mathbb{J}^i{}_l ~.
\end{align}

The superconformal algebra is then obtained by extending the translation generator $P_{a}$ to $P_A$ and the special conformal generator $K^{a}$ to $K^A$. The commutation relations involving the $Q$-supersymmetry generators with the bosonic ones are:
\begin{subequations} 
	\bea
	\big[M_{ab}, Q_\g^i \big] &=& (\s_{ab})_\g{}^\d Q_\d^i ~,\quad 
	\big[M_{ab}, \bar Q^\gd_i \big] = (\tilde{\s}_{ab})^\gd{}_\dd \bar Q^\dd_i~,\\
	\big[\mathbb{D}, Q_\a^i \big] &=& \hf Q_\a^i ~, \quad
	\big[\mathbb{D}, \bar Q^\ad_i \big] = \hf \bar Q^\ad_i ~, \\
	\big[\mathbb{Y}, Q_\a^i \big] &=&  Q_\a^i ~, \quad
	\big[\mathbb{Y}, \bar Q^\ad_i \big] = - \bar Q^\ad_i ~, \label{2.19c} \\
	\big[\mathbb{J}^i{}_j, Q_\a^k \big] &=&  - \d^k_j Q_\a^i + \frac{1}{\mathcal{N}} \d^i_j Q_\a^k ~, \quad
	\big[\mathbb{J}^i{}_j, \bar Q^\ad_k \big] = \d^i_k \bar Q^\ad_j - \frac{1}{\mathcal N} \d^i_j \bar Q^\ad_k ~,  \\
	\big[K^a, Q_\b^i \big] &=& -\ri (\s^a)_\b{}^\bd \bar{S}_\bd^i ~, \quad 
	\big[K^a, \bar{Q}^\bd_i \big] = 
	-\ri ({\s}^a)^\bd{}_\b S^\b_i ~.
	\eea
\end{subequations}
As noted above, when $\cN=4$ the chiral parameter $\r[\xi]$ is constant. At the level of the superconformal algebra, this means that $\mathbb{Y}$ is a central charge commuting with all elements of the superconformal algebra, hence the relations \eqref{2.19c} should be omitted.

The commutation relations involving the $S$-supersymmetry generators 
with the bosonic operators are: 
\begin{subequations}
	\bea
	\big [M_{ab} , S^\g_i \big] &=& - (\s_{ab})_\b{}^\g S^\b_i ~, \quad
	\big[M_{ab} , \bar S_\gd^i \big] = - (\ts_{ab})^\bd{}_\gd \bar S_\bd^i~, \\
	\big[\mathbb{D}, S^\a_i \big] &=& -\hf S^\a_i ~, \quad
	\big[\mathbb{D}, \bar S_\ad^i \big] = -\hf \bar S_\ad^i ~, \\
	\big[\mathbb{Y}, S^\a_i \big] &=&  -S^\a_i ~, \quad
	\big[\mathbb{Y}, \bar S_\ad^i \big] =  \bar S_\ad^i ~,  \label{2.20c}\\
	\big[\mathbb{J}^i{}_j, S^\a_k \big] &=&  \d^i_k S^\a_j - \frac{1}{\mathcal{N}} \d^i_j S^\a_k ~, \quad
	\big[\mathbb{J}^i{}_j, \bar S_\ad^k \big] = - \d_j^k \bar S_\ad^i + \frac{1}{\mathcal N} \d^i_j \bar S_\ad^k ~,  \\
	\big[ S^\a_i , P_b \big] &=& \ri (\s_b)^\a{}_\bd \bar{Q}^\bd_i ~, \quad 
	\big[\bar{S}_\ad^i , P_b \big] = 
	\ri ({\s}_b)_\ad{}^\b Q_\b^i ~.
	\eea
\end{subequations}
We emphasise that for $\cN=4$, the commutators \eqref{2.20c} should be omitted. 

Finally, the anti-commutation relations of the fermionic generators are: 
\begin{subequations}
	\bea
	\{Q_\a^i , \bar{Q}^\ad_j \} &=& - 2 \ri \d^i_j (\s^b)_\a{}^\ad P_b=- 2 \ri \d^i_j  P_\a{}^\ad~, \\
	\{ S^\a_i , \bar{S}_\ad^j \} &=& 2 \ri  \d_i^j (\s^b)^\a{}_\ad K_b=2 \ri \d_i^j  K^\a{}_\ad
	~, \\
	\{ S^\a_i , Q_\b^j \} &=& \d_i^j \d^\a_\b \Big(2 \mathbb{D} + \frac{\mathcal{N}-4}{\mathcal{N}}\mathbb{Y} \Big) - 4 \d_i^j  M^\a{}_\b 
	+ 4 \d^\a_\b  \mathbb{J}^j{}_i ~, \\
	\{ \bar{S}_\ad^i , \bar{Q}^\bd_j \} &=& \d_j^i \d^\bd_\ad \Big(2 \mathbb{D} - \frac{\mathcal{N}-4}{\mathcal{N}}\mathbb{Y} \Big) + 4 \d_j^i  \bar{M}_\ad{}^\bd 
	- 4 \d_\ad^\bd  \mathbb{J}^i{}_j  ~, \label{2.21d}
	\eea
\end{subequations}
Note that all remaining (anti-)commutators not listed above vanish identically. These graded commutation relations
describe the $\cN$-extended superconformal algebra, ${\mathfrak{su}}(2,2|\mathcal{N})$. 

To conclude, we give the superconformal transformation laws for primary superfields. Given a conformal Killing supervector field $\xi$, the corresponding infinitesimal superconformal transformation of primary tensor superfield $U$ (with its indices suppressed) is
\bea
\d_\x U &=&\Big\{
\x + \hf K[\xi]^{ab} M_{ab} 
+\D_U \s[\x] + \ri q_U \r[\xi] + \L[\xi]^i{}_j \mathbb{J}^j{}_i\Big\}U~,
\eea
The parameters $\D_U$ and $q_U$ are called the dimension (or Weyl weight) and $\sU(1)_R$ charge of $U$, respectively.

\section{Conformal geometry in $d\geq3$ dimensions} \label{Chapter2.2}

This subsection is devoted to a brief review of 
conformal gravity in $d \geq 3$ dimensions\footnote{The $d=2$ case is somewhat special, and will be described in further detail in section \ref{Chapter5.2}.} as the gauge theory of the conformal group $\sSO(d,2)$. This approach was pioneered in four dimensions in \cite{KTvN1}.
We closely follow the presentations given in  \cite{BKNT-M1,BKNT}.

The Lie algebra of the conformal
group, $\mathfrak{so}(d,2)$, whose generators we collectively denote $X_{\tilde a}$, is spanned by the translation ($P_a$), Lorentz ($M_{ab}$), dilatation ($\mathbb{D}$) and special conformal generators ($K^a$). We recall their non-vanishing commutation relations from \eqref{2.17}. It is convenient to group these generators into the two disjoint subsets $P_a$ and $X_{\underline a}$:
\begin{align}
	X_{\tilde a} = (P_a, X_{\underline a}) ~, \qquad X_{\underline a} = (M_{ab},\mathbb{D},K^a)~.
\end{align}
Then, the conformal algebra may be rewritten as follows
\bsubeq
\begin{align}
	[X_{\underline{a}} , X_{\underline{b}} ] &= -f_{\underline{a} \underline{b}}{}^{\underline{c}} X_{\underline{c}} \ , \\
	[X_{\underline{a}} , P_{{b}} ] &= -f_{\underline{a} { {b}}}{}^{\underline{c}} X_{\underline{c}}
	- f_{\underline{a} { {b}}}{}^{ {c}} P_{ {c}} \label{A.2}
	~,
\end{align}
\esubeq
where $f_{\underline{a} \underline{b}}{}^{\underline{c}}$, $f_{\underline{a} { {b}}}{}^{\underline{c}}$ and $f_{\underline{a} { {b}}}{}^{ {c}}$ are the structure coefficients of the conformal algebra.

\subsection{Gauging the conformal algebra in $d \geq 3 $ dimensions}

Let $\mathcal{M}^d$ be a $d$-dimensional spacetime ($d \geq 3$) parametrised by the local coordinates $x^m$, $m = 0,1, \dots, d-1.$ To gauge the conformal algebra it is necessary to associate each non-translational generator $X_{\underline{a}}$ with a connection one-form $\o^{\underline{a}} = (\o^{ab},b,\mathfrak{f}_a)=\rd x^{m} \o_m{}^{\underline{a}}$ and with $P_a$ a vielbein one-form $e^a = \rd x^m e_m{}^a$, where it is assumed that $e:={\rm det}(e_m{}^a) \neq 0$, hence there exists a unique inverse vielbein $e_a{}^m$:
\begin{align}
	e_a{}^m e_m{}^b = \d_a{}^b~, \qquad e_m{}^a e_a{}^n=\d_m{}^n~.
\end{align}
The latter may be used to construct
the vector fields $e_a = e_a{}^m \pa_m $, which constitute a basis for the tangent space at each
point of $\mathcal{M}^{d}$. It may then be used to express the connection in the vielbein basis as
$\omega^{\underline{a}} =e^b\omega_b{}^{\underline{a}}$, 
where $\omega_b{}^{\underline{a}}=e_b{}^m\omega_m{}^{\underline{a}}$. 

The covariant derivatives
have the form
\be
\nabla_a =  e_a{}^m \partial_m - \hf \omega_a{}^{bc} M_{bc} - b_a \mathbb D - \mathfrak{f}_{ab} K^b ~.
\ee
We note that the translation generators $P_a$ do not appear in the above expression. Instead, we assume that they are replaced by $\nabla_a$ and obey the commutation relations:
\begin{align}
	[X_{\underline{a}} , \nabla_{{b}} ] &= -f_{\underline{a} { {b}}}{}^{\underline{c}} X_{\underline{c}}
	- f_{\underline{a} { {b}}}{}^{ {c}} \nabla_{ {c}} 
	\ .
\end{align}

By definition, the gauge group of conformal gravity is generated by local transformations of the form
\begin{subequations}\label{CGgentransmations}
	\bea
	\delta_{\mathscr K} \nabla_a &=& [\mathscr{K},\nabla_a] \ , \\
	\mathscr{K} &=& \xi^b \nabla_b +  \L^{\underline{b}} X_{\underline{b}}
	=  \xi^b \nabla_b + \hf K^{bc} M_{bc} + \s \mathbb{D} + \L_b K^b ~,
	\eea
\end{subequations}
where  the gauge parameters satisfy natural reality conditions. These gauge transformations act
on a conformal tensor field $U$ (with indices suppressed) as 
\bea 
\label{CGgenMattertfs}
\d_{\mathscr K} U = {\mathscr K} U ~.
\eea
Further, we will say that $U$ is primary if it satisfies
\bea
K^a U = 0~, \quad \mathbb D U = \D_U U~,
\label{conformalproperties}
\eea
where $\D_U$ is called the dimension (or Weyl weight) of $U$.

Amongst themselves, the covariant derivatives satisfy the following commutation relations
\be
[\nabla_a , \nabla_b] = -\cT_{ab}{}^c \nabla_c - \hf \cR(M)_{ab}{}^{cd} M_{cd}
- \cR(\mathbb{D})_{ab} \mathbb D - \cR(K)_{abc} K^c \ ,
\ee
where we have made the definitions:
\begin{subequations}
	\begin{align}
		\mathcal{T}_{ab}{}^c&=-\mathscr{C}_{ab}{}^{c}+2{\o}_{[ab]}{}^c+2{b}_{[a}\delta_{b]}{}^{c}~,\\
		\mathcal{R}(M)_{ab}{}^{cd}&=R_{ab}{}^{cd}+8\mathfrak{f}_{[a}{}^{[c}\delta_{b]}{}^{d]}~,\label{LorCur}\\ 
		\mathcal{R}(K)_{abc}&=-\mathscr{C}_{ab}{}^d\mathfrak{f}_{dc}-2{\o}_{[a|c|}{}^{d}\mathfrak{f}_{b]d}-2{b}_{[a}\mathfrak{f}_{b]c}+2e_{[a}\mathfrak{f}_{b]c}~, \label{2.19cc}\\
		\mathcal{R}(\mathbb{D})_{ab}&= -\mathscr{C}_{ab}{}^{c}{b}_c+4\mathfrak{f}_{[ab]}+2e_{[a}{b}_{b]}~,\\
		R_{ab}{}^{cd}&=-\mathscr{C}_{ab}{}^{f}{\o}_{f}{}^{cd}+2e_{[a}{\o}_{b]}{}^{cd}-2{\o}_{[a}{}^{cf}{\o}_{b]f}{}^{d}~. \label{RiemannB}
	\end{align}
\end{subequations}
Here $R_{ab}{}^{cd}$ is the curvature tensor\footnote{It should be emphasised that our curvature tensor does not satisfy the Bianchi identity $R_{[abc]d}=0$ unless \eqref{CC} is imposed and $b_a=0$.} constructed from the Lorentz connection $\omega_a{}^{bc}$ and we have introduced the anholonomy coefficients $\mathscr{C}_{ab}{}^{c}$
\begin{align}
	[e_a,e_b] = \mathscr{C}_{ab}{}^{c} e_c ~.
\end{align}

In order for the above geometry to describe conformal gravity, it is necessary to impose certain covariant constraints such that the only independent geometric fields are the vielbein and dilatation connection. They are as follows:
\begin{subequations}
	\label{CC}
	\begin{align}
		\cT_{ab}{}^c &=0 ~, \label{CCa}\\
		\eta^{bd} \cR(M)_{abcd} &=0~. \label{CCb}
	\end{align}
\end{subequations}
The first constraint determines $\o_{a}{}^{bc}$ in terms of the vielbein and dilatation connection, while the second determines $\mathfrak{f}_{ab}$ to be
\be
\label{SCconn}
\mathfrak{f}_{ab} = - \hf P_{ab} = - \frac{1}{2 (d - 2)} R_{ab} + \frac{1}{4 (d - 1) (d - 2)} \eta_{ab} R ~,
\ee
where $P_{ab}$ is the Schouten tensor and we have defined
\be 
R_{ac} = \eta^{bd} R_{abcd} \ , \quad R = \eta^{ab} R_{ab} ~.
\ee
Inserting \eqref{SCconn} into \eqref{LorCur} leads to the result that $\cR(M)_{ab}{}^{cd}$ is exactly the Weyl tensor
\be
\mathcal{R}(M)_{ab}{}^{cd} = C_{ab}{}^{cd} = R_{ab}{}^{cd} - 4 P_{[a}{}^{[c} \d_{b]}{}^{d]} ~,
\ee
which is a primary field of dimension 2
\begin{align}
	K_e C_{abcd} = 0 ~, \qquad \mathbb{D} C_{abcd} = 2 C_{abcd}~.
\end{align} 
It should be emphasised that, since $C_{abcd}$ is primary, it is independent of the dilatation connection $b_a$.

Next, it is necessary to analyse the Jacobi identity 
\be
[\nabla_a , [\nabla_b, \nabla_c]] + [\nabla_c , [\nabla_a, \nabla_b]] + [\nabla_b , [\nabla_c, \nabla_a]] = 0~,
\ee
which leads to the constraints
\begin{subequations}
	\begin{align}
		\cR(\mathbb{D})_{ab} &= 0 ~, \\
		\cR(K)_{[abc]} &= 0 \ , \label{CGDBI1} \\ 
		C_{[abc]}{}^d &= 0 \ , \label{CGDBI2} \\ 
		\nabla_{[a} \cR(K)_{bc]}{}^d &= 0 \ , \label{CGDBI3} \\
		\nabla_{[a} C_{bc ]}{}^{de} - 4 \cR(K)_{[ab}{}^{[d} \d^{e]}_{c]} &= 0 \label{CGDBI4} \ . 
	\end{align}
\end{subequations}
In particular, eq. \eqref{CGDBI4} implies the important identity
\be 
\label{A.17}
\hf \nabla_c C_{ab}{}^{c e} + (d - 3) \cR(K)_{ab}{}^e - 2 \cR(K)_{c [a}{}^c \d^e_{b]} = 0 \ .
\ee
Thus, to continue our analysis it is necessary to consider the cases $d > 3$ and $d=3$ separately.

For $d > 3$, it follows from \eqref{A.17} that the special conformal curvature takes the form
\begin{align}
	\cR(K)_{abc} = \frac{1}{2(d-3)} \nabla^d C_{abcd}~,
\end{align}
which implies that the algebra of covariant derivatives is
\begin{align}
	[\nabla_a , \nabla_b] = - \frac{1}{2} C_{abcd} M^{cd} - \frac{1}{2(d-3)} \nabla^d C_{abcd} K^c ~.
\end{align}
To conclude our discussion of this case, we list  the algebraic properties 
of $C_{abcd}$:
\bea
C_{abcd}= C_{[ab] [cd] } = C_{cd ab} ~, \qquad C_{a[bcd]} =0 ~,\qquad \eta^{bc} C_{abcd}=0~.
\eea

The $d = 3$ case is special because the Weyl tensor identically vanishes, $C_{abcd} = 0$. As a result, the algebra of covariant derivatives takes the form
\be
[\nabla_a , \nabla_b] = -\cR(K)_{abc} K^c ~,
\ee
and the conformal geometry of spacetime is controlled by the primary field $\cR(K)_{abc}$. One can show that this field takes the form
\begin{align}
	\cR(K)_{abc} = - \cD_{[a} P_{b]c} = - \hf W_{abc} ~, \qquad \cD_a = e_a{}^m \partial_m - \o_{a}{}^{bc} M_{bc} ~,
\end{align}
where we have introduced the Lorentz covariant derivative\footnote{This definition of $\cD_a$ is valid for generic spacetime dimensions.} $\cD_a$ and the Cotton tensor $W_{abc}$. The latter proves to be a primary field of dimension $3$,
\begin{align}
	K_d W_{abc} = 0 ~, \qquad \mathbb{D} W_{abc} = 3 W_{abc}~.
\end{align}
It is useful to introduce its dual
\begin{align}
	W_{ab} = \hf \ve_{acd} W^{cd}{}_b ~,
\end{align}
which proves to be symmetric and traceless,
\begin{align}
	W_{ab} = W_{ba} ~, \qquad \eta^{ab} W_{ab} = 0~,
\end{align}
and also satisfies the conservation equation
\begin{align}
	\nabla^b W_{ab} = 0~.
\end{align}

\subsection{Conformal action principle}

In order to formulate conformal field theories, an action principle is required. 
We look for a real scalar field $\cL$ such that the action
\be
\cS = \int\text{d}^{d}x \, e \, \mathcal{L} ~,
\ee
is locally conformal. Requiring this functional to be invariant under conformal gravity gauge transformations, \eqref{CGgenMattertfs} and \eqref{CGgentransmations},
we obtain the following condition on $\cL$
\be
K^a \cL =0~, \qquad {\mathbb D} \cL = d \cL~.
\ee
Thus, the Lagrangian $\cL$ must be a primary field of dimension $d$.

\subsection{Degauging to Lorentzian geometry}

As mentioned above, the only independent geometric fields in this geometry are the vielbein and dilatation gauge field. Actually, the latter proves to be a purely gauge degree of freedom. Specifically, it transforms algebraically under special conformal transformations \eqref{CGgentransmations}
\begin{align}
	\mathscr{K}(\L) = \L_a K^a \qquad \implies \qquad \d_{\mathscr{K}(\L)} b_a = - 2 \L_a~.
\end{align}
Hence, it is possible to impose the gauge condition $b_a = 0$ at the cost of breaking special conformal symmetry.\footnote{This process is known as `degauging.'} As a result, the connection $\mathfrak{f}_{ab}$ is no longer required for the covariance of $\nabla_a$ under the residual gauge freedom and it should be separated
\begin{align}
	\nabla_a = \cD_a - \mathfrak{f}_{ab} K^b = \cD_a + \hf P_{ab} K^b ~.
\end{align}
It may then be shown that the Lorentz covariant derivatives $\cD_a$ satisfy the algebra
\begin{align}
	[\cD_a , \cD_b] = - \hf \big( C_{ab}{}^{cd} + 4 P_{[a}{}^{c} \d_{b]}{}^{d} \big) M_{cd}~.
\end{align}

Next, it is important to describe the supergravity gauge freedom of this geometry, which corresponds to the residual gauge transformations of \eqref{CGgentransmations} in the gauge $b_a = 0$. These include local $\mathcal{K}$-transformations of the form
\begin{subequations}
	\label{A.30}
	\bea
	\delta_{\mathcal K} \cD_A &=& [\mathcal{K},\cD_A] \ , \\
	\mathcal{K} &=& \xi^b \cD_b + \hf K^{bc} M_{bc} ~,
	\eea
	which act on tensor superfields ${U}$ (with their indices suppressed) as
	\begin{align}
		\d_\cK U = \cK U ~.
	\end{align}
\end{subequations}

The gauge transformations \eqref{A.30} prove to not be the most general conformal gravity gauge transformations preserving the gauge $b_a=0$.
Specifically, it may be shown that the following transformation also enjoys this property
\begin{align}
	\mathscr{K}(\s) = \s \mathbb{D} + \frac{1}{2} \nabla_b \s K^b \quad \Longrightarrow \quad \d_{\mathscr{K}(\s)} b_a = 0~,
\end{align}
where $\s$ is real but otherwise unconstrained. 
As a result, it is necessary to consider how this transformation manifests in the degauged geometry
\begin{align}
	\d_{\mathscr{K}(\s)} \nabla_a \equiv \d_{\s} \nabla_a = \d_\s \cD_a - \d_\s \mathfrak{f}_{ab} K^b~.
\end{align}
By a routine computation, we obtain
\begin{subequations}
	\begin{align}
		\d_\s \cD_a &= \s \cD_a + \cD^b \s M_{ba} ~, \\
		\d_\s C_{abcd} &= 2 \s C_{abcd} ~, \\
		\d_\s P_{ab} &= 2 \s P_{ab} - \cD_a \cD_b \s~,
	\end{align}
\end{subequations}
which are the standard Weyl transformations.

\section{Gauging the superconformal algebra} \label{Chapter2.3}

Conformal superspace is a gauge theory of the superconformal algebra. It can be identified with a pair $(\cM^{4|4 \mathcal{N}}, \nabla)$. Here $\mathcal{M}^{4|4 \mathcal{N}}$ denotes a supermanifold parametrised by local coordinates 
$z^M = (x^m, \q^\m_\imath, \bar \q_{\dot \m}^\imath)$, and $\nabla$ is a covariant derivative associated with the superconformal algebra. We recall that the generators $ X_{\tilde A}$ of the superconformal algebra are given by eq. \eqref{2.15}. They can be grouped into two disjoint subsets,
\bea 
X_{\tilde A} = (P_A, X_{\underline{A}} )~, \qquad
X_{\underline{A}} =
( M_{ab} ,{\mathbb D}, \mathbb{Y} , \mathbb{J}^i{}_j, K^A)~,
\eea  
each of which constitutes a superalgebra:
\bsubeq\label{4DRigidAlgebra}
\begin{align}
	[P_{ {A}} , P_{ {B}} \} &= -f_{{ {A}} { {B}}}{}^{{ {C}}} P_{ {C}}
	\ , \\
	[X_{\underline{A}} , X_{\underline{B}} \} &= -f_{\underline{A} \underline{B}}{}^{\underline{C}} X_{\underline{C}} \ , \\
	[X_{\underline{A}} , P_{{B}} \} &= -f_{\underline{A} { {B}}}{}^{\underline{C}} X_{\underline{C}}
	- f_{\underline{A} { {B}}}{}^{ {C}} P_{ {C}}
	~ . \label{4DMixing}
\end{align}
\esubeq
Here the structure constants $f_{{ {A}}{ {B}}}{}^{ {C}}$ contain only one non-zero component, which is
\be
f_{\a}^i{}^\bd_j{}^c = 2 \ri \d^i_j \,(\s^c)_\a{}^\bd~.
\ee

To define the covariant derivatives, $\nabla_A= (\nabla_a, \nabla_\a^i, \bar \nabla^\ad_i)$, we associate with each generator $X_{\underline{A}} =
( M_{ab} ,{\mathbb D}, \mathbb{Y} , \mathbb{J}^i{}_j, K^A)$
a connection one-form 
$\Omega^{\underline{A}} = (\O^{ab},B, \Phi, \Theta^j{}_i ,\frak{F}_{A})= \rd z^M \Omega_M{}^{\underline{A}}$,  
and with $P_{ {A}}$ a supervielbein one-form
$E^{ {A}} = (E^a, E^\a,\bar{E}_\ad) = \rd z^{ {M}} E_M{}^A$. 
It is assumed that the supermatrix $E_M{}^A$ is nonsingular, $E:= {\rm Ber} (E_M{}^A) \neq 0$, 
and hence there exists a unique inverse supervielbein. The latter is given by 
the supervector fields $E_A = E_A{}^M \pa_M $, which constitute a new basis for the tangent space at every 
point. The supermatrices $E_A{}^M $ and $E_M{}^A$ satisfy the 
properties $E_A{}^ME_M{}^B=\d_A{}^B$ and $E_M{}^AE_A{}^N=\d_M{}^N$.
With respect to  the basis $E^A$,  the connection is expressed as 
$\Omega^{\underline{A}} =E^B\Omega_B{}^{\underline{A}}$, 
where $\Omega_B{}^{\underline{A}}=E_B{}^M\Omega_M{}^{\underline{A}}$. 
The covariant derivatives are given by 
\bea\label{4Dnabla}
\nabla_A 
&=& E_A  - \O_A{}^{\underline B} X_{\underline B}=
E_A -  \hf \Omega_A{}^{bc} M_{bc} - B_A \mathbb{D} - \ri \Phi_A \mathbb{Y} - \Theta_A{}^{i}{}_j \mathbb{J}^{j}{}_i
- \mathfrak{F}_{AB} K^B ~.
\eea
The translation generators $P_B$ do not show up in \eqref{4Dnabla}. It is assumed that the operators $\nabla_A$ replace $P_A$ and obey the graded commutation relations
\be
[ X_{\underline{B}} , \nabla_A \} = -f_{\underline{B} A}{}^C \nabla_C
- f_{\underline{B} A}{}^{\underline{C}} X_{\underline{C}} ~ ,
\ee
compare with \eqref{4DMixing}.

By definition, the gauge group of conformal supergravity  is generated by local transformations of the form
\begin{subequations}
	\label{2.63}
	\bea
	\delta_{\mathscr K} \nabla_A &=& [{\mathscr K},\nabla_A] ~ , \\
	{\mathscr K} &=& \xi^B \nabla_B +  \L^{\underline{B}} X_{\underline{B}}
	=  \xi^B \nabla_B+ \hf K^{bc} M_{bc} + \s \mathbb{D} + \ri \rho \mathbb{Y} + \chi^{i}{}_j \mathbb{J}^{j}{}_i
	+ \L_B K^B ~ ,~~~~
	\eea
\end{subequations}
where  the gauge parameters satisfy natural reality conditions. The conformal supergravity gauge group acts on a conformal tensor superfield $U$ (with its indices suppressed) as 
\bea 
\label{2.64}
\d_{\mathscr K} U = {\mathscr K} U ~.
\eea
Of special significance are primary superfields. 
The superfield $U$ is said to be primary if it is characterised by the properties 
\bea
K^A U = 0~, \quad \mathbb D U = \D_U U~,  \quad {\mathbb Y} U = q_U U~,
\label{2.67}
\eea
for some real constants $\D_U$ and $q_U$ which are called the dimension (or Weyl weight)  and $\sU(1)_R$ charge of $U$, respectively.

Amongst themselves, the covariant derivatives $\nabla_A$ obey the graded commutation relations
\be
\label{2.66}
[ \nabla_A , \nabla_B \} = - \mathcal{T}_{AB}{}^C \nabla_C - \mathcal{R}(X)_{AB}{}^{\underline{C}} X_{\underline{C}} 
\ ,
\ee
where $ {\cal{T}}_{AB}{}^C $ and $ \mathcal{R}(X)_{AB}{}^{\underline{C}} $ denote the torsion and the curvature, respectively. They obey Bianchi identities which follow from:
\be 
(-1)^{\ve_A \ve_C} [ \nabla_A , [ \nabla_B , \nabla_C \} \} +
\text{(two cycles)} = 0~.
\label{2.65}
\ee

This framework defines a geometric set-up to obtain a multiplet of conformal supergravity containing the metric. However, in general, the resulting multiplet is reducible. Thus, it is necessary to impose constraints on the torsion and curvatures appearing in eq.~\eqref{2.66}. This is a standard task in geometric superspace approaches to supergravity, and it is pedagogically reviewed in \cite{BK,GGRS}. One beautiful feature of the construction of \cite{ButterN=1} is the simplicity of the superspace constraints needed to obtain the Weyl multiplet of conformal supergravity. In fact, to obtain a sufficient set of constraints, one requires the algebra \eqref{2.66} to have a Yang-Mills structure.

\subsection{$\mathcal{N}=1$ conformal superspace}
\label{section2.3.1}
This subsection is devoted to a review of $\mathcal{N}=1$ conformal superspace in four dimensions. It was introduced by Butter in the seminal work \cite{ButterN=1}.
\subsubsection{The algebra of covariant derivatives}
As discussed above, to describe conformal supergravity, the algebra of conformally covariant derivatives $\nabla_A$ must be constrained to have a Yang-Mills structure. Specifically, we require
\bsubeq
\label{2.68}
\bea
&	\{ \nabla_{\a} , \nabla_{\b} \}  =  0 ~, \quad \{ \bar{\nabla}_{\ad} , \bar{\nabla}_{\bd} \} = 0 ~, \quad \{\nabla_{\a} , \bar{\nabla}_{\ad} \} = - 2 \ri \nabla_{\a \ad} ~, 
\\
&\big[ \nabla_{\a} , \nabla_{\b \bd} \big] =  2\ri \ve_{\a \b} \bar{\mathscr{W}}_\bd
~,\quad
\big[ \bar{\nabla}_{\ad} , \nabla_{\b \bd} \big] =  - 2\ri \ve_{\ad \bd} \mathscr{W}_{\b}
~,
\eea
\esubeq
where the operator $\bar{\mathscr{W}}_{\ad}$ is the complex conjugate of $\mathscr{W}_{\a}$. The latter takes the form
\bea
{\mathscr{W}}_{\a}
&=&
\frac{1}{2}{\mathscr{W}}(M)_{\a}{}^{bc} M_{bc}
+{\mathscr{W}}(\mathbb D)_{\a}\mathbb D
+\ri {\mathscr{W}}(\mathbb{Y})_{\a} \mathbb{Y}
\non\\
&&
+{\mathscr{W}}(S)_{\a}{}_\b S^\b
+{\mathscr{W}}(\bar{S})_{\a}{}^\bd \bar{S}_\bd
+{\mathscr{W}}(K)_{\a}{}_b K^b
~.
\eea
Having imposed the constraints \eqref{2.68}, the Bianchi identities \eqref{2.65} become non-trivial and now play the role of consistency conditions which we use to determine the torsion and curvature. Their solution is as follows
\begin{subequations}
	\label{2.70}
	\bea
	\big[ \nabla_{\a} , \nabla_{\b \bd} \big] & = & \ri \ve_{\a \b} \Big( 2 \bar{W}_{\bd \gd \dd} \bar{M}^{\gd \dd} - \frac{1}{2} \bar{\nabla}^{\ad} \bar{W}_{\ad \bd \gd} \bar{S}^{\gd} + \frac{1}{2} \nabla_{\g}{}^{\ad} \bar{W}_{\ad \bd \gd} K^{\g \gd} \Big) ~, \\
	\big[ \bar{\nabla}_{\ad} , \nabla_{\b \bd} \big] & = & - \ri \ve_{\ad \bd} \Big( 2 W_{\b}{}^{\g \d} M_{\g \d} + \frac{1}{2} \nabla^{\a} W_{\a \b \g} S^{\g} + \frac{1}{2} \nabla^{\a \gd} W_{\a \b}{}^{\g} K_{\g \gd} \Big) ~,~~~~~~
	\\
	\big[ \nabla_{\a \ad} , \nabla_{\b \bd} \big] & = & \ve_{\ad \bd} \psi_{\a \b} + \ve_{\a \b} \bar{\psi}_{\ad \bd} ~, 
	\eea
	where the operator $\psi_{\a\b} = \psi_{\b \a}$ and its conjugate $\bar{\psi}_{\ad\bd}$ are given by
	\bea
	\psi_{\a \b} & = & W_{\a \b}{}^{\g} \nabla_{\g} + \nabla^{\g} W_{\a \b}{}^{\d} M_{\g \d} - \frac{1}{8} \nabla^{2} W_{\a \b \g} S^{\g} + \frac{\ri}{2} \nabla^{\g \gd} W_{\a \b \g} \bar{S}_{\gd} \non \\
	&& + \frac{1}{4} \nabla^{\g \dd} \nabla_{(\a} W_{\b) \g}{}^{\d} K_{\d \dd} + \frac{1}{2} \nabla^{\g} W_{\a \b \g} \mathbb{D} - \frac{3}{4} \nabla^{\g} W_{\a \b \g} \mathbb{Y} ~, \\
	\bar{\psi}_{\ad \bd} & = & - \bar{W}_{\ad \bd}{}^{\gd} \bar{\nabla}_{\gd} - \bar{\nabla}^{\gd} \bar{W}_{\ad \bd}{}^{\dd} \bar{M}_{\gd \dd} + \frac{1}{8} \bar{\nabla}^{2} \bar{W}_{\ad \bd \gd} \bar{S}^{\gd} + \frac{\ri}{2} \nabla^{\g \gd} \bar{W}_{\ad \bd \gd} S_{\g} \non \\
	&& - \frac{1}{4} \nabla^{\d \gd} \bar{\nabla}_{(\ad} \bar{W}_{\bd) \gd}{}^{\dd} K_{\d \dd} - \frac{1}{2} \bar{\nabla}^{\gd} \bar{W}_{\ad \bd \gd} \mathbb{D} - \frac{3}{4} \bar{\nabla}^{\gd} \bar{W}_{\ad \bd \gd} \mathbb{Y} ~.
	\eea
\end{subequations}

We note that the conformal superspace algebra is expressed in terms of a single superfield $W_{\alpha \beta \gamma}= W_{(\a\b\g)}$, its conjugate $\bar W_{\ad \bd \gd}$, and their covariant derivatives. It defines an $\mathcal{N}=1$ extension of the Weyl tensor; it is known as the ($\cN=1$) super-Weyl tensor. Further, it is a primary covariantly chiral superfield of dimension 3/2 
\bea
\label{SuperWeyl}
K^D W_{\a \b \g} =0~, \quad \bar \nabla^\dd W_{\a\b\g}=0 ~, \quad
{\mathbb D} W_{\a\b\g} = \frac 32 W_{\a\b\g}~,
\eea
and it obeys the Bianchi identity
\bea
\cB_{\a\ad} :=  \ri \nabla^\b{}_{\ad} \nabla^\g W_{\a\b\g}
=\ri \nabla_{\a}{}^{ \bd} \bar \nabla^\gd \bar W_{\ad\bd\gd}
= \bar \cB_{\a\ad}~.
\label{super-Bach}
\eea
Here we have defined the real superfield $\cB_{\a\ad}$, which is
the $\cN=1$ supersymmetric generalisation of the Bach tensor introduced in \cite{BK88} (see also \cite{KMT}).
The super-Bach tensor, $\cB_{\a\ad}$, proves to be primary, $K^B \cB_{\a\ad} =0$, carries weight $3$, ${\mathbb D} \cB_{\a\ad} = 3 \cB_{\a\ad}$, and satisfies the conservation equation
\bea
\nabla^\a \cB_{\a\ad}=0 ~~~\Longleftrightarrow~~~
\bar \nabla^\ad \cB_{\a\ad} =0
~.
\eea

The structure of the algebra \eqref{2.68} and \eqref{2.70} leads to highly non-trivial implications. In particular, its consistency with eq. \eqref{2.21d} implies that a primary covariantly chiral superfield can carry only undotted spinor indices
\begin{align}
	K^{B} \F =0 ~, \quad \bar{\nabla}^{\ad} \F = 0 \quad \implies \quad \F \equiv \F_{\a(n)}~,
\end{align}
and that its $\sU(1)_R$ charge is determined in terms of its dimension, 
\bea
K^B \F_{\a(n)} =0~, \quad \bar \nabla^\bd \F_{\a(n)} =0 \ ,\quad {\mathbb D} \F_{\a(n)} = \D_\F \F_{\a(n)} ~~
\implies ~~ q_\Phi = -\frac 23  \D_\Phi~.
\label{N=1Chiral} 
\eea
There is a regular procedure to construct such constrained multiplets. Given a complex tensor superfield $\J_{\a(n)}$ with the superconformal properties
\bea
K^B \J_{\a(n)} =0~,\quad
{\mathbb D} \J_{\a(n)} = (\D-1) \J_{\a(n)}
~,\quad
\mathbb{Y} \J_{\a(n)} =  \Big(2-\frac 23  \D\Big) \J_{\a(n)} 
~
\eea
its descendant 
\bea
\label{2.77}
\F_{\a(n)}= - \frac 14 \bar \nabla^2 \Psi_{\a(n)}
\eea
proves to be primary and  covariantly chiral. Here $\F_{\a(n)}$ is invariant under gauge transformations of the form $\d \J_{\a(n)} = \bar \nabla^\bd \l_{\a(n) \bd} $, where the gauge parameter $\l_{\a(n)\bd} $ is primary.

\subsubsection{Superconformal action principles}

In order to formulate locally superconformal field theories, an action principle is required. 
Such actions can be constructed in two different ways:  either as integrals over the full superspace or over its chiral subspace. Here we consider these two options. 

First, we look for a scalar superfield $\cL$ such that the action
\be
\label{2.78}
\cS = \int\text{d}^{4|4}z \, E \, \mathcal{L} ~, \qquad \rd^{4|4}z = \rd^{4}x \, \rd^{2} \q \, \rd^{2} \bar{\q}~,
\ee
is locally superconformal.  
Performing a  gauge transformation, eqs. \eqref{2.63} and \eqref{2.64}, we arrive at the variation
\bea
\label{2.79}
\d_{\mathscr{K}} \cS = \int\text{d}^{4|4}z \, E \, \bigg ( (-1)^{\ve_A}\Big[  \nabla_A( \xi^A \mathcal{L}) &+&  
\xi^B   \mathcal{T}_{BA}{}^A  \mathcal{L} \Big]\non \\ 
\qquad + \L^{\underline{B}}\Big[  (-1)^{\ve_A}  f_{\underline{B} A}{}^{A} \mathcal{L} 
&+& 
X_{\underline{B}} \mathcal{L}\Big] \bigg) ~, 
\eea
which must vanish.  Requiring the contributions containing $\L^{\underline{B}}$ to vanish gives
\be
\label{2.80}
X_{\underline{B}} \mathcal{L} = - (-1)^{\ve_A} f_{\underline{B} A}{}^{A} \mathcal{L}
\quad \Longleftrightarrow \quad K^B \cL =0~, \quad {\mathbb D} \cL = 2 \cL~, 
\quad \mathbb{Y} \cL =0~.
\ee
Once the conditions 
\eqref{2.80} are satisfied, it is readily seen that the remaining $\x$-dependent contributions in \eqref{2.79} cancel out. In summary, given a primary real dimension-2 scalar Lagrangian $\cL$, the action \eqref{2.78} is locally superconformal. 

Given a  primary chiral scalar Lagrangian $\cL_{\rm c}$ of weight $3$, 
\bea
K^B \cL_{\rm c} =0~, \quad \bar \nabla_\ad \cL_{\rm c} =0 \ ,\quad 
{\mathbb D} \cL_{\rm c} = 3 \cL_{\rm c} ~,
\label{2.81}
\eea
the  {\rm chiral} action
\bea
\cS_{\rm c}=\int\rd^4x\rd^2\q\, \cE \,\cL_{\rm c} 
\label{2.82}
\eea
is locally superconformal. Here $\cE$ is a chiral density. The precise definition of $\cE$ requires the use of a prepotential formulation for supergravity, see \cite{KRTM1,BK}. Actions over the full and chiral superspaces may be related by the rule
\begin{align}
	\cS = \int\rd^{4|4}z\,  E\, \cL = - \frac{1}{4} \int \rd^4 x \, \rd^2 \q \,\cE \, \bar{\nabla}^2 \cL ~.
\end{align}

There is an alternative definition of the chiral action that follows from the superform approach to 
the construction of supersymmetric invariants \cite{Ectoplasm,GGKS}.
It is based on the use of the  following  super 4-form 
\bea
\Xi_4&=&
2\ri \bar{E}_\dd \wedge \bar{E}_\gd\wedge  E^b\wedge  E^a(\ts_{ab})^{\gd\dd}
\cL_{\rm c}
+\frac{\ri}{6}\ve_{abcd}\bar{E}_\dd\wedge  E^c\wedge E^b\wedge  E^a (\ts^d)^{\dd\d}\nabla_\d \cL_{\rm c}
\non\\
&&
-\frac{1}{96}\ve_{abcd}E^d\wedge E^c\wedge E^b\wedge E^a 
\nabla^2\cL_{\rm c}~,
\label{Sigma_4}
\eea
which was constructed by Bin\'etruy {\it et al.} 
\cite{Binetruy:1996xw} and independently by Gates {\it et al.} \cite{GGKS} in the GWZ superspace. This superform is closed, 
\bea
\rd \, \X_4 =0~,
\eea
and it proves to be primary\footnote{The superform may be degauged to the $\sU(1)$ and GWZ superspaces described in section \ref{section2.4.1}. Then the condition \eqref{2.89} is equivalent to the super-Weyl invariance of $\X_4$. The latter property was proven in \cite{KT-M17} in the GWZ geometry.} 
\bea
K^B \X_4 =0~.
\label{2.89}
\eea
The chiral action \eqref{2.82} can be recast
as an integral of $\Xi_4$ over a spacetime $\cM^4$,
\begin{subequations}\label{2.90}
	\bea
	\cS_{\rm c} &=& \int_{\cM^4} \Xi_4
	~, \label{2.90a}
	\eea
	where $\cM^4$ is the bosonic body of the curved superspace  $\cM^{4|4}$
	obtained by switching off  the Grassmann variables. 
	It turns out that 
	\eqref{2.90a} leads to the representation 
	\bea
	\cS_{\rm c} 
	=
	\int {\rm d}^4x\, e \,\Big(
	-\frac{1}{4}\nabla^2
	+\frac{\ri}{2}(\tilde{\s}^{a})^{\ad\a}\bar{\psi}_a{}_\ad\nabla _\a
	-(\tilde{\s}^{ab})^{\ad\bd}\bar{\psi}_a{}_\ad\bar{\psi}_b{}_\bd\Big){\cL}_{\rm c}\Big|_{\theta^\a=\bar{\q}_\ad=0} ~,~~~~~~ 
	\label{2.90b}
	\eea
\end{subequations} 
which is the simplest way to reduce the action from superfields to components. 
Here we have introduced the vielbein and gravitino fields
\begin{subequations}
	\begin{align}
		e_m{}^a &= E_{m}{}^a|_{\theta^\a=\bar{\q}_\ad=0} ~, \qquad e= {\rm det}(e_m{}^a)~, \\
		\psi_{m}{}^\a &= 2 E_m{}^\a|_{\theta^\a=\bar{\q}_\ad=0}~, \quad \psi_a{}^\a = e_a{}^m \psi_m{}^\a~.
	\end{align}
\end{subequations}

\subsection{$\mathcal{N}=2$ conformal superspace}
This subsection is devoted to a review of $\cN=2$ conformal superspace in four dimensions. It was introduced in \cite{ButterN=2} as a generalisation of the $\cN=1$ construction \cite{ButterN=1} described above.\footnote{Owing to the fact that $\cN=2$ supersymmetry in four dimensions involves eight supercharges, the geometry of this superspace obeys almost identical constraints to that of the $\cN=(1,0)$ conformal superspace in six dimensions \cite{BKNT}, see appendix \ref{Appendix2C} for a review. }

\subsubsection{The algebra of covariant derivatives}
The defining constraints on the covariant derivatives $\nabla_A$ in $\cN=2$ conformal superspace are\footnote{Specific to the $\cN=2$ case is presence of the totally antisymmetric $\sSU(2)_R$ invariant tensor $\ve_{ij}$,  $\ve^{\underline{1}\underline{2}} = \ve_{\underline{2}\underline{1}} = 1$. It may be used to raise and lower isospinor indices by the rule: $\psi^i = \ve^{ij} \psi_j$, $\psi_i = \ve_{ij} \psi^j$.}
\bsubeq
\label{N=2CSSAlgebra-0}
\bea
\{ \nabla_{\a}^i , \nabla_{\b}^j \}  &=&  -2 \ve^{ij} \ve_{\a \b} \bar{\mathscr W} ~, \quad \{ \bar{\nabla}^{\ad}_i , \bar{\nabla}^{\bd}_j \} = 2 \ve_{ij} \ve^{\ad \bd} \mathscr{W} ~, \\ &&\quad \;\; \{\nabla_{\a}^i , \bar{\nabla}^{\bd}_{j} \} = - 2 \ri \d^i_j \nabla_{\a}{}^{\bd} ~,
\eea
\esubeq
where the operator $\bar{\mathscr{W}}$ is the complex conjugate of $\mathscr{W}$. The latter takes the form
\bea
{\mathscr{W}}
&=&
\frac{1}{2}{\mathscr{W}}(M)^{ab} M_{ab}
+\ri {\mathscr{W}}(\mathbb Y)\mathbb Y
+\mathscr{W}(\mathbb{J})^{ij} \mathbb{J}_{ij}
+{\mathscr{W}}(\mathbb D)\mathbb D
\non\\
&&
+{\mathscr{W}}(S)_\a^i S^\a_i
+{\mathscr{W}}(S)_{i}^\ad \bar{S}_\ad^i
+{\mathscr{W}}(K)_a K^a
~.
\eea
Having imposed the constraints \eqref{N=2CSSAlgebra-0}, the Bianchi identities \eqref{2.65} become non-trivial and now play the role of consistency conditions which may be used to determine the torsion and curvature. Their solution, up to mass dimension-3/2,
is as follows
\begin{subequations}
	\label{2.91}
	\begin{align}
		\{ \nabla_\a^i , \nabla_\b^j \} &= 2 \ve^{ij} \ve_{\a\b} \big( \bar{W}_{\gd\dd} \bar{M}^{\gd\dd} + \frac 1 4 \bar{\nabla}_{\gd k} \bar{W}^{\gd\dd} \bar{S}^k_\dd - \frac 1 4 \nabla_{\g\dd} \bar{W}^\dd{}_\gd K^{\g \gd} \big)~, \\
		\{ \bar{\nabla}^\ad_i , \bar{\nabla}^\bd_j \} &= -2 \ve_{ij} \ve^{\ad\bd} \big( W^{\g\d} M_{\g\d} - \frac 1 4 \nabla^{\g k} W_{\g\d} S_k^\d + \frac 1 4 \nabla^{\g\gd} W_\g^\d K_{\d \gd} \big)~, \\
		\{ \nabla_\a^i , \bar{\nabla}^\bd_j \} &= - 2 \ri \d_j^i \nabla_\a{}^\bd~, \\
		[\nabla_{\a\ad} , \nabla_\b^i ] &= - \ri \ve_{\a\b} \bar{W}_{\ad\bd} \bar{\nabla}^{\bd i} - \frac{\ri}{2} \ve_{\a\b} \bar{\nabla}^{\bd i} \bar{W}_{\ad\bd} \mathbb{D} - \frac{\ri}{4} \ve_{\a\b} \bar{\nabla}^{\bd i} \bar{W}_{\ad\bd} {\mathbb Y} + \ri \ve_{\a\b} \bar{\nabla}^\bd_j \bar{W}_{\ad\bd} \mathbb{J}^{ij}
		\\ & \quad
		- \ri \ve_{\a\b} \bar{\nabla}_\bd^i \bar{W}_{\gd\ad} \bar{M}^{\bd \gd} - \frac{\ri}{4} \ve_{\a\b} \bar{\nabla}_\ad^i \bar{\nabla}^\bd_k \bar{W}_{\bd\gd} \bar{S}^{\gd k} + \frac{1}{2} \ve_{\a\b} \nabla^{\g \bd} \bar{W}_{\ad\bd} S^i_\g
		\\ & \quad
		+ \frac{\ri}{4} \ve_{\a\b} \bar{\nabla}_\ad^i \nabla^\g{}_\gd \bar{W}^{\gd \bd} K_{\g \bd}~, \\
		[ \nabla_{\a\ad} , \bar{\nabla}^\bd_i ] &=  \ri \d^\bd_\ad W_{\a\b} \nabla^{\b}_i + \frac{\ri}{2} \d^\bd_\ad \nabla^{\b}_i W_{\a\b} \mathbb{D} - \frac{\ri}{4} \d^\bd_\ad \nabla^{\b}_i W_{\a\b} {\mathbb Y} + \ri \d^\bd_\ad \nabla^{\b j} W_{\a\b} \mathbb{J}_{ij}
		\\ & \quad
		+ \ri \d^\bd_\ad \nabla^{\b}_i W^\g{}_\a M_{\b\g} + \frac{\ri}{4} \d^\bd_\ad \nabla_{\a i} \nabla^{\b j} W_\b{}^\g S_{\g j} - \hf \d^\bd_\ad \nabla^\b{}_\gd W_{\a\b} \bar{S}^{\gd}_i
		\\ & \quad
		+ \frac{\ri}{4} \d^\bd_\ad \nabla_{\a i} \nabla^\g{}_\gd W_{\b\g} K^{\b\gd} ~.
	\end{align}
\end{subequations}

We note that the conformal superspace algebra is expressed in terms of a single superfield $W_{\alpha \beta }= W_{\b \a}$, its conjugate $\bar W_{\ad \bd}$, and their covariant derivatives. This superfield is an $\mathcal{N}=2$ extension of the Weyl tensor, and is called the super-Weyl tensor. It proves to be a primary chiral superfield of dimension 1,
\bea
K^C W_{\a \b } =0~, \quad \bar \nabla^\gd_k W_{\a\b}=0 ~, \quad
{\mathbb D} W_{\a\b} = W_{\a\b}~,
\eea
and it obeys the Bianchi identity
\bea
\cB &:=&  \nabla_{\a \b} W^{\a\b}
= \bar \nabla^{\ad \bd} \bar W_{\ad \bd}
= \bar \cB~,
\eea
where we have defined the second-order operators:
\begin{subequations}
	\begin{align}
		\nabla_{\a \b} = \nabla_{(\a}^i \nabla_{\b) i}~,	\qquad \bar{\nabla}^{\ad \bd} = \bar{\nabla}^{(\ad}_i \nabla^{\bd) i}~.
	\end{align}
	It is also useful to introduce the following:
	\begin{align}
		\nabla^{ij} = \nabla^{\a(i} \nabla_\a^{ j)}  ~, \qquad 
		\bar \nabla_{ij} = \bar{\nabla}_{\ad (i} \bar{\nabla}^\ad_{j)} ~.
	\end{align}
\end{subequations}
The superfield $\cB$ is
the $\cN=2$ supersymmetric generalisation of the Bach tensor.
This super-Bach multiplet proves to be primary, $K^A \cB =0$, carries weight $2$, ${\mathbb D} \cB= 2 \cB$, and satisfies the conservation equation \cite{BdeWKL}
\bea 
\nabla^{ij} \cB&=&0 \quad \Longleftrightarrow \quad
\bar \nabla_{ij} \cB =0
~.
\eea

The structure of the algebra \eqref{2.91} leads to highly non-trivial implications. In particular, its consistency with eq. \eqref{2.21d} implies that 
primary covariantly chiral superfields can carry neither isospinor nor dotted spinor indices
\begin{align}
	\bar{\nabla}^{\ad}_i \F = 0 \quad \implies \quad \F \equiv \F_{\a(n)}~.
\end{align}
Given such a superfield, eq. \eqref{2.21d} further implies that its $\sU(1)_R$ charge is determined in terms of its dimension, 
\bea
K^B \F_{\a(n)} =0~, \quad \bar \nabla_j^\bd \F_{\a(n)} =0 \ ,\quad {\mathbb D} \F_{\a(n)} = \D_\F \F_{\a(n)} ~~
\implies ~~ q_\F = - 2  \D_\F~.
\label{2.97} 
\eea
There is a regular procedure to construct primary chiral multiplets and their conjugate antichiral ones. It is based on the use of operators
\bea
\nabla^4 :=\nabla^{\a\b}\nabla_{\a\b} =-\nabla^{ij}\nabla_{ij}~,\quad
\bar \nabla^4:= \bar \nabla_{\ad\bd} \bar \nabla^{\ad\bd} = -\bar \nabla_{ij} \bar \nabla^{ij}~.
\eea
Let us consider a $\sSU(2)_R$ neutral rank-$n$ spinor superfield $\J_{\a(n)}$ with the following superconformal properties:
\bea
K^B  \J_{\a(n)} = 0~, \quad {\mathbb D} \J_{\a(n)} = (\D-2)  \J_{\a(n)}~, 
\quad {\mathbb Y}  \J_{\a(n)} = 2(2-\D)  \J_{\a(n)}~.
\eea
Then its descendant 
\bea
\label{2.100}
\F_{\a(n)} = -\frac{1}{48} \bar \nabla^4  \J_{\a(n)}
\eea
is a primary covariantly chiral superfield of the type \eqref{2.97}.

\subsubsection{Superconformal action principles} 

To construct supergravity-matter systems, a locally superconformal action principle is required. 
Here we review three types of superconformal actions in $\cN=2$ supergravity that have played important roles in the literature. 

The simplest locally superconformal action involves a full superspace integral:
\begin{align}
	\cS = \int\rd^{4|8}z\,  E\, \cL~,\qquad\rd^{4|8}z=\rd^4x\,\rd^4\q \,\rd^4 \bar \q~,
\end{align}
where $\cL$ is a primary real dimensionless scalar Lagrangian,
\bea
K^A \cL =0~, \qquad \bar \cL = \cL~, \qquad {\mathbb D}\cL=0~.
\eea

More general  is the chiral action, which involves an integral
over the chiral subspace
\bea  
\cS_{\rm c} = \int  \rd^4 x \, \rd^4 \q \,\cE \, \cL_{\rm c} \ .
\label{chiralAc}
\eea
Here  $\cE$ is a
suitably chosen chiral measure,
and $\cL_{\rm c}$ is a primary covariantly chiral Lagrangian of dimension $2$,
\bea
K^A \cL_{\rm c}=0~, \qquad
\bar\nabla^\ad_i \cL_{\rm c} = 0~, \qquad {\mathbb D}\cL_{\rm c} = 2 \cL_{\rm c}~.
\label{ChiralLagrangian}
\eea
The precise definition of $\cE$ in conformal superspace is somewhat technical\footnote{In $\sSU(2)$ superspace, which will be described in section \ref{section2.4.2}, $\cE$ was obtained by making use of normal coordinates 
	\cite{KT-M2009}.}
\cite{ButterN=2}.
Actions over the full and chiral superspaces may be related by the rule
\begin{align}
	\cS = \int\rd^{4|8}z\,  E\, \cL = - \frac{1}{48} \int \rd^4 x \, \rd^4 \q \,\cE \, \bar{\nabla}^4 \cL ~.
\end{align}

There is an alternative definition of the chiral action that follows from the superform approach to 
the construction of supersymmetric invariants \cite{Castellani,Ectoplasm,GGKS}.
It is based on the use of the  following  super 4-form \cite{GKT-M}
\bea
\Xi_4&=&
-4\bar{E}_\bd^{j}\wedge \bar{E}^\bd_{j}\wedge \bar{E}_\ad^{i}\wedge \bar{E}^\ad_{i}\,
\cL_{\rm c}
-2\bar{E}_\bd^{j}\wedge \bar{E}^{\bd}_{ j}\wedge \bar{E}_\ad^{i}\wedge E^a\, 
(\ts_a)^{\ad\a}\nabla_{\a i}\cL_{\rm c}
\non\\
&&
-\frac{\ri}{2}\bar{E}_\bd^{j}\wedge \bar{E}_\ad^{i}\wedge E^b\wedge  E^a\,(\ts_{ab})^{\ad\bd}\nabla^{ij}\cL_{\rm c}
\non\\
&&
-\frac{\ri}{4}\bar{E}_\ad^{i}\wedge \bar{E}^\ad_{i}\wedge E^b\wedge  E^a\Big(
(\s_{ab})_{\a\b}\nabla^{\a\b}
-8({\ts}_{ab})_{\ad\bd}\bar{W}^{\ad\bd}\Big)\cL_{\rm c}
\non\\
&&
-\frac{\ri}{36}\ve_{abcd}\bar{E}_\ad^{i}\wedge E^c\wedge E^b\wedge  E^a\Big(
(\ts^d)^{\ad\a}\nabla_\a^{j}\nabla_{ij}
-6(\ts^d)^{\bd\a}\bar{W}_{\ad\bd}\nabla_{\a i}\Big){\cL}_{\rm c}
\non\\
&&
+\frac{1}{24}\ve_{abcd}E^d\wedge E^c\wedge E^b\wedge E^a
\Big(
\nabla^4+\bar{W}^{\ad\bd}\bar{W}_{\ad\bd}\Big)\cL_{\rm c}
~.
\label{Sigma_4}
\eea
This superform is closed, 
\bea
\rd \, \X_4 =0~.
\eea
It proves to be primary\footnote{The superform may be degauged to the $\sU(2)$ and $\sSU(2)$ superspaces described in section \ref{section2.4.2}. Then the condition \eqref{337} is equivalent to the super-Weyl invariance of $\X_4$. } 
\bea
K^B \X_4 =0~.
\label{337}
\eea
The chiral action \eqref{chiralAc} can be recast
as an integral of $\Xi_4$ over a spacetime $\cM^4$,
\begin{subequations}\label{2.24}
	\bea
	\cS_{\rm c} &=& \int_{\cM^4} \Xi_4
	~, \label{2.24.a}
	\eea
	where $\cM^4$ is the bosonic body of the curved superspace  $\cM^{4|4}$
	obtained by switching off  the Grassmann variables. 
	It turns out that 
	\eqref{2.24.a} leads to the following representation 
	\cite{ButterN=2} (see also \cite{Butter:2012xg}):
	\bea
	\cS_{\rm c} 
	&=&
	\int {\rm d}^4x\, e \,\bigg(
	~
	\frac{1}{48}\nabla^4 
	+ \bar{W}^{\ad\bd} \bar{W}_{\ad\bd} 
	-  \frac{\ri}{12} \bar{\psi}_d{}^l_\dd \Big((\tilde{\s}^d)^{\dd \a} \nabla_\a^q \nabla_{lq} -6(\s^d)_{\a \ad} \bar{W}^{\ad \dd} \nabla^\a_l \Big)
	\non\\
	&&
	+\frac{1}{4} \bar{\psi}_c{}_\gd^k \bar{\psi}_d{}^l_\dd  \Big( 
	(\tilde{\s}^{cd})^{\gd \dd} \nabla_{kl}
	-\hf \ve^{\gd \dd} \ve_{kl} (\s^{cd})_{\b \g} \nabla^{\b \g} 
	- 4\ve^{\gd\dd} \ve_{kl} (\tilde{\s}^{cd})_{\ad\bd} \bar{W}^{\ad\bd} \Big)
	\non\\
	&&
	- \frac{1}{4} \ve^{abcd} (\tilde{\s}_a)^{\bd \a} \bar{\psi}_b{}_\bd^j \bar{\psi}_c{}_\gd^k \bar{\psi}_d{}^\gd_k \nabla_{\a j} 
	- \frac{\ri}{4} \ve^{abcd} \bar{\psi}_a{}_{\ad}^i \bar{\psi}_b{}^{\ad}_i \bar{\psi}_c{}_{\bd}^j \bar{\psi}_d{}^{\bd}_j \bigg) {\cL}_{\rm c}\Big|_{\theta^\a_i=\bar{\q}_\ad^i=0} ~,~~~~~~~~~~~
	\label{2.24.b}
	\eea
\end{subequations}
where we have introduced the vielbein $e_m{}^a$ and gravitini $\psi_{m}{}^\a_i$ fields as follows:
\begin{align}
	e_m{}^a = E_{m}{}^a|_{\theta^\a_i=\bar{\q}_\ad^i=0} ~, \qquad \psi_{m}{}^\a_i = 2 E_m{}^\a_i|_{\theta^\a_i=\bar{\q}_\ad^i=0}~.
\end{align}

\subsection{$\mathcal{N}>2$ conformal superspace with vanishing curvature}
With the exception of the $\cN=4$ case \cite{ButterN=4}, the $\cN>2$ conformal superspaces with non-vanishing super-Weyl tensor have not yet appeared in the literature. However, even in the absence of curvature, such superspaces are a powerful setting to study superconformal field theories in conformally-flat backgrounds. Thus, the goal of this subsection is to sketch $\cN > 2$ conformal superspace with vanishing curvature.
\subsubsection{The algebra of covariant derivatives}
In the absence of curvature, the algebra of covariant derivatives for $\cN>2$ conformal superspace is flat. Specifically, they satisfy the anti-commutation relations:
\bea
\label{2.109}
\{ \nabla_{\a}^i , \nabla_{\b}^j \}  =  0 ~, \qquad \{ \bar{\nabla}^{\ad}_i , \bar{\nabla}^{\bd}_j \} = 0 ~, \qquad \{ \nabla_{\a}^i , \bar{\nabla}^\bd_j \} = - 2 \ri \d^i_j \nabla_\a{}^\bd ~.
\eea

Consistency of this algebra with eq. \eqref{2.21d} implies that 
primary covariantly chiral superfields can carry neither isospinor nor dotted spinor indices
\begin{align}
	\bar{\nabla}^{\ad}_i \F = 0 \quad \implies \quad \F \equiv \F_{\a(n)}~.
\end{align}
Given such a superfield, eq. \eqref{2.21d} further implies that its $\sU(1)_R$ charge\footnote{We remind the reader that, for $\cN=4$, the $\sU(1)_R$ charge of a superfield is ill-defined.} is determined in terms of its dimension, 
\bea
\label{2.111}
K^B \F_{\a(n)} =0~, \quad \bar \nabla_j^\bd \F_{\a(n)} =0 \ ,\quad {\mathbb D} \F_{\a(n)} = \D_\F \F_{\a(n)} ~~
\implies ~~ q_\F = \frac{2\cN}{\cN-4} \D_\F~.~~~~
\eea
There is a regular procedure to construct primary chiral multiplets and their conjugate antichiral ones. It is based on the use of the operators
\bea
\nabla^{2\cN} := \nabla^{\a(\N)}\nabla_{\a(\N)}~,\quad
\bar \nabla^{2\cN}:= \bar \nabla_{\ad(\cN)} \bar \nabla^{\ad(\cN)}~,
\eea
where we have introduced the rank-$\cN$ operators
\bea
\nabla_{\a(\cN)}= \ve_{i_1 \dots i_\cN} \nabla_{(\a_1}^{i_1} \dots \nabla_{\a_\cN)}^{i_\cN} ~, \quad \bar{\nabla}^{\ad(\cN)} = \ve^{i_1 \dots i_\cN} \bar{\nabla}^{(\ad_1}_{i_1} \dots \bar{\nabla}^{\ad_\cN)}_{i_\cN}~,
\eea
where the totally antisymmetric $\sSU(\cN)_R$ invariant tensors $\ve^{i_1 \dots i_\cN}$ and $\ve_{i_1 \dots i_\cN}$ are normalised as $\ve^{\underline{1} \dots \underline{\cN}} = - \ve_{\underline{1} \dots \underline{\cN}}  = 1$.
Let us consider a $\sSU(\cN)_R$ neutral rank-$n$ spinor superfield $\J_{\a(n)}$ with the following superconformal properties:
\bea
K^B  \J_{\a(n)} = 0~, \quad {\mathbb D} \J_{\a(n)} = (\D-\cN)  \J_{\a(n)}~, 
\quad {\mathbb Y}  \J_{\a(n)} = \cN \Big( 1 + \frac{2 \D}{\cN-4} \Big)  \J_{\a(n)}~.~~~
\eea
Then its descendant 
\bea
\F_{\a(n)} = \bar \nabla^{2\cN} \J_{\a(n)}
\eea
is a primary covariantly chiral superfield of the type \eqref{2.111}.

\subsubsection{Superconformal action principles} 

The simplest locally superconformal action in $\cN$-extended conformal superspace involves a full superspace integral:
\begin{align}
	\cS = \int\rd^{4|4\cN}z\,  E\, \cL~,\qquad\rd^{4|4\cN}z=\rd^4x\,\rd^{2 \cN}\q\, \rd^{2\cN} \bar \q~,
\end{align}
where $\cL$ possesses the superconformal properties:
\bea
K^A \cL =0~, \qquad \bar \cL = \cL~, \qquad {\mathbb D}\cL= (4 - 2\cN) \cL~.
\eea

More general  is the chiral action, which involves an integral
over the chiral subspace
\bea  
\cS_{\rm c} = \int  \rd^4 x \, \rd^{2\cN} \q \,\cE \, \cL_{\rm c} ~ .
\label{2.118}
\eea
Here  $\cE$ is an appropriately defined chiral measure,
and $\cL_{\rm c}$ is constrained by
\bea
K^A \cL_{\rm c}=0~, \qquad
\bar\nabla^\ad_i \cL_{\rm c} = 0~, \qquad {\mathbb D}\cL_{\rm c} = (4 - \cN) \cL_{\rm c}~.
\eea
Actions over the full and chiral superspaces may be related by the rule
\begin{align}
	S = \int\rd^{4|4\cN}z\,  E\, \cL = - \frac{1}{2^\cN \cN!(\cN+1)!} \int \rd^4 x \, \rd^{2\cN} \q \,\cE \, \bar{\nabla}^{2\cN} \cL ~.
\end{align}

To the best of our knowledge, the rule to recast the chiral action \eqref{2.118} as an integral over the bosonic body of $\cM^{4|4\cN}$ has not yet appeared in the literature. However, if one sets the gravitini fields to zero, $\psi_{m}{}^\a_i = 2 E_m{}^\a_i|_{\theta^\a_i=\bar{\q}_\ad^i=0}=0$, the rule is simple:
\bea  
S_{\rm c} = - \frac{1}{2^\cN \cN!(\cN+1)!} \int \rd^4 x \,e \, \nabla^{2\N}\cL_{\rm c}|_{\theta^\a_i=\bar{\q}_\ad^i=0} ~ .
\eea

\section{Degauging to Lorentzian supergeometry} \label{Chapter2.4}

According to \eqref{2.63}, under an infinitesimal special superconformal gauge transformation $\mathscr{K} = \Lambda_{B} K^{B}$, the dilatation connection transforms as follows
\bea
\d_{\mathscr{K}} B_{A} = - 2 \Lambda_{A} ~.
\eea
Thus, it is possible to choose a gauge condition
$B_{A} = 0$, which completely fixes 
the special superconformal gauge freedom.\footnote{There is a class of residual gauge transformations preserving the gauge $B_{A}=0$. These generate the super-Weyl transformations of the degauged geometry, see below.} As a result, the corresponding connection is no longer required for the covariance of $\nabla_A$ under the residual gauge freedom and
should be separated
\bea
\nabla_{A} &=& \mathcal{D}_{A} - \mathfrak{F}_{AB} K^{B} ~.
\eea
Here the operator $\mathcal{D}_{A} $ involves only the Lorentz, $\sU(1)_R$ and $\sSU(\cN)_R$ connections.

The residual gauge freedom of \eqref{2.63} in the gauge $B_A=0$ includes $\mathcal{K}$ gauge transformations of the form
\begin{subequations}
	\label{2.127}
	\bea
	\delta_{\mathcal K} \mathcal{D}_A &=& [{\mathcal K}, \mathcal{D}_A] ~ , \\
	{\mathcal K} &=& \xi^B \mathcal{D}_B + \hf K^{bc} M_{bc} + \ri \rho \mathbb{Y} + \chi^{i}{}_j \mathbb{J}^{j}{}_i ~ ,~~~~
	\eea
\end{subequations}
where the gauge parameters satisfy natural reality conditions. The corresponding transformation of a given tensor superfield $U$ (with its indices suppressed) is:
\bea 
\d_{\mathcal K} U = {\mathcal K} U ~.
\eea

The next step is to relate the special superconformal connection
$\mathfrak{F}_{AB}$  to the torsion tensor of the degauged geometry. To do this, it is necessary to make use of the relation
\bea
\label{2.112}
[ \nabla_{A} , \nabla_{B} \} &=& [ \mathcal{D}_{A} , \mathcal{D}_{B} \} - \big(\mathcal{D}_{A} \mathfrak{F}_{BC} - (-1)^{\ve_A \ve_B} \mathcal{D}_{B} \mathfrak{F}_{AC} \big) K^C - \mathfrak{F}_{AC} [ K^{C} , \nabla_B \} \non \\
&& + (-1)^{\ve_A \ve_B} \mathfrak{F}_{BC} [ K^{C} , \nabla_A \} + (-1)^{\ve_B \ve_C} \mathfrak{F}_{AC} \mathfrak{F}_{BD} [K^D , K^C \} ~.
\eea
It should be noted that such an analysis has only appeared in the literature for $\cN=1$ and $2$. As a result, the $\cN > 2$ case will be omitted below.

\subsection{The $\sU(1)$ and GWZ superspace geometries}
\label{section2.4.1}

In section \ref{section2.3.1} we introduced the $\cN=1$ conformal superspace as a powerful approach to off-shell $\cN=1$ conformal supergravity. Here, we will instead study some other important superspace formulations, namely: (i) $\sU(1)$ superspace \cite{Howe}; and (ii) the GWZ superspace \cite{GWZ}. They differ by their structure groups, which are $\sSL( 2, {\mathbb C}) \times \sU(1)_R$ and $\sSL( 2, {\mathbb C}) $, respectively. Both of them can be derived from conformal superspace by making use of the degauging procedure described above. We elaborate on this in more detail below.

\subsubsection{Degauging to the $\sU(1)$ superspace geometry}

By making use of the conformal superspace algebra, \eqref{2.68} and \eqref{2.70}, the relation \eqref{2.112} leads to a set on consistency conditions equivalent to the Bianchi identities of $\sU(1)$ superspace \cite{Howe}. Their solution expresses the components of $\mathfrak{F}_{AB}$ in terms of the torsion 
tensor of $\sU(1)$ superspace \cite{ButterN=1}. Specifically, one obtains
\begin{subequations} 
	\label{2.113}
	\bea
	\mathfrak{F}_{\a \b} & = & \frac{1}{2} \ve_{\a \b} \bar{R} ~, \quad \bar{\mathfrak{F}}_{\ad \bd} = -\frac{1}{2} \ve_{\ad \bd} R ~, \quad
	\mathfrak{F}_{\a \bd} 
	= -\bar{\mathfrak{F}}_{\bd \a} 
	= \frac{1}{4} G_{\a \bd} ~, 
	\\
	\mathfrak{F}_{\a , \b \bd} & = & - \frac{\ri}{4} \mathcal{D}_{\a} G_{\b \bd} - \frac{\ri}{6} \ve_{\a \b} \bar{X}_{\bd}  = \mathfrak{F}_{\b \bd , \a} ~, \\
	\bar{\mathfrak{F}}_{\ad , \b \bd} &=& \frac{\ri}{4} \bar{\mathcal{D}}_{\ad} G_{\b \bd} + \frac{\ri}{6} \ve_{\ad \bd} X_{\b} =\mathfrak{F}_{\b \bd , \a} ~, \\
	\mathfrak{F}_{\a \ad , \b \bd} & = & - \frac{1}{8} \big[ \mathcal{D}_{\a} , \bar{\mathcal{D}}_{\ad} \big] G_{\b \bd} - \frac{1}{12} \ve_{\ad \bd} \mathcal{D}_{\a} X_{\b} + \frac{1}{12} \ve_{\a \b} \bar{\mathcal{D}}_{\ad} \bar{X}_{\bd}  \non \\
	&& 
	+ \frac{1}{2} \ve_{\a \b} \ve_{\ad \bd} \bar{R} R
	+ \frac{1}{8} G_{\a \bd} G_{\b \ad} ~,
	\eea
\end{subequations}
where $R$ and $X_{\a}$ are complex chiral
\begin{subequations}
	\bea
	\bar{\mathcal{D}}_{\ad} R &=& 0 ~, \qquad \mathbb{Y} R = -2 R~, 
	\\
	\bar{\mathcal{D}}_{\ad} X_{\a} &=& 0 ~, \qquad \mathbb{Y} X_\a = - X_\a~,
	\eea
	while $G_{\a \ad}$ is a real vector superfield. These are related via
	\bea
	X_{\a} &=& \mathcal{D}_{\a}R - \bar{\mathcal{D}}^{\ad}G_{\a \ad} ~. \label{Bianchi1}
	\eea
\end{subequations}

We now pause and comment on the geometry described by $\mathcal{D}_A$. In particular, by employing \eqref{2.112} one arrives at the following anticommutation relation
\be
\label{2.115}
\{ \mathcal{D}_\a , \bar{\mathcal{D}}_{\ad} \} = - 2 \ri \mathcal{D}_{\aa} - G^{\b}{}_{\ad} M_{\a \b} + G_\a{}^{\bd} \bar{M}_{\ad \bd} + \frac{3}{2} G_\aa \mathbb{Y} ~.
\ee
It is convenient to work in a geometry where right hand side of \eqref{2.115} contains no curvature-dependent terms. To this end, we perform the following redefinition
\begin{align}
	\mathcal{D}_{\a \ad} ~\longrightarrow ~ \cD_{\a \ad} + \frac{\ri}{2} G^{\b}{}_{\ad} M_{\a \b} - \frac{\ri}{2} G_{\a}{
	}^{\bd} \bar{M}_{\ad \bd} - \frac{3 \ri}{4} G_{\a \ad} \mathbb{Y} ~.
\end{align}

It may be shown that the resulting algebra of covariant derivatives takes the form
\begin{subequations} \label{U(1)algebra}
	\bea
	\{ \cD_{\a}, \cD_{\b} \} &=& -4{\bar R} M_{\a \b}~, \qquad
	\{\cDB_{\ad}, \cDB_{\bd} \} =  4R {\bar M}_{\ad \bd}~, \label{U(1)algebra.a}\\
	&& {} \qquad \{ \cD_{\a} , \cDB_{\ad} \} = -2{\rm i} \cD_{\a \ad} ~, 
	\label{U(1)algebra.b}	\\
	\big[ \cD_{\a} , \cD_{ \b \bd } \big]
	& = &
	{\rm i}
	{\ve}_{\a \b}
	\Big({\bar R}\,\cDB_\bd + G^\g{}_\bd \cD_\g
	- (\cD^\g G^\d{}_\bd)  M_{\g \d}
	+2{\bar W}_\bd{}^{\gd \dot{\d}}
	{\bar M}_{\gd \dot{\d} }  \Big) \non \\
	&&
	+ {\rm i} (\cDB_{\bd} {\bar R})  M_{\a \b}
	-\frac{\ri}{3} \ve_{\a\b} \bar X^\gd \bar M_{\gd \bd} - \frac{\ri}{2} \ve_{\a\b} \bar X_\bd \mathbb{Y}
	~, \label{U(1)algebra.c}\\
	\big[ {\bar \cD}_{\ad} , \cD_{\b\bd} \big]
	& = &
	- {\rm i}
	\ve_{\ad\bd}
	\Big({R}\,\cD_{\b} + G_\b{}^\gd \cDB_\gd
	- (\cDB^{\gd} G_{\b}{}^{\dd})  \bar M_{\gd \dd}
	+2{W}_\b{}^{\g \d}
	{M}_{\g \d }  \Big) \non \\
	&&
	- {\rm i} (\cD_\b R)  {\bar M}_{\ad \bd}
	+\frac{\ri}{3} \ve_{\ad \bd} X^{\g} M_{\g \b} - \frac{\ri}{2} \ve_{\ad\bd} X_\b \mathbb{Y}
	~, \label{U(1)algebra.d}
	\eea
	which lead to 
	\bea
	\left[ \cD_{\a \ad} , \cD_{\b \bd} \right] & = & \ve_{\a \b} \bar \chi_{\ad \bd} + \ve_{\ad \bd} \chi_{\a \b} ~, \\
	\chi_{\a \b} & = & - \ri G_{ ( \a }{}^{\gd} \cD_{\b ) \gd} + \frac{1}{2} \cD_{( \a } R \cD_{\b)} + \frac{1}{2} \cD_{ ( \a } G_{\b )}{}^{\gd} \cDB_{\gd} + W_{\a \b}{}^{\g} \cD_{\g} \non \\
	&& + \frac{1}{6} X_{( \a} \cD_{ \b)} + \frac{1}{4} (\cD^{2} - 8R) {\bar R} M_{\a \b} + \cD_{( \a} W_{ \b)}{}^{\g \d} M_{\g \d} \non \\
	&& - \frac{1}{6} \cD_{( \a} X^{\g} M_{\b) \g} - \frac{1}{2} \cD_{ ( \a} \cDB^{\gd} G_{\b)}{}^{\dd} {\bar M}_{\gd \dd} + \frac{1}{4} \cD_{( \a} X_{\b)} \mathbb{Y} ~, \\
	{\bar \chi}_{\ad \bd} & = & \ri G^{\g}{}_{( \ad} \cD_{\g \bd)} - \frac{1}{2} \cDB_{( \ad } {\bar R} \cDB_{\bd)} - \frac{1}{2} \cDB_{ ( \ad } G^{\g}{}_{\bd)} \cD_{\g} - {\bar W}_{\ad \bd}{}^{\gd} \cDB_{\gd} \non \\
	&& - \frac{1}{6} {\bar X}_{( \ad} \cDB_{ \bd)} + \frac{1}{4} (\cDB^{2} - 8{\bar R}) R {\bar M}_{\ad \bd} - \cDB_{( \ad} {\bar W}_{ \bd)}{}^{\gd \dd} {\bar M}_{\gd \dd} \non \\
	&& + \frac{1}{6} \cDB_{( \ad} {\bar X}^{\gd} {\bar M}_{\bd) \gd} + \frac{1}{2} \cDB_{ ( \ad} \cD^{\g} G^{\d}{}_{\bd)} M_{\g \d} + \frac{1}{4} \cDB_{( \ad} {\bar X}_{\bd)} \mathbb{Y} ~.
	\eea
\end{subequations}
These relations should be supplemented with the following Bianchi identities:
\begin{subequations}
	\bea
	\cD^{\a} X_{\a} &=& \cDB_{\ad} {\bar X}^{\ad} ~, \label{Bianchi2}\\
	\bar{\cD}_{\ad} W_{\a \b \g} &=& 0 ~, \\
	\cD^{\g} W_{\a \b \g} &=& {\rm i} \cD_{(\a}{}^{\gd} G_{\b ) \gd} - \frac{1}{3} \cD_{(\a} X_{\b)} ~.
	\label{4.10c}
	\eea
\end{subequations}
In particular, we note that \eqref{Bianchi2} implies that $X_{\a}$ is the chiral field strength of a $\sU(1)$ vector multiplet. The geometry described above is the $\sU(1)$ superspace geometry \cite{Howe,GGRS} in the form described in \cite{BK11-2}.

To conclude our discussion of $\sU(1)$ superspace , we make two comments. 
Firstly, one may check that degauging the relation \eqref{2.77} gives
\bea
\F_{\a(n)}= - \frac 14 \big(\bar \cD^2 -4R \big)\Psi_{\a(n)}~, \qquad 
\bar \cD_\bd \F_{\a(n)}=0~.
\eea
Secondly, integration by parts is remarkably simple in $\sU(1)$ superspace; given an arbitrary supervector $\cV^A$, one may show that the following holds
\bea
\int\text{d}^{4|4}z \, E \,  (-1)^{\ve_A} \cD_A \cV^A =0~.
\eea

\subsubsection{The super-Weyl transformations of $\sU(1)$ superspace}

Above we made use of the special conformal gauge freedom to degauge from conformal to $\sU(1)$ superspace. Here we will show that the residual dilatation symmetry manifests in the latter as super-Weyl transformations.

Specifically, to preserve the gauge $B_{A}=0$, every local dilatation transformation with parameter $ \s $ should be accompanied by a compensating special conformal one
\begin{align}
	\mathscr{K}(\sigma) = \L_{B} (\sigma)K^{B} + \s \mathbb{D} \quad \implies \quad \d_{\mathscr{K}( \s )} B_A = 0~.
\end{align}
We then arrive at the following constraints
\bea
\L_{A}(\s) = \hf \nabla_A \s~.
\eea
As a result, we define the following transformation
\bea
\label{2.139}
\d_{\mathscr{K}(\s)} \nabla_A \equiv \d_{\s} \nabla_A = \d_\s \cD_A - \d_\s \mathfrak{F}_{AB} K^B~.
\eea

By making use of \eqref{2.113} and \eqref{2.139}, we arrive at the following transformation laws for the covariant derivatives of $\sU(1)$ superspace 
\begin{subequations}
	\label{superWeylTf}
	\bea
	\delta_{\s}\cD_{\a} & = & \frac{1}{2} \s \cD_{\a} + 2 \cD^{\b} \sigma M_{\b \a} - \frac{3}{2} \cD_{\a} 
	\sigma \mathbb{Y} ~, \\
	\d_{\s} \cDB_{\ad} & = & \frac{1}{2} \s \cDB_{\ad} + 2 \cDB^{\bd} \s {\bar M}_{\bd \ad} +
	\frac{3}{2} \cDB_{\ad} \s \mathbb{Y} ~, \\
	\d_{\s} \cD_{\a \ad} & = & \s \cD_{\a \ad} + {\rm i} \cD_{\a} \s \cDB_{\ad} 
	+ {\rm i} \cDB_{\ad} \s \cD_{\a}  + {\rm i} \cDB_{\ad} \cD^{\b} \s  M_{\b \a} \non \\
	&& + {\rm i} \cD_{\a} \cDB^{\bd} \s { \bar M}_{\bd \ad} + \frac{3 \ri}{4} \left[ \cD_{\a} , \cDB_{\ad} \right]\s \mathbb{Y} ~, \label{superWeylTf.c}
	\eea
\end{subequations}
while the torsion superfields arising from the degauged torsion $\mathfrak{F}_{AB}$ transform as follows
\begin{subequations}
	\label{superWeylTfTorsions}
	\bea
	\d_{\s} R & = & \s R + \frac{1}{2} \cDB^{2} \s ~, \\
	\d_{\s} G_{\a \ad} & = &  \s G_{\a \ad} + [ \cD_{\a} , \cDB_{\ad} ] \s ~, \\
	\d_{\s} X_{\a} & = & \frac{3}{2} \s X_{\a} - \frac{3}{2} (\cDB^{2} - 4 R) \cD_{\a} \s ~. \label{sWX}
	\eea
	Finally, as the super-Weyl tensor is conformally covariant, its transformation law is readily obtained via
	\bea
	\d_{\s} W_{\a \b \g} = \big(\L_{B}(\s) K^{B} + \s \mathbb{D}\big) W_{\a \b \g} = \frac{3}{2} \s W_{\a \b \g}~.
	\eea
\end{subequations}

The relations  \eqref{superWeylTf} and \eqref{superWeylTfTorsions} 
give the super-Weyl transformations 
in $\sU(1)$ superspace
\cite{GGRS,Howe} (see also \cite{BK11-2}).
The conditions \eqref{2.67}, which define a primary superfield $U$, are equivalent to the following 
\bea
\label{2.142}
\d_\s U = \D_U \s U~, \qquad  {\mathbb Y} U = q_U U~,
\eea
in $\sU(1)$ superspace.

\subsubsection{The Grimm-Wess-Zumino formulation}

As pointed out above, the covariantly chiral spinor $X_\a$ is the field strength 
of an Abelian vector multiplet. It follows from \eqref{sWX} that the super-Weyl freedom allows us to choose the gauge 
\bea
X_\a =0~.
\eea
In this gauge the $\sU(1)_{R}$ curvature vanishes, in accordance with  
\eqref{U(1)algebra}, and therefore the $\sU(1)_{R}$ connection may be gauged away.
The resulting algebra of covariant derivatives \eqref{U(1)algebra} 
reduces to that of the  GWZ geometry \cite{GWZ}.

Equation \eqref{sWX}  tells us that imposing the condition $X_\a=0$ does not completely fix the super-Weyl freedom. The residual transformations are generated 
by parameters of the form 
\bea 
\s=\hf \big(\S +\bar \S \big) ~, \qquad \bar \cD_\ad \S =0~. 
\eea
However, in order to preserve the gauge of vanishing $\sU(1)_{R}$ connection, 
every residual super-Weyl transformation must be accompanied by a 
compensating $\sU(1)_{R}$ transformation with parameter
\bea
\r = \frac{3 \ri}{4} \big( \S - \bar \S\big)~.
\eea
This leads to the transformations of the covariant derivatives \cite{HT}:
\begin{subequations} 
	\label{superweyl}
	\bea
	\d_\S \cD_\a &=& \Big( {\bar \S} - \hf \S \Big)  \cD_\a + \cD^\b \S M_{\a \b}  ~, \\
	\d_\S \bar \cD_\ad & = & \Big(  \S -  \hf {\bar \S} \Big)
	\bar \cD_\ad +  \bar \cD^\bd  {\bar \S}  {\bar M}_{\ad \bd} ~,\\
	\d_\S \cD_{\a\ad} &=& \hf( \S +\bar \S) \cD_{\a\ad} 
	+\frac{\ri}{2} \bar \cD_\ad \bar \S \cD_\a + \frac{\ri}{2} \cD_\a  \S \bar \cD_\ad \non \\
	&& + \cD^\b{}_\ad \S M_{\a\b} + \cD_\a{}^\bd \bar \S \bar M_{\ad \bd}~.
	\eea
\end{subequations}
At the same time, the torsion tensors transform as follows:
\begin{subequations} 
	\bea
	\d_\S R &=& 2\S R +\frac{1}{4} (\bar \cD^2 -4R ) \bar \S ~, \\
	\d_\S G_{\a\ad} &=& \hf (\S +\bar \S) G_{\a\ad} +\ri \cD_{\a\ad} ( \S- \bar \S) ~, 
	\label{s-WeylG}\\
	\d_\S W_{\a\b\g} &=&\frac{3}{2} \S W_{\a\b\g}~.
	\label{s-WeylW}
	\eea
\end{subequations} 
The conditions \eqref{2.142}, which define a primary superfield $U$, turn into 
\bea
\d_\S U = (\mf{p}_U\S +\mf{q}_U\bar \S) U~, \qquad \mf{p}_U+\mf{q}_U =\D_U, \quad \mf{p}_U-\mf{q}_U = - \frac 32 q_U~,
\eea
in the GWZ formulation.

\subsection{The $\sU(2)$ and $\sSU(2)$ superspace geometries}
\label{section2.4.2}

Here we consider two covariant formulations for $\cN=2$ conformal supergravity that have found applications in recent years, specifically:
(i) $\sU(2)$ superspace 
\cite{Howe,KLRT-M2}; and (ii) $\sSU(2)$ superspace 
\cite{Grimm,KLRT-M1}. They differ by their structure groups, which are $\sSL( 2, {\mathbb C}) \times \sU(2)_R$ and $\sSL( 2, {\mathbb C}) \times \sSU(2)_R$, respectively.

\subsubsection{Degauging to the $\sU(2)$ superspace geometry}

In conjunction with \eqref{2.91}, the identity \eqref{2.112} leads to a set of consistency conditions that are equivalent to the Bianchi identities of $\sU(2)$ superspace \cite{Howe}. Their solution expresses the components of $\mathfrak{F}_{AB}$ in terms of the torsion 
tensor of $\sU(2)$ superspace. Here we will present results only up to mass dimension 3/2. The outcome of the analysis is as follows:
\begin{subequations} 
	\label{2.133}
	\bea
	\mathfrak{F}_\a^i{}_\b^j  
	&=&
	-\hf\ve_{\a\b}S^{ij}
	+\hf\ve^{ij}Y_{\a\b}
	~,
	\\
	\bar{\mathfrak{F}}^\ad_i{}^\bd_j  
	&=&
	-\hf\ve^{\ad\bd}\bar{S}_{ij}
	+\hf\ve_{ij}\bar{Y}^{\ad\bd}
	~,\\
	\mathfrak{F}_\a^i{}^\bd_j
	&=&
	- \bar{\mathfrak{F}}^\bd_j{}_\a^i
	=
	-\d^i_jG_\a{}^\bd
	-\ri G_\a{}^\bd{}^i{}_j
	~,
	\\
	\mathfrak{F}_{\a}^{i}{}_{,\bb}
	&=&
	\frac{\ri}{4}\ve_{\a\b}\bar{\mathcal{D}}^{\gd i} \bar{W}_{\bd\gd}  
	-\frac{1}{6}\ve_{\a\b}\mathcal{D}^{\g}_{j}G_{\g\bd}{}^{ij}
	+\frac{\ri}{12}\ve_{\a\b}\bar{\mathcal{D}}_{\bd j}S^{ij}
	\non\\
	&&
	-\frac{\ri}{4}\bar{\mathcal{D}}_\bd^iY_{\a\b}
	+\frac{1}{3}\mathcal{D}_{(\a j}G_{\b)\bd}{}^{ij}
	~,
	\\
	\bar{\mathfrak{F}}^{\ad}_{i}{}^{,\bb}
	&=&
	\frac{\ri}{4}\ve^{\ad\bd}{\mathcal{D}}_{\g i}W^{\b\g}  
	+\frac{1}{6}\ve^{\ad\bd}\bar{\mathcal{D}}_{\gd}^{ j}G^{\b\gd}{}_{ij}
	+\frac{\ri}{12}\ve^{\ad\bd}{\mathcal{D}}^{\b j}\bar{S}_{ij}
	\non\\
	&&
	-\frac{\ri}{4}\mathcal{D}^\b_{i}\bar{Y}^{\ad\bd}
	-\frac{1}{3}\bar{\mathcal{D}}^{(\ad  j}G^{\b\bd)}{}_{ij}
	~,
	\\
	\mathfrak{F}_{\aa,}{}_{\b}^j 
	&=&
	-\frac{\ri}{12}\ve_{\a\b}\bar{\mathcal{D}}_{\bd k}S^{kj}
	-\frac{\ri}{4}\bar{\mathcal{D}}_\ad^jY_{\a\b}
	+\frac{1}{3}\mathcal{D}_{\a k}G_{\b\ad}{}^{jk}
	~,
	\\
	\mathfrak{F}^{\aa,}{}^\bd_j 
	&=&
	-\frac{\ri}{12}\ve^{\ad\bd}{\mathcal{D}}^{\b k}\bar{S}_{kj}
	-\frac{\ri}{4}{\mathcal{D}}^{\a}_{j}\bar{Y}^{\ad\bd}
	-\frac{1}{3}\bar{\mathcal{D}}^{\ad k}G^{\a\bd}{}_{jk}
	~.
	\eea
\end{subequations}
The dimension-1 superfields introduced above have the following symmetry properties:  
\bea
S^{ij}=S^{ji}~, \qquad Y_{\a\b}=Y_{\b\a}~, 
\qquad W_{\a\b}=W_{\b\a}~, \qquad G_{\a\ad}{}^{ij}=G_{\a\ad}{}^{ji}~,
\eea
and satisfy the reality conditions:
\bea
\overline{S^{ij}} =  \bar{S}_{ij}~,\quad
\overline{W_{\a\b}} = \bar{W}_{\ad\bd}~,\quad
\overline{Y_{\a\b}} = \bar{Y}_{\ad\bd}~,\quad
\overline{G_{\b\ad}} = G_{\a\bd}~,\quad
\overline{G_{\b\ad}{}^{ij}} = ~G_{\a\bd}{}_{ij}
~.~~~~~~
\eea
Their ${\sU}(1)_R$ charges are (real superfields are omitted):
\bea
{\mathbb Y} \,S^{ij}=2S^{ij}~,\qquad
{\mathbb Y}  \,Y_{\a\b}=2Y_{\a\b}~, \qquad
{\mathbb Y} \, W_{\a\b}=-2W_{\a\b}~.
\eea

At the same time, the algebra obeyed by ${\mathcal{D}}_A$ takes the form:
\begin{subequations} \label{U(2)algebra}
	\bea
	\{ \mathcal{D}_\a^i , \mathcal{D}_\b^j \}
	&=&
	4 S^{ij}  M_{\a\b} 
	+2\ve_{\a\b}\ve^{ij}Y^{\g\d}  M_{\g\d}  
	+2\ve^{ij} \ve_{\a\b}  \bar{W}_{\gd\dd} \bar{M}^{\gd\dd} 
	\non\\
	&&
	+2\ve_{\a\b}\ve^{ij}S^{kl}  \mathbb{J}_{kl}
	+ 4Y_{\a\b}  \mathbb{J}^{ij}
	~,
	\label{U(2)algebra.a}
	\\
	\{ \mathcal{D}_\a^i , \bar{\mathcal{D}}^\bd_j \}
	&=&
	- 2 \ri \d_j^i\mathcal{D}_\a{}^\bd
	+4\Big(
	\d^i_jG^{\g\bd}
	+\ri G^{\g\bd}{}^i{}_j
	\Big) 
	M_{\a\g} 
	+4\Big(
	\d^i_jG_{\a\gd}
	+\ri G_{\a\gd}{}^i{}_j
	\Big)  
	\bar{M}^{\bd\gd}
	\non\\
	&&
	+8 G_\a{}^\bd \mathbb{J}^i{}_j
	-4\ri\d^i_j G_\a{}^\bd{}^{kl}  \mathbb{J}_{kl}
	-2\Big(
	\d^i_jG_\a{}^\bd
	+\ri G_\a{}^\bd{}^i{}_j
	\Big)
	\mathbb{Y} 
	~,
	\label{U(2)algebra.b}
	\\
	{[} \mathcal{D}_a, \mathcal{D}_\b^j{]}&=& 
	-\ri 
	(\ts_a)^{\ad\g}\Big(
	\d^j_kG_{\b\ad}
	+ \ri G_{\b\ad}{}^{j}{}_k\Big)
	\mathcal{D}_\g^k
	\non\\
	&&
	+{\frac\ri 2}\Big(({\s}_a)_{\b\gd}S^{jk}
	-\ve^{jk}({\s}_a)_\b{}^{\dd}\bar{W}_{\dd\gd}
	-\ve^{jk}({\s}_a)^{\a}{}_\gd Y_{\a\b}\Big)\bar{\mathfrak D}^\gd_k
	\non\\
	&&
	-\hf \mathfrak{R}_a{}_\b^j{}^{cd}M_{{c}{d}}
	-\mathfrak{R}_a{}_\b^j{}^{kl}\mathbb{J}_{kl}
	-\ri \mathfrak{R}_a{}_\b^j\,{\mathbb Y}
	~,
	\label{U(2)algebra.c}
	\eea
	\esubeq
	where the dimension-3/2 components of the curvature appearing in (\ref{U(2)algebra.c}) 
	are
	\begin{subequations}
		\bea
		\mathfrak{R}_a{}_\b^j{}_{cd}&=&
		-\ri(\s_d)_{\b}{}^{\dd} \mathfrak{T}_{ac}{}_\dd^j
		+\ri(\s_a)_{\b}{}^{\dd} \mathfrak{T}_{cd}{}_\dd^j
		-\ri(\s_c)_{\b}{}^{\dd} \mathfrak{T}_{da}{}_\dd^j
		~,
		\label{3/2curvature-1}
		\\
		\mathfrak{R}_{\aa,}{}_{\b}^j{}^{kl}
		&=&
		{\ri}\ve^{j(k}\bar{\mathcal{D}}_\ad^{l)}Y_{\a\b}
		+{\ri}\ve_{\a\b}\ve^{j(k}\bar{\mathcal{D}}^{\dd l)}\bar{W}_{\ad\dd}
		+{\frac\ri 3}\ve_{\a\b}\ve^{j(k}\bar{\mathcal{D}}_{\ad q}S^{l)q}
		\non\\
		&&-{\frac43}\ve^{j(k}\mathcal{D}_{(\a q}G_{\b)\ad}{}^{l)q}
		-{\frac23}\ve_{\a\b}\ve^{j(k}\mathcal{D}^\d_{q}G_{\d\ad}{}^{l)q}
		~,
		\\
		\mathfrak{R}_{\aa,}{}_{\b}^j&=&
		\mathcal{D}_{\b}^jG_{\a\ad}
		-{\frac\ri 3}\mathcal{D}_{(\a k}G_{\b)\ad}{}^{jk}
		-{\frac\ri 2}\ve_{\a\b}\mathcal{D}^{\g}_kG_{\g\ad}{}^{jk}
		~.
		\eea
	\end{subequations}
	The right-hand side of  (\ref{3/2curvature-1}) involves the dimension-3/2 components 
	of the torsion, which take the form
	\begin{subequations}
		\bea
		&&\mathfrak{T}_{ab}{}_\gd^k\equiv(\s_{ab})^{\a\b}\mathfrak{T}_{\a\b}{}_{\gd}^{k}
		-(\ts_{ab})^{\ad\bd}\mathfrak{T}_{\ad\bd}{}_{\gd}^{k}~,~~~
		\\
		&&\mathfrak{T}_{\a\b}{}_{\gd}^{k}
		=
		{\frac14}\bar{\mathcal{D}}_{\gd}^{k}Y_{\a\b}
		-{\frac\ri 3}\mathcal{D}_{(\a}^lG_{\b)\gd}{}^{k}{}_l
		~,
		\\
		&&\mathfrak{T}_{\ad\bd}{}_{\gd}^{ k}
		=
		{\frac14}\bar{\mathcal{D}}_{\gd}^k\bar{W}_{\ad\bd}
		+{\frac16}\ve_{\gd(\ad}\bar{\mathcal{D}}_{\bd)l}S^{kl}
		+{\frac\ri 3}\ve_{\gd(\ad}\mathcal{D}^{\d}_qG_{\d\bd)}{}^{kq}
		~.
		\eea
	\end{subequations}
	There are several consistency conditions arising from solving \eqref{2.112} and the constraints satisfied by $W_{\a\b}$ in conformal superspace. They lead to the following set of dimension-3/2 Bianchi identities:
	\begin{subequations}\label{BI-U2}
		\bea
		\mathcal{D}_{\a}^{(i}S^{jk)}&=&0~,
		\\
		\mathcal{D}_\a^i\bar{W}_{\bd\gd}&=&0~,\\
		\mathcal{D}_{(\a}^{i}Y_{\b\g)}&=&0~,
		\\
		\mathcal{D}_{(\a}^{(i}G_{\b)\bd}{}^{jk)}&=&0~, 
		\\
		\bar{\mathcal{D}}_{\ad}^{(i}S^{jk)} &=& \ri\mathcal{D}^{\b (i}G_{\b\ad}{}^{jk)}~,
		\\
		\mathcal{D}_{\a}^{i}S_{ij}
		&=&
		-\mathcal{D}^{\b}_{j}Y_{\b\a}
		~,
		\\
		\mathcal{D}_\a^iG_{\b\bd}&=&
		- \frac{1}{ 4}\bar{\mathcal{D}}_\bd^iY_{\a\b}
		+ \frac{1}{ 12}\ve_{\a\b}\bar{\mathcal{D}}_{\bd j}S^{ij}
		- \frac{1}{ 4}\ve_{\a\b}\bar{\mathcal{D}}^{\gd i}\bar{W}_{\gd\bd}
		\non\\
		&&
		- \frac{\ri }{ 3}\ve_{\a\b}\mathcal{D}^{\g}_j G_{\g \bd}{}^{ij}
		~,
		\eea
	\end{subequations}
	and the dimension-2 constraint
	\bea
	\big( \mathcal{D}_{\a \b}
	-4Y_{\a\b} \big) W^{\a\b}
	&=& \big( \bar{\mathcal{D}}^{\ad \bd}
	-4\bar{Y}^{\ad\bd} \big) \bar{W}_{\ad\bd}
	~.
	\label{BI-U2-2}
	\eea
	Here we have made the definitions
	\begin{align}
		\mathcal{D}_{\a \b} = \mathcal{D}^i_{(\a} \mathcal{D}_{\b) i} ~, \qquad \bar{\mathcal{D}}_{\ad \bd} = \bar{\mathcal{D}}_i^{( \ad }\bar{\mathcal{D}}^{\bd ) i}~,
	\end{align}
	and it is useful to also define
	\begin{align}
		\mathcal{D}^{ij} = \mathcal{D}^{\a (i} \mathcal{D}_{\a}^{j)} ~, \qquad \bar{\mathcal{D}}_{i j} = \bar{\mathcal{D}}_{\ad (i}\bar{\mathcal{D}}^{\ad}_{j)}~.
	\end{align}
	
	In closing, we note that, upon degauging, relation \eqref{2.100} takes the form \cite{KT-M2009,Muller}
	\begin{align}
		\F_{\a(n)} &= - \frac{1}{96}\Big(\bar{\mathcal{D}}^{ij}\bar{\mathcal{D}}_{ij} - \bar{\mathcal{D}}_{\ad \bd}\bar{\mathcal{D}}^{\ad \bd} +16 \bar{S}^{ij} \bar{\mathcal{D}}_{ij} +16 \bar{Y}_{\ad \bd} \bar{\mathcal{D}}^{\ad \bd} \Big) \Psi_{\a(n)} ~.
	\end{align}

	\subsubsection{The super-Weyl transformations of $\sU(2)$ superspace}
	
	In the previous subsection we made use of the special conformal gauge freedom to degauge from conformal to $\sU(2)$ superspace. The goal of this subsection is to show that the residual dilatation symmetry manifests in the latter as super-Weyl transformations.
	
	To preserve the gauge $B_{A}=0$, every local dilatation transformation with parameter $ \s $ should be accompanied by a compensating special conformal one
	\begin{align}
		\mathscr{K}(\sigma) = \L_{B} (\sigma)K^{B} + \s \mathbb{D} \quad \implies \quad \d_{\mathscr{K}( \s )} B_A = 0~.
	\end{align}
	We then arrive at the following constraints
	\bea
	\L_{A}(\s) = \hf \nabla_A \s~.
	\eea
	As a result, we define the following transformation
	\bea
	\d_{\mathscr{K}(\s)} \nabla_A \equiv \d_{\s} \nabla_A = \d_\s \cD_A - \d_\s \mathfrak{F}_{AB} K^B~.
	\eea
	
	By making use of \eqref{2.133}, one can obtain the following transformation laws for the $\sU(2)$ superspace covariant derivatives
	\begin{subequations}
		\label{2.148}
		\bea
		\d_{\s} \mathcal{D}_\a^i&=&\hf\s\mathcal{D}_\a^i+2\mathcal{D}^{\g i}\s M_{\g\a}-2\mathcal{D}_{\a k}\s \mathbb{J}^{ki}
		-\hf \mathcal{D}_\a^i\s {\mathbb Y} 
		~,
		\label{Finite_D}\\
		\d_{\s}\bar{\mathcal{D}}_{\ad i}&=&\hf\s\bar{\mathcal{D}}_{\ad i}
		+2\bar{\mathcal{D}}^{\gd}_{i}\s \bar{M}_{\gd\ad}
		+2\bar{\mathcal{D}}_{\ad}^{k}\s \mathbb{J}_{ki}
		+\hf \bar{\mathcal{D}}_{\ad i}\s {\mathbb Y}~,
		\label{Finite_Db} 
		\\
		\d_{\s}\mathcal{D}_{\a\ad}
		&=&
		\s\mathcal{D}_{\a\ad}
		+\ri \bar{\mathcal{D}}_{\ad k}\s\mathcal{D}_\a^k
		+\ri \mathcal{D}_\a^k\s \bar{\mathcal{D}}_{\ad k}
		\non\\
		&&
		+\mathcal{D}^\g{}_\ad \s M_{\g\a}
		+\mathcal{D}_\a{}^\gd \s \bar{M}_{\gd\ad}
		~.
		\label{Finite_D_c}
		\eea
	\end{subequations}
	The dimension-1 components of the torsion transform as
	\begin{subequations}
		\label{2.149}
		\bea
		\d_{\s} W_{\a\b}&=&\s{W}_{\a \b}~,
		\label{Finite_W}
		\\
		\d_{\s} Y_{\a\b}&=&\s Y_{\a\b}
		-\hf\mathcal{D}_{\a \b}\s
		\label{Finite_Y}~,
		\\
		\d_{\s} S_{ij}&=&\s S_{ij}
		-\hf\mathcal{D}_{ij} \s
		\label{Finite_S}~,
		\\
		\d_{\s} G_{\a\ad}&=&
		\s G_{\a\ad}
		-{\frac18}[\mathcal{D}_\a^k,\bar{\mathcal{D}}_{\ad k}]\s
		~,
		\label{Finite_G}
		\\
		\d_{\s} G_{\a\ad}{}^{ij}&=&\s G_{\a\ad}{}^{ij}
		+{\frac\ri 4}[\mathcal{D}_\a^{(i},\bar{\mathcal{D}}_\ad^{j)}]\s
		~.
		\label{Finite_Gij}
		\eea
	\end{subequations}
	
	Relations \eqref{2.148} and \eqref{2.149} 
	give the super-Weyl transformations 
	in $\sU(2)$ superspace \cite{Howe}.
	The conditions \eqref{2.67}, which define a primary superfield $U$, are equivalent to the following 
	\bea
	\d_\s U = \D_U \s U~, \qquad  {\mathbb Y} U = q_U U~,
	\label{primaryU(2)}
	\eea
	in $\sU(2)$ superspace.
	
	\subsubsection{$\sSU(2)$ superspace}
	
	The torsion $G_{\a\ad}{}^{ij}$ of $\sU(2)$ superspace may be shown to be a pure gauge degree of freedom \cite{Howe,KLRT-M2}. Specifically, one can use the super-Weyl freedom \eqref{Finite_Gij} to set
	\bea
	G_{\a\bd}{}^{ij}=0~.
	\label{G2}
	\eea
	In this gauge, it is natural to redefine $\cD_a$ in accordance with the shift
	\bea
	\label{2.169}
	\cD_a ~\longrightarrow ~\mathcal{D}_a -\ri G_a \,{\mathbb Y}~.
	\eea
	Making use of \eqref{U(2)algebra}, we find that they obey the graded commutation relations
	\begin{subequations} 
		\label{4.21}
		\bea
		\{\cD_\a^i,\cD_\b^j\}&=&
		4S^{ij}M_{\a\b}
		+2\ve^{ij}\ve_{\a\b}Y^{\g\d}M_{\g\d}
		+2\ve^{ij}\ve_{\a\b}\bar{W}^{\gd\dd}\bar{M}_{\gd\dd}
		\non\\
		&&
		+2 \ve_{\a\b}\ve^{ij}S^{kl}\mathbb{J}_{kl}
		+4 Y_{\a\b}\mathbb{J}^{ij}~, 
		\label{acr1} \\
		\{\cD_\a^i,\cDB^\bd_j\}&=&
		-2\ri\d^i_j \cD_\a{}^\bd
		+4\d^{i}_{j}G^{\d\bd}M_{\a\d}
		+4\d^{i}_{j}G_{\a\gd}\bar{M}^{\gd\bd}
		+8 G_\a{}^\bd \mathbb{J}^{i}{}_{j}~,~~~~~~~~~
		\\
		{[}\cD_a,\cD_\b^j{]}&=&
		\ri(\s_a)_{(\b}{}^{\bd}G_{\g)\bd}\cD^{\g j}
		\non\\
		&&
		+{\frac{\ri}2}\Big(({\s}_a)_{\b\gd}S^{jk}
		-\ve^{jk}({\s}_a)_\b{}^{\dd}\bar{W}_{\dd\gd}
		-\ve^{jk}({\s}_a)^{\a}{}_\gd Y_{\a\b}\Big)\cDB^\gd_k
		\non\\
		&&
		+{\frac{\ri}2}\Big((\ts_a)^{\gd\g}\ve^{j(k}\cDB_\gd^{l)}Y_{\b\g}
		-(\s_a)_{\b\gd}\ve^{j(k}\cDB_{\dd}^{l)}\bar{W}^{\gd\dd}
		-{\frac12}(\s_a)_\b{}^{\gd}\cDB_{\gd}^{j}S^{kl}\Big)
		\mathbb{J}_{kl}
		\non\\
		&&
		+{\frac{\ri}2}\Big((\s_a)_{\b}{}^{\dd} \hat{\cT}_{cd}{}_\dd^j
		+(\s_c)_{\b}{}^{\dd} \hat{\cT}_{ad}{}_\dd^j
		-(\s_d)_{\b}{}^{\dd} \hat{\cT}_{ac}{}_\dd^j\Big)M^{{c}{d}}
		~,
		\eea
		where 
		\bea
		\hat{\cT}_{ab}{}_\gd^k&=&-{\frac14}(\s_{ab})^{\a\b}\cDB_\gd^{ k}Y_{\a\b}
		+{\frac14}(\ts_{ab})^{\ad\bd}\cDB_{\gd}^{ k}\bar{W}_{\ad\bd}
		-{\frac16}(\ts_{ab})_{\gd\dd}\cDB^{\dd}_{l}S^{kl}~.
		\eea
	\end{subequations}
	The various torsion tensors in \eqref{4.21} obey the Bianchi identities \eqref{BI-U2} and \eqref{BI-U2-2} upon imposing \eqref{G2} and applying the shift \eqref{2.169}. By examining equations \eqref{4.21} we see that the $\sU(1)_R$ curvature has been eliminated
	and therefore the corresponding connection 
	is flat. Hence, by performing an appropriate local $\sU(1)_R$ transformation the latter may be gauged away.
	As a result, the gauge group reduces to $\sSL( 2, {\mathbb C}) \times \sSU(2)_R$ and the superspace geometry is the so-called $\sSU(2)$ superspace of \cite{Grimm,KLRT-M1}.
	
	It turns out that the gauge condition \eqref{G2} allows for residual super-Weyl transformations, which are described
	by a parameter $\s$ constrained by
	\be
	[\cD_\a^{(i},\bar{\cD}_\ad^{j)}]\s=0~.
	\label{Ucon}
	\ee
	The general solution of this condition is \cite{KLRT-M1}
	\bea
	\s = \frac{1}{2} (\S+\bar{\S})~, \qquad {\bar \cD}^\ad_i \S =0~,
	\qquad {\mathbb Y}\, \S =0~,
	\eea
	where the parameter $\S$ is covariantly chiral, with zero $\sU(1)_R$ charge, but otherwise arbitrary.
	To preserve the gauge of vanishing $\sU(1)_R$ connection, every super-Weyl transformation, see (\ref{Finite_D}) and (\ref{Finite_Db}), must be accompanied by a following compensating $\sU(1)_R$ transformation \eqref{2.127} with parameter $\r = \frac{\ri}{4} (\S - \bar{\S})$. As a result, the $\sSU(2)$ geometry is left invariant by the following set of super-Weyl transformations \cite{KLRT-M1}:
	\bsubeq
	\bea
	\d_{\S} \cD_\a^i&=&\hf\bar{\S}\cD_\a^i+(\cD^{\g i}\S)M_{\g\a}-(\cD_{\a k}\S)\mathbb{J}^{ki}~, 
	\\
	\d_{\S} \cDB_{\ad i}&=&\hf\S\cDB_{\ad i}+(\cDB^{\gd}_{i}\bar{\S})\bar{M}_{\gd\ad}
	+(\cDB_{\ad}^{k}\bar{\S})\mathbb{J}_{ki}~, 
	\label{super-Weyl1} \\
	\d_{\S} \cD_\aa&=&
	\hf(\S+\bar{\S})\cD_\aa
	+\frac{\ri}{2} \cDB_{\ad k} \bar{\S} \cD_\a^k 
	+\frac{\ri}{2} \cD_{\a}^k {\S} \cDB_{\ad k}\non \\
	&\phantom{=}&
	+ \hf \cD^\g{}_\ad (\S + \bar{\S}) M_{\a \g}
	+ \hf \cD_\a{}^{\gd} (\S + \bar{\S}) \bar{M}_{\ad \gd}
	~,
	\\ 
	\d_{\S} S^{ij}&=&\bar{\S} S^{ij}-{\frac14}\cD^{ij} \S~, 
	\label{super-Weyl-S} \\
	\d_{\S} Y_{\a\b}&=&\bar{\S} Y_{\a\b}-{\frac14}\cD_{\a \b}\S~,
	\label{super-Weyl-Y} \\
	\d_{\S} {W}_{\a \b}&=&\S {W}_{\a \b }~,\\
	\d_{\S} G_{\a\bd} &=&
	\hf(\S+\bar{\S})G_{\a\bd} -{\frac{\ri}4}
	\cD_{\a \bd} (\S-\bar{\S})~.
	\label{super-Weyl-G} 
	\eea
\end{subequations}
The conditions \eqref{primaryU(2)}, which define a primary superfield $U$, turn into 
\bea
\d_\S U = (\mf{p}_U\S +\mf{q}_U\bar \S) U~, \qquad \mf{p}_U+\mf{q}_U =\D_U, \quad \mf{p}_U-\mf{q}_U = - 2 q_U~,
\eea
in $\sSU(2)$ superspace.

Due to these transformations, $\sSU(2)$ superspace provides a geometric description of the Weyl multiplet of $\cN=2$ conformal supergravity \cite{KLRT-M1}. It should be emphasised that the algebra of covariant derivatives \eqref{4.21} was derived 
originally
by Grimm \cite{Grimm}. However, no discussion of super-Weyl transformations was given in \cite{Grimm}

\begin{subappendices}

\section{Spinors in four dimensions} \label{Appendix2A}

Our four-dimensional notation and conventions coincide with those of \cite{BK}, which are similar to the ones employed in \cite{WB}. In particular, we work with the `mostly plus' Minkowski metric
$\eta_{ab} = \textrm{diag}(-1,1,1,1)$, and normalise the Levi-Civita tensor $\ve_{abcd}$ as $\ve_{0123} = -\ve^{0123} = -1$.

We almost exclusively use two-component $\sSL(2,\mathbb{C})$ spinors when working in four dimensions. There are two inequivalent representations of $\sSL(2,\mathbb{C})$, namely the fundamental and complex conjugate representations. An object $\psi_\a$ transforming in the former is known as a two-component left-handed Weyl spinor and its indices are raised and lowered using the antisymmetric $\sSL(2,\mathbb{C})$-invariant tensors
\begin{subequations}
	\begin{align}
		\ve_{\a \b} &= - \ve_{\b \a}~, \quad \ve_{1 2} = -1 ~, \\
		\ve^{\a \b} &= - \ve^{\b \a}~, \quad \ve^{1 2} = 1 ~, \\
		&\quad~ \ve^{\a \b} \ve_{\b \g} = \d^\a_\g~,
	\end{align}
\end{subequations}
in the following way:
\begin{align}
	\psi^\a = \ve^{\a \b} \psi_\b ~, \qquad \psi_\a = \ve_{\a \b} \psi^\b ~.
\end{align}
On the other hand, an object $\psi_\ad$ transforming in the complex conjugate representation is known as a two-component right-handed Weyl spinor and its indices are raised and lowered by making use of the invariant tensors
\begin{subequations}
	\begin{align}
		\ve_{\ad \bd} &= - \ve_{\bd \ad}~, \quad \ve_{\dot{1} \dot{2}} = -1 ~, \\
		\ve^{\ad \bd} &= - \ve^{\bd \ad}~, \quad \ve^{\dot{1} \dot{2}} = 1 ~, \\
		&\quad~ \ve^{\ad \bd} \ve_{\bd \gd} = \d^\ad_\gd~,
	\end{align}
\end{subequations}
as follows:
\begin{align}
	\psi^\ad = \ve^{\ad \bd} \psi_\bd ~, \qquad \psi_\ad = \ve_{\ad \bd} \psi^\bd ~.
\end{align}

A vector $V_a$ is in one-to-one correspondence with an $\sSL(2,\mathbb{C})$ spin tensor $V_\aa$ via the bijection
\be 
V_{\a\ad} = (\s^a)_{\a\ad} V_a \quad \Longleftrightarrow  \quad V_a = -\frac{1}{2} (\tilde{\s}_a)^{\a\ad} V_{\a\ad} ~,
\ee
where we have made the definitions
\begin{align}
	\s_a = (\mathds{1},\vec{\s}) ~, \qquad (\tilde{\s}_a)^{\ad \a} = \ve^{\ad \bd} \ve^{\a \b} (\s_a)_{\bb}~.
\end{align}
Here $\vec{\s}$ are the Pauli matrices. The tensors $\s_a$ and $\tilde{\s}_a$ satisfy the important identities:
\begin{subequations}
	\begin{align}
		\s_{(a} \tilde{\s}_{b)} &= - \eta_{a b} \mathds{1} ~, \qquad \tilde{\s}_{(a} {\s}_{b)} = - \eta_{a b} \mathds{1} ~, \\
		\text{Tr}(\s_a \tilde{\s}_b) &= -2 \eta_{a b} ~, \qquad (\s^a)_\aa (\tilde{\s}_a)^{\bb} = -2 \d_\a^\b \d_\ad^\bd~.
	\end{align}
\end{subequations}
Additionally, they may be used to construct the antisymmetric matrices
\begin{align}
	\label{2.182}
	\s^{a b} = - \hf \s^{[a} \tilde{\s}^{b]} ~, \qquad \tilde{\s}^{a b} = - \hf \tilde{\s}^{[a} {\s}^{b]}~,
\end{align}
which allow one to relate a real antisymmetric matrix $X_{ab}=-X_{ba}$ to a pair of symmetric $\sSL(2,\mathbb{C})$ tensors via the prescription
\begin{subequations}
	\begin{align}
		X_{\a \b} &= \hf (\s^{ab})_{\a \b} X_{ab} ~, \qquad \bar{X}_{\ad \bd} = - \hf (\tilde{\s}^{ab})_{\ad \bd} X_{ab}~, \\
		\Longleftrightarrow& \quad 
		X_{a b} = (\s_{a b})_{\a \b} X^{\a \b} - (\tilde{\s}_{a b})_{\ad \bd} X^{\ad \bd}~.
	\end{align}
\end{subequations}

In closing, we note that we will occasionally make use of the convention that indices denoted by the same symbol should be symmetrised over, e.g.
\begin{align}
	U_{\a(n)} V_{\a(m)} = U_{(\a_1 . . .\a_n} V_{\a_{n+1} . . . \a_{n+m})} =\frac{1}{(n+m)!}\big(U_{\a_1 . . .\a_n} V_{\a_{n+1} . . . \a_{n+m}}+\cdots\big)~,
\end{align}
with a similar convention for dotted spinor indices. 

\section{Conformal compensators} \label{Appendix2B}

In order to describe Einstein (super)gravity\footnote{In the supersymmetric case, this is often referred to as Poincar\'e supergravity.} in four dimensions, 
the multiplet of conformal (super)gravity has to be coupled to some compensator(s) $\X$. In general, the latter are primary Lorentz scalars, and at least one of them must have a non-zero dimension
\bea
K^A \X = 0 ~, \qquad \mathbb{D}\X=\D_\X\X~, \qquad \D_\X \neq 0~.
\label{3.22}
\eea
Further, they are required to be nowhere vanishing such that 
$|\X|^2 $ is strictly positive. In the supersymmetric case there are several choices\footnote{For instance, in the $\cN=1$ case, a chiral compensator leads to old minimal supergravity, while a real linear compensator yields the new minimal theory, see \cite{KRTM1} for a recent review.} of conformal compensator $\Xi$, which
yield inequivalent off-shell supergravity theories. The (super)space corresponding to Einstein (super)gravity 
is identified with a triple $(\cM^{4|4\cN}, \nabla,\X)$. 
Two curved (super)spaces $(\cM^{4|4\cN}, \widetilde \nabla, \widetilde \X)$ and 
$(\cM^{4|4\cN}, \nabla,\X)$ are conformally related if their covariant derivatives
and the compensators $\widetilde \X$ and $\X$ 
are connected by the same finite dilatation transformation, 
\bea
\widetilde{\X} = \re^{\s \mathbb{D}} \X~, \qquad \widetilde{\nabla}_A = \re^{\s \mathbb{D}} \nabla_A \re^{- \s \mathbb{D}}~.
\eea
Once $\X$ has been fixed, 
the off-shell (super)gravity multiplet is completely described in terms of the following data:
(i) a superspace geometry for conformal supergravity; and (ii) the conformal compensators. 

\section{Conformal supergravity in six dimensions} \label{Appendix2C}

This appendix is devoted to a review of aspects of $\cN=(1,0)$ conformal supergravity in six dimensions relevant to this thesis. Specifically, we begin by describing our six-dimensional spinor conventions, which are similar to those of \cite{LTM}, and utilise them to describe the $\cN=(1,0)$ superconformal algebra. Building on this, we review the salient details of the $\cN=(1,0)$ conformal superspace of \cite{BKNT}.

\subsection{Spinors in six dimensions}

When working in six dimensions we will exclusively present our results in terms of four component spinors, though it is useful to relate such objects to their eight component cousins. Thus, we will employ the latter as our starting point. We recall that the $8 \times 8$ Dirac matrices $\Gamma^a$ and the charge conjugation matrix
$C$ obey the relations
\begin{gather}
	\{\G_a, \G_b\} = -2 \eta_{a b} \mathbbm{1}~, \quad
	(\G^a)^\dag = -\G_a~, \quad
	\eta_{ab} = \textrm{diag}(-1,1,1,1,1,1)
	\eol
	C \G_a C^{-1} = -\G_a^T~, \quad
	C^\dag C = \mathbbm{1}~, \qquad C = C^{\rm T} = C^*~.
\end{gather}
In particular, $\Gamma_a C^{-1}$ is antisymmetric. The chirality matrix
$\Gamma_*$ is defined by
\begin{align}
	\Gamma_{[a} \Gamma_b \Gamma_c \Gamma_d \Gamma_e \Gamma_{f]} = \ve_{abcdef} \Gamma_*~,
\end{align}
where $\ve_{abcdef}$ is the Levi-Civita tensor and is normalised as $\ve_{012345} = -\ve^{012345} = 1$.
As a consequence of the above conditions, one can show that
\begin{align}\label{eq:GammaReality}
	\Gamma^a = B (\Gamma^a)^* B^{-1}~, \qquad B = \Gamma_* \Gamma_0 C^{-1}~.
\end{align}
The charge conjugate $\Psi^c$ of a Dirac spinor is conventionally defined by
\begin{align}
	\bar\Psi \equiv \Psi^\dag \Gamma_0 =: (\Psi^c)^T C \qquad \implies \quad
	\Psi^c = - \Gamma_0 C^{-1} \Psi^* = -\Gamma_* B \Psi^*~.
\end{align}
Because $B^* B = -\mathbbm{1}$,
charge conjugation is an involution only for objects with an even number of spinor indices,
so it is not possible to have Majorana spinors in six dimensions.
One can instead have a symplectic Majorana condition when the spinors possess an $\sSU(2)$ index.
Conventionally this is denoted
\begin{align}\label{eq:SympMaj}
	(\Psi_i)^c = \Psi^i \quad \implies \quad
	\Psi^i = -\Gamma_0 C^{-1} (\Psi_i)^* = -\Gamma_* B (\Psi_i)^*
\end{align}
for a spinor of either chirality. We raise and lower $\sSU(2)$ indices $i=\1,\2$
using the conventions
\begin{align}
	\Psi^i = \ve^{i j} \Psi_j~, \qquad \Psi_i = \ve_{i j} \Psi^j~, \qquad \ve^{\1\2} = \ve_{\2\1} = 1~.
\end{align}

We employ a Weyl basis for the gamma matrices so that
an eight-component Dirac spinor $\Psi$ decomposes into a four-component
left-handed Weyl spinor $\psi^\alpha$ and a four-component right-handed spinor $\chi_\alpha$
so that
\begin{align}\label{eq:ChiralDecomp1}
	\Psi =
	\begin{pmatrix}
		\psi^\alpha \\
		\chi_\alpha
	\end{pmatrix}~, \qquad
	\Gamma_* =
	\begin{pmatrix}
		\delta^\a{}_\b & 0 \\
		0 & -\delta_\a{}^\b
	\end{pmatrix}~, \qquad \alpha, \beta =1,\cdots, 4~.
\end{align}
The spinors $\psi^\alpha$ and $\chi_\alpha$ are valued in the
two inequivalent fundamental representations 
of $\mathfrak{su}^*(4) \cong \mathfrak{so}(5,1)$.
We further take
\begin{align}
	\Gamma^a =
	\begin{pmatrix}
		0 & (\tilde\gamma^a)^{\alpha\beta} \\
		(\gamma^a)_{\alpha\beta} & 0 
	\end{pmatrix}~,\qquad
	C =
	\begin{pmatrix}
		0 & \delta_\alpha{}^\beta \\
		\delta^\alpha{}_\beta & 0
	\end{pmatrix}~.
\end{align}
The Pauli-type matrices $\gamma^a$ 
and $\tilde\gamma^a$ are antisymmetric and related by
\begin{align}
	(\tilde\gamma^a)^{\alpha\beta} = \frac{1}{2} \ve^{\a\b\g\d} (\gamma^a)_{\g\d}~, \qquad
	(\gamma^a)^* = \tilde\gamma^a~,
\end{align}
where $\ve^{\a\b\g\d}$ is the canonical antisymmetric symbol of $\mathfrak{su}^*(4)$.
They obey
\bsubeq
\bea (\g^a)_{\a\b} (\tilde{\g}^b)^{\b\g}
+ (\g^b)_{\a\b} (\tilde{\g}^a)^{\b\g} &=& - 2 \eta^{ab} \d^\g_\a \ , \\
(\tilde{\g}^a)^{\a\b} (\g^b)_{\b\g}
+ (\tilde{\g}^b)^{\a\b} (\g^a)_{\b\g} 
&=& - 2 \eta^{ab} \d^\a_\g \ ,
\eea
\esubeq
and as a consequence of \eqref{eq:GammaReality},
\begin{subequations}
	\begin{align}
		(\gamma^a)_{\alpha\beta} = B_{\alpha}{}^\gd B_\beta{}^{\dd}
		\big((\gamma^a)_{\gamma \delta}\big)^*~, \quad
		(\tilde\gamma^a)^{\alpha\beta} = B^{\alpha}{}_\gd B^\beta{}_{\dd}
		\big((\tilde\gamma^a)^{\gamma \delta}\big)^*~,
	\end{align}
	where we have made the definition
	\begin{align}
		B=
		\begin{pmatrix}
			B^{\alpha}{}_\bd & 0 \\
			0 & B_{\alpha}{}^\bd
		\end{pmatrix}~.
		\label{eq:PauliReality}
	\end{align}
\end{subequations}
A dotted index denotes the complex conjugate representation in $\mathfrak{su}^*(4)$.
It is natural to use the $B$ matrix to define bar conjugation on a
four component spinor via
\begin{align}
	\bar\psi^\alpha = B^\alpha{}_\bd (\psi^\beta)^*~, \qquad
	\bar \chi_\alpha = B_\alpha{}^\bd (\chi_\beta)^*~.
	\label{A.12}
\end{align}
For example, $\overline{(\gamma^a)_{\alpha\beta}} = (\gamma^a)_{\alpha\beta}$
using \eqref{eq:PauliReality} and similarly for $\tilde\gamma^a$.
We also note that, as a consequence of $B^* B = -\mathbbm{1}$,
\be
\overline{\overline{\psi^{\a}}} = -\psi^\a.
\ee
A symplectic Majorana spinor $\Psi_i$, decomposed as in \eqref{eq:ChiralDecomp1}
and obeying \eqref{eq:SympMaj}, has Weyl components that obey
\begin{align}\label{eq:SympMaj4c}
	\overline{\psi^{\alpha i}} = \psi^\alpha_{i}~, \qquad
	\overline{\chi_{\alpha i}} = \chi_\alpha^{i}~.
\end{align}

We define the antisymmetric products of two or three Pauli-type matrices as
\bsubeq
\begin{alignat}{2}
	\g_{ab} &:= \g_{[a} \tilde{\g}_{b]} := \hf (\g_a \tilde{\g}_b - \g_b \tilde{\g}_a) \ , &\quad
	\tilde{\g}_{ab} &:= \tilde{\g}_{[a} \g_{b]}  = -(\g_{ab})^T\ , \\
	\g_{abc} &:= \g_{[a} \tilde{\g}_b \g_{c]} \ , &\quad \tilde{\g}_{abc} &:= \tilde{\g}_{[a} \g_b \tilde{\g}_{c]} \ .
\end{alignat}
\esubeq
Note that $\g_{ab}$ and $\tilde\g_{ab}$ are traceless, whereas $\g_{abc}$ and
$\tilde\g_{abc}$ are symmetric. Making use of the completeness relations
\begin{subequations}
	\begin{align}
		(\gamma^a)_{\a\b} (\tilde\gamma_{a})^{\g\d} &= 4\, \delta_{[\a}{}^\g \delta_{\b]}{}^{\d}~, \\
		(\gamma^{ab})_\a{}^\b (\gamma_{ab})_\g{}^\d &= - 8\,\delta_{\a}{}^\d \delta_{\g}{}^{\b}
		+ 2\, \delta_{\a}{}^\b \delta_{\g}{}^{\d}~, \\
		(\gamma^{abc})_{\a\b} (\tilde\gamma_{abc})^{\g\d} &= 48\, \delta_{(\a}{}^\g \delta_{\b)}{}^{\d}~, \\
		(\gamma^{abc})_{\a\b} (\tilde\gamma_{abc})_{\g\d} &= (\gamma^{abc})^{\a\b} (\tilde\gamma_{abc})^{\g\d} = 0~,
	\end{align}
\end{subequations}
it is straightforward to establish natural isomorphisms between tensors of $\mathfrak{so}(5,1)$
and matrix representations of $\mathfrak{su}^*(4)$.
Vectors $V^a$ and antisymmetric matrices $V_{\a\b} = - V_{\b\a}$ 
are related by
\be 
V_{\a\b} := (\g^a)_{\a\b} V_a \quad \Longleftrightarrow  \quad V_a = \frac{1}{4} (\tilde{\g}_a)^{\a\b} V_{\a\b} \ .
\ee
Antisymmetric rank-two tensors $F_{ab}$ are related to traceless matrices $F_\a{}^\b$ 
via
\bea 
F_\a{}^\b := - \frac{1}{4} (\g^{ab})_\a{}^\b F_{ab} \ , \quad F_\a{}^\a = 0 \quad
\Longleftrightarrow 
\quad F_{ab} = \hf (\g_{ab})_\b{}^\a F_\a{}^\b = - F_{ba} \ .
\label{A.18}
\eea
Self-dual and anti-self-dual rank-three antisymmetric tensors $T^{(\pm)}_{abc}$,
\be \frac{1}{3!} \ve^{abcdef} T_{def}^{(\pm)} = \pm T^{(\pm)abc} \ ,
\ee
are related to symmetric matrices $T_{\a\b}$ and $T^{\a\b}$ 
via
\bsubeq
\bea
T_{\a\b} &:=& \frac{1}{3!} (\g^{abc})_{\a\b} T_{abc} = T_{\b\a} \quad \Longleftrightarrow \quad 
T_{abc}^{(+)} = \frac{1}{8} (\tilde{\g}_{abc})^{\a\b} T_{\a\b} \ , \\
T^{\a\b} &:=& \frac{1}{3!} (\tilde{\g}^{abc})^{\a\b} T_{abc} = T^{\b\a} \quad 
\Longleftrightarrow \quad
T^{(-)}_{abc} = \frac{1}{8} (\g_{abc})_{\a\b} T^{\a\b} \ .
\eea
\esubeq

\subsection{The $\mathcal{N}=(1,0)$ superconformal algebra} \label{AppendixA.2}

The bosonic sector of the $\cN = (1, 0)$ superconformal algebra contains the translation ($P_{a}$), Lorentz ($M_{ab}$), 
special conformal ($K_{a}$), dilatation ($\mathbb{D}$) and $\sSU(2)_R$ generators ($\mathbb{J}^{ij}$).
Amongst themselves, they obey the algebra
\bsubeq
\begin{align} [M_{a b} , M_{c d}] &= 2 \eta_{c [a} M_{b] d} - 2 \eta_{d [ a} M_{b ] c} \ , \\
	[M_{a b}, P_c] &= 2 \eta_{c [a} P_{b]} \ , \quad [\mathbb{D} , P_a] = P_a \ , \\
	[M_{a b} , K_c] &= 2 \eta_{c [a} K_{b]} \ , \quad [\mathbb{D} , K_a] = - K_{a} \ , \\
	[K_a , P_b] &= 2 \eta_{a b} {\mathbb D} + 2 M_{ab} \ , \\
	[\mathbb{J}^{ij} , \mathbb{J}^{kl}] &= \ve^{k(i} \mathbb{J}^{j) l} + \ve^{l (i} \mathbb{J}^{j) k} \ ,
\end{align}
\esubeq

To obtain the $\cN=(1,0)$ superconformal algebra, we extend the 
translation generator to $P_A = (P_a , Q_\a^i)$ and the special conformal generator to
$K^A = (K^a , S^\a_i)$.
The fermionic generator $Q_\a^i$ obeys the algebra
\bsubeq \label{SCA1}
\begin{align} \{ Q_\a^i , Q_\b^j \} &= - 2 \ri \ve^{ij} (\g^{c})_{\a\b} P_{c} \ , \quad [Q_\a^i , P_a ] = 0 \ , \quad [{\mathbb D} , Q_\a^i ] = \hf Q_\a^i \ , \\
	[M_{ab} , Q_\g^k ] &= - \hf (\g_{ab})_\g{}^\d Q_\d^k \ , \quad [\mathbb{J}^{ij} , Q_\a^k ] = \ve^{k (i} Q_\a^{j)} \ ,
\end{align}
\esubeq
while the generator $S^\a_i$ obeys the algebra
\bsubeq\label{SCA2}
\begin{align} \{ S^\a_i , S^\b_j \} &= - 2 \ri \ve_{ij} (\tilde{\g}^{c})^{\a\b} K_{c} \ , \quad [S^\a_i , K_a ] = 0 \ , \quad [{\mathbb D} , S^\a_i ] = - \hf S^\a_i~, \\
	[M_{ab} , S^\g_k ] &= \hf (\g_{ab})_\d{}^\g S^\d_k  \ , \quad [\mathbb{J}^{ij} , S^\a_k ] = \d_k^{(i} S_\a^{j)} \ ,
\end{align}
\esubeq
Finally, the (anti-)commutators of $K^{A}$ and $P_B$ are
\bsubeq\label{SCA3}
\begin{align} [ K_a , Q_\a^i ] &= - \ri (\g_a)_{\a\b} S^{\b i} \ , \quad [S^\a_i , P_a ] = - \ri (\tilde{\g}_a)^{\a\b} Q_{\b i} \ , \\
	\{ S^\a_i , Q_\b^j \} &= 2 \d^\a_\b \d^j_i \mathbb D - 4 \d^j_i M_\b{}^\a + 8 \d^\a_\b \mathbb{J}_i{}^j \ .
\end{align}
\esubeq

\subsection{$\cN=(1,0)$ conformal superspace}

In the approach of \cite{BKNT}, conformal superspace is identified with a pair $(\cM^{6|8},\nabla)$, where $\cM^{6|8}$ denotes a supermanifold parametrised by local coordinates $z^M = (x^m, \q^\m_\imath)$ and $\nabla_A = (\nabla_a , \nabla_\a^i)$ is a covariant derivative associated with the superconformal algebra. The latter takes the form
\bea
\label{6DCD}
\nabla_A
= E_A - \hf  {\Omega}_A{}^{bc} M_{bc} - \Phi_A{}^{jk} \mathbb{J}_{jk} - B_A \mathbb D
-  {\mathfrak{F}}_A{}_B K^B ~.
\eea
The translation generators $P_A$ do not show up in \eqref{6DCD}. It is assumed that the operators $\nabla_A$ replace $P_A$ and obey the following graded commutation relations
\be
[ X_{\underline{B}} , \nabla_A \} = -f_{\underline{B} A}{}^C \nabla_C
- f_{\underline{B} A}{}^{\underline{C}} X_{\underline{C}}
\quad
\iff
\quad
[ X_{\underline{B}} , P_A \} = -f_{\underline{B} A}{}^C P_C
- f_{\underline{B} A}{}^{\underline{C}} X_{\underline{C}} ~,
\ee
where $X_{\underline{B}} = (M_{ab},\mathbb{J}_{ij}, \mathbb{D},K^B)$ collectively denotes the non-translational generators of the $\cN=(1,0)$ superconformal algebra described above.

By definition, the gauge group of conformal supergravity is generated by local transformations of the form
\begin{subequations}
	\bea
	\delta_{\mathscr K} \nabla_A &=& [{\mathscr K},\nabla_A] ~ , \\
	{\mathscr K} &=& \xi^B \nabla_B +  \L^{\underline{B}} X_{\underline{B}}
	=  \xi^B \nabla_B+ \hf K^{bc} M_{bc} + \s \mathbb{D} + \chi^{ij} \mathbb{J}_{ij}
	+ \L_B K^B ~ ,
	\eea
\end{subequations}
where the gauge parameters satisfy natural reality conditions. The conformal supergravity gauge group acts on a conformal tensor superfield $U$ (with its indices suppressed) as 
\bea 
\d_{\mathscr K} U = {\mathscr K} U ~.
\eea
Of special significance are primary superfields. 
The superfield $U$ is said to be primary if it is characterised by the properties 
\bea
K^A U = 0~, \quad \mathbb D U = \D_U U~,
\eea
for some real constant $\D_U$ which is called the dimension (or Weyl weight) of $U$.

To describe the $\mathcal{N} = (1,0)$ Weyl multiplet in conformal superspace,
we require the algebra of covariant derivatives to have a Yang-Mills structure. 
Specifically, we require that the latter takes the form
\bea
\{ \nabla_\a^i , \nabla_\b^j \} = - 2 \ri \ve^{ij} (\g^a)_{\a\b} \nabla_a \ , ~~~~~~
\left[ \nabla_a , \nabla_\a^i \right] = (\g_a)_{\a\b} \mathscr{W}^{\b i} \ ,
\label{algb-000}
\eea
where $\mathscr{W}^{\a i}$ is a primary dimension-$3/2$ operator 
\begin{align}
	[\mathbb{D},\mathscr{W}^{\a i}] = \frac 3 2 \mathscr{W}^{\a i} ~, \quad [K^B,\mathscr{W}^{\a i}] = 0~,
\end{align}
which is valued in the superconformal algebra
\begin{align}
	\mathscr{W}^{\a i} &= \mathscr{W}(\nabla)^{\a i, B} \nabla_B + \hf \mathscr{W}(M)^{\a i,bc} M_{bc} + \mathscr{W}(\mathbb{J})^{\a i,jk} \mathbb{J}_{jk} \non \\
	&+ \mathscr{W}(\mathbb{D})^{\a i} \mathbb{D} + \mathscr{W}(K)^{\a i}{}_B K^B ~. 
\end{align}

To obtain further information regarding the structure of $\mathscr{W}^{\a i}$, we study the Jacobi identity
\be 
(-1)^{\ve_A \ve_C} [ \nabla_A , [ \nabla_B , \nabla_C \} \} +
\text{(two cycles)} = 0~.
\label{6DJI}
\ee
Specifically, as a consequence of \eqref{algb-000}, one finds
\bea [\nabla_a , \nabla_b] = - \frac{\ri}{8} (\g_{ab})_\a{}^\b \{ \nabla_\b^k , \mathscr{W}^\a_k \}~,
\eea
and the additional constraints
\be \{ \nabla_\a^{(i} , \mathscr{W}^{\b j)} \} = \frac{1}{4} \d^\b_\a \{ \nabla_\g^{(i} , \mathscr{W}^{\g j)} \} \label{WalbeBI}
~ , \quad \{ \nabla_\g^k , \mathscr{W}^\g_k \} = 0 ~ .
\ee
The operator $\mathscr{W}^{\a i}$ is then constrained to be
\bea
\mathscr{W}^{\a i} &=& 
W^{\a\b} \nabla_\b^i
+ \nabla_\g^i W^{\a\b} M_\b{}^\g
- \frac{1}{4} \nabla_\g^i W^{\b\g} M_\b{}^\a
+ \frac{1}{2} \nabla_{\b j} W^{\a\b} \mathbb{J}^{ij}
+ \frac{1}{8} \nabla_\b^i W^{\a\b} \mathbb D \non\\
&&
- \frac{1}{16} \nabla_\b^j \nabla_\g^i W^{\a \g} S^\b_j
+ \frac{\ri}{2} \nabla_{\b\g} W^{\g\a} S^{\b i} \non\\
&&- \frac{1}{12} (\g^{ab})_\b{}^\g \nabla_b \big( \nabla_\g^i W^{\b \a} 
- \hf \d_\g^\a \nabla_\d^i W^{\b\d} \big) K_a~.
\label{def-W}
\eea

We see from eq. \eqref{def-W} that the geometry of conformal superspace is controlled by the superfield $W^{\a \b}$, which is an $\cN=(1,0)$ supersymmetric extension of the Weyl tensor. It is characterised by the superconformal properties
\bea
K^C W^{\a\b} = 0 ~,
\quad \mathbb D W^{\a\b} = W^{\a\b} ~,
\eea
and satisfies the Bianchi identities
\bsubeq\label{WBI}
\bea
\nabla_\a^{(i} \nabla_{\b}^{j)} W^{\g\d} &=& - \d^{(\g}_{[\a} \nabla_{\b]}^{(i} \nabla_{\r}^{j)} W^{\d) \r} \ , \\
\nabla_\a^k \nabla_{\g k} W^{\b\g} - \frac{1}{4} \d^\b_\a \nabla_\g^k \nabla_{\d k} W^{\g\d}
&=& 8 \ri \nabla_{\a \g} W^{\g \b} \ .
\eea
\esubeq

An alternative formulation of $\cN=(1,0)$ conformal supergravity, namely the $\sSU(2)$ superspace of \cite{LTM}, may be obtained from the present geometry via a certain gauge fixing procedure. In the original work \cite{BKNT}, this was achieved by coupling the background to a conformal compensator, though the same result may be obtained by utilising the degauging procedure described in section \ref{Chapter2.4}, see \cite{KLRTM} for the technical details. Additionally, the component reduction of this superspace formulation was described in \cite{BNTM} and yields the superconformal tensor calculus of \cite{BSvP}.

\end{subappendices}

\chapter{Symmetries of curved (super)space and supersymmetric field theories} \label{Chapter3}

This chapter is, in part, devoted to the study of geometric symmetries of curved (super)spaces, specifically their (conformal) isometries. In addition to characterising the background (super)space, they are fundamentally important in the study of symmetries of field theories defined on the background. Specifically, conformal isometries of the background induce symmetries of every (super)conformal action, and consequently preserve the equations of motion.\footnote{Similarly, isometries induce symmetries of the action of each field theory on the background, though they are inconsistent with (super)conformal invariance.} 

Building on this property, the latter half of this chapter is devoted to the study of higher-derivative symmetries of several important (super)conformal kinetic operators and their massive counterparts. These are known in the literature as higher symmetries and play a fundamental role in the context of (super)conformal higher-spin theories, which will be described in the next chapter. In particular, the algebra of higher symmetries for super(conformal) kinetic operators yield geometric realisations of (super)conformal higher-spin algebras.\footnote{See \cite{Bekaert} for a proof of this statement in the non-supersymmetric case.}

This chapter is based on the publications \cite{KR19,KLRTM,KPR22,KR23} and is organised as follows. The conformal isometries of curved (super)space are described in section \ref{Chapter3.1}. Following this, in section \ref{Chapter3.2} we fix a (super)gravity background and study its isometries. In section \ref{Chapter3.4} we introduce the conformal Killing tensor (super)fields of a background superspace as a generalisation of its conformal Killing vector (super)fields. They are employed in section \ref{Chapter3.3} to derive higher symmetries of several important (super)conformal kinetic operators in conformally-flat backgrounds. Higher symmetries of kinetic operators for massive multiplets are then studied in section \ref{Chapter3.5}. The results of this chapter are summarised in section \ref{Chapter3.6}. The main body of this chapter is accompanied by a single technical appendix, namely appendix \ref{Appendix3A}, which is devoted to symmetries of $\cN=(1,0)$ supergravity backgrounds and higher symmetries of the Sohnius operator in six dimensions.

\section{Conformal isometries} \label{Chapter3.1}

The conformal isometries of four-dimensional curved $\cN \leq 1$ (super)space have been studied extensively, see e.g. \cite{BK, K12}. The objective of this subchapter is to lift the known non-supersymmetric and $\cN=1$ results to conformal superspace and present, for the first time, the $\cN\geq2$ story.

A universal description of the conformal isometries of a given curved background was given in \cite{K15}. In the context of four-dimensional conformal (super)space it can be summarised as follows. A real (super)vector field $\x= \x^B E_B$ on $(\cM^{4 |4 \cN}, \nabla)$ is said to be conformal Killing if 
\bea
\label{3.1}
\d_\mathscr{K} \nabla_A = [\xi^B \nabla_B + \L^{\underline{B}} X_{\underline {B}}, \nabla_A]  = 0~.
\eea
As will be verified below, the following general properties hold:

\begin{itemize}
	\item The parameters $\L^{\underline{B}}$ are uniquely determined 
	in terms of $\x^B$; $\L^{\underline{B}} = \L^{\underline{B}}[\xi]$.
	\item For $\cN \geq 1$, the spinor parameters $\x^{\b}_j$ and $\bar{\x}_\bd^j$ are uniquely determined in terms of $\x^b$.
	\item The vector parameter $\x^b$ obeys a closed-form equation containing complete information regarding the  conformal Killing (super)vector field.
	\item The set of conformal Killing (super)vector fields forms a finite-dimensional Lie (super)algebra with respect to the standard Lie bracket. This is the (super)conformal algebra of $(\cM^{4 |4 \cN}, \nabla)$.
\end{itemize}
Equivalent properties to those above have been established for diverse dimensions and off-shell supersymmetries in several publications \cite{BK,KLRST-M,BIL,KNT-M,LO},
though the $\cN\geq2$ case in four dimensions remains largely unexplored. This section is aimed at bridging this gap.

\subsection{Curved spacetime}
In this subsection we recast the known results for conformal isometries of gravitational backgrounds in the language of conformal geometry. As discussed above, in this framework conformal isometries may be computed by analysing equation \eqref{3.1} with $A=a$. A routine analysis leads us to the differential constraints:
\begin{subequations}
	\label{3.3}
	\bea
	\nabla_a \xi^b &=& K_a{}^b[\xi] + \d_a^b \s[\xi] ~, \\
	\nabla_a K^{bc}[\xi] &=& 2 \d_a^{[b} \L^{c]}[\xi] - \hf \xi^d C_{ad}{}^{bc} ~, \\
	\nabla_a \s[\xi] &=& 2 \L_a[\xi] ~, \\
	\nabla_a \L_b[\xi] &=& \hf \xi^c \nabla^d C_{acbd}~.
	\eea
\end{subequations}
Their solution is given by
\bea
\label{N=0ConfIsoParameters}
K^{\a \b}[\xi] = \nabla^{[a} \xi^{b]} ~, \qquad
\s [\xi] = \frac 1 4 \nabla_a \xi^a ~, \qquad
\L_a [\xi] = \frac 1 8 \nabla_a \nabla_b \xi^b ~,
\eea
where $\xi^a$ obeys the conformal Killing vector equation
\bea
\label{CKV}
\nabla_{(a} \xi_{b)} = \frac 1 4 \eta_{ab} \nabla_c \xi^c ~,
\eea
hence it is a primary field of dimension $-1$
\bea
\label{CKVConfProps}
K^b \xi^a = 0 ~, \qquad \mathbb{D} \xi^a = - \xi^a ~.
\eea
We emphasise that \eqref{CKVConfProps} determines the conformal properties of the parameters \eqref{N=0ConfIsoParameters}.

\subsection{$\cN=1$ curved superspace}
The conformal isometries of $\mathcal{N}=1$ curved superspace have been studied extensively, see e.g. \cite{BK, K12}. Here, we will uplift the known results to conformal superspace by studying equation \eqref{3.1}. Its implications are as follows:
\begin{subequations}
	\label{3.8}
	\bea
	\nabla_\a \xi^b &=& 2 \ri (\s^b)_{\aa} \bar \xi^{\ad} ~, \\
	\nabla_\a \xi^{\b} &=& K_{\a}{}^\b [\xi] + \Big( \frac 1 2 \s [\xi] + \ri \r [\xi] \Big) \d_\a^\b ~, \qquad
	\nabla_\a \bar \xi_{\bd} = 0 ~, \\
	\nabla_\a K^{\b \g} [\xi] &=& - 4 \d_\a^{(\b} \L^{\g)} [\xi] ~, \qquad
	\nabla_\a \bar{K}^{\bd \gd} [\xi] = - \ri \xi_{\a\ad} \bar{W}^{\ad \bd \gd} ~, \\
	\nabla_\a \s [\xi] &=& 2 \L_\a [\xi] ~,  \qquad
	\nabla_\a \r [\xi] = 3 \ri \L_\a [\xi] ~, \\
	\nabla_\a \L_\b [\xi] &=& 0 ~, \qquad
	\nabla_\a \bar{\L}^{\bd} [\xi] = \ri \L_{\a}{}^{\bd} [\xi] - \frac \ri {4}  \xi_{\a \bd} \bar{\nabla}_{\bd} \bar{W}^{\ad \bd(2)}  ~,   \\
	\nabla_\a \L_\bb [\xi] &=& -2\ri \xi_{\a}{}^{\ad} \nabla_\b{}^{\ad} \bar{W}_{\ad(2) \bd} ~.
	\eea
\end{subequations}
The solution to the differential constraints \eqref{3.8} is as follows:
\begin{subequations}
	\label{N=1ConfIsoParameters}
	\bea
	\xi^{\a} &=& - \frac{\ri}{8} \bar \nabla_{\ad} \xi^\aa ~, \qquad
	K^{\a \b}[\xi] = \frac 1 4 \nabla^{(\a \ad} \xi^{\b)}{}_{\ad} ~, \\
	\s [\xi] &=& - \frac 1 8 \nabla_\aa \xi^\aa ~, \qquad
	\r [\xi] = - \frac 1 {32} [ \nabla_\a , \bar \nabla_\ad ] \xi^\aa ~, \\
	\L_\a [\xi] &=& - \frac {1} {16} \nabla_\a \nabla_\bb \xi^\bb ~, \qquad
	\L_\aa [\xi] = - \frac{\ri}{16} \nabla_\a \bar \nabla_\ad \nabla_\bb \xi^\bb + \frac{1}{4} \xi_{\a}{}^{\bd} \bar{\nabla}^{\bd} \bar{W}_{\ad \bd(2)} ~,~
	\eea
\end{subequations}
where $\xi^\aa$ obeys the conformal Killing vector equation
\bea
\label{N=1CKV}
\nabla_{(\a} \xi_{\b) \bd} = 0 \quad \Longleftrightarrow \quad \bar{\nabla}_{(\ad} \xi_{\b \bd)} = 0~,
\eea
hence it is a primary superfield of dimension $-1$
\bea
\label{N=1ConfProps}
K^B \xi^\aa = 0 ~, \qquad \mathbb{D} \xi^\aa = - \xi^\aa ~.
\eea
These, in turn, determine the superconformal properties of the parameters \eqref{N=1ConfIsoParameters}. An important corollary of \eqref{N=1CKV} is
\bea
\nabla_{(\a (\ad} \xi_{\b) \bd)} = 0 \quad \Longleftrightarrow  \quad\nabla_{(a} \xi_{b)} = \frac 1 4 \eta_{ab} \nabla_c \xi^c ~.
\eea

\subsection{$\cN=2$ curved superspace}
We now turn to studying the conformal isometries of $\cN=2$ curved superspace.\footnote{In appendix \ref{Appendix3A.1}, the conformal isometries of curved $\cN=(1,0)$ superspaces in six dimensions are studied. It is expected that these symmetries may be related to the present ones via an appropriate dimensional reduction.} To this end, we begin by fixing $A = (_\a^i)$ in eq. \eqref{3.1}. It implies the following differential constraints
\begin{subequations}
	\label{3.13}
	\bea
	\nabla_\a^i \xi^b &=& 2 \ri (\s^b)_{\aa} \bar \xi^{\ad i} ~, \label{2.3a}\\
	\nabla_\a^i \xi^{\b}_j &=& K_{\a}{}^\b [\xi] \d^i_j - \chi^{i}{}_j [\xi] \d_\a^\b + \Big( \frac 1 2 \s [\xi] + \ri \r [\xi] \Big) \d_\a^\b \d^i_j ~, \label{2.3b}\\
	\nabla_\a^i \bar \xi_{\bd}^j &=& - \frac{\ri}{2} \ve^{ij} \xi_\a{}^{\ad} \bar{W}_{\ad \bd} ~, \label{2.3c}\\
	\nabla_\a^i K^{\b \g} [\xi] &=& - 4 \d_\a^{(\b} \L^{\g) i} [\xi] ~, \label{2.3d}\\
	\nabla_\a^i \bar{K}^{\bd \gd} [\xi] &=& - \frac{\ri}{2} \xi_\a{}^\ad \bar{\nabla}^{(\bd i} \bar{W}_{\ad}{}^{\gd)} + 2 \xi_{\a}^i \bar{W}^{\bd \gd} ~, \label{2.3e}\\
	\nabla_\a^i \chi^{jk} [\xi] &=& - \frac \ri 2 \ve^{i(j} \big( \xi_\a{}^{\ad} \bar \nabla^{\bd k)} \bar W_{\ad \bd} + 8 \ri \L_\a^{k)} [\xi] \big) ~, \label{2.3f}\\
	\nabla_\a^i \s [\xi] &=& - 2 \L_\a^i [\xi] - \frac \ri 4 \xi_{\a}{}^{\ad} \bar \nabla^{\bd i} \bar{W}_{\ad \bd} ~, \label{2.3g} \\
	\nabla_\a^i \r [\xi] &=& \ri \L_\a^i [\xi] - \frac 1 8 \xi_\a{}^{\ad} \bar \nabla^{\bd i} \bar W_{\ad \bd} ~, \label{2.3h}\\
	\nabla_\a^i \L_\b^j [\xi] &=& \frac{1}{4} \ve^{ij} \xi_{\a}{}^{\ad} \nabla_{\b}{}^{\bd} \bar{W}_{\ad \bd} ~, \label{2.3i}\\
	\nabla_\a^i \bar{\L}^{\bd}_j [\xi] &=& \ri \L_{\a}{}^{\bd} [\xi] \d^i_j - \frac \ri {16}  \xi_\a{}^{\ad} \big( \d^i_j \bar \nabla_{\ad \gd} \bar W^{\bd \gd} + \bar \nabla^{i}{}_j \bar W_{\ad}{}^{\bd} \big) + \frac 1 2 \xi_{\a}^i \bar{\nabla}_{\ad j} \bar W^{\ad \bd} ~,  \label{2.3j} \\
	\nabla_\a^i \L_\bb [\xi] &=& \frac \ri 4 \xi_\a{}^\gd \bar \nabla_\gd^i \nabla_\b{}^\ad \bar W_{\ad \bd} + \xi_\a^i \nabla_{\b \gd} \bar W_{\bd}{}^{\gd}\label{2.3k} ~.
	\eea
\end{subequations}
Equations \eqref{3.13} allow us to express all parameters of $\mathscr{K[\xi]}$ solely in terms of the real vector superfield $\xi^{\aa}$. Explicitly:
\begin{subequations}
	\label{N=2ConfIsoParameters}
	\bea
	\xi^{\a i} &=& - \frac{\ri}{8} \bar \nabla_{\ad}^i \xi^\aa ~, \qquad
	K^{\a \b}[\xi] = \frac 1 4 \nabla^{(\a \ad} \xi^{\b)}{}_{\ad} ~, \\
	\chi^{ij}[\xi] &=& - \frac{\ri}{16} \nabla_\a^{(i} \bar \nabla_{\ad}^{j)} \xi^{\aa} ~, \qquad
	\r [\xi] = - \frac 1 {64} [ \nabla_\a^i , \bar \nabla_{\ad i} ] \xi^\aa ~, \\
	\s [\xi] &=& - \frac 1 8 \nabla_\aa \xi^\aa ~, \qquad
	\L_\a^i [\xi] = \frac 1 {16} \nabla_{\a}^i \nabla_\bb \xi^\bb ~, \\
	\L_\aa [\xi] &=& - \frac{\ri}{32} \nabla_\a^i \bar \nabla_{\ad i} \nabla_\bb \xi^{\bb} + \frac{1}{4} \xi_{\a \bd} \bar{\nabla}^{\bd \gd} \bar{W}_{\ad \gd} + \frac{1}{32} \bar{\nabla}^{\bd i} \xi_{\a \bd} \bar{\nabla}_i^\gd \bar{W}_{\ad \gd}~.
	\eea
\end{subequations}
Further, \eqref{2.3a} implies that $\xi^\aa$ obeys the conformal Killing vector equation
\bea
\label{N=2CKV}
\nabla_{(\a}^i \xi_{\b) \bd} = 0 \quad \Longleftrightarrow \quad \bar{\nabla}_{(\ad}^i \xi_{\b \bd)} = 0~,
\eea
which is a primary superfield of dimension $-1$
\bea
\label{N=2ConfProps}
K^B \xi^\aa = 0 ~, \quad \mathbb{D} \xi^\aa = - \xi^\aa ~.
\eea
These, in turn, determine the superconformal properties of the parameters \eqref{N=2ConfIsoParameters}. We note that \eqref{N=2CKV} implies $\nabla_{(\a (\ad} \xi_{\b) \bd)} = 0$.

\subsection{Conformally-flat $\cN>2$ superspace}
For completeness, we now investigate the conformal isometries of conformally-flat $\cN>2$ curved superspace. As above, we fix $A = (_\a^i)$ in \eqref{3.1} and determine its implications:
\begin{subequations}
	\label{3.18}
	\bea
	\label{3.18a}
	\nabla_\a^i \xi^b &=& 2 \ri (\s^b)_{\aa} \bar \xi^{\ad i} ~, \qquad \nabla_\a^i \bar \xi_{\bd}^j = 0~,\\
	\nabla_\a^i \xi^{\b}_j &=& K_{\a}{}^\b [\xi] \d^i_j + K^{i}{}_j [\xi] \d_\a^\b + \Big( \frac 1 2 \s [\xi] + \ri \r [\xi] \Big) \d_\a^\b \d^i_j ~,\\
	\nabla_\a^i K^{\b \g} [\xi] &=& - 4 \d_\a^{(\b} \L^{\g) i} [\xi] ~, \qquad	\nabla_\a^i \bar{K}^{\bd \gd} [\xi] = 0 ~,\\
	\nabla_\a^i \chi[\xi]^j{}_k &=& 4 \Big(\d^i_k \L_\a^j[\xi] - \frac 1 \cN \d^j_k \L_\a^i [\xi] \Big) ~, \\
	\nabla_\a^i \s [\xi] &=& - 2 \L_\a^i [\xi] ~, \qquad
	\nabla_\a^i \r [\xi] = - \frac{\ri(\cN-4)}{\cN} \L_\a^i [\xi] ~,\\
	\nabla_\a^i \L_\b^j [\xi] &=& 0 ~, \qquad \nabla_\a^i \bar{\L}^{\bd}_j [\xi] = \ri \L_{\a}{}^{\bd} [\xi] \d^i_j  ~, \\
	\nabla_\a^i \L_\bb [\xi] &=& 0~.
	\eea
\end{subequations}
Equations \eqref{3.18} allow us to express all parameters of $\mathscr{K[\xi]}$ solely in terms of the real vector superfield $\xi^{\aa}$:
\begin{subequations}
	\label{N>2ConfIsoParameters}
	\bea
	\xi^{\a i} &=& - \frac{\ri}{8} \bar \nabla_{\ad}^i \xi^\aa ~, \qquad
	K^{\a \b}[\xi] = \frac 1 4 \nabla^{(\a \ad} \xi^{\b)}{}_{\ad} ~, \\
	\chi^{i}{}_j [\xi] &=& - \frac{\ri}{32} \Big ( [\nabla_\a^{i} , \bar \nabla_{\ad j}] - \frac{1}{\cN} \d^i_j [\nabla_\a^{k} , \bar \nabla_{\ad k}] \Big ) \xi^{\aa} ~, \\
	\s [\xi] &=& - \frac 1 8 \nabla_\aa \xi^\aa ~, \qquad
	\r [\xi] = - \frac 1 {32 \cN} [ \nabla_\a^i , \bar \nabla_{\ad i} ] \xi^\aa ~, \\
	\L_\a^i [\xi] &=& \frac 1 {16} \nabla_{\a}^i \nabla_\bb \xi^\bb ~, \qquad
	\L_\aa [\xi] = - \frac{1}{16} \nabla_\aa \nabla_\bb \xi^{\bb} ~.
	\eea
\end{subequations}
Further, \eqref{3.18a} implies that $\xi^\aa$ obeys the conformal Killing vector equation
\bea
\label{N>2CKV}
\nabla_{(\a}^i \xi_{\b) \bd} = 0 \quad \Longleftrightarrow \quad \bar{\nabla}_{(\ad}^i \xi_{\b \bd)} = 0~.
\eea
It is conformally invariant provided that $\xi^\aa$ is a primary superfield of dimension $-1$
\bea
\label{N>2ConfProps}
K^B \xi^\aa = 0 ~, \qquad \mathbb{D} \xi^\aa = - \xi^\aa ~.
\eea
This equation \eqref{N>2ConfProps} determines the superconformal properties of the parameters \eqref{N>2ConfIsoParameters}. Further, \eqref{N>2CKV} has the important corollary $\nabla_{(\a (\ad} \xi_{\b) \bd)} = 0$.

\subsection{Algebra of conformal isometries}

An important consequence of equation \eqref{3.1} is that the commutator of two conformal isometries results in another transformation of the same type, 
\begin{subequations}
	\bea
	\left[ \d_{\mathscr{K}[\xi_{2}]} , \d_{\mathscr{K}[\xi_{1}]} \right] \nabla_{A} &=& \d_{\mathscr{K}[\xi_3]} \nabla_{A} = 0 ~,
	\eea
	where we have defined
	\bea
	\label{2.12b}
	\mathscr{K}[\xi_3] &:=& \left[ \mathscr{K}[\xi_{2}] , \mathscr{K}[\xi_{1}] \right] ~.
	\eea
\end{subequations}
Thus, the set of all conformal Killing (super)vector fields forms a Lie (super)algebra, which is the (super)conformal algebra of $(\mathcal{M}^{4|4\cN}, \nabla) $. 

By employing the bracket \eqref{2.12b}, we may readily compute $\x^\aa_3$ in terms of $\x_1^{\aa}$ and $\x_2^\aa$. We obtain:
\bea
\label{3.20}
\x^{\a\ad}_3 &=& [\xi_1 , \xi_2 ]^{\aa} := -\hf  \x^{\b\bd}_1 \nabla_{\b \bd} \x^{\a\ad}_2 
- \frac{\ri}{16}  \bar \nabla_{\bd i} \x^{\a\bd}_1 \nabla_{\b}^i \x^{\b \ad}_2 ~-~ \big(1 \leftrightarrow 2\big)
~,
\eea
where it should be emphasised that the second term is not present for $\cN=0$.
It may be readily shown that $\x^{\aa}_3$ is conformal Killing.
We emphasise that \eqref{3.20} is the curved uplift of the bracket \eqref{2.10b}.

Next, we prove that the superconformal algebra of $(\mathcal{M}^{4|4\cN}, \nabla)$ is finite dimensional, and its dimension does not exceed that of the (super)conformal group $\sSU(2,2|\cN)$. To prove this claim, we introduce the set of parameters: 
\bea
\label{conformalparameters}
{\bm \Xi} := \Big\{ \xi^{A} , \,K^{ab} [ \xi ] , \,\s [ \xi ] , \, \r [ \xi ] , \, \chi^i{}_j[\xi] , \,\L_A[\xi] \Big\}. 
\eea
Our claim is then equivalent to the requirement that $\nabla_A \bm \Xi $ is 
a linear combination of the elements of $\bm \Xi$. The non-supersymmetric proof trivially follows from \eqref{3.3}, while in the $\cN=1$, $\cN=2$ and conformally-flat $\cN > 2$ cases this holds for $A = (_\a^i)$ by virtue of \eqref{3.8}, \eqref{3.13} and \eqref{3.18}, respectively. The general case immediately follows, proving the claim.

\section{Isometries} \label{Chapter3.2}

We recall from appendix \ref{Appendix2B} that (super)gravity backgrounds may be described in conformal (super)space by coupling to some compensating multiplets, where at least one compensator satisfies the conditions \eqref{3.22}. Given such a background, one may describe its isometries as those conformal isometries \eqref{3.1} which preserve the conformal compensators. Specifically, a conformal Killing supervector field $\x = \x^B E_B$ on $(\cM^{4|4\cN},\nabla,\Xi)$ is said to be Killing if
\bea
\mathscr{K}[\xi] \X =0~.
\label{KillingCondition4d}
\eea
It immediately follows that the set of Killing vectors on $(\cM^{4 | 4 \cN}, \nabla,\X)$
is a Lie superalgebra.

Making use of $\X$, we can always construct a primary scalar
(super)field ${\bm \X}$ with the properties: 
(i) it is an algebraic function of $\X$; 
(ii)  it is nowhere vanishing; and 
(iii) it has non-zero dimension $\D_{\bm \X}$.
It follows from \eqref{KillingCondition4d} that isometries of the background must satisfy the condition
\bea
(\x^{B} \nabla_{B} + \D_{\bm \X} \s[\x] + \ri q_{\bm \X} \r[\x] ) {\bm \X} =0~.
\label{2.23CC}
\eea
The dilatation and $\sU(1)_R$ freedom (if $q_{\bm \Xi} \neq 0$) may be used to impose the gauge condition 
\bea
{\bm \X}=1 \ ,
\label{Xgauge1}
\eea
Then, eq. \eqref{2.23CC} reduces to
\begin{subequations}
	\label{Killing}
	\bea
	\s [\x] &=& 0 \quad \implies \quad \nabla_a \xi^a = 0~, \\
	\r[\x] &=& 0 \quad \implies \quad [\nabla_\a^i , \bar{\nabla}_{\ad i}] \xi^{\aa} = 0~. \label{3.25b}
	\eea
\end{subequations}
where \eqref{3.25b} should be omitted if $q_{\bm \Xi} = 0$. Equations \eqref{Killing} define the Killing conditions for a vector (super)field and are preserved by the bracket \eqref{3.20}. Further in the $\cN=1$ and $\cN=2$ cases we may degauge these equations to the GWZ and $\sSU(2)$ superspaces, respectively:
\begin{subequations}
	\begin{align}
		\cN=1:& \qquad \cD_a \xi^a = 0~, \\
		\cN=2: & \qquad [\cD_\a^i , \bar{\cD}_{\ad i}] \xi^{\aa} = 8 G_\aa \xi^\aa~.
	\end{align}
\end{subequations}

\section{Conformal Killing tensor (super)fields} \label{Chapter3.4}

In the previous section a central role was played by the conformal Killing vector (super)fields of a given background. Here, we will study their higher-rank extensions, known as conformal Killing tensors. They will play a central role in the following section, where it will be shown that they also parametrise higher symmetries for the kinetic operators of certain matter multiplets. Additionally, as will be discussed in the following chapter, they also arise in the study of (super)conformal higher spin theories. Thus, it is of interest to further analyse their properties.

A primary tensor (super)field $\xi_{\a(m) \ad(n)}$, $m,n\geq0$ is said to be a conformal Killing if it satisfies
\bea
\label{N=2CKT}
\nabla_{\a}^i \xi_{\a(m) \ad(n)} = 0 ~, \qquad \bar \nabla_{\ad i} \xi_{ \a(m) \ad(n)} = 0 \quad \implies \quad \nabla_{\aa} \xi_{\a(m) \ad(n)} = 0~.
\eea
These constraints generalise the notion of conformal Killing vector (super)fields and (super)conformal Killing-Yano tensors \cite{HL3}. It may be shown that equations \eqref{N=2CKT} are conformally invariant provided
\be
\mathbb{D} \xi_{\a(m) \ad(n)} = - \hf (m+n) \xi_{\a(m) \ad(n)} ~, \qquad
{\mathbb Y} \xi_{\a(m) \ad(n)} = -\frac{\cN(m-n)}{\cN-4} \xi_{\a(m) \ad(n)} ~,
\ee
where we emphasise that the latter condition should be omitted if $\cN=4$.
If $m=n$, $\xi_{\a(m)\ad(n)}$ is inert under $\sU(1)_R$ transformations and thus it may be consistently restricted to be real. This case was first studied in a supersymmetric setting in \cite{HL1}.

Given two conformal Killing tensors $\xi^1_{\a(m_1) \ad(n_1)}$ and $\xi^2_{\a(m_2) \ad(n_2)}$ defined on $(\mathcal{M}^{4|4\cN},\nabla)$, it may be shown that their symmetric product
\bea
\hat{\xi}_{\a(m_1+m_2) \ad(n_1+n_2)} = \xi^1_{\a(m_1) \ad(n_1)} \xi^2_{\a(m_2) \ad(n_2)} ~,
\eea
is a solution to \eqref{N=2CKT}. This allows one to construct new conformal Killing tensors from existing ones. We may also endow the set of
conformal Killing tensor superfields with an additional algebraic structure. Specifically, we will construct the bracket of conformal Killing tensor superfields, generalising the vector case \eqref{3.20}. This bracket is valid only for real conformal Killing tensors, thus we will require $m_1 = n_1 = p$ and $m_2 = n_2 = q$. Their bracket is necessarily a real, primary superfield satisfying \eqref{N=2CKT}
\bea
&& [ \xi^{1} , \xi^{2} ]_{\a(p+q-1) \ad(p+q-1)} =  - \frac{p}{2} \xi^{1}_{\a(p-1)}{}^{\b}{}_{\ad(p-1) }{}^{\bd} \nabla_{\b \bd } \xi^{2}_{\a(q) \ad(q)} + \frac{q}{2} \xi^{2}_{\a(q-1) }{}^{\b}{}_{\ad(q-1)} {}^{\bd} \nabla_{\b \bd } \xi^{1}_{\a(p) \ad(p)} \non \\&& - \frac{\ri p q}{4(p+1)(q+1)} \Big \{ \bar{\nabla}_{\bd i} \xi^{1}_{\a(p)}{}_{\ad(p-1)}{}^{\bd} \nabla_{\b}^i \xi^{2}_{\a(q-1)}{}^{\b}{}_{\ad(q)} - \bar{\nabla}_{\bd i} \xi^{2}_{\a(q)}{}_{\ad(q-1)}{}^{\bd} \nabla_{\b}^i \xi^{1}_{\a(p-1)}{}^{\b}{}_{\ad(p)} \Big \} ~. ~~~~~~~~
\label{SNB}
\eea
As a result, the set of real conformal Killing tensor superfields $\xi_{\a(n) \ad(n)}$ on a fixed supergravity background forms a superalgebra with respect to \eqref{SNB}. This bracket was introduced in \cite{HL1} where it was called the ``supersymmetric even Schouten-Nijenhuis bracket."

\section{Symmetries of (super)conformal field theories} \label{Chapter3.3}

Consider some (super)conformal field theory describing the dynamics of a primary matter multiplet $\Psi$ (with its indices suppressed)
\begin{align}
	K^A \Psi = 0 ~, \qquad \mathbb{D} \Psi = \D_\Psi \J ~, \qquad \mathbb{Y} \Psi = q_{\J} \J ~,
\end{align}
where the last condition should be omitted for $\cN=0$ and $\cN=4$. We collectively denote its defining differential constraints in terms of the operator $\mathfrak{L}$
\begin{align}
	\label{3.23}
	\mathfrak{L} \J = 0~.
\end{align}

A linear differential operator $\mathfrak{D}$ is said to be a symmetry of $\mathfrak{L}$, $\mathfrak{D} \in \rm{Symm}(\mathfrak{L})$, if it preserves \eqref{3.23}
\begin{subequations} 
	\label{3.28}
	\bea
	\mathfrak{L} \mathfrak{D} \J = 0~,
	\eea
	and all (super)conformal properties of $\Psi$
	\bea
	K^A \mathfrak{D} \J =0~, \qquad \mathbb D \mathfrak{D} \J =  \D_{\J} \mathfrak{D} \J ~, \qquad \mathbb{Y} \mathfrak{D} \J = q_{\J}  \mathfrak{D} \J~.
	\label{3.22b}
	\eea
\end{subequations}
Condition \eqref{3.22b} means that $\mathfrak D$ is a conformal dimensionless and uncharged operator. Given $\mathfrak{D}_1, \mathfrak{D}_2 \in \rm{Symm}(\mathfrak{L})$, it is clear that their product also satisfies \eqref{3.28}, $\mathfrak{D}_1 \mathfrak{D}_2 \in \rm{Symm}(\mathfrak{L})$.

The set $\rm{Symm}(\mathfrak{L})$ naturally forms an associative algebra. In this algebra, it is natural to 
introduce the equivalence relation
\bea
\label{SymmEQR}
\mathfrak{D}_1 \sim \mathfrak{D}_2 
\quad \Longleftrightarrow \quad
\big( \mathfrak{D}_1 - \mathfrak{D}_2 \big)\J = 0~,
\eea
to quotient out trivial symmetries. In what follows we will often attach a superscript to $\mathfrak{D}$ to denote its order; $\mathfrak{D}^{(n)}$ denotes an order-$n$ symmetry operator. For $n>1$, we will say that $\mathfrak{D}^{(n)}$ defines a higher symmetry of $\mathfrak{L}$.

The simplest non-trivial symmetries of $\mathfrak{L}$ are the conformal isometries \eqref{3.1}; $\mathfrak{D}^{(1)} = \mathscr{K}[\xi]$, where $\xi^a$ is a conformal Killing vector (super)field. This is easily verified by making use of the identities
\begin{align}
	[\mathscr{K}[\xi], \nabla_A] = 0 ~, \qquad [\mathscr{K}[\xi], X_{\underline{A}}] = 0~.
\end{align}
An immediate corollary of this is that, for any conformal Killing
vector (super)fields $\x^a_1, \x^a_2, \dots \x^a_n$, the operator
\bea
\mathfrak{D}^{(n)} =  \mathscr{K}[\xi_1] \mathscr{K}[\xi_2]
\dots  \mathscr{K}[\xi_n] \in \rm{Symm}(\mathfrak{L})
\eea
is a higher symmetry of $\mathfrak{L}$. Therefore, $\rm{Symm}(\mathfrak{L})$ includes the universal enveloping algebra of the (super)conformal algebra of the background. Importantly, this means that $\rm{Symm}(\mathfrak{L})$ is a (super)conformal higher-spin algebra \cite{FL-algebras,Bekaert}.

In this section we will study in further detail the higher symmetries of some important $\cN \leq 2$ (super)conformal kinetic operators. Additionally, in the interest of tractability of calculations, we will restrict our attention to conformally-flat backgrounds.

\subsection{Higher symmetries of the conformal d'Alembertian}\label{Chapter3.3.1}

Over the past three decades, there has been an extensive study of the higher symmetries of the conformal d'Alembertian in dimensions $d>2$, including the important publications \cite{ShSh,Eastwood}. Here, we will classify all such symmetries in conformally-flat backgrounds.

We recall the action for a free massless complex scalar field $\varphi$ propagating on a general gravitational background
\begin{align}	
	\mc{S}[\varphi,\bar{\varphi}]= \int\text{d}^{4}x \, e \,  \bar{\varphi}\, \square \, \varphi~, \qquad \square := \nabla^a \nabla_a ~.\label{CSaction}
\end{align}
This action is known to be conformal if $\varphi$ is a primary field of unit dimension
\begin{align}
	K^a \varphi = 0~, \qquad \mb{D}\varphi=\varphi. \label{3.30}
\end{align}
Varying \eqref{CSaction} with respect to $\bar{\varphi}$ yields the equation of motion
\begin{align}
	\square \varphi = 0~.
	\label{3.29}
\end{align}

Keeping in mind the general construction above, a linear differential operator $\mathfrak{D}$ is called a symmetry of the 
conformal d'Alembertian, $\square$, if it obeys: 
\begin{subequations} 
	\bea
	\square \mathfrak{D} \vf &=& 0~, \label{5.4a}\\
	K^a \mathfrak{D} \vf &=&0~, \qquad \mathbb D \mathfrak{D} \vf =  \mathfrak{D} \vf ~. \label{5.4b}
	\eea
\end{subequations}
Utilising the equivalence relation \eqref{SymmEQR}, it is possible to show that every symmetry operator
of order $n$ can be reduced to the canonical form
\bea
\label{HSdAlembertianCanonicalForm}
\mathfrak{D}^{(n)} = \sum_{k=0}^{n}\xi^{\a(k) \ad(k)} (\nabla_\aa)^{k} ~.
\eea
Requiring $\mathfrak{D}^{(n)}$ to preserve condition \eqref{3.29}, we find that the top parameter $\xi^{\a(n) \ad(n)}$ is a conformal Killing tensor
\begin{subequations}
	\label{3.33}
	\begin{align}
		\nabla^\aa \xi^{\a(n) \ad(n)} = 0~,
	\end{align}
	and the remaining parameters satisfy the constraints 
	\begin{align}
		\Box \xi^{\a(k+1) \ad(k+1)} - \nabla^\aa \xi^{\a(k) \ad(k)} = 0 ~, \qquad 0 \leq k \leq n ~,
	\end{align}
\end{subequations}
where one should keep in mind that $\xi^{\a(n+1)\ad(n+1)}=0$.

Further, we require that the transformed field is primary and of unit dimension \eqref{5.4b}, which implies that the coefficient of lowest weight (highest rank) is primary, 
\begin{subequations}
	\label{3.35}
	\begin{align}
		K^\bb \xi^{\a(n) \ad(n)} = 0~,\qquad \mathbb{D} \xi^{\a(n) \ad(n)}=-n\xi^{\a(n) \ad(n)}~,
	\end{align}
	and the remaining coefficients satisfy
	\begin{align}
		K^{\b \dot{\b}} \xi^{\a(k) \ad(k)}  = 4(k+1)^2 \xi^{\a(k) \b \ad(k) \bd}~,  \qquad 0 \leq k \leq n-1 ~.
	\end{align}
\end{subequations}
The unique ansatz for $\xi^{\a(k) \ad(k)}$ compatible with constraints \eqref{3.33} and \eqref{3.35} is
\begin{align}
	\xi^{\a(k) \ad(k)} = A_k (\nabla_{\bb})^{n-k} \xi^{\a(k) \b(s-k) \ad(k) \bd(s-k)} ~, \qquad 0 \leq k \leq n ~,
\end{align}
for undetermined coefficients $A_k \in \mathbb{C}$. They may be readily computed (up to an overall normalisation) by making use of \eqref{3.33} and \eqref{3.35}. One finds
\begin{align}
	A_k= \binom{n}{k}^2\binom{2n+2}{n-k}^{-1} ~, \qquad 0 \leq k \leq n~.
\end{align}
Thus, in conformally-flat backgrounds, the higher symmetries of the conformal d'Alembertian are defined by the operators
\begin{align}
	\label{HSCD}
	\mathfrak{D}^{(n)} &= \sum_{k=0}^{n}\binom{n}{k}^2\binom{2n+2}{n-k}^{-1} (\nabla_{\bb})^{n-k} \xi^{\a(k) \b(n-k) \ad(k) \bd(n-k)} (\nabla_\aa)^k~.
\end{align}

\subsection{Higher symmetries of the massless Wess-Zumino operator} \label{Chapter3.3.2}
In this subsection we study the higher symmetries of the massless Wess-Zumino operator. As is well-known, the latter is a superconformal extension of the conformal d'Alembertian and massless Dirac operator, see e.g. \cite{BK} for more details.

The action for the massless, non-interacting Wess-Zumino model is:
\bea
\label{WZAction}
\cS[\F, \bar \F] &=&\int \rd^{4|4}z  \,E \,\bar \F \F ~,
\eea
where $\F$ is a primary superfield of unit dimension,
\begin{align}
	\mathbb{D} \F = \F~, \qquad  K^A \F = 0~,
\end{align}
subject to the off-shell constraint
\bea
\label{4.2}
\bar{\nabla}_{\ad} \F = 0 ~.
\eea
We note that this constraint uniquely fixes the $\sU(1)_R$ charge of $\F$, $\mathbb{Y} \Phi = - \frac 2 3 \Phi$. Varying the action \eqref{WZAction} with respect to $\bar \Phi$ yields the equation of motion
\bea
\label{EoMmasslessWZ}
\nabla^2 \F = 0~.
\eea

In keeping with the general prescription detailed at the start of this section, a linear differential operator $\mathfrak D$ will be called a symmetry of the Wess-Zumino operator if it
obeys the conditions
\begin{subequations}
	\bea
	\bar \nabla_\ad {\mathfrak D} \F &=& 0~, \qquad
	\nabla^2 {\mathfrak D} \F =0~, \label{phiConditions1}\\
	K^A \mathfrak{D} \F &=& 0 ~, \qquad \mathbb{D} \mathfrak{D} \F = \mathfrak{D} \F ~, \qquad \mathbb{Y} \mathfrak{D} \F = -\frac{2}{3} \mathfrak{D} \F \label{phiConditions2}~.
	\eea
\end{subequations}
for every on-shell chiral scalar $\F$. Making use of the equivalence relation \eqref{SymmEQR}, it is possible to reduce every $n$th-order symmetry operator to the canonical form
\bea
\label{canonicalForm}
{\mathfrak D}^{(n)} = \sum_{k=0}^{n} \xi^{\a(k) \ad(k)} (\nabla_{\aa})^k  + \sum_{k=1}^{n} \xi^{\a(k) \ad(k-1)} (\nabla_\aa)^{k-1} \nabla_{\a}  ~.
\eea

To begin with, we impose \eqref{phiConditions1}, which implies that $\xi^{\a(n) \ad(n)}$ is a conformal Killing tensor superfield
\bea
\nabla_{\a} \xi_{\a(n) \ad(n)} = 0 ~, \quad
\bar \nabla_{\ad} \xi_{\a(n) \ad(n)} =0 ~,
\eea
in addition to:
\begin{subequations}
	\label{WZparametersDiffConstraints}
	\bea
	\bar \nabla_{\bd} \xi^{\a(k) \a(k-1)} &=& 0~, \quad 1 \leq k \leq n ~, \\
	\bar \nabla_{\bd} \xi^{\a(k) \ad(k)} + 2 \ri \d_{\bd}^{\ad} \xi^{\a(k) \ad(k-1)} &=& 0~, \quad 0 \leq k \leq n ~, \\
	\nabla^2 \xi^{\a(k) \ad(k)} &=& 0~, \quad 0 \leq k \leq n ~, \\
	\nabla^2 \xi^{\a(k) \ad(k-1)} + 2 \nabla^{\a} \xi^{\a(k-1) \ad(k-1)} &=& 0 ~, \quad 1 \leq k \leq n ~.
	\eea
\end{subequations}
Further, imposing \eqref{phiConditions2} we see that $\xi^{\a(n) \ad(n)}$ is a primary superfield of dimension $-n$
\bea
K^B \xi_{\a(n) \ad(n)} = 0 ~, \qquad \mathbb{D} \xi_{\a(n) \ad(n)} = -n \xi_{\a(n) \ad(n)}~, \qquad \mathbb{Y} \xi_{\a(n) \ad(n)} = 0~,
\eea
and that the remaining parameters obey
\begin{subequations}
	\label{WZparametersConfConstraints}
	\bea
	S^{\b} \xi^{\a(k) \ad(k)} - 4(k+1) \xi^{\a(k) \b \ad(k)} &=& 0 ~, \quad 0 \leq k \leq n-1 ~, \\
	S^{\b} \xi^{\a(k) \ad(k-1)} &=& 0 ~, \quad 1 \leq k \leq n ~,  \\
	\bar{S}_\bd \xi^{\a(k) \ad(k)} &=& 0 ~, \quad 0 \leq k \leq n -1 ~,  \\
	\bar{S}_\bd \xi^{\a(k) \ad(k-1)} - 2 \ri k \xi^{\a(k) \ad(k-1)}{}_{\bd} &=& 0 ~, \quad 1 \leq k \leq n ~.
	\eea
\end{subequations}

In order to solve these equations, we introduce the most general ansatz consistent with the requirements above. This is:
\begin{subequations}
	\label{WZparametersAnsatz}
	\bea
	\xi^{\a(k) \ad(k-1)} &=& A_k (\nabla_\bb)^{n-k} \bar \nabla_{\bd} \xi^{\a(k) \b(n-k) \ad(k-1) \bd(n-k+1)} ~, \quad 1 \leq k \leq n ~, \\
	\xi^{\a(k) \ad(k)} &=& B_k (\nabla_\bb)^{n-k} \xi^{\a(k) \b(n-k) \ad(k) \bd(n-k)} \non \\ &\phantom{=}&+ C_k (\nabla_\bb)^{n-k-1} \nabla_\b \bar \nabla_\bd \xi^{\a(k) \b(n-k) \ad(k) \bd(n-k)} ~, \quad 0 \leq k \leq n ~,
	\eea
\end{subequations}
where $A_k, B_k, C_k \in \mathbb{C}$. Inserting \eqref{WZparametersAnsatz} into the constraints \eqref{WZparametersDiffConstraints} and \eqref{WZparametersConfConstraints} yields:
\begin{subequations}
	\bea
	A_k = \frac{\ri n}{2(n+1)}  {n \choose k} {n-1 \choose k-1}{{2n+1 \choose n-k}}^{-1} ~, \quad 1 \leq k \leq n~, \\
	B_k = {n \choose k}^2  {2n+1 \choose n-k}^{-1} ~, \quad 0 \leq k \leq n ~, \\
	C_k = - \frac{\ri (n-k)}{2(n+1)} {n \choose k}^2  {2n+1 \choose n-k}^{-1} ~, \quad 0 \leq k \leq n ~.
	\eea
\end{subequations}

In summary, we state that
\bea
\label{WZHSsoln}
\mathfrak{D}^{(n)} \Phi &=& \sum_{k=0}^n  {n \choose k}^2  {2n+1 \choose n-k}^{-1} \Big \{ (\nabla_\bb)^{n-k} - \frac{\ri (n-k)}{2(n+1)} (\nabla_\bb)^{n-k-1} \nabla_\b \bar{\nabla}_\bd \Big \} \xi^{\a(k) \b(n-k) \ad(k) \bd(n-k)} \non \\
&\phantom{=}& \times (\nabla_\aa)^k \Phi + \sum_{k=1}^n \frac{\ri n}{2(n+1)}  {n \choose k} {n-1 \choose k-1}{{2n+1 \choose n-k}}^{-1} (\nabla_\bb)^{n-k} \bar{\nabla}_\bd \xi^{\a(k) \b(n-k) \ad(k-1) \bd(n-k+1)} \non \\
&& \times   (\nabla_\aa)^{k-1} \nabla_\a \Phi ~,
\eea
is the unique (modulo overall normalisation) $n$th order higher symmetry of the massless Wess-Zumino operator in conformally-flat superspaces. 

\subsection{Higher symmetries of the massless Sohnius operator}\label{Chapter3.3.3}

Building on the non-supersymmetric and $\cN=1$ studies undertaken above, in this subsection we study higher-derivative symmetries 
of the massless Sohnius operator.\footnote{Higher symmetries of the six-dimensional Sohnius operator are studied in appendix \ref{Appendix3A.2}. It is expected that the results of the latter may be related to the present case via a dimensional reduction.} The latter defines the on-shell equations for a massless hypermultiplet.

We consider an on-shell hypermultiplet, which is described by a primary isospinor $q^i$ (and its conjugate $\bar q_i$) subject to the constraints\footnote{It may be shown that $q^i$ encodes two massless chiral multiplets, defined by $\F_{+} = \bar{q}_{\underline 1}|_{\q_{\underline 2}^\a = \bar{\q}_\ad^{\underline 2} = 0}$ and $\F_{-} = {q}^{\underline 2}|_{\q_{\underline 2}^\a = \bar{\q}_\ad^{\underline 2} = 0}$. Making use of \eqref{qEoM}, it is easily verified that they obey \eqref{4.2} and \eqref{EoMmasslessWZ}.}
\bea
\label{qEoM}
\nabla_{\a}^{(i} q^{j)} = 0~, \qquad \bar \nabla_{\ad}^{(i} q^{j)} = 0~.
\eea
These are consistent with the superconformal algebra provided
\bea
\mathbb{D} q^i = q^i~, \qquad \mathbb{Y} q^i = 0~.
\eea

A linear differential operator $\mathfrak D$ is said to be a symmetry of the massless Sohnius operator if it obeys the conditions
\begin{subequations}
	\label{HMSymmProps}
	\bea
	\nabla_\a^{(i} {\mathfrak D} q^{j)} &=& 0 ~, \qquad \bar \nabla_\ad^{(i} {\mathfrak D} q^{j)} = 0~, \\
	K^A \mathfrak{D} q^i &=& 0~, \qquad \mathbb{D} \mathfrak{D} q^i = \mathfrak{D} q^i ~, \qquad \mathbb{Y} \mathfrak{D} q^i = 0 ~.
	\eea
\end{subequations}
Modulo the equivalence \eqref{SymmEQR}, every $n$th-order symmetry operator may be brought to a canonical form given by
\bea
\label{B5}
\mathfrak{D}^{(n)} &=&  \sum_{k=0}^{n} \xi^{\a(k) \ad(k)} (\nabla_{\aa})^k + \sum_{k=0}^{n-1} \Big \{ \xi^{\a(k+1) \ad(k)}{}_j (\nabla_{\aa})^k \nabla_{\a}^j + \xi^{\a(k) \ad(k+1)}{}_j (\nabla_{\aa})^k \bar{\nabla}_{\ad}^j \non \\
&\phantom{=}& + \xi^{\a(k) \ad(k) j(2)} (\nabla_{\aa})^k \mathbb{J}_{j(2)} \Big \}
\eea

We now impose the conditions \eqref{HMSymmProps}. For brevity we will not spell out their implications and instead simply report the results. All coefficients appearing in \eqref{B5} may be expressed in terms of $\xi^{\a(n) \ad(n)}$, which proves to be a conformal Killing tensor superfield
\begin{subequations}
	\label{3.58}
	\begin{align}
		\nabla_\a^i \xi_{\a(n) \ad(n)} &= 0 ~, \qquad \bar{\nabla}_\ad^i \xi_{\a(n) \ad(n)} = 0 ~, \\
		K^B \xi_{\a(n) \ad(n)} &= 0 ~, \qquad \mathbb{D} \xi_{\a(n) \ad(n)} = - n \xi_{\a(n) \ad(n)} ~, \qquad \mathbb{Y} \xi_{\a(n) \ad(n)} = 0~.
	\end{align}
\end{subequations}
Specifically, we find:
\begin{subequations}
	\label{B.6}
	\bea
	\xi^{\a(k) \ad(k)} &=&  \bigg [ \frac{n((n+k+2)(k+1)+n(n-k))}{2(n+1)^3} \binom{n+k}{2k} \binom{2k}{k} \binom{2n}{n-1}^{-1}(\nabla_\bb)^{n-k} \non \\
	&\phantom{=}& + \frac{(-1)^k (n-1)^2}{32(n+1)^2(2n-1)} \binom{n+k}{k} \binom{n-2}{k} \binom{2n-2}{n}^{-1} (\nabla_\bb)^{n-k-2} \big \{ \nabla_{\b(2)} , \bar{\nabla}_{\bd(2)} \big \} \bigg ] \non \\
	&\phantom{=}& \times \xi^{\a(k) \b(n-k) \ad(k) \bd(n-k)} ~, \\
	\xi^{\a(k+1) \ad(k) i} &=& \bigg [ \frac{\ri n^2(k+2)}{2(n+1)(k+1)} \binom{n+k+1}{k} \binom{n-1}{k} \binom{2n}{n-1}^{-1} (\nabla_\bb)^{n-k-1} \bar{\nabla}_\bd^i \non \\
	&\phantom{=}& - \frac{4(-1)^k(n+k+1)}{k+1} \binom{n+k+1}{k} \binom{n-2}{k} \binom{2n-2}{n}^{-1} (\nabla_\bb)^{n-k-2} \nabla_\b^i \bar{\nabla}_{\bd(2)} \bigg ] \non \\
	&\phantom{=}& \times \xi^{\a(k+1) \b(n-k-1) \ad(k) \bd(n-k)} ~, \\
	\xi^{\a(k) \ad(k+1) i} &=& \bigg [ - \frac{\ri n^2(k+2)}{2(n+1)(k+1)} \binom{n+k+1}{k} \binom{n-1}{k} \binom{2n}{n-1}^{-1} (\nabla_\bb)^{n-k-1} \nabla_\b^i \non \\
	&\phantom{=}& - \frac{4(-1)^k(n+k+1)}{k+1} \binom{n+k+1}{k} \binom{n-2}{k} \binom{2n-2}{n}^{-1} (\nabla_\bb)^{n-k-2} \bar{\nabla}_\bd^i {\nabla}_{\b(2)} \bigg ] \non \\
	&\phantom{=}& \times \xi^{\a(k) \b(n-k) \ad(k+1) \bd(n-k-1)} ~, \\
	\xi^{\a(k) \ad(k) j(2)} &=& \bigg [ \frac{\ri n^2}{2(n+1)^2} \binom{n+k+1}{k} \binom{n-1}{k} \binom{2n}{n-1}^{-1} (\nabla_\bb)^{n-k-1} \nabla_\b^{(j_1} \bar{\nabla}_\bd^{j_2)} \bigg ] \non \\
	&\phantom{=}& \times \xi^{\a(k) \b(n-k) \ad(k) \bd(n-k)} ~.
	\eea
\end{subequations}

Thus, in conformally-flat superspace the higher symmetries of the massless Sohnius operator take the form \eqref{B5}, where the coefficents are given in the expressions \eqref{B.6}. As a result, such symmetries are determined in terms of a single superfield subject to the constraints \eqref{3.58}.

\section{Symmetries of massive field theories} \label{Chapter3.5}

In the previous section we successfully described higher symmetries for the kinetic operators of three massless matter multiplets propagating in conformally-flat backgrounds. Such symmetries were required to preserve both the equations of motion and all (super)conformal properties. Here, by endowing each multiplet with a mass, we break (super)conformal symmetry. Consequently, the corresponding constraint on its symmetries should be relaxed.
As will be described below, the higher symmetries of the resulting non-(super)conformal kinetic operators may be obtained from their cousins derived above, provided that the conformal Killing tensor obeys an additional constraint, the Killing condition.

\subsection{Higher symmetries of the massive Klein-Gordon operator}

We recall that a massive scalar field propagating in curved spacetime is described by the action
\bea
\cS[\vf, \bar \vf] &=&\int \rd^4x \,e \, \Big \{ \bar \varphi \square \varphi - \frac{m}{2} \Xi^2 \bar \varphi \varphi \Big \} ~,
\eea
where $m=\bar m $ is the mass of $\varphi$ and $\X$ is a conformal compensator
\begin{align}
	K^a \X = 0 ~, \qquad \mathbb{D} \X = \X ~.
\end{align}
In what follows we will make use of the local dilatation freedom to impose the gauge $\X=1$.\footnote{It follows that $\nabla_a \X = -b_a$ and we must make use of the degauging procedure to eliminate $b_a$.} 
In this gauge the equation of motion is the massive Klein-Gordon equation
\bea
(\square - m^2) \varphi = 0~.
\eea

In the massive case, the requirement that the symmetry operator $\mathfrak{D}^{(n)}$, see eq. \eqref{HSCD}, preserves the equation of motion
\bea
(\square - m^2) \mathfrak{D}^{(n)} \varphi = 0~.
\eea
leads to the well-known Killing condition
\be
\nabla_{\bb} \xi^{\a(n-1) \b \ad(n-1) \bd} = 0 \quad \implies \quad \cD_{\bb} \xi^{\a(n-1) \b \ad(n-1) \bd} = 0~,
\ee
where the latter constraint arises upon degauging. It should be emphasised that, given two Killing tensors, their bracket \eqref{SNB}, is also Killing.

\subsection{Higher symmetries of the massive Wess-Zumino operator}
\label{section3.5.2}

The propagation of a massive chiral supermultiplet in curved superspace 
is described by the action
\bea
S[\F, \bar \F] &=&\int \rd^4x\rd^2\q\rd^2\qb  \,E \,\bar \F \F 
+ \Big\{ \frac{m}{2}  \int\rd^4x\rd^2\q\, \cE \,S_0\F^2 +{\rm c.c.} \Big\}~,
\qquad \bar \nabla_\ad \F =0~, ~~
\eea
with $m=\bar m $ a mass parameter. Here $S_0$ is the chiral compensator of old minimal supergravity, with the properties
\begin{align}
	\bar{\nabla}_\ad S_0 = 0~, \qquad K^A S_0 = 0 ~, \qquad \mathbb{D} S_0 = S_0  \quad \implies \quad \mathbb{Y} S_0 = -\frac{2}{3} S_0~.
\end{align}
In what follows we will make use of the local dilatation and $\sU(1)_R$ freedom to impose the gauge $S_0=1$.\footnote{Note that $\nabla_A S_0 = - B_A + \frac{2\ri}{3} \F_A =0$ in this gauge and the resulting geometry is that of the GWZ superspace.} 
Then the equation of motion is:
\bea
- \frac 1 4 \nabla^2 \Phi + m \bar{\Phi} = 0~.
\eea

We now determine what additional conditions must be imposed upon the $n$th-order operator \eqref{WZHSsoln} so that we obtain a symmetry of the massive equations of motion. Since it has been shown that all coefficients
are expressed in terms of the top component, we expect that this condition may be written as a closed
form equation in $\xi_{\a(n) \ad(n)}$.

In the massive case, the requirement that $\mathfrak{D}^{(n)}$ preserves the equation of motion
\bea
- \frac 1 4 \nabla^2 \mathfrak{D}^{(n)} \Phi + m \overline{ \left( \mathfrak{D}^{(n)} \Phi  \right) } = 0 ~,
\eea
leads to the following nontrivial constraint
\be
\nabla_{\b} \bar{\nabla}_{\bd} \xi^{\b \a(n-1) \bd \ad(n-1)} = \frac{2 \ri (n+1)}{n} \left( \xi^{\a(n-1) \ad(n-1)}  - \bar{\xi}^{\a(n-1) \ad(n-1)} \right) ~.
\ee
Upon degauging to the GWZ geometry and inserting the expression for $\x^{\a(n-1) \ad(n-1)}$, we obtain the Killing condition 
\bea
\label{KillingTensorN=1}
(\cD_{\b} \cDB_{\bd} - 2 n ( n + 1 ) G_{\b \bd} ) \xi^{\b \a(n-1) \bd \ad(n-1)} = 0 ~,
\eea
which implies 
\bea
\cD^{\b \bd} \xi_{\b \a(n-1) \bd \ad(n-1)} = 0 ~.
\eea
In the case of AdS superspace AdS${}^{4|4}$, 
$G_{\a \ad} = 0$ and the Killing condition \eqref{KillingTensorN=1} reduces to the one
originally described in \cite{GKS}. Further, it may be proven that, given two Killing tensor superfields \eqref{KillingTensorN=1}, their bracket, defined by \eqref{SNB}, is also Killing.

\subsection{Higher symmetries of the massive Sohnius operator}
\label{section3.5.3}

It is possible to endow a hypermultiplet with a mass by coupling it to a `frozen' vector multiplet \`a la \cite{BK97}. This approach is most useful to deal with off-shell hypermultiplets without intrinsic central charge, such as the $q^+$ hypermultiplet \cite{GIKOS} and the polar hypermultiplet (see \cite{G-RRWLvU,K2010} and references therein).
It can also be used for the Fayet-Sohnius hypermultiplet \cite{Fayet, Sohnius78}
which suffices for our goals. 
To this end, we introduce the gauge covariant derivatives
\bea
\mathscr{D}_A = (\mathscr{D}_a,\mathscr{D}_\a^i,\bar{\mathscr{D}}^{\ad}_i) = \nabla_A - V_A \D ~,
\eea
where $\D$ is a central charge operator, $[\D, \mathscr{D}_A] = 0$. The covariant derivatives $\mathscr{D}_A$ are characterised by the anti-commutation relations
\begin{subequations}\label{VMAlgebra}
	\bea
	&\{ \mathscr{D}_\a^i , \mathscr{D}_\b^j \} = - 2 \ve^{ij} \ve_{\a\b}  \bar{W}_0 \D ~, 
	\quad \{{\bar {\mathscr{D}}}_{\dot \a i} \, ,  {\bar {\mathscr{D}}}_{\dot  \b j} \} = - 2 
	\ve_{ij}\, \ve_{\dot \a \dot \b} {W}_0 \D  ~, 
	\\
	& \{ \mathscr{D}_\a^i , \bar{\mathscr{D}}^\bd_j \} = - 2 \ri \d_j^i \mathscr{D}_\a{}^\bd ~,
	\eea
\end{subequations}
where ${W}_0$ is a field strength describing the vector multiplet. The latter is subject to the off-shell constraints
\begin{align}
	\bar{\mathscr{D}}_\ad^i {W}_0 = 0 ~, \qquad \mathscr{D}^{ij} {W}_0 = \bar{\mathscr{D}}^{ij} \bar{{W}}_0~.
\end{align}
and its superconformal properties are as follows
\begin{align}
	K^A {W}_0 = 0~, \qquad \mathbb{D} {W}_0 = {W}_0 ~, \qquad \mathbb{Y} {W}_0 = -2 {W}_0~.
\end{align}
In what follows we employ the dilatation and $\sU(1)_R$ freedom to impose the gauge ${W}_0 = \ri$.\footnote{An immediate consequence of this gauge fixing is $\mathscr{D}_A {W}_0 = -\ri B_A -2\ri \F_A=0$ and the resulting geometry is that of the $\sSU(2)$ superspace.}

To describe a massive hypermultiplet,  the constraints \eqref{qEoM} should be replaced with
\be
\label{HMVM}
\mathscr{D}_{\a}^{(i} q^{j)} = 0~, \quad \bar{\mathscr{D}}_{\ad}^{(i} q^{j)} = 0~.
\ee
An important consequence of \eqref{HMVM} is\footnote{This is just a different way of looking at the Fayet-Sohnius hypermultiplet \cite{Fayet, Sohnius78}.}
\bea
(\mathscr{D}^a \mathscr{D}_a + \D^2 ) {q}^i = 0 ~.
\eea
Thus, requiring that $q^i$ is an eigenvector of $\D$ (with non-zero eigenvalue)
\bea
\D q^i = \ri m q^i ~, \qquad m \in \mathbb{R}\setminus\{0\}~,
\eea
is equivalent to endowing the hypermultiplet with a mass,
\bea
(\mathscr{D}^a \mathscr{D}_a - m^2 ) q^i = 0~.
\eea

As the gauge group now includes central charge transformations generated by $\D$, the higher symmetries \eqref{B5} should be modified such they preserve the massive equations of motion \eqref{HMVM}. It turns out that the necessary modification is as follows:
\begin{align}
	\label{3.85}
	\bold{\hat{\mathfrak{D}}}^{(n)} = \mathfrak{D}^{(n)} + \sum_{k=0}^{n-1} \chi^{\a(k) \ad(k)} (\mathscr{D}_{\aa})^k \D~,
\end{align}
for some tensor superfields $\chi^{\a(k)\ad(k)}$. Imposing the conditions
\begin{align}
	\mathscr{D}_\a^{(i} \bold{\hat{\mathfrak{D}}}^{(n)} q^{j)} = 0~, \qquad \bar{\mathscr{D}}_\ad^{(i} \bold{\hat{\mathfrak{D}}}^{(n)} q^{j)} = 0~,
\end{align}
we obtain the constraints
\begin{align}
	\label{3.87}
	\bm{\mathscr{D}}^{\b i} \chi^{\a(k) \ad(k)} = 2 \ri \xi^{\b \a(k) \ad(k) i} ~.
\end{align}
We emphasise that this implies that $\chi^{\a(k) \ad(k)}$ is, in general, a non-local function of $\xi^{\a(n) \ad(n)}$.
Equation \eqref{3.87} implies the following non-trivial condition
\begin{align}
	\label{B.8}
	{\mathscr{D}}^i_\b \bar{\mathscr{D}}_{\bd i} \xi^{\a(n-1) \b \ad(n-1) \bd} = 0 \quad \implies \quad \mathscr{D}_\bb \xi^{\a(n-1) \b \ad(n-1) \bd} = 0~.
\end{align}
Upon degauging to the $\sSU(2)$ superspace, this takes the form
\begin{align}
	\label{3.89}
	({\cD}^i_\b \bar{\cD}_{\bd i} - 4(n+1) G_{\bb}) \xi^{\a(n-1) \b \ad(n-1) \bd} = 0 \quad \implies \quad {\cD}_\bb \xi^{\a(n-1) \b \ad(n-1) \bd} = 0~.
\end{align}
We note that, given two solutions to \eqref{3.89}, their bracket, defined by \eqref{SNB}, also satisfies \eqref{3.89}.

\section{Summary of results} \label{Chapter3.6}

This chapter was devoted to the study of geometric symmetries of curved (super)space backgrounds and the higher symmetries of several important kinetic operators. We initated this analysis in section \ref{Chapter3.1} by computing conformal isometries of curved backgrounds in conformal (super)space, where they arise as those structure group transformations preserving the covariant derivatives \eqref{3.1}. An important result of our analysis is that conformal isometries are parametrised by a single real vector (super)field, known as a conformal Killing vector. The latter proves to be primary, of dimension $-1$, and satisfies a first-order differential constraint. It is also proven that the conformal isometries of a given curved background form a finite-dimensional (super)algebra with respect to the bracket \eqref{3.20}.\footnote{The conformal isometries of $\cN=(1,0)$ conformal supergravity backgrounds were also described in appendix \ref{Appendix3A.1}. It is expected that these results may be related to the four-dimensional story via a dimensional reduction.} Building on these results, in section \ref{Chapter3.2} we studied a special class of geometric symmetries known as isometries. They were defined as those conformal isometries preserving the conformal compensator(s) of the (super)gravity theory considered, and thus are incompatible with (super)conformal symmetry. Our analysis lead to the conditions \eqref{Killing}, which characterise a Killing vector (super)field. The latter naturally parametrise the isometries of a given background.

Following this, in section \ref{Chapter3.4}, we introduced the conformal Killing tensor (super)fields of a given curved background as higher-rank generalisations of its conformal Killing vectors and studied their properties. Such (super)fields proved to be fundamentally important in section \ref{Chapter3.3}, where we studied the higher symmetries of the kinetic operators for several important $\cN \leq 2$ matter multiplets in conformally-flat backgrounds. In particular, sections \ref{Chapter3.3.1}, \ref{Chapter3.3.2} and \ref{Chapter3.3.3} were devoted to higher symmetries of the conformal d'Alembertian, massless Wess-Zumino operator and massless Sohnius operator, respectively.\footnote{Higher symmetries of the six-dimensional Sohnius operator were also described in appendix \ref{Appendix3A.2}. We expect that these results may be related to the four-dimensional story by performing a dimensional reduction.} In each case, it was shown that such symmetries are parametrised by conformal Killing tensors.

Finally, in section \ref{Chapter3.5}, we analysed higher symmetries of massive extensions for the kinetic operators studied in section \ref{Chapter3.3}. By requiring that the higher symmetries derived in section \ref{Chapter3.3} preserve the massive equations of motion, we arrived at new conditions on the conformal Killing tensor parameter. Such constraints are, by construction, incompatible with (super)conformal symmetry and are higher-rank extensions of the Killing conditions for vector (super)fields \eqref{Killing}.

\begin{subappendices}
	
\section{Symmetries in six dimensions} \label{Appendix3A}

This appendix is devoted, in part, to studying the conformal isometries of curved $\cN=(1,0)$ superspace in six spacetime dimensions as an extension of the four-dimensional analysis undertaken in section \ref{Chapter3.1}. Building on this, we then proceed to studying higher symmetries for the Sohnius operator in general conformally flat backgrounds. All results will be presented in the conformal superspace geometry reviewed in appendix \ref{Appendix2C}.

\subsection{Conformal isometries}
\label{Appendix3A.1}

Keeping in mind the general discussion of conformal isometries in section \ref{Chapter3.1}, a real supervector field $\x= \x^B E_B$ on $(\cM^{6 |8}, \nabla)$ is said to be conformal Killing if 
\bea
\label{CSSConfIso}
\d_{\mathscr{K}[\xi]} \nabla_A =[\mathscr{K}[\xi], \nabla_A] = [\xi^B  {\nabla}_B + \L^{\underline {B}} X_{\underline{B}} , \nabla_A]  = 0 
~.
\eea
The solution to \eqref{CSSConfIso} is:
\begin{subequations} \label{3.8parameters}
	\bea
	\xi^{\a}_{i} &=& \frac{\ri}{12} \nabla_{\b}{}_{i} \xi^{\b \a}=-\frac{\ri}{12}(\tilde{\g}^a)^{\a\b} \nabla_{\b}{}_{i} \xi_a~, \\
	\L_{\a}{}^{\b} [\xi] &=& \frac{1}{2} \big( \nabla_{\a}^{i} \xi^{\b}_{i} - \frac{1}{4} \d_{\a}^{\b} \nabla_{\g}^{i} \xi^{\g}_{i} \big) + \xi_{\a \g} W^{\b \g}  = - \frac{1}{4} (\g^{ab})_{\a}{}^{\b} \nabla_{a} \xi_b  + \xi_{\a \g} W^{\b \g} \\
	\L_{ij} [\xi] &=&
	\frac{1}{4} \nabla_{\a (i} \xi^{\a}_{j)}
	=-\frac{\ri}{48}(\tilde{\g}^a)^{\a\b} \nabla_{\a (i} \nabla_{\b}{}_{j)} \xi_a 
	~, \\
	\s [\xi] &=& \frac{1}{4} \nabla_{\a}^{i} \xi_{i}^{\a} = \frac{1}{6} \nabla_{a} \xi^{a}~, \\
	\L_{\a}^i [\xi] &=& \frac{1}{2} \nabla_{\a}^{i} \s[\xi] - \frac{1}{16} \xi_{\a \b} \nabla_{\g}^{i} W^{\b \g} = \frac{1}{12} \nabla_{\a}^i \nabla^a \xi_a - \frac{1}{16} \xi_{\a \b} \nabla_{\g}^i W^{\b \g} ~, \\
	\L_{a} [\xi] &=& 2 (\tilde{\g}_a)^{\a \b} \nabla_{\a}^{i} \L_{\b i} = \frac{4}{3} \nabla_{a} \nabla^b \xi_b - \frac{1}{8} (\tilde{\g}_a)^{\a \b} \nabla_{\a}^i (\xi_{\b \g} \nabla_{\d i} W^{\g \d}) ~,
	\eea
\end{subequations}
where $\xi^{a}$ obeys the conformal Killing vector equation
\bea
\label{CKVCSS}
\nabla_{\a}^{i} \xi^{a} = - \frac{1}{5} (\g^{ab})_{\a}{}^{\b} \nabla_{\b}^{i} \xi_{b} ~.
\eea
This equation is conformally invariant provided $\xi^{a}$ is primary and of dimension
$-1$, 
\bea
K^B \x^a = 0 ~,
\qquad \mathbb D \x^a = -\x^a~.
\label{3.10}
\eea
These relations
determine the superconformal properties of the parameters 
in \eqref{3.8parameters}.
An important corollary of \eqref{3.10} is
\bea
\nabla_{(a} \xi_{b)} = \frac{1}{6} \eta_{a b} \nabla_{c} \xi^{c} ~.
\eea

It is clear from \eqref{CSSConfIso} that the commutator of superconformal transformations must result in another such transformation
\begin{subequations}
	\bea
	\left[ \d_{\mathscr{K}[\xi_{2}]} , \d_{\mathscr{K}[\xi_{1}]} \right] \nabla_{A} &=& \d_{\mathscr{K}[\xi_3]} \nabla_{A} = 0 ~,\\
	\mathscr{K}[\xi_3] &:=& \Big[ \mathscr{K}[\xi_{2}] , \mathscr{K}[\xi_{1}] \Big] ~,
	\eea
\end{subequations}
From this we may extract the form of $\xi_3$, which is determined solely in terms of its vector component
\bea
\x^{a}_3 &=& \xi^{b}_1 \nabla_{b} \xi^{a}_2 - \xi^{b}_2 \nabla_{b} \xi^{a}_1 - \frac{\ri}{48} (\tilde{\g}^a)^{\a\b} \nabla_{\a}^{i} \xi_1^b \nabla_{\b i} \xi_{2b} + \frac{\ri}{48} (\tilde{\g}^{a}{}_{bc})^{\a \b} \nabla_{\a}^i \xi_1 ^b \nabla_{\b i} \xi_2 ^c
~. \label{superLieBracket}
\eea
By a straightforward computation, it may be shown that $\xi_3^a$ satisfies the conformal Killing vector equation \eqref{CKVCSS}. This analysis implies that the conformal Killing supervector fields generate a finite dimensional Lie superalgebra, which is the superconformal algebra of $(\mathcal{M}^{6|8},\nabla)$.

\subsection{Higher symmetries of the Sohnius operator}
\label{Appendix3A.2}

In six dimensions, the off-shell formulation for a hypermultiplet coupled to conformal supergravity is given
in  \cite{LTM}. 
On the mass shell, the former is described by an isospinor superfield $q^{i}$ satisfying 
the equation
\bea
\label{HM}
\nabla_{\a}^{(i} q^{j)} = 0~.
\eea
The constraint is conformally invariant provided $q^i$ is a primary superfield of dimension 2
\bea
\label{HMSW}
K^{B} q^i = 0~, \qquad
\mathbb{D} q^i = 2 q^i~.
\eea
Additionally, \eqref{HM} yields the useful corollary
\bea
\label{HMCorollary}
\nabla_{\a}^{i} \nabla_{\b}^{j} q_{j} &=& - 4 \ri \nabla_{\a \b} q^i ~.
\eea

In keeping with the general prescription given in section \ref{Chapter3.3}, a linear differential operator $\mathfrak{D}$ is said to be a symmetry of the (six-dimensional) Sohnius operator if it obeys the constraints
\begin{subequations}
	\bea
	\label{HigherSymmConstraint}
	\nabla_{\a}^{(i} \mathfrak{D}q^{j)} &=& 0 ~, \\
	\label{HMSymmSW}
	K^{A} \mathfrak{D} q^i &=& 0~, \qquad
	\mathbb{D}  \mathfrak{D} q^i = 2 q^i~.
	\eea
\end{subequations}
Modulo the equivalence \eqref{SymmEQR}, every $n$th-order symmetry operator may be brought to a canonical form given by
\bea
\label{HSCanonicalForm}
\mathfrak{D}^{(n)} &=& \sum_{k=0}^{n} \xi^{a(k)} (\nabla_{a})^k + \sum_{k=0}^{n-1} \xi^{a(k) \b}{}_i (\nabla_a)^k \nabla_{\b}^{i} + \sum_{k=0}^{n-1} \xi^{a(k)ij} (\nabla_a)^k \mathbb{J}_{ij} ~.
\eea
Here all parameters are symmetric and traceless in their vector indices, $\xi^{a(k)\b}{}_i$ is gamma-traceless, $(\g_b)_{\a \b}\xi^{a(k-1)b\b}{}_i = 0$, and $\xi^{a(k)ij}$ is symmetric in its isospinor indices.

Equation \eqref{HigherSymmConstraint} yields several constraints on the parameters of $\mathfrak{D}^{(n)}$, including
\begin{subequations}
	\bea
	\label{HSHMImplications}
	\nabla_{\a}^{i} \xi^{a(n)} &=& \frac{n}{n+4} (\g^{b (a_1)})_{\a}{}^{\b} \nabla_{\b}^i \xi^{a_2 \dots a_n)}{}_{b} ~, \\
	\xi^{a(n-1) \b i} &=& \frac{\ri}{4(n+2)} \nabla_{\a}^i \xi^{a(n-1)b} (\tilde{\g}_b)^{\a \b}~, \\
	\xi^{a(n-1) ij} &=& - \frac{\ri n (n+1)}{8(n+2)(n+3)} \nabla^{(i} \tilde{\gamma}_{b} \nabla^{j)} \xi^{a(n-1) b} ~.
	\eea
\end{subequations}
Hence, we obtain expressions for $\xi^{a(n-1) \b i}$ and $\xi^{a(n-1) ij}$ in terms of $\xi^{a(n)}$, which obeys the closed form equation \eqref{HSHMImplications}. The latter implies that $\xi^{a(n)}$ is a conformal Killing tensor.

If the background admits a conformal Killing vector superfield $\x^a$, it may be shown that the corresponding conformal isometry \eqref{CSSConfIso} yields the unique first-order symmetry operator
\bea
\label{3.104}
\nabla_{\a}^{(i} \mathfrak{D}^{(1)} q^{j)} = \nabla_{\a}^{(i} \mathscr{K}[\xi] q^{j)} = \nabla_{\a}^{(i} \Big[ \xi^C  {\nabla}_C 	+  {\L}^{kl} [\xi] \mathbb{J}_{kl} + 2 \s[\xi] \Big] q^{j)} = 0 ~.
\eea
We emphasise that \eqref{3.104} is a symmetry of the Sohnius operator on general curved backgrounds.

Next, we turn to the evaluation of $\mathfrak{D}^{(2)}$ in general conformally flat backgrounds.\footnote{By this we mean that the super-Weyl tensor vanishes, $W^{\a \b} = 0$.} When acting on $q^i$ it takes the form
\bea
\label{Rank2HS}
\mathfrak{D}^{(2)} q^i &=& \xi^{a b} \nabla_{a} \nabla_{b} q^i - \hf \xi^{a \a i} \nabla_{a} \nabla_{\a j} q^j + \xi^{a i}{}_{j} \nabla_{a} q^j + \xi^{a} \nabla_{a} q^i - \hf \xi^{\a i} \nabla_{\a j} q^j \non \\
&+& \xi^{i}{}_{j} q^j + \xi q^i ~.
\eea
The unique solution compatible with \eqref{HigherSymmConstraint} and \eqref{HMSymmSW} is
\begin{subequations}
	\label{Rank2CSSSoln}
	\bea
	\xi^{a \a i } &=& - \frac{\ri}{16} (\tilde{\g}_b)^{\a \b} \nabla_{\b}^i \xi^{ab} ~, \\
	\xi^{a ij} &=& - \frac{3 \ri}{80}  \nabla^{(i} \tilde{\g}_b \nabla^{j)} \xi^{ab} ~, \\
	\xi^{a} &=& \frac{3}{4} \nabla_{b} \xi^{ab} ~, \\
	\xi^{\a i} &=& -\frac{\ri}{20} (\tilde{\gamma}_a)^{\a \b} \nabla_{\b}^i \nabla_b \xi^{ab} + \frac{1}{800} (\tilde{\g}_a)^{\a \b} (\tilde{\g}_b)^{\g \d} \nabla_{\g}^i \nabla_{(\b}^j \nabla_{\d) j} \xi^{ab} ~, \\
	\xi^{ij} &=& - \frac{\ri}{80} \nabla^{(i} \tilde{\gamma}_{a} \nabla^{j)} \nabla_b \xi^{ab} ~, \\
	\xi &=& - \frac{3}{20} \nabla_{a} \nabla_{b} \xi^{ab} + \frac{1}{800} (\tilde{\g}_a)^{\a \b} (\tilde{\g}_b)^{\g \d} \nabla_{(\a}^i \nabla_{\b)i} \nabla^j_{(\g} \nabla_{\d)j} \xi^{ab} ~.
	\eea
\end{subequations}
In particular, we find that all parameters are expressed solely in terms of the conformal Killing tensor $\xi^{ab}$.

\end{subappendices}

\chapter{(Super)conformal higher-spin theory} \label{Chapter4}

In this chapter we study models for $\cN$-extended (super)conformal higher-spin gauge multiplets in four dimensions. Their corresponding (super)fields are higher-spin generalisations of those describing conformal (super)gravity. Further, as will be shown below, they are dual to conformal (super)current multiplets via the Noether procedure. By making use of the (super)conformal covariance inherent to conformal (super)space, the classification of such (super)currents, and consequently their dual gauge multiplets, is significantly simplified. Further, such symmetry principles allow us to show the existence of a unique (super)conformal kinetic action for each gauge (super)field and become vital tools when studying models for the latter with non-trivial dynamics.

This chapter is based on the publications \cite{HKR,KR19,KPR,KPR2,KR21,KR21-2,KPR22} and is organised as follows. The conformal higher-spin (super)current multiplets are determined in section \ref{Chapter4.1}. This allows us to compute their dual gauge (super)fields and corresponding gauge-invariant actions in conformally-flat background in section \ref{Chapter4.2}. Building on this construction, section \ref{Chapter4.3} is devoted to the study of $\sU(1)$ duality invariant models for (S)CHS gauge multiplets. In section \ref{Chapter4.4} we employ the Noether procedure to construct cubic interactions between matter multiplets and an infinite tower of (S)CHS fields. The results of this chapter are summarised in section \ref{Chapter4.5}. The main body of this chapter is accompanied by two technical appendices. In appendix \ref{Appendix4A} we review the off-shell formulation for the $\cN=2$ vector multiplet and also study the $\cN=2$ superconformal gravitino multiplet.  Appendix \ref{Appendix4B} is devoted to the study of $\sU(1)$ duality rotations for complex CHS fields.

\section{Conformal higher-spin (super)current multiplets} \label{Chapter4.1}

The objective of this section is to identify all possible conformal higher-spin (super)current multiplets in general curved backgrounds. This classification will allow us, in the following section, to elucidate the structure of their dual gauge (super)fields via the the Noether coupling \eqref{1.12}.

\subsection{Conformal higher-spin currents}
Let $m$ and $n$ be positive integers. A primary tensor field $j^{\a(m) \ad(n)}$ defined on the background spacetime is said to be a conformal current if it satisfies the transverse condition
\bea
\label{4.1}
\nabla_{\bb} j^{\b \a(m-1) \bd \ad(n-1)} = 0~.
\eea
This constraint uniquely fixes the dimension of $j^{\a(m) \ad(n)}$
\bea
\mathbb{D} j^{\a(m) \ad(n)} = \hf (m+n+4) j   ^{\a(m) \ad(n)} ~.
\eea
For $m=n=s$, $j^{\a(s) \ad(s)}$ may be restricted to be real.

Given a conformal current $j^{\a(m)\ad(n)}$ and a conformal Killing tensor $\z_{\a(p) \ad(q)}$ \eqref{N=2CKT}, with $m > p \geq 1$ and $n > q \geq 1$, it may be shown that
\bea
\mathfrak{j}^{\a(m-p) \ad(n-q)} = j^{\a(m-p) \b(p) \ad(n-q) \bd(q)} \z_{\b(p) \bd(q)} ~,
\eea
is also a conformal current.

\subsection{Conformal higher-spin supercurrents}
Let us now extend the above analysis to the supersymmetric case. Specifically, we recall the appropriate superspace constraints on a conformal supercurrent $J$ which yield a multiplet of conserved currents \eqref{4.1} at the component level.\footnote{It should be emphasised that, in general, some lower-spin supercurrents carry isospinor indices. In the $\cN=2$ case such supercurrents, and their corresponding gauge prepotentials, are studied in appendix \ref{Appendix4A}. The $\cN>2$ story will not be considered here.}  A component analysis of the supercurrent multiplets described below is given in section \ref{section4.1.3}.

Let $m$ and $n$ be positive integers. A primary tensor superfield $J^{\a(m) \ad(n)}$ defined on the background superspace will be called a conformal supercurrent if it obeys
\begin{subequations}
	\label{41}
	\bea
	\nabla_\b^i J^{\b \a(m-1) \ad(n)} &=& 0 \quad \Longrightarrow \quad \nabla^{ij} J^{\a(m) \ad(n)} = 0 ~, \\
	\bar{\nabla}_{\bd i} J^{\a(m) \bd \ad(n-1)} &=& 0 \quad \Longrightarrow \quad \bar{\nabla}_{ij} J^{\a(m) \ad(n)} = 0 ~.
	\eea
\end{subequations}
These constraints uniquely fix the superconformal properties of $J^{\a(m) \ad(n)}$
\bea
\label{eq4.4}
\mathbb{D} J^{\a(m) \ad(n)} = \hf (m+n+4) J^{\a(m) \ad(n)} ~, \qquad \mathbb{Y} J^{\a(m) \ad(n)} = \frac{\cN (m-n)}{\cN - 4} J^{\a(m) \ad(n)} ~.
\eea
For $m=n=s$, $J^{\a(s) \ad(s)}$ is invariant under $\sU(1)_R$ transformations and thus it may be restricted to be real. This special case was first described in Minkowski superspace in \cite{HST}. In the $\cN=1$ case, $J^\aa$ corresponds to the ordinary conformal supercurrent \cite{FWZ}. Additionally, the description of the higher-spin supercurrents $J^{\a(s) \ad(s)}$ was extended to $\text{AdS}^{4|4}$ in \cite{BHK18}.

When $m \geq 1$ and $n = 0$, the constraints \eqref{41} should be replaced with
\begin{subequations}
	\label{42}
	\bea
	\nabla_\b^i J^{\b \a(m-1)} &=& 0 \quad \Longrightarrow \quad \nabla^{ij} J^{\a(m) } = 0 ~, \\
	\bar{\nabla}_{ij} J^{\a(m)} &=& 0 ~.
	\eea
\end{subequations}
Consistency of \eqref{42} with the superconformal algebra implies:
\bea
\mathbb{D} J^{\a(m)} = \hf (m + 4) J^{\a(m)} ~, \qquad \mathbb{Y} J^{\a(m)} = \frac{\cN m}{\cN - 4} J^{\a(m)} ~.
\eea
For $\cN=1$, the spinor $J^\a$ naturally originates from the reduction of the $\cN=2$ conformal supercurrent to $\cN=1$ superspace \cite{KT,Sohnius79}. 

Finally, for the special case $m=n=0$, we replace \eqref{42} with
\bea
\label{43}
\nabla^{ij} J = 0 ~, \qquad \bar{\nabla}_{ij} J = 0 ~.
\eea
These imply:
\bea
\mathbb{D} J = 2 J ~, \qquad \mathbb{Y} J= 0 ~.
\eea
It is clear that $J$ may be restricted to be real and for $\cN=1$ it corresponds to the flavour current supermultiplet \cite{FWZ}, while for $\cN=2$ it describes ordinary conformal supercurrent
\cite{HST,Sohnius,KT}.

The conformal Killing tensors described in the previous chapter, see eq.\eqref{N=2CKT}, may be utilised to construct new conserved conformal currents from existing ones. Specifically, given a conformal supercurrent $J^{\a(m)\ad(n)}$ and a conformal Killing tensor $\z_{\a(p) \ad(q)}$, with $m \geq p$ and $n \geq q$, it may be shown that
\bea
\mathfrak{J}^{\a(m-p) \ad(n-q)} = J^{\a(m-p) \b(p) \ad(n-q) \bd(q)} \z_{\b(p) \bd(q)} ~,
\eea
is also a conformal supercurrent.

\subsection{Superspace and component reduction}
\label{section4.1.3}
To conclude our discussion of conformal higher-spin supercurrents, it is instructive to comment on the superspace and component reductions of the conformal (super)currents described above. This analysis requires us to work in conformally flat backgrounds, thus we set the super-Weyl tensor to zero.

First, we consider conformal supercurrents in $\cN = N \geq 2$ superspace. Let $\Nabla_\a^{\hat i} $, $\bar{\Nabla}^\ad_{\hat i} $ and $\Nabla_{\aa} = \frac{\ri}{2(N-1)} \{\Nabla_\a^{\hat i} , \bar \Nabla_{\ad \hat{i}} \}$, $\hat{i} = \underline{1},~ \dots ,~ \underline{N - 1}$, be the covariant derivatives of $\cN=N-1$ conformal superspace. We will define them as follows: $ \Nabla_\a^{\hat{i}} {\bm U} =  \nabla_\a^{\hat{i}} U|_{\theta^\a_{\underline {N}} = \bar{\theta}_\ad^{ \underline {N}} = 0}$ and $\bar{\Nabla}^\ad_{\hat i} {\bm U}= \bar{\nabla}^{\ad}_{\hat{i}} U|_{\theta^\a_{\underline {N}} = \bar{\theta}_\ad^{ \underline {N}} = 0}$, where ${\bm U} :=U|_{\theta^\a_{\underline {N}} = \bar{\theta}_\ad^{ \underline {N}} = 0}$
is the $\cN = N-1$ projection of $U$.

Returning to the reduction procedure, we first consider the supercurrent $J^{\a(m)\ad(n)}$ \eqref{eq4.4}, with $m,n \geq 1$. It contains four independent $\cN=N-1$ conserved current supermultiplets defined by:
\begin{subequations}
	\label{410}
	\bea
	\bm{J}^{\a(m) \ad(n)} &=& J^{\a(m) \ad(n)} |_{\theta^\a_{\underline {N}} = \bar{\theta}_\ad^{ \underline {N}} = 0} ~, \\
	\bm{J}^{\a(m+1) \ad(n)} &=& \nabla^{\a \underline N} J^{\a(m) \ad(n)} |_{\theta^\a_{\underline {N}} = \bar{\theta}_\ad^{ \underline {N}} = 0} ~, \\
	\bm{J}^{\a(m) \ad(n+1)}{} &=& \bar \nabla^{\ad}_{\underline N} J^{\a(m) \ad(n)} |_{\theta^\a_{\underline {N}} = \bar{\theta}_\ad^{ \underline {N}} = 0} ~, \\
	\bm{J}^{\a(m+1) \ad(n+1)} &=& \hf \big[ \nabla^{\a \underline{N}} , \bar \nabla_{\underline{N}}^{\ad} \big] J^{\a(m) \ad(n)} |_{\theta^\a_{\underline {N}} = \bar{\theta}_\ad^{ \underline {N}} = 0} 
	- \frac{1}{2(m+n+N+1)} \big[ \Nabla^{\a {\hat{i}}} , \bar \Nabla^{\ad}_{\hat{i}} \big] \bm{J}^{\a(m) \ad(n)} \non \\
	&\phantom{=}  & - \frac{\ri (m-n)}{m+n+N+1} \Nabla^{\aa} \bm{J}^{\a(m) \ad(n)} ~.
	\eea
\end{subequations}
Similarly, if we fix $n=0$, $J^{\a(m)}$ \eqref{42} is composed of four $\cN=N-1$ supercurrents
\begin{subequations}
	\bea
	\bm{J}^{\a(m)} &=& J^{\a(m)} |_{\theta^\a_{\underline {N}} = \bar{\theta}_\ad^{ \underline {N}} = 0} ~, \\
	\bm{J}^{\a(m+1)} &=& \nabla^{\a \underline N} J^{\a(m)} |_{\theta^\a_{\underline {N}} = \bar{\theta}_\ad^{ \underline {N}} = 0} ~, \\
	\bm{J}^{\a(m) \ad} &=& \bar \nabla^{\ad}_{\underline N} J^{\a(m)} |_{\theta^\a_{\underline {N}} = \bar{\theta}_\ad^{ \underline {N}} = 0} ~, \\
	\bm{J}^{\a(m+1) \ad} &=& \hf \big[ \nabla^{\a \underline N} , \bar \nabla_{\underline{N}}^{\ad} \big] J^{\a(m)} |_{\theta^\a_{\underline {N}} = \bar{\theta}_\ad^{ \underline {N}} = 0} 
	- \frac{1}{2(m+N+1)} \big[ \Nabla^{\a {\hat{i}}} , \bar \Nabla^{\ad}_{\hat{i}} \big] \bm{J}^{\a(m)} \non \\
	&\phantom{=}& - \frac{\ri m}{m+N+1} \Nabla^{\a \ad} \bm{J}^{\a(m)} ~.
	\eea
\end{subequations}
Finally, upon reduction of $J$ \eqref{43} we obtain three $\cN=N-1$ current multiplets
\begin{subequations}
	\bea
	\bm{J}^{} &=& J^{} |_{\theta^\a_{\underline {N}} = \bar{\theta}_\ad^{ \underline {N}} = 0} ~, \\
	\bm{J}^{\a} &=& \nabla^{\a \underline N} J |_{\theta^\a_{\underline {N}} = \bar{\theta}_\ad^{ \underline {N}} = 0} ~, \\
	\bm{J}^{\a \ad} &=& \hf \big[ \nabla^{\a \underline N} , \bar \nabla_{\underline{N}}^{\ad} \big] J |_{\theta^\a_{\underline {N}} = \bar{\theta}_\ad^{ \underline {N}} = 0} - \frac{1}{2(N+1)} \big[ \Nabla^{\a \hat{i}} , \bar \Nabla^{\ad}_{\hat{i}} \big] \bm{J} ~.
	\eea
\end{subequations}

The superspace reduction procedure described above provides a systematic method to reduce any conformal supercurrent to $\cN=1$ superspace. Thus, to complete this story, it is also necessary to describe the reduction of the $\cN=1$ conformal supercurrents to spacetime. Given such a superfield $J^{\a(m) \ad(n)}$, with $m,n \geq 1$, it contains four independent component fields, which can be chosen as follows
\begin{subequations}
	\bea
	\bm{j}^{\a(m) \ad(n)} & = & J^{\a(m) \ad(n)} |_{\theta^\a = \bar{\theta}_\ad = 0} ~, \\
	\bm{j}^{\a(m+1) \ad(n)} & = & \nabla^{\a} J^{\a(m) \ad(n)} |_{\theta^\a = \bar{\theta}_\ad = 0} ~, \\
	\bm{j}^{\a(m) \ad(n+1)} & = & \bar{\nabla}^{\ad} J^{\a(m) \ad(n)} |_{\theta^\a = \bar{\theta}_\ad = 0} ~, \\
	\bm{j}^{\a(m+1) \ad(n+1)} & = & [ \nabla^{\a} , \bar{\nabla}^{\ad} ] J^{\a(m) \ad(n)} |_{\theta^\a = \bar{\theta}_\ad = 0} - \frac{\ri (m - n)}{m+n+2} \bm{\nabla}^{\a \ad} \bm{j}^{\a(m) \ad(n)} ~.
	\label{6.31d}
	\eea
\end{subequations}
Each of the component fields defined above satisfy \eqref{4.1}, hence they constitute conformal currents.

Next, if $m \geq 1$ and $n=0$, the conformal supercurrent takes the form $J^{\a(m)}$. At the component level, it contains two conserved currents in its multiplet:
\begin{subequations}
	\bea
	\bm{j}^{\a(m) \ad} &=& \bar{\nabla}^{\ad} J^{\a(m)} |_{\theta^\a = \bar{\theta}_\ad = 0} ~, \\
	\bm{j}^{\a(m+1) \ad} &=& [ \nabla^{\a} , \bar{\nabla}^{\ad} ] J^{\a(m)} |_{\theta^\a = \bar{\theta}_\ad = 0} - \frac
	{\ri m }{m+2} \bm{\nabla}^{\a \ad} \bm{j}^{\a(m) \ad(n)} ~.
	\eea
\end{subequations}

The final case of interest is that of the flavour current $J$. It contains a single current at the component level, 
\bea
\bm{j}^{\a \ad} = [ \nabla^{\a} , \bar{\nabla}^{\ad} ] J |_{\theta^\a = \bar{\theta}_\ad = 0}~.
\eea

\section{Conformal higher-spin (super)fields} \label{Chapter4.2}

In the previous section, we described the conformal higher-spin (super)currents in general curved backgrounds. Now, by making use of the method of supercurrent multiplets advocated in \cite{HST,OS,BdeRdeW}, we determine their associated gauge (super)fields by requiring that the Noether couplings
\begin{subequations}
	\label{NoetherCoupling}
	\begin{align}
		\label{NoetherCoupling-a}
		\cN = 0:& \qquad \mathcal{S}^{(m,n)}_{\rm N.C.} = \int \rd^4x \, e \, h_{\a(m) \ad(n)} j^{\a(m) \ad(n)}~, \\
		\label{NoetherCoupling-b}
		\cN > 0:& \qquad \mathcal{S}^{(m,n)}_{\rm N.C.} = \int \text{d}^{4|4\cN}z \, E \, H_{\a(m) \ad(n)} J^{\a(m) \ad(n)} ~,
	\end{align}
\end{subequations}
are locally (super)conformal and gauge-invariant.

\subsection{Conformal higher-spin gauge models} \label{section4.1}

We recall from the previous section that conformal currents are primary tensor fields $j^{\a(m) \ad(n)}$, with $m,n \geq 1$, subject to the conservation equation \eqref{4.1}. It then follows from \eqref{NoetherCoupling-a} that their duals, $h_{\a(m) \ad(n)}$, possess the gauge freedom \cite{KMT,KP}
\bea
\label{CHS}
\d_{\ell} h_{\a(m) \ad(n)} = \nabla_{\aa} \ell_{\a(m-1) \ad(n-1)} ~,
\eea
where $\ell_{\a(m-1) \ad(n-1)}$ is, in general, complex unconstrained. We note that the special cases $(m,n) = (1,1),~(2,1),~(2,2)$ and $(3,1)$ describe the Maxwell field, conformal gravitino, conformal graviton and conformal hook field \cite{Curtright2}, respectively. Additionally, the $m=n=s$ and $m=n+1=s+1$ cases describe a conformal spin-$s$ and $s+\hf$ field, respectively \cite{FT}. Further, in former case, we are able to choose $h_{\a(s) \ad(s)}$ to be real.

Requiring that both $h_{\a(m) \ad(n)}$ and $\ell_{\a(m-1) \ad(n-1)}$ are primary,
\begin{subequations}
	\label{CHSprimary}
	\bea
	K^{\bb} h_{\a(m) \ad(n)} = 0 ~, \qquad K^{\bb} \ell_{\a(m-1) \ad(n-1)} = 0 ~,
	\eea
	uniquely fixes the dimension of the gauge field
	\bea
	\mathbb{D} h_{\a(m) \ad(n)} &=& \hf \big(4 -(m+n)\big) h_{\a(m) \ad(n)} ~.
	\eea
\end{subequations}

From $h_{\a(m)\ad(n)}$ and its conjugate $\bar{h}_{\a(n)\ad(m)}$, we construct the higher-derivative descendants \cite{KMT,KP}
\begin{subequations}
	\label{4.20}
	\bea
	\hat{\mathfrak{C}}_{\a(m+n)} (h) = (\nabla_{\a}{}^{\bd})^n h_{\a(m) \bd(n)} ~, \\
	\check{\mathfrak{C}}_{\a(m+n)} (\bar{h}) = (\nabla_{\a}{}^{\bd})^m \bar{h}_{\a(n) \bd(m)} ~.
	\eea
\end{subequations}
We note that when the gauge field is real $(m=n=s)$, the above descendants coincide
\begin{align}
	m=n=s \quad \Longrightarrow \quad   {\mathfrak{C}}_{\a(2s)} (h) \equiv \hat{\mathfrak{C}}_{\a(2s)} (h) = \check{\mathfrak{C}}_{\a(2s)} (h) ~.
\end{align}
Additionally, we point out that $\mathfrak{C}_{\a(2)}(h)$ is Maxwell's field strength and $\mathfrak{C}_{\a(4)}(h)$ is the linearised Weyl tensor.
The descendants \eqref{4.20} have the following dimensions:
\begin{subequations}
	\bea
	\hat{\mathfrak{C}}_{\a(m+n)} (h) &=&  \hf (n-m +4) 	\hat{\mathfrak{C}}_{\a(m+n)} (h) ~, \\
	\check{\mathfrak{C}}_{\a(m+n)} (\bar{h}) &=& \hf (m-n + 4) \check{\mathfrak{C}}_{\a(m+n)} (\bar{h}) ~.
	\eea
\end{subequations}
Further, it may be shown that they are primary,
\bea
K^\bb \hat{\mathfrak{C}}_{\a(m+n)} (h) = 0 ~, \quad K^\bb \check{\mathfrak{C}}_{\a(m+n)} (\bar{h}) = 0 ~.
\eea

These properties imply that the action \cite{KMT,KP}
\bea
\label{CHSaction}
{\cS}^{(m,n)}[h,\bar{h}] = \ri^{m+n} \int \rd^4x \, e \, \hat{\mathfrak{C}}^{\a(m+n)} (h) \check{\mathfrak{C}}_{\a(m+n)} ({\bar h}) + \text{c.c.}
\eea
is locally superconformal. The overall factor of $\ri^{m+n}$ above has been chosen due to the identity
\bea
\ri^{m+n+1} \int \rd^4x \, e \, \hat{\mathfrak{C}}^{\a(m+n)} (h) \check{\mathfrak{C}}_{\a(m+n)} ({\bar h}) + \text{c.c.} = 0~,
\eea
which holds up to a total derivative for any conformally flat background. 

We now restrict our attention to conformally flat spacetimes. In these geometries it may be shown that the descendants \eqref{4.20} are gauge-invariant field strengths
\bea
\d_{\ell}\hat{\mathfrak{C}}_{\a(m+n)} (h) = 0 ~, \quad \d_{\ell} \check{\mathfrak{C}}_{\a(m+n)} ({\bar h}) = 0 ~.
\eea
As a result, it follows that the action \eqref{CHSaction} is also gauge-invariant.

\subsection{Superconformal higher-spin gauge models} \label{section4.2.2}

Having described the CHS fields $h_{\a(m)\ad(n)}$ and computed their free actions above, we now study their supersymmetric extensions by utilising the Noether coupling \eqref{NoetherCoupling-b}. We recall that the first family of conformal supercurrents take the form $J^{\a(m) \ad(n)}$, $m,n \geq 1$, and are subject to the conservation equation \eqref{41}. It then follows that their duals, $H_{\a(m) \ad(n)}$, are defined modulo:
\begin{subequations}
	\label{4.27}
	\bea
	\label{SCHS1}
	\d_{\z,\l} H_{\a(m) \ad(n)} = \bar{\nabla}_{\ad i} \z_{\a(m) \ad(n-1)}{}^i + \nabla_{\a}^i \l_{\a(m-1) \ad(n) i} ~,
	\eea
	where $\z_{\a(m) \ad(n-1)}{}^i$ and $\l_{\a(m-1) \ad(n) i}$ are complex unconstrained. For $m=n$, the flat analogue of the tranformation law \eqref{SCHS1} first appeared in \cite{HST}.
	Further, in the $\cN=1$ case, $H_{\aa}$ describes the conformal supergravity multiplet \cite{FZ2} and the gauge transformations \eqref{SCHS1} were proposed in \cite{KMT}, see
	also \cite{KP}.
	
	When $m \geq 1,~ n=0$, the supercurrents take the form $J^{\a(m)}$ and satisfy eq. \eqref{42}. It is clear that their corresponding prepotentials, $H_{\a(m)}$, are characterised by the gauge transformation law
	\bea
	\label{SCHS2}
	\d_{\z,\l} H_{\a(m)} =\bar{\nabla}_{ij} \z_{\a(m)}{}^{ij} + \nabla_{\a}^i \l_{\a(m-1) i} ~,
	\eea
	where $\z_{\a(m)}{}^{ ij }$ and $\l_{\a(m-1) i}$ are complex unconstrained. For $\cN=1$, the prepotential $H_\a$ describes the conformal gravitino \cite{GS} multiplet.
	
	Finally, for $m=n=0$, we obtain the real scalar supercurrent $J$. It is constrained by \eqref{43} and its dual, $H$, is real and defined modulo the gauge transformations:
	\bea
	\label{SCHS3}
	\d_{\z} H = \bar \nabla_{ij} \z^{ij} + \text{c.c.}
	\eea
\end{subequations}
The gauge parameter $\z^{ij}$ is complex unconstrained. It should be noted that, for $\cN=1$ and $\cN=2$, $H$ describes a vector multiplet, see e.g. \cite{BK}, and the conformal supergravity multiplet \cite{Sohnius79,HST,KT}, respectively.

\subsubsection{The $H_{\a(m) \ad(n)}$ prepotentials}

First, we study the prepotentials $H_{\a(m) \ad(n)}$, $m,n\geq1$, which possess the gauge freedom \eqref{SCHS1}. Requiring that both this multiplet and its corresponding gauge parameters, $\z_{\a(m) \ad(n-1)}{}^i$ and $\l_{\a(m-1) \ad(n) i}$, are primary,
\bea
K^B H_{\a(m) \ad(n)} = 0 ~, \quad K^B \z_{\a(n) \ad(n-1)}{}^i = 0 ~, \quad K^B \l_{\a(m-1) \ad(n) i} = 0 ~, 
\eea
uniquely fixes the dimension and $\sU(1)_R$ charge of $H_{\a(m) \ad(n)}$
\begin{subequations}
	\bea
	\mathbb{D} H_{\a(m) \ad(n)} &=& - \hf (m+n+4\cN - 4) H_{\a(m) \ad(n)} ~, \\
	\mathbb{Y} H_{\a(m) \ad(n)} &=& - \frac{\cN (m-n)}{\cN - 4} H_{\a(m) \ad(n)} ~.
	\eea
\end{subequations}
This implies that for $m = n = s$, we are able to choose $H_{\a(s) \ad(s)}$ to be real, in which case \eqref{SCHS1} reduces to
\bea
\label{421}
\d_\z H_{\a(s) \ad(s)} = \bar{\nabla}_{\ad i} \z_{\a(s) \ad(s-1)}{}^i + \text{c.c.}
\eea

From the prepotential $H_{\a(m) \ad(n)}$ and its conjugate $\bar{H}_{\a(n) \ad(m)}$, we construct the higher-derivative chiral descendants\footnote{In the flat superspace limit, these chiral descendants reduce to those introduced in \cite{SG,GGRS}. In the $\cN=1$ case their $\text{AdS}^{4|4}$ superspace extension was given in \cite{KMT}.}
\begin{subequations}
	\label{linSW1}
	\bea
	\hat{\mathfrak{W}}_{\a(m+n+\N)} (H) &=& - \frac{1}{2^\cN \cN!(\cN+1)!} \bar \nabla^{2\cN} (\nabla_{\a}{}^{\bd})^n \nabla_{\a(\cN)} H_{\a(m) \bd(n)} ~, \\
	\check{\mathfrak{W}}_{\a(m+n+\cN)} (\bar H) &=& - \frac{1}{2^\cN \cN!(\cN+1)!} \bar \nabla^{2\cN} (\nabla_{\a}{}^{\bd})^m \nabla_{\a(\N)} \bar H_{\a(n) \bd(m)} ~.
	\eea
\end{subequations}
We note that when the gauge prepotential is real $(m=n=s)$, they coincide
\begin{align}
	m=n=s \quad \Longrightarrow \quad   {\mathfrak{W}}_{\a(2s+\N)} (H) \equiv \hat{\mathfrak{W}}_{\a(2s+\N)} (H) = \check{\mathfrak{W}}_{\a(2s+\N)} (H) ~.
\end{align}
These chiral descendants have the following dimensions:
\begin{subequations}
	\bea
	\mathbb{D} \hat{\mathfrak{W}}_{\a(m+n+\cN)} (H) &=&  \hf (n-m -\N +4) \hat{\mathfrak{W}}_{\a(m+n+\N)} (H) ~, \\
	\mathbb{D} \check{\mathfrak{W}}_{\a(m+n+\cN)} (\bar H) &=& \hf (m-n -\N + 4) \check{\mathfrak{W}}_{\a(m+n+\N)}  (\bar H) ~.
	\eea
\end{subequations}
Further, it may be shown that they are primary,
\bea
K^B \hat{\mathfrak{W}}_{\a(m+n+\N)} (H) = 0 ~, \quad K^B \check{\mathfrak{W}}_{\a(m+n+\N)} (\bar H) = 0 ~.
\eea

These properties imply that the action
\bea
\label{SCHSaction1}
\cS^{(m,n)}[H,\bar{H}] = \ri^{m+n} \int \rd^4x \rd^{2\N} \q \, \cE\, \hat{\mathfrak{W}}^{\a(m+n+\cN)} (H) \check{\mathfrak{W}}_{\a(m+n+\cN)} (\bar H) + \text{c.c.}
\eea
is locally superconformal.\footnote{We point out that, in the $\cN=1$ case, action \eqref{SCHSaction1} was proposed in \cite{KMT,KP}.} The overall factor of $\ri^{m+n}$ in \eqref{SCHSaction1} has been chosen due to the identity
\bea
\ri^{m+n+1} \int \rd^4x \rd^{2\N} \q \, \cE\, \hat{\mathfrak{W}}^{\a(m+n+\cN)} (H) \check{\mathfrak{W}}_{\a(m+n+\cN)} (\bar H) + \text{c.c.} = 0~,
\eea
which holds up to a total derivative for any conformally flat background. 

We restrict our attention to conformally flat superspaces. In these geometries it may be shown that the chiral descendants \eqref{linSW1} are gauge-invariant
\bea
\d_{\z,\l} \hat{\mathfrak{W}}_{\a(m+n+\N)} (H) = 0 ~, \quad \d_{\z,\l} \check{\mathfrak{W}}_{\a(m+n+\N)} (\bar H) = 0 ~.
\eea
As a result, the action \eqref{SCHSaction1} is also gauge-invariant.

\subsubsection{The $H_{\a(m)}$ prepotentials}

The gauge prepotentials $H_{\a(m)}$, $m \geq 1$ are defined modulo the transformations \eqref{SCHS2}. Similar to the $n\neq0$ case, requiring that both the prepotentials and gauge parameters are superconformally primary,
\bea
K^B H_{\a(m)} = 0 ~, \quad K^C \z_{\a(n)}{}^{ij} = 0 ~, \quad K^B \l_{\a(m-1) i} = 0 ~,
\eea
uniquely determines the dimension and $\sU(1)_R$ charge of $H_{\a(m)}$
\bea
\mathbb{D} H_{\a(m)} = -\hf (m + 4\cN - 4) H_{\a(m)} ~, \quad \mathbb{Y} H_{\a(m) } = - \frac{\cN m}{\cN - 4} H_{\a(m)} ~.
\eea

Associated with the prepotential $H_{\a(m)}$ (and its conjugate $\bar{H}_{\ad(m)}$) are the chiral descendants:
\begin{subequations}
	\label{linSW2}
	\bea
	\hat{\mathfrak{W}}_{\a(m+\N)} (H) &=& - \frac{1}{2^\cN \cN!(\cN+1)!} \bar \nabla^{2\N} \nabla_{\a(\N)} H_{\a(m)} ~, \\
	\check{\mathfrak{W}}_{\a(m+\N)} (\bar H) &=& - \frac{1}{2^\cN \cN!(\cN+1)!} \bar \nabla^{2\N} (\nabla_{\a}{}^{\bd})^m \nabla_{\a(\N)} \bar H_{\bd(m)} ~.
	\eea
\end{subequations}
It may readily be shown that their dimensions are
\begin{subequations}
	\bea
	\mathbb{D} \hat{\mathfrak{W}}_{\a(m+\cN)} (H) &=& - \hf (m + \N - 2)  \hat{\mathfrak{W}}_{\a(m+\N)} (H) ~,	\\
	\mathbb{D} \check{\mathfrak{W}}_{\a(m+\N)} (\bar H) &=& \hf(m - \N + 2) \check{\mathfrak{W}}_{\a(m+\N)} (\bar H) ~.
	\eea
\end{subequations}
and that they are primary
\bea
K^B \hat{\mathfrak{W}}_{\a(m+\N)} (H) = 0 ~, \quad K^B \check{\mathfrak{W}}_{\a(m+\N)} (\bar H) = 0 ~.
\eea
As a result, the following action
\bea
\label{SCHSaction2}
\cS^{(m,0)}[H,\bar{H}] = \ri^{m} \int \rd^4x \rd^{2\N} \q \, \cE\, \hat{\mathfrak{W}}^{\a(m+\N)} (H) \check{\mathfrak{W}}_{\a(m+\N)} (\bar H) + \text{c.c.}
\eea
is locally superconformal. The overall coefficient of $\ri^{m}$ has been chosen since on conformally flat backgrounds
\bea
\ri^{m+1} \int \rd^4x \rd^{2\N} \q \, \cE\, \hat{\mathfrak{W}}^{\a(m+\cN)} (H) \check{\mathfrak{W}}_{\a(m+\cN)} (\bar H) + \text{c.c.} = 0~,
\eea
i.e. it is a total derivative.

In backgrounds with vanishing super-Weyl tensor, a routine calculation allows us to show that \eqref{linSW2} are gauge-invariant field strengths
\bea
\d_{\z,\l} \hat{\mathfrak{W}}_{\a(m+\cN)} (H) = 0 ~, \quad \d_{\z,\l} \check{\mathfrak{W}}_{\a(m+\cN)} (\bar H) = 0 ~.
\eea
Thus, the action \eqref{SCHSaction2} proves to be gauge-invariant.

\subsubsection{The $H$ prepotential}

The scalar gauge prepotential $H = \bar H $ possesses the gauge freedom \eqref{SCHS3}. The requirement that both $H$ and $\z^{ij}$ are superconformally primary,
\bea
K^B H = 0 ~, \quad K^B \z^{ ij } = 0 ~,
\eea
leads to
\bea
\mathbb{D} H = - 2(\N-1) H ~, \quad \mathbb{Y} H = 0 ~.
\eea
From the prepotential $H$, we may construct the single chiral descendant\footnote{It should be noted that for $\cN=1$ and $\cN=2$, the superfield \eqref{linSW3} coincides with the field strength of a $\sU(1)$ vector multiplet and the linearised super-Weyl tensor, respectively.}
\bea
\label{linSW3}
\mathfrak{W}_{\a(\N)} (H) &=& - \frac{1}{2^\cN \cN!(\cN+1)!} \bar \nabla^{2\N} \nabla_{\a(\N)} H ~.
\eea
It has the following superconformal properties
\bea
\mathbb{D} \mathfrak{W}_{\a(\N)} (H) = \hf(4 - \cN) \mathfrak{W}_{\a(\N)} (H) ~, \quad K^B \mathfrak{W}_{\a(\N)} (H) = 0 ~,
\eea
which imply that the following action 
\bea
\label{SCHSaction3}
\cS^{(0,0)}[H] = \int \rd^4x \rd^{2\N} \q \, \cE\, \mathfrak{W}^{\a(\N)} (H) \mathfrak{W}_{\a(\N)} ( H) + \text{c.c.}
\eea
is locally superconformal.
We note that 
\bea
\ri \int \rd^4x \rd^{2\N} \q \, \cE\, \mathfrak{W}^{\a(\N)} (H) \mathfrak{W}_{\a(\N)} ( H) + \text{c.c.} = 0 ~,
\eea
is a total derivative in backgrounds with vanishing super-Weyl tensor.

When the background geometry is conformally flat, it is easily shown that
\bea
\d_{\z} \mathfrak{W}_{\a(\N)} (H) = 0 ~.
\eea
Hence, \eqref{SCHSaction3} is gauge-invariant. Actually, in the $\cN=1$ case, $\mathfrak{W}_\a(H)$ proves to be gauge-invariant on arbitrary backgrounds.

\subsection{Superspace and component reduction}
\label{section4.2.3}

This subsection is devoted to a discussion of the component content of the gauge superfields described above in conformally flat backgrounds.\footnote{In the presence of conformal curvature, the gauge transformations of its constituent fields may entangle to each other, see \cite{KPR} for more details.} To do this, we will first reduce all $\cN = N \geq 2$ prepotentials to $\cN = 1$ superspace and then complete the picture by performing a component reduction of each $\cN=1$ gauge multiplet.

We recall that a superconformal gauge multiplet $H_{\a(m) \ad(n)}$, $m,n \geq 0$, is defined modulo the transformations \eqref{4.27}. Utilising this freedom, we construct a Wess-Zumino gauge on $H_{\a(m) \ad(n)}$ such that the only non-vanishing $\N=N-1$ superfields in its multiplet are:
\begin{subequations}
	\label{4.16}
	\begin{align}
		\bm{H}_{\a(m+1) \ad(n+1)} &= \frac{1}{2} [\nabla_{\a}^{\underline{N}} , \bar{\nabla}_{\ad \underline{N}}] H_{\a(m) \ad(n)} |_{\theta^\a_{\underline {N}} = \bar{\theta}_\ad^{ \underline {N}} = 0} ~, \\
		\bm{\Psi}_{\a(m+1) \ad(n)} &= - \frac{1}{4} \nabla_{\a}^{\underline{N}} (\bar{\nabla}_{\underline{N}})^2 H_{\a(m) \ad(n)} |_{\theta^\a_{\underline {N}} = \bar{\theta}_\ad^{ \underline {N}} = 0} ~, \\
		\bm{\Phi}_{\a(m) \ad(n+1)} &= - \frac{1}{4} \bar{\nabla}_{\ad \underline{N}} ({\nabla}^{\underline{N}})^2 H_{\a(m) \ad(n)} |_{\theta^\a_{\underline {N}} = \bar{\theta}_\ad^{ \underline {N}} = 0} ~, \\
		\bm{G}_{\a(m) \ad(n)} &= 
		\frac{1}{32}  \{ (\nabla^{\underline{N}})^2 , (\bar{\nabla}_{\underline{N}})^2 \}  H_{\a(m) \ad(n)} |_{\theta^\a_{\underline {N}} = \bar{\theta}_\ad^{ \underline {N}} = 0} \non \\
		&\phantom{=}+ \frac{1}{8(m+n+3)} [\Nabla^{\b \underline{N}} , \bar{\Nabla}^{\bd}_{\underline{N}}] \bm{H}_{\a(m) \b \ad(n) \bd} 
		\non \\
		&\phantom{=}
		- \frac{\ri(m-n)}{4(m+n+3)} \Nabla^{\bb} \bm{H}_{\a(m) \b \ad(n) \bd} ~.
	\end{align}
\end{subequations}
We emphasise that these superfields have been defined such that they are primary
\begin{subequations}
	\bea
	K^B \bm{H}_{\a(m+1) \ad(n+1)} &=& 0 ~, \qquad K^B \bm{\Psi}_{\a(m+1) \ad(n)} = 0 ~, \\
	K^B \bm{\Phi}_{\a(m) \ad(n+1)} &=& 0 ~, \qquad K^B \bm{G}_{\a(m) \ad(n)} = 0 ~.
	\eea
\end{subequations}
Further, owing to the residual gauge freedom associated with this Wess-Zumino gauge, each prepotential in \eqref{4.16} is defined modulo the appropriate gauge transformations \eqref{4.27}.

This superspace reduction procedure allows us to systematically reduce any superconformal higher-spin gauge multiplet to its constituent $\cN=1$ superfields. We now complete our analysis by imposing an appropriate Wess-Zumino gauge on some $\cN=1$ gauge superfield $H_{\a(m)\ad(n)}$, $m,n \geq 0$, and then reduce it to components. In such a gauge the only non-vanishing component fields are
\begin{subequations}
	\label{455}
	\begin{align}
		\bm{h}_{\a(m+1) \ad(n+1)} &= \frac{1}{2} [\nabla_{\a} , \bar{\nabla}_{\ad}] H_{\a(m) \ad(n)} |_{\theta^\a = \bar{\theta}_\ad = 0} ~, \\
		\bm{\psi}_{\a(m+1) \ad(n)} &= - \frac{1}{4} \nabla_{\a} \bar{\nabla}^2 H_{\a(m) \ad(n)} |_{\theta^\a = \bar{\theta}_\ad = 0} ~, \\
		\bm{\phi}_{\a(m) \ad(n+1)} &= - \frac{1}{4} \bar{\nabla}_{\ad} {\nabla}^2 H_{\a(m) \ad(n)} |_{\theta^\a = \bar{\theta}_\ad = 0} ~, \\
		\bm{g}_{\a(m) \ad(n)} &= 
		\frac{1}{32} \{ \nabla^2 , \bar{\nabla}^2 \}  H_{\a(m) \ad(n)} |_{\theta^\a = \bar{\theta}_\ad = 0} - \frac{\ri(m-n)}{4(m+n+3)} \Nabla^{\bb} \bm{h}_{\a(m) \b \ad(n) \bd} ~.
	\end{align}
\end{subequations}
Each component field above has been defined such that it is primary
\begin{subequations}
	\bea
	K^\bb \bm{h}_{\a(m+1) \ad(n+1)} &=& 0 ~, \qquad K^\bb \bm{\psi}_{\a(m+1) \ad(n)} = 0 ~, \\
	K^\bb \bm{\phi}_{\a(m) \ad(n+1)} &=& 0 ~, \qquad K^\bb \bm{g}_{\a(m) \ad(n)} = 0 ~.
	\eea
\end{subequations}
Additionally, due to the residual gauge freedom associated with this Wess-Zumino gauge, each component field in \eqref{455} is defined modulo the gauge transformations \eqref{CHS}, provided that it possesses at least one undotted and one dotted spinor index. The component fields \eqref{455} carrying only (un)dotted spinor indices constitute conformal non-gauge fields \cite{KP}. These are known to play a key role in the description of gauge-invariant models for CHS fields in backgrounds with Weyl curvature, see e.g. \cite{Kuzenko:2019eni}.

\section{Duality-invariant models for (S)CHS fields} \label{Chapter4.3}

Having identified the (S)CHS gauge multiplets and computed their corresponding gauge-invariant free actions above, we now construct models for these (super)fields with non-trivial dynamics. In particular, building on the success of the so-called ``ModMax" theory \cite{BLST}, see also \cite{Kosyakov}, and its $\cN=1$ extension \cite{BLST2,K21}, we formulate the theory of $\sU(1)$ duality invariant models for (S)CHS fields in conformally flat backgrounds.

\subsection{Duality-invariant CHS models} \label{section4.3.1}

Consider a dynamical system describing the propagation of a conformal spin-$s$ field
$h_{\a(s) \ad(s)}$, with $s\geq 1$, 
in a conformally flat spacetime. Its action functional 
$\cS^{(s)}[\mathfrak{C},\bar{\mathfrak C}]$ is assumed to depend 
only on the field strength $\mathfrak{C}_{\a(2s)}$\footnote{For notational simplicity we use the shorthand $\mathfrak{C}_{\a(2s)} \equiv \mathfrak{C}_{\a(2s)}(h)$.} (and its conjugate 
$\bar{\mathfrak{C} }_{\ad (2s) }$), defined in \eqref{4.20}.
It is important to note that the latter obeys the Bianchi identity
\be
\label{4.55}
(\nabla^{\b}{}_{\ad})^s \mathfrak{C}_{\a(s) \b(s)} 
=
(\nabla_{\a}{}^{\bd})^s \bar{\mathfrak{C}}_{\ad(s) \bd(s)}  ~.
\ee

Now, we assume that  $\cS^{(s)}[\mathfrak{C},\bar{\mathfrak C}]$ is extended  to be a functional of an unconstrained field $\mathfrak{C}_{\a(2s)}$ and its conjugate. We then introduce the dual field
\be
\ri \mathfrak{M}_{\a(2s)} := \frac{\d \cS^{(s)}[\mathfrak{C},\bar{\mathfrak{C}}]}{\d \mathfrak{C}^{\a(2s)}} ~,
\ee
where we have defined
\be
\d \cS^{(s)}[\mathfrak{C},\bar{\mathfrak{C}}] = \int \rd^4x\, e \, \d \mathfrak{C}^{\a(2s)} \frac{\d \cS^{(s)}[\mathfrak{C},\bar{\mathfrak{C}}]}{\d \mathfrak{C}^{\a(2s)}}  + \text{c.c.}
\ee
Varying $\cS^{(s)}[\mathfrak C , \bar{\mathfrak C} ]$ with respect to $h_{\a(s) \ad(s)}$ yields the equation of motion
\be
\label{4.58}
(\nabla^{\b}{}_{\ad})^s \mathfrak{M}_{\a(s) \b(s)} 
=
(\nabla_{\a}{}^{\bd})^s \bar{\mathfrak{M}}_{\ad(s) \bd(s)}  ~.
\ee

A crucial outcome of our analysis above is that the functional form of the equation of motion \eqref{4.58} mirrors that of the Bianchi identity \eqref{4.55}. As a result, it is clear that their union is invariant under the $\sSO(2)\cong \sU(1)$ duality transformations:
\be
\label{4.59}
\d_\l \mathfrak{C}_{\a(2s)} = \l \mathfrak{M}_{\a(2s)} ~, \quad \d_\l \mathfrak{M}_{\a(2s)} = - \l \mathfrak{C}_{\a(2s)} ~,
\ee
where $\l$ is a constant, real parameter. One may then obtain two equivalent expressions for the variation of $\cS^{(s)}[\mathfrak C , \bar{\mathfrak C}]$ with respect to \eqref{4.59}
\be
\label{4.60}
\d_\l \cS^{(s)}[\mathfrak{C},\bar{\mathfrak{C}}] = \frac{\ri \l}{4} \int \rd^4x \, e \, \Big \{ \mathfrak{C}^2 - \mathfrak{M}^2 \Big \} + \text{c.c.} = -\frac{\ri \l}{2} \int \rd^4x \, e \, \mathfrak{M}^2 + \text{c.c.}~,
\ee
as a generalisation of similar derivations in nonlinear electrodynamics 
\cite{GZ2,GZ3,KT2}.\footnote{In \eqref{4.60} we have employed the notational shorthand $T^2 = T^{\a(m)} T_{\a(m)}$ (similarly $\bar{T}^2 = \bar{T}_{\ad(m)} \bar{T}^{\ad(m)}$).}
This implies the self-duality equation
\be
\label{4.61}
\text{Im} \int \rd^4x \, e \, \Big \{ \mathfrak{C}^{2}
+ \mathfrak{M}^{2} \Big \}  = 0 ~,
\ee
which must hold for an unconstrained field $\mathfrak{C}_{\a(2s)}$ and its conjugate.
The simplest solution of \eqref{4.61} is the
free CHS model \eqref{CHSaction}.\footnote{For the free CHS 
	model \eqref{CHSaction}, one can also consider scale transformations in addition to 
	the $\sU(1)$ duality ones \eqref{4.59}, which is similar to the case of electrodynamics
	discussed, e.g., in \cite{KT2}.} The self-duality equation for the CHS fields
$h_{\a(m) \ad(n)}$, with $m,n \geq 1$, is derived in appendix \ref{Appendix4A}.

In the $s=1$ case, the self-duality equation \eqref{4.61} was originally 
derived by Bialynicki-Birula \cite{B-B} and independently re-discovered  
by Gibbons and Rasheed in 1995 \cite{GR1}. Two years later, it was again re-derived by  Gaillard and Zumino  \cite{GZ2} with the aid of their formalism 
developed in 1981 \cite{GZ1} but originally applied only in the linear case. 

If we allow for the action  $\cS^{(s)}[\mathfrak{C},\bar{\mathfrak C}]$  to depend on a dimension-$1$ conformal compensator $\Xi$,
then the  family of $\sU(1)$ duality-invariant theories is very large. 
For instance, the following $\sU(1)$ duality-invariant model 
\bea
\label{HSBI}
\cS^{(s)}_\text{BI}[\mathfrak{C},\bar{\mathfrak{C}} ; \Xi] = - \int \rd^4x \, e \, \Xi^4 \bigg \{ 1 - \bigg (1 + (-1)^s \frac{\mathfrak{C}^2 + \bar{\mathfrak{C}}^2}{\Xi^4} + \frac{(\mathfrak{C}^2 - \mathfrak{C}^2)^2}{4 \Xi^8} \bigg )^{\frac{1}{2}} \bigg \} 
\eea
is a higher-spin generalisation of Born-Infeld electrodynamics \cite{BI}.\footnote{
	The latter is obtained from \eqref{HSBI} for $s=1$ by making use of local dilatation transformations to impose a gauge condition $\Xi^2 = g^{-1} = {\rm const}$.}
Owing to the dependence of $\cS^{(s)}_\text{BI}[\mathfrak{C},\bar{\mathfrak{C}} ; \Xi] $ on $\Xi$, it is clear that \eqref{HSBI} is not conformal. 
Another important solution to \eqref{4.61} is the following one-parameter duality-invariant extension of \eqref{HSBI}
\begin{align}
	\label{HSBIgen}
	\cS^{(s)}_\text{BIgen}[\mathfrak{C},\bar{\mathfrak{C}} ; \Xi] &= - \int \rd^4x \, e \, \Xi^4 \bigg \{ 1 - \bigg (  1 + \frac{2}{\Xi^4} \bigg [ \frac{(-1)^s}{2} \cosh \g (\mathfrak{C}^2 + \bar{\mathfrak{C}}^2) + \sinh \g (\mathfrak{C}^2 \bar{\mathfrak{C}}^2)^{\frac 1 2} \bigg ] \non \\
	& \qquad \qquad \qquad \qquad + \frac{(\mathfrak{C}^2 - \mathfrak{C}^2)^2}{4 \Xi^8} \bigg)^{\frac 1 2} \bigg \}~, \qquad \g \in \mathbb{R}~.
\end{align}
For $s=1$ this model was introduced in \cite{Bandos:2020hgy}.

\subsubsection{Self-duality under Legendre transformation} \label{section2.2}

In the case of nonlinear (super) electrodynamics,  $\sU(1)$ duality invariance implies self-duality under Legendre transformations, see \cite{KT2} for a review. This remarkable property proves to extend to the higher-spin case, as will be shown below.

We start by describing a Legendre transformation for a generic theory with action 
$\cS^{(s)}[\mathfrak{C},\bar{\mathfrak C}]$.
For this we introduce the parent action
\begin{align}
	\label{parent}
	\cS^{(s)} [\mathfrak C,\bar{\mathfrak C},
	\mathfrak C^{\rm D},\bar{\mathfrak C}^{\rm D}] = \cS^{(s)}[\mathfrak C,\bar{\mathfrak C}] + \int \rd^4x \, e \, \Big ( \frac \ri 2 \mathfrak C^{\a(2s)} \mathfrak C^{\rm D}_{\a(2s)} + \text{c.c.} \Big ) ~.
\end{align}
Here $\mathfrak{C}_{\a(2s)}$ is an unconstrained field and $\mathfrak C^{\rm D}_{\a(2s)}$ takes the form
\begin{align}
	\mathfrak C^{\rm D}_{\a(2s)} = (\nabla_{\a}{}^{\bd})^s h^{\rm D}_{\a(s) \bd(s)} ~,
\end{align}
where $h^{\rm D}_{\a(s) \ad(s)}$ is a Lagrange multiplier field. Indeed, upon varying \eqref{parent} with respect to 
$h^{\rm D}_{\a(s) \ad(s)}$ 
one obtains the Bianchi identity \eqref{4.55}, 
and its general solution is given by eq. \eqref{4.20}, for some real field $h_{\a(s)\ad(s)}$. 
As a result the second term in \eqref{parent} becomes a total derivative, and we end up with the original action 
$\cS^{(s)}[\mathfrak{C},\bar{\mathfrak C}]$.
Alternatively, if we first vary  \eqref{parent} with respect to $\mathfrak{C}^{\a(2s)}$, the equation of motion is
\begin{align}
	\mathfrak{M}_{\a(2s)} = - \mathfrak{C}^{\rm D}_{\a(2s)}~,
\end{align}
which we may solve to express $\mathfrak{C}_{\a(2s)}$ as a function of 
$\mathfrak{C}^{\rm D}_{\a(2s)}$ and its conjugate. 
Inserting this solution into \eqref{parent}, we obtain the dual model
\begin{align}
	\label{dualmodel}
	\cS^{(s)}_{\rm D}[\mathfrak C^{\rm D},\bar{\mathfrak C}^{\rm D}] 
	:= \Big [ \cS^{(s)}[\mathfrak C,\bar{\mathfrak C}] +  \int \rd^4x \, e \, \Big ( \frac \ri 2 \mathfrak C^{\a(2s)} \mathfrak C^{\rm D}_{\a(2s)} + \text{c.c.} \Big ) \Big ]\Big |_{\mathfrak{C} = \mathfrak{C}( \mathfrak{C}^{\rm D},  \bar{\mathfrak{C}}^{\rm D})}~.
\end{align}

Now, given an action $\cS^{(s)}[\mathfrak C , \bar{\mathfrak C} ]$ satisfying \eqref{4.61}, our aim is to show that it satisfies
\begin{align}
	\label{legendre}
	\cS^{(s)}_{\rm D}[\mathfrak C,\bar{\mathfrak C}] = \cS^{(s)} [\mathfrak C,\bar{\mathfrak C}]~,
\end{align}
which means that the corresponding Lagrangian is invariant under Legendre transformations. A routine calculation allows one to show that the following functional
\begin{align}
	\label{invariant}
	\cS^{(s)}[\mathfrak C,\bar{\mathfrak C}] +  \int \rd^4x \, e \, \Big ( \frac \ri 4 \mathfrak C^{\a(2s)} \mathfrak M_{\a(2s)} + \text{c.c.} \Big )
\end{align}
is invariant under \eqref{4.59}. The latter may be exponentiated to obtain the finite $\sU(1)$ duality transformations
\begin{align}
	\mathfrak{C}'_{\a(2s)} = \text{cos} \l \, \mathfrak{C}_{\a(2s)} + \text{sin} \l \, \mathfrak{M}_{\a(2s)} ~, \quad
	\mathfrak{M}'_{\a(2s)} = - \text{sin} \l \, \mathfrak{C}_{\a(2s)} + \text{cos} \l \, \mathfrak{M}_{\a(2s)}  ~.
\end{align}
Performing such a transformation with $\l = \frac \pi 2$ on \eqref{invariant} yields 
\begin{align}
	\cS^{(s)}[\mathfrak C, \bar{\mathfrak C}] 
	= \cS^{(s)}[\mathfrak C^{\rm D}, \bar{\mathfrak C}^{\rm D}] 
	- \int \rd^4x \, e \, \Big( \frac \ri 2 \mathfrak{C}^{\a(2s)} \mathfrak{C}^{\rm D}_{\a(2s)} + \text{c.c.} \Big) ~. 
\end{align}
Upon inserting this expression into \eqref{dualmodel}, we obtain \eqref{legendre}.

In the above analysis, we made use of the fact that the  general solution of  the Bianchi identity \eqref{4.55} is given by \eqref{4.20}. To justify this claim, it suffices to work in Minkowski space. Let $\mathfrak{C}_{\a(2s)} $ be a field subject to the equation \eqref{4.55}, with $\nabla_a=\pa_a$. Introduce its descendant defined by
\begin{subequations}
	\bea
	h^{\perp}_{\a(s) \ad(s)} := (\pa^{\b}{}_{\ad})^s \mathfrak{C}_{\a(s) \b(s)}
	\label{perp}
	\eea
	which is automatically transverse, 
	\bea
	\pa^{\b\bd} h^{\perp}_{\b\a(s-1) \bd \ad(s-1)} &=&0~.
	\eea
\end{subequations}
The Bianchi identity \eqref{4.55} tells us that $h^{\perp}_{\a(s) \ad(s)} $ is real, 
$\overline{h^{\perp}_{\a(s) \ad(s)}  }= h^{\perp}_{\a(s) \ad(s)} $.
Now we can express $\mathfrak{C}_{\a(2s)} $ in terms of $h^{\perp}_{\a(s) \ad(s)}$,
\bea
\mathfrak{C}_{\a(2s)} =  \square^{-s} (\pa_{\a}{}^{\bd})^s h^{\perp}_{\a(s) \bd(s)} 
=(\pa_{\a}{}^{\bd})^s h_{\a(s) \bd(s)} ~.
\eea
In the final relation the real field $h_{\a(s)\ad(s)}$ is not assumed to be transverse.
This field is related to $\square^{-s} h^{\perp}_{\a (s) \ad(s)} $ by a gauge transformation
\bea
\d_\ell h_{\a(s) \ad(s)} = \pa_{\aa} \ell_{\a(s-1) \ad(s-1)}~,
\eea
with a real gauge parameter $\ell_{\a(s-1) \ad(s-1)}$.  Our consideration may be extended to the supersymmetric case considered in section \ref{section4.3.2}.

\subsubsection{Auxiliary variable formulation} \label{section2.3}

As a generalisation of the Ivanov-Zupnik \cite{IZ_N3,IZ1,IZ2} approach, here we will introduce a powerful formalism to generate duality-invariant models that makes use of auxiliary variables. 

Consider the following action functional
\begin{align}
	\label{auxiliaryCHS}
	\cS^{(s)}[\mathfrak{C},\bar{\mathfrak C}, \r, \bar \r] &= (-1)^s \int \rd^4x 
	\, e \, \Big \{ 2 \r \mathfrak{C} - 
	\r^2- \frac{1}{2}\mathfrak{C}^2 \Big \} + \text{c.c.} + \mathfrak{S}^{(s)}_{\text{int}} [\r , \bar{\r}] ~.
\end{align}
Here we have introduced the auxiliary variable $\r_{\a(2s)}$,
which is an unconstrained primary dimension-2 field, 
\bea
K^\bb \mathcal{\r}_{\a(2s)} =0~, \qquad {\mathbb D} \mathcal{\r}_{\a(2s)} = 2 \mathcal{\r}_{\a(2s)}~.
\eea
The functional 
$\mathfrak{S}^{(s)}_{\text{int}} [\r , \bar{\r}]$, by definition, contains cubic and higher powers of $\r_{\a(2s)}$ and its conjugate. The equation of motion for $\r^{\a(2s)}$ is 
\be
\label{etaEoM}
\r_{\a(2s)} = \mathfrak{C}_{\a(2s)} + \frac{(-1)^s}{2} \frac{\d \mathfrak{S}^{(s)}_{\text{int}} [\r , \bar{\r}]}{\d \r^{\a(2s)}}~.
\ee
Equation \eqref{etaEoM} allows one to express $\r_{\a(2s)}$ as a functional of $\mathfrak{C}_{\a(2s)}$ and its conjugate. This means that \eqref{auxiliaryCHS} is equivalent to a CHS theory with action
\begin{align}
	\label{dualTheory}
	\cS^{(s)}[\mathfrak{C},\bar{\mathfrak C}] &= \frac{(-1)^s}{2} \int \rd^4x 
	\, e \, \mathfrak{C}^2 + \text{c.c.} + {\cS}^{(s)}_{\text{int}} [\mathfrak{C} , \bar{\mathfrak C}] ~.
\end{align}
Thus, \eqref{auxiliaryCHS} and \eqref{dualTheory} provide two equivalent realisations of the same model.

The power of this formulation is most evident when the self-duality equation \eqref{4.61} is applied. A routine computation reveals that this constraint is equivalent to
\be
\label{dualCondition}
\text{Im} \int \rd^4x \, e \, \r^{\a(2s)} \frac{\d \mathfrak{S}^{(s)}_{\text{int}} [\r , \bar{\r}]}{\d \r^{\a(2s)}} = 0 ~.
\ee
Thus, self-duality of the action \eqref{auxiliaryCHS} is equivalent to the requirement that $\mathcal{S}^{(s)}_{\text{int}}[\r,\bar{\r}]$ is invariant under rigid $\sU(1)$ phase transformations
\be
\mathfrak{S}^{(s)}_{\text{int}} [\re^{\ri \varphi} \r , \re^{- \ri \varphi} \bar{\r}] 
= \mathfrak{S}^{(s)}_{\text{int}} [\r , \bar{\r}] ~, \quad \varphi \in \mathbb{R} ~.
\label{dualCondition2}
\ee

As an example of a solution to \eqref{dualCondition2}, we can consider the model 
\bea 
\mathfrak{S}^{(s)}_{\text{int}} [\r , \bar{\r} ; \X]  = 
\int \rd^4x \, e \, \X^4 {\mathfrak F} \Big( 
\frac{ \mathcal{\r}^2 \bar{\mathcal{\r}}^2 }{\X^8} \Big)~,
\eea
where ${\mathfrak F}(x) $ is a real analytic function of a real variable. However, such models are not conformal if the action depends on $\X$. Thus condition of conformal invariance imposes additional nontrivial restrictions.

\subsubsection{Conformal $\sU(1)$ duality-invariant models}

In the $s=1$ case, there is a unique conformal $\sU(1)$ duality-invariant extension of free electrodynamics proposed in \cite{BLST} (see also \cite{Kosyakov}). This was called  ``ModMax electrodynamics'' in \cite{BLST}. It turns out that for $s>1$, families of conformal $\sU(1)$ duality-invariant models exist.

As a warm-up example, let us consider the following nonlinear conformal action
\begin{align}
	\cS^{(s)}[\mathfrak{C},\bar{\mathfrak C}] = \frac{ (-1)^s \a}{2} \int \rd^4x \, e \, \Big \{ \mathfrak{C}^2 + \bar{\mathfrak{C}}^2 \Big \} + \b \int \rd^4x \, e \, \sqrt{\mathfrak{C}^2 \bar{\mathfrak{C}}^2} ~, \quad \a,\b \in \mathbb{R}.
\end{align}
Requiring this action to obey the self-duality equation \eqref{4.61}, we obtain the constraint
\be
\a^2 - \b^2 = 1 \quad \Longrightarrow \quad \a = \text{cosh} \, \g ~, \quad \b = \text{sinh} \, \g ~, \quad \g \in \mathbb{R}~.
\ee
Thus, the nonlinear theory
\begin{align}
	\label{4.89}
	\cS^{(s)}[\mathfrak{C},\bar{\mathfrak C}] = \frac{(-1)^s \text{cosh} \, \g}{2} \int \rd^4x \, e \, \Big \{ \mathfrak{C}^2 + \bar{\mathfrak{C}}^2 \Big \} + \text{sinh} \, \g \int \rd^4x \, e \, \sqrt{\mathfrak{C}^2 \bar{\mathfrak{C}}^2} ~, \quad \g \in \mathbb{R},
\end{align}
is a one-parameter conformal $\sU(1)$ duality-invariant  extension of the free CHS action \eqref{4.20}.
In the $s=1$ case our model coincides with  ModMax electrodynamics. It should also be noted that this action may also be obtained from \eqref{auxiliaryCHS} by choosing the interaction
\begin{align}
	\mathfrak{S}_{\rm int}^{(s)}[\r,\bar{\r}] = \b \int \rd^4x \, e \, \sqrt{\r^2 \bar{\r}^2}~, \qquad \b \in \mathbb{R}~.
\end{align}
Upon eliminating the auxiliary variables, it may be shown that the resulting action coincides with \eqref{4.89} upon making the identification
\begin{align}
	\cosh \g = \frac{1+(\b/2)^2}{1-(\b/2)^2} \quad \Longleftrightarrow \quad \sinh \g = \frac{\b}{1-(\b/2)^2}~.
\end{align}

In order to construct more general models, it is advantageous to make use of the auxiliary variable formulation described above. We introduce algebraic invariants of the symmetric rank-$(2s)$ spinor $\r_{\a(2s)} $, which has the same algebraic properties as $\mathfrak{C}_{\a(2s)}$:
\bea
\r^2 := (-1)^s \r_{\a(s)}{}^{\b(s)} \r_{\b(s)}{}^{\a(s)} ~, \qquad 
\r^3 := \r_{\a(s)}{}^{\b(s)} \r_{\b(s)}{}^{\g(s)} \r_{\g(s)}{}^{\a(s)}~, \qquad \dots
\label{invariants}
\eea
If $s$ is odd, all invariants $\r^{2n +1}$, with $n$ a non-negative integer, vanish.

For simplicity, we restrict our analysis to the conformal graviton, $s=2$. 
In this case the family of invariants \eqref{invariants} contains only two functionally independent invariants \cite{PenroseR}, $\r^2 $ and $\r^3$. In particular, one may show that 
\bea
s=2: \qquad \r^4 = \hf (\r^2)^2~.
\eea
Now we choose the self-interaction in \eqref{auxiliaryCHS} to be of the form 
\bea
\label{2.35}
\mathfrak{S}^{(2)}_{\text{int}} [\r , \bar{\r}] 
=\int \rd^4 x\, e\, \Big\{ \b \big( \r^2 \bar \r^2\big)^{\hf} + \k \big( \r^3 \bar \r^3\big)^{\frac 13} \Big\}~,
\eea
where $\b$ and $\k$ are real coupling constants. The resulting model is clearly 
conformal and $\sU(1)$ duality-invariant. For $\k \neq 0$, elimination of the auxiliary variables $\r_{\a(4)}$ and $\bar \r_{\ad(4)}$ does not 
result 
in a simple action like \eqref{4.89}. In particular, such an elimination, to quadratic order in the couplings, yields the following self-dual model
\bea
\label{confgravitonNLAction}
\cS^{(2)}[\mathfrak{C},\bar{\mathfrak{C}}] &=& \int \rd^4 x\, e\, \bigg\{ 
\frac 1 2 \Big (1 + \frac 1 2 \b^2 \Big) (\mathfrak{C}^2 +\bar{\mathfrak{C}}^2) 
+ \b (\mathfrak{C}^2 \bar{\mathfrak{C}}^2)^{\frac 1 2} + \kappa (\mathfrak{C}^3 \bar{\mathfrak{C}}^3)^{\frac 1 3} \non \\
&& \quad + \frac{1}{2} \b \k \frac{(\mathfrak{C}^3)^2 \bar{\mathfrak{C}}^2 + (\bar{\mathfrak{C}^3})^2 {\mathfrak{C}}^2 }{(\mathfrak{C}^3 \bar{\mathfrak{C}}^3)^{\frac 2 3} (\mathfrak{C}^2 \bar{\mathfrak{C}}^2)^{\frac 1 2}}
+ \frac{1}{12} \kappa^2 \frac{(\mathfrak{C}^2)^2 + (\bar{\mathfrak{C}}^2)^2}{(\mathfrak{C}^3 \bar{\mathfrak{C}}^3)^{\frac 1 3}} \non \\
&& \quad - \frac{1}{24} \kappa^2 \frac{(\mathfrak{C}^3)^2 (\bar{\mathfrak{C}}^2)^2 + (\bar{\mathfrak{C}^3})^2 ({\mathfrak{C}}^2)^2}{(\mathfrak{C}^3 \bar{\mathfrak{C}}^3)^{\frac 4 3}} + \dots \bigg \} ~.
\eea
The ellipsis in \eqref{confgravitonNLAction} denotes additional contributions to the full nonlinear theory which are cubic or higher order in the coupling constants.
We emphasise that for the special case $\k=0$ the above action yields 
\eqref{4.89}.

For $s> 2$ the number of algebraic invariants of $\r_{\a(2s)}$ grows, and therefore one can define families of conformal $\sU(1)$ duality-invariant models.

\subsection{Duality-invariant SCHS models} \label{section4.3.2}

In this subsection we describe $\sU(1)$ duality-invariant models for SCHS multiplets. Such analyses require us to employ a supersymmetric generalisation of the formalism described above. For the case of nonlinear $\cN=1$ and $\cN=2$ super electrodynamics, such a formalism was developed in the rigid supersymmetric case in \cite{KT1,KT2} and extended to supergravity in \cite{KMcC,K12-2}. Below, we will describe its extension to the case of higher-spins.

To begin, we consider a dynamical system describing the propagation of a superconformal gauge multiplet $H_{\a(s) \ad(s)}$, $s \geq 0$, in $\cN$-extended conformally flat superspace. We assume that its action $\cS^{(s)}[\mathfrak{W},\bar{\mathfrak{W}}]$ depends only on the chiral field strength $\mathfrak{W}_{\a(2s+\cN)}$\footnote{As in the non-supersymmetric case, we make use of the shorthand $\mathfrak{W}_{\a(2s+\cN)} \equiv \mathfrak{W}_{\a(2s+\cN)}(H)$.}  (and its conjugate $\bar{\mathfrak{W}}_{\ad(2s+\cN)}$), defined in \eqref{linSW1} ($s>0$) and \eqref{linSW3} ($s=0$). It is important to note that the field strengths obey the Bianchi identity
\be
\label{4.87}
(\nabla^{\b}{}_{\ad})^s \nabla^{\b(\N)} \mathfrak{W}_{\a(s) \b(s+\N)} 
= (-1)^\cN
(\nabla_{\a}{}^{\bd})^s \bar{\nabla}^{\bd(\cN)} \bar{\mathfrak{W}}_{\ad(s) \bd(s + \N)}  ~.
\ee 
Here the real superfield $(\nabla^{\b}{}_{\ad})^s \nabla^{\b(\N)} \mathfrak{W}_{\a(s) \b(s+\N)}$ proves to be primary.

We assume that  $\cS^{(s)}[\mathfrak{W} , \bar{\mathfrak{W}}]$ is consistently defined as a functional of a general chiral superfield $\mathfrak{W}_{\a(2s+\cN)}$ and its conjugate. This allows us to introduce the dual superfield
\be
\ri \mathfrak{M}_{\a(2s+\N)} := 2 \frac{\d \cS^{(s)}[\mathfrak{W},\bar{\mathfrak{W}}]}{\d \mathfrak{W}^{\a(2s+\N)}} ~,
\ee
where the variational derivative is defined by 
\be
\d \cS^{(s)}[\mathfrak{W},\bar{\mathfrak{W}}] = \int \rd^4x \rd^{2\N} \q \, \mathcal{E}\, \d \mathfrak{W}^{\a(2s+\N)} \frac{\d \cS^{(s)}[\mathfrak{W},\bar{\mathfrak{W}}]}{\d \mathfrak{W}^{\a(2s+\N)}}  + \text{c.c.}
\ee
It follows that $\mathfrak{M}_{\a(2s+\cN)}$ is a primary chiral superfield,
\bea
K^B \mathfrak{M}_{\a(2s+\cN)} =0~, \qquad \bar \nabla^\bd_i \mathfrak{M}_{\a(2s+\cN)} =0~,
\qquad 
\mathbb{D} \mathfrak{M}_{\a(2s+\cN)} = \frac{4-\cN}{2} \mathfrak{M}_{\a(2s+\cN)}~.
\eea
Varying the action with respect to the prepotential $H_{\a(s) \ad(s)}$ yields
the equation of motion
\be
\label{4.91}
(\nabla^{\b}{}_{\ad})^s \nabla^{\b(\N)} \mathfrak{M}_{\a(s) \b(s+\N)} 
= (-1)^\cN
(\nabla_{\a}{}^{\bd})^s \bar{\nabla}^{\bd(\cN)} \bar{\mathfrak{M}}_{\ad(s) \bd(s + \N)} ~.
\ee
We emphasise that the real superfield $(\nabla^{\b}{}_{\ad})^s \nabla^{\b(\N)} \mathfrak{M}_{\a(s) \b(s+\N)}$ is primary.

The analysis above indicates that the functional form of \eqref{4.87} coincides with that of \eqref{4.91}. It then follows that their union is invariant under the $\sU(1)$ duality rotations:
\be
\label{4.99}
\d_\l \mathfrak{W}_{\a(2s+\cN)} = \l \mathfrak{M}_{\a(2s+\cN)} ~, \quad \d_\l \mathfrak{M}_{\a(2s+\cN)} = - \l \mathfrak{W}_{\a(2s+\cN)} ~,
\ee
where $\l$ is a constant, real parameter. One may then obtain two equivalent expressions for the variation of $\cS^{(s)}[\mathfrak{W} , \bar{\mathfrak{W}}]$ with respect to \eqref{4.99}
\bea
\d_\l \cS^{(s)}[\mathfrak{W} , \bar{\mathfrak{W}}] &=& \frac{\ri \l}{4} \int \rd^4x \, \rd^{2\N}\q \, \cE \, \Big \{ \mathfrak{W}^2 - \mathfrak{M}^2 \Big \} + \text{c.c.} \non \\
&=& -\frac{\ri \l}{2} \int \rd^4x \, \rd^{2\cN}\q \, \cE \, \mathfrak{M}^2 + \text{c.c.}~,
\label{315}
\eea
as a generalisation of similar derivations in nonlinear $\cN=1$ and $\cN=2$ supersymmetric electrodynamics
\cite{KT1,KT2,KMcC,K12-2}.
This implies the self-duality equation
\be
\label{4.94}
\text{Im} \int \rd^4x \rd^{2\N} \q \, \mathcal{E}\, \Big \{ \mathfrak{W}^2 + \mathfrak{M}^{2} \Big \}  = 0 ~,
\ee
which must hold for a general chiral superfield $\mathfrak{W}_{\a(2s+\cN)}$  and its conjugate. Every solution $\cS^{(s)}[\mathfrak{W} , \bar{\mathfrak{W}}]$ of the self-duality equation describes a $\sU(1)$ duality-invariant theory. For the $s=0$ case, equation \eqref{4.94} was originally derived in the $\cN=1$ ($\cN=2$) case in  \cite{KT1} (\hspace{-0.01cm}\cite{KT2}) in Minkowski superspace and extended to supergravity in \cite{KMcC} (\hspace{-0.01cm}\cite{K12-2}). The simplest solutions of the self-duality equation \eqref{4.94} are the free SCHS models \eqref{SCHSaction1} ($s>0$) and \eqref{SCHSaction3} ($s=0$).

The above results allow one to prove, in complete analogy with the non-supersymmetric analysis conducted in section \ref{section2.2}, that $\sU(1)$ duality-invariant theories are  self-dual under Legendre transformations.

\subsubsection{Auxiliary variable formulation} \label{section3.2}

As a generalisation of the auxiliary variable formalism sketched in section \ref{section2.3},
here we will develop a reformulation of the supersymmetric $\sU(1)$ duality-invariant theories introduced above. In the $s=0$ case, it will reduce 
to the auxiliary superfield approach for $\sU(1)$ duality-invariant supersymmetric electrodynamics introduced in \cite{K13,ILZ}.

Consider the action functional
\begin{align}
	\label{4.95}
	\cS^{(s)}[\mathfrak{W},\bar{\mathfrak{W}}, \eta, \bar \eta] &= (-1)^{s} \int \rd^4x 
	\rd^{2\N} \q \, \mathcal{E}\, \Big \{ \eta \mathfrak{W} - 
	\frac{1}{2} \eta^{2}- \frac{1}{4}\mathfrak{W}^2 \Big \} + \text{c.c.} 
	+ \mathfrak{S}^{(s)}_{\text{int}} [\eta , \bar{\eta}] ~.
\end{align}
Here we have introduced the new multiplet $\eta_{\a(2s+\cN)}$, which is required to be primary and covariantly chiral, 
\bea
K^B \eta_{\a(2s+\N)} =0~, \qquad
\bar{\nabla}^\bd_i \eta_{\a(2s+\N)} = 0~, \qquad {\mathbb D} \eta_{\a(2s+\N) } = \frac{4-\cN}{2} \eta_{\a (2s+\N)}
~.
\eea
By definition, the functional $\mathfrak{S}^{(s)}_{\text{int}} [\eta , \bar{\eta}]$ contains cubic and higher powers of $\eta_{\a(2s+\cN)}$ and its conjugate.

The equation of motion for $\eta^{\a(2s+\N)}$ is 
\be
\label{4.97}
\eta_{\a(2s+\N)} = \mathfrak{W}_{\a(2s+\N)}
+ (-1)^{s} \frac{\d \mathfrak{S}^{(s)}_{\text{int}} [\eta , \bar{\eta}]}{\d \eta^{\a(2s+\N)}}~.
\ee
Employing perturbation theory, equation \eqref{4.97} allows one to express $\eta_{\a(2s+\N)}$ as a functional of $\mathfrak{W}_{\a(2s+\N)}$ and its conjugate. This means that \eqref{4.95} is dual to a SCHS theory with action
\begin{align}
	\label{4.98}
	\cS^{(s)}[\mathfrak{W},\bar{\mathfrak{W}}] &= \frac{(-1)^s}{4} \int \rd^4x 
	\rd^{2\cN} \q \, \mathcal{E}\, \mathfrak{W}^2 + \text{c.c.} 
	+ {\cS}^{(s)}_{\text{int}} [\mathfrak{W} , \bar{\mathfrak{W}}] ~.
\end{align}
Thus, \eqref{4.95} and \eqref{4.98} provide two equivalent realisations of the same model.

The power of this formulation is most evident when the self duality equation \eqref{4.94} is applied. A routine computation reveals that this constraint is equivalent to
\be
\label{4.4}
\text{Im} \int \rd^4x \rd^{2\N} \q \, \mathcal{E}\, \eta^{\a(2s+\N)} 
\frac{\d \mathfrak{S}^{(s)}_{\text{int}} [\eta , \bar{\eta}]}{\d \eta^{\a(2s+\N)}} = 0 ~.
\ee
Thus, self-duality of the action \eqref{4.95} is equivalent to the requirement that $\mathfrak{S}^{(s)}_{\text{int}}[\eta,\bar{\eta}]$ is invariant under rigid $\sU(1)$ phase transformations
\be
\label{4.5}
\mathfrak{S}^{(s)}_{\text{int}} [\re^{\ri \varphi} \eta , \re^{- \ri \varphi} \bar{\eta}] = \mathfrak{S}^{(s)}_{\text{int}} [\eta , \bar{\eta}] ~, \quad \varphi \in \mathbb{R} ~.
\ee

Having formulated the general theory for $\cN$-extended duality invariant models above, for the remainder of this subsection we employ this formalism to derive some interesting supersymmetric duality-invariant higher-spin models.

\subsubsection{$\cN=1$ duality-invariant models}

Here we will construct some interesting duality-invariant models for $\cN=1$ superconformal gauge multiplets.
To this end, we permit the action \eqref{4.95} to depend on a dimension-2 conformal compensator $\X$\footnote{A simple realisation for the conformal compensator is $\X=S_0 \bar{S}_0$, where $S_0$ is the chiral compensator of old minimal supergravity, see section \ref{section3.5.2}.}, allowing us to generate families of theories satisfying the self-duality constraint \eqref{4.5}. First, we consider the following interaction
\begin{subequations}
	\bea
	\label{4.101a}
	\mathfrak{S}^{(s)}_{\text{int}} [\eta , \bar{\eta}; \X] &=& \frac{1}{2(2s+2)!} \int \rd^4x \rd^2 \q \rd^2 \bar{\q} \, E \, \frac{ (\eta^{2}\bar{\eta}^{2})^{s+1}}{\X^{3s+2}} 
	\mathfrak{F}^{(s)} (v, \bar v) ~, 
	\eea
	where 
	\bea
	v: = \frac{1}{8} \nabla^2 \big[ {\eta^{2}}{\X^{-2}} \big]~.
	\label{4.101c}
	\eea
\end{subequations}
Applying \eqref{4.5}, we find that our model is $\sU(1)$ duality invariant provided
\bea
\label{4.7}
\mathfrak{F}^{(s)} (v, \bar v) = \mathfrak{F}^{(s)} (v \bar v) ~.
\eea

The present model is superconformal if the action is independent of $\X$. This uniquely fixes the functional form of $\mathfrak{F}(v \bar v)$ modulo a single real parameter
\be
\mathfrak{F}^{(s)}_{\text{SC}}(v \bar v) = \frac{\kappa}{(v \bar v)^{\frac{1}{4}(3s+2)}} ~, \quad \kappa \in \mathbb{R} ~.
\ee
Employing \eqref{4.97}, we arrive at
\bea
\mathfrak{W}_{\a(2s+1)} = \eta_{\a(2s+1)} \bigg \{ 1 &+& \frac{(-1)^s \kappa}{8(2s+2)!} \bar{\nabla}^2 \Big[ (2s+2) \frac{\eta^{2s} \bar{\eta}^{2s+2}}{\X^{3s+2}} \mathfrak{F}^{(s)}_{\text{SC}} \non \\
&&\qquad + \frac{\bar{\eta}^{2s+2}}{4 \X^{3s+2}} \nabla^2 \Big( \eta^{2s+2} \partial_v \mathfrak{F}^{(s)}_{\text{SC}} \Big) \Big] \bigg \} ~,
\eea
which, along with its conjugate, allows us to integrate out the auxiliary variables present in \eqref{4.101a}. 
The final result, for $s=0$, is the model for superconformal $\sU(1)$ duality-invariant electrodynamics introduced in \cite{K21,BLST2}
\bea
\label{N=1MM}
\cS^{(0)} [\mathfrak{W},\bar {\mathfrak{W} }] &=&
\frac{1}{4} \cosh \g \int  \rd^4 x \rd^2 \q  \,\cE \, \mathfrak{W}^2 +{\rm c.c.}
\non \\
&& + \frac{1}{4} \sinh \g   \int \rd^4 x \rd^2 \q \rd^2\bar \q \,E \,
\frac{\mathfrak{W}^2\,{\mathfrak{W}}^2}{\X^2\sqrt{u\bar u} }~,
\eea
where $\g$ is a real coupling constant. 
For $s>0$ the resulting model is
\begin{align} \label{scAction}
	\cS^{(s>0)}[\mathfrak{W},\bar{\mathfrak{W}}] &= \frac{(-1)^s}{4} \int \rd^4x 
	\rd^2 \q \, \mathcal{E}\, \mathfrak{W}^2 + \text{c.c.} \non \\
	&\qquad + \frac{\kappa}{2(2s+2)!} \int \rd^4x \rd^2 \q \rd^2 \bar{\q} \, E \, \frac{( \mathfrak{W}^{2}\bar{\mathfrak{W}}^{2})^{s+1}}{\X^{3s+2} (u \bar u)^{\frac 14 (3s+2)}} ~.
\end{align}
In both cases we have made use of the shorthand 
\begin{align}
	u = \frac{1}{8} \nabla^2 \big[ \mathfrak{W}^2 \X^{-2} \big] ~.
\end{align}
We note that \eqref{scAction} is invariant under rescalings of the conformal compensator 
\begin{align}
	\label{ccscale}
	\X \rightarrow \re^{2\s} \X~,
\end{align}
which implies that its dependence on $\X$ is superficial. Thus, \eqref{scAction} is superconformal.

It is important to note that, at the component level, the purely bosonic sector of the interaction (the $\k$-term) present in \eqref{scAction} identically vanishes.\footnote{The
	numerator in the $\k$-term in \eqref{scAction} contains a product of $4(s+1)$ fermionic superfields. Thus, we do not have enough spinor derivatives to convert all 
	fermionic superfields into bosonic ones for $s>0$.
} 
Thus, these actions describe different duality-invariant models than those presented in the previous section.
However, one may construct a supersymmetric duality-invariant model
that contains, for instance, the bosonic theory  \eqref{4.89}
at the component level.

Consider the following supersymmetric duality-invariant model 
\begin{align}
	\label{4.115}
	\cS^{(s)} [\mathfrak{W}, \bar{\mathfrak{W}}, \eta , \bar{\eta}; \X] &= (-1)^{s} \int \rd^4x 
	\rd^2 \q \, \mathcal{E}\, \Big \{ \eta \mathfrak{W} - 
	\frac{1}{2} \eta^2- \frac{1}{4}\mathfrak{W}^2 \Big \} + \text{c.c.} \non \\
	&\qquad + \frac{\b}{8} \int \rd^4x \rd^2 \q \rd^2 \bar{\q} \, E \, \frac{ \eta^{2}\bar{\eta}^{2}}{\X^{2} \sqrt{v \bar{v}}} ~, \qquad \b \in \mathbb{R}~.
\end{align}
It does not enjoy invariance under \eqref{ccscale}, hence it is not superconformal. However,
at the component level, this model contains certain conformal duality-invariant 
actions.
For this we restrict our attention
to the bosonic sector of \eqref{4.115}. The bosonic field strengths contained in 
$\mathfrak{W}_{\a(2s+1)}$  are:\footnote{These field strengths can be related to those of eq. \eqref{4.20} by a rescaling.}
\begin{subequations}
	\begin{align}
		\mathfrak{C}_{\a( 2s+2)} &= \ri \nabla_{\a } \mathfrak{W}_{\a(2s+1) } |_{\theta^\a=\bar{\q}_\ad=0} ~, \qquad \mathfrak{C}_{\a(2s)} = - \frac{4(s+1)}{2s+1} \nabla^\b \mathfrak{W}_{\b \a(2s) } |_{\theta^\a=\bar{\q}_\ad=0} ~. 
	\end{align}
	Similarly, the bosonic component fields of  $\eta_{\a(2s+1)}$ are:
	\begin{align}
		\r_{\a( 2s+2)} &= \ri \nabla_{\a } \eta_{\a(2s+1) } |_{\theta^\a=\bar{\q}_\ad=0} ~, \qquad \r_{\a(2s)} = - \frac{4(s+1)}{2s+1} \nabla^\b \eta_{\b \a(2s) } |_{\theta^\a=\bar{\q}_\ad=0} ~.
	\end{align}
\end{subequations}
Action \eqref{4.115} can be reduced to components by making use of the reduction rule \eqref{2.90b}.
Since we are interested only in the bosonic sector, it suffices to approximate
\bea
v|_{\theta^\a=\bar{\q}_\ad=0} \approx - \frac{1}{\chi^2} \bigg \{ \frac{128 (s+1)^3}{(2s+1)^3} \r^{\a(2s+2)} \r_{\a(2s+2)} 
+ \frac{1}{2} \r^{\a(2s)} \r_{\a(2s)}  \bigg\}~,
\eea 
where we have made the definition $\chi := \Xi|_{\theta^\a=\bar{\q}_\ad=0}$.
Further, it may be checked that  the interaction (the $\b$-term)  in \eqref{4.115} does not contain contributions  which are (i) linear in $\r_{\a(2s)}$ and its conjugate; and (ii)  linear in $\r_{\a(2s+2)}$ and its conjugate.
Therefore, if 
we switch off the spin-$s$ field, 
\begin{subequations}
	\label{turnoff}
	\bea
	\label{switchoffGF}
	\mathfrak{C}_{\a(2s)} = 0~,
	\eea
	then 
	\bea
	\r_{\a(2s)} = 0
	\label{switchoffRF}
	\eea
\end{subequations}
is a solution of the corresponding  equation of motion.
Under these conditions, 
the resulting bosonic action proves coincides with 
$\cS^{(s+1)}[\mathfrak{C},\bar{\mathfrak C}, \r, \bar \r]$, which may be
obtained from \eqref{auxiliaryCHS} by performing the shift $s \rightarrow s+1$.
We emphasise that the compensator $\X$ does not contribute to this action, and therefore it
is locally conformal.
Instead of considering the branch  \eqref{switchoffGF},
we may switch off the spin-$(s+1) $ field, $\mathfrak{C}_{\a(2s+2)}=0$.
Then it follows that $\r_{\a(2s+2)}=0$ is a solution of the corresponding equation of motion, and the resulting bosonic  action proves to coincide with 
the conformal duality-invariant action \eqref{auxiliaryCHS}.

To conclude this section, we derive a supersymmetric extension of the cubic interaction present in \eqref{2.35}. For this purpose, we will fix $s=1$ in \eqref{4.95}. The relevant interaction is constructed in terms of the following scalar descendants of $\eta_{\a(3)}$: 
\begin{align}
	\L = \frac{\ri}{2} \eta^{\a(2) \b} \eta^{\a(2)}{}_\b \nabla_{(\a_1} \eta_{\a_2 \a_3 \a_4)} ~, \qquad
	w = \nabla^2 \big( \L \X^{-3} \big) ~,
\end{align}
and the conformal compensator $\X$. Using these, we construct the following action
\begin{align}
	\label{4.120}
	\cS^{(1)} [\mathfrak{W}, \bar{\mathfrak{W}}, \eta , \bar{\eta}; \X] &= - \int \rd^4x 
	\rd^2 \q \, \mathcal{E}\, \Big \{ \eta \mathfrak{W} - 
	\frac{1}{2} \eta^2- \frac{1}{4}\mathfrak{W}^2 \Big \} + \text{c.c.} \non \\
	&\qquad + 16 \kappa \int \rd^4x \rd^2 \q \rd^2 \bar{\q} \, E \, \frac{ \L \bar{\L} }{\X^{4} (w \bar{w})^{\frac 23}} ~.
\end{align}
We note that it is not superconformal as it lacks invariance under \eqref{ccscale}. However, we will show that, at the component level, it contains the cubic interaction present in \eqref{2.35}. To demonstrate this, we restrict our attention to the bosonic sector of \eqref{4.120}, which allows us to approximate
\begin{align}
	\label{w|}
	\quad w |_{\theta^\a=\bar{\q}_\ad=0} \approx - \frac{8192\sqrt{3}}{243 \chi^3}\Big( \r^{\a(2)}{}_{\b(2)} \r^{\b(2)}{}_{\g(2)} \r^{\g(2)}{}_{\a(2)} - \frac{9}{32} \r^{\a(2)} \r^{\a(2)} \r_{\a(4)} \Big) ~.
\end{align}
Further, the interaction (the $\k$-term) present in \eqref{4.120} contains no terms linear in either $\r_{\a(4)}$ or $\r_{\a(2)}$. Hence, if we switch off the spin-$1$ field, $\mathfrak{C}_{\a(2)} = 0$, then $\r_{\a(2)} = 0$ solves the corresponding equations of motion. Under these conditions, the resulting bosonic action takes the form
\begin{align}
	\cS^{(2)}[\mathfrak{C},\bar{\mathfrak C}, \r, \bar \r] &= \int \rd^4x 
	\, e \, \Big \{ 2 \r \mathfrak{C} - 
	\r^2- \frac{1}{2}\mathfrak{C}^2 \Big \} + \text{c.c.} + \kappa \int \rd^4x 
	\, e \, \big( \r^3 \bar \r^3\big)^{\frac 13} ~.
\end{align}
In particular, we note that it is locally conformal as it is independent of the compensator. Thus, \eqref{4.120} is a supersymmetric extension of the cubic interaction of \eqref{2.35}. In closing, we emphasise that if one instead attempts to consider the branch $\mathfrak{C}_{\a(4)} = 0$ and $\mathfrak{C}_{\a(2)} \neq 0$, it is not legal to set $\r_{\a(4)} = 0$. Such a choice would imply $w|_{\theta^\a=\bar{\q}_\ad=0} \approx 0$, making the corresponding component action undefined.

\subsubsection{$\cN=2$ duality-invariant models}

Building on the $\cN=1$ story above, here we derive some $\sU(1)$ duality-invariant models in $\cN=2$ superspace. As in the $\cN=1$ case, we begin by allowing \eqref{4.95} to depend on a dimension-2 conformal compensator $\Xi$.\footnote{One may, for instance, choose $\Xi=W_0 \bar{W}_0$, where $W_0$ is the field strength of a compensating vector multiplet, see section \ref{section3.5.3}.} This greatly simplifies the procedure of constructing possible interactions satisfying condition \eqref{4.5}, for instance
\bea
\mathfrak{S}^{(s)}_{\text{int}} [{\eta} , \bar{{\eta}} ; \X] =
\int \rd^4x \rd^4 \q \rd^4 \bar \q \, {E}\, {\mathfrak F} \Big( 
\frac{ {\eta}^2 \bar{{\eta}}^{2} }
{ \X^2 } \Big)~.
\eea
Here $ {\mathfrak F} (x)$ is a real analytic function of a real variable. 

It is not difficult to construct explicit examples of $\cN=2$ superconformal 
and $\sU(1)$ duality-invariant higher-spin theories, as a generalisation of the $\cN=2$ vector multiplet models presented in \cite{K21}. We consider the self-interaction
\bea
\label{4.124}
\mathfrak{S}^{(s)}_{\text{int}} [{\eta} , \bar{{\eta}}] =
c \int \rd^4x \rd^4 \q \rd^4 \bar \q \,  \ln 
{\eta}^2
\ln   \bar{{\eta}}^2~,
\eea
with $c$ a coupling constant. It is clearly  invariant under rigid $\sU(1)$ phase transformations, eq. \eqref{4.5}, and therefore it generates a $\sU(1)$ duality-invariant model. Additionally, \eqref{4.124} is $\cN=2$ superconformal. The above functional is well defined provided the auxiliary variables ${\eta}_{\a(2s+2)}$ are chosen to belong to the open domain ${\eta}^2 \neq 0$.

\section{(S)CHS fields interacting with matter} \label{Chapter4.4}

The theory of an infinite tower of interacting CHS fields is most easily defined as an induced action, see \cite{Segal, Tseytlin, BJM}. For example, given the gauge-invariant action $\cS[\vf,\bar{\vf},h]$ for a massless complex scalar $\vf$ coupled to background 
CHS fields $h_{\a(s) \ad(s)}$, $s \geq 1$, an effective action may be defined according to
\begin{align}
	e^{\ri \Gamma_{\text{eff}}[h]}=\int\mc{D}\varphi\mc{D}\bar{\varphi}\, e^{\ri \mc{S}[\varphi, \bar{\varphi}, h] }~. \label{EffAct}
\end{align}
As $\mc{S}[\varphi,\bar{\vf},h]$ is bilinear in $\varphi$, the latter may be integrated out in \eqref{EffAct}, and the CHS action may be identified with the logarithmically divergent part of $\Gamma_{\text{eff}}[h]$, which is local and invariant under the gauge transformations of the higher-spin fields $h_{\a(s) \ad(s)}$. The $\cN=1$ supersymmetric extension of this story is described in \cite{KMT,KP23}.

Hence, it is clear that the gauge-invariant action for some (super)conformal matter multiplets coupled to background (S)CHS fields plays a central role in this construction.
Therefore, it is the objective of this section to derive such gauge-invariant models where the (super)conformal symmetry is manifest.\footnote{In the non-supersymmetric case, the action for a complex scalar coupled to background bosonic CHS fields was considered in \cite{Segal,BJM}, though conformal symmetry was neither discussed nor manifest in their construction.}

\subsection{Complex scalar field coupled to bosonic CHS fields} \label{secCS}
The results of ``Closing the gauge algebra" and ``Inclusion of auxiliary gauge fields" below were obtained by Michael Ponds \cite{KPR22}.

This subsection is devoted to the study of interactions between a massless complex scalar $\vf$ and a tower of background CHS fields $h_{\a(s) \ad(s)}$, $s \geq 1$. Further, by requiring off-shell gauge-invariance of the corresponding action, we deduce the non-abelian extension of the gauge transformations of $h_{\a(s) \ad(s)}$.

\subsubsection{Differential operators on the space of primary scalar fields} \label{secCSLinVar}

As a first step in building interactions involving $\varphi$, we analyse the possible transformation rules consistent with its kinematic properties. 
To begin, we propose the following linear transformation rule for $\vf$
\begin{subequations}
	\label{4.127}
	\begin{align}
		\delta \varphi=-{\mc{U}}\varphi~,\qquad {\mc{U}}=\sum_{s=0}^{\infty}{\mc{U}}^{(s)}, \label{CSgt}
	\end{align} 
	where ${\mc{U}}^{(s)}$ is a differential operator of maximal order $s$ 
	\begin{align}
		{\mc{U}}^{(s)}=\sum_{k=0}^{s}\zeta_{(s)}^{\a(k)\ad(k)}(\nabla_{\a\ad})^k ~. \label{GenOpphi}
	\end{align}
\end{subequations}
At this stage, the coefficients $\zeta_{(s)}^{\a(k)\ad(k)}$ are complex. Operator ${\mc{U}}$ must preserve the off-shell properties of $\vf$ \eqref{3.30}, and hence so too must each individual ${\mc{U}}^{(s)}$. As we now show, this implies that the coefficients of each operator \eqref{GenOpphi} may be expressed in terms of a single parameter, which may ultimately be identified with $\ell_{\a(s)\ad(s)}$, the gauge parameter associated with $h_{\a(s+1) \ad(s+1)}$. Keeping in mind the analysis of section \ref{Chapter3.3.1}, it is clear that ${\mc{U}}^{(s)}$ takes the form
\begin{align}
	{\mc{U}}^{(s)} &= \ri^s \sum_{k=0}^{s}\binom{s}{k}^2\binom{2s+2}{s-k}^{-1} (\nabla_{\bb})^{s-k} \z^{\a(k) \b(s-k) \ad(k) \bd(s-k)} (\nabla_\aa)^k~, \label{CSOp}
\end{align}
where we have identified $\z_{(s)}^{\a(s) \ad(s)}\equiv \z^{\a(s)\ad(s)}$, which is an unconstrained complex parameter with the conformal properties:
\begin{align}
	\label{4.129}
	\mathbb{D} \z^{\a(s) \ad(s)} = -s \z^{\a(s) \ad(s)} ~, \qquad K^{\bb} \z^{\a(s) \ad(s)} = 0~.
\end{align}

The operator \eqref{CSOp} is the unique one (modulo normalisation) of the functional form \eqref{GenOpphi} preserving all off-shell properties of $\varphi$. However, it is not the most general operator with this property, which would include terms involving powers of the conformal d'Alembertian in the ansatz \eqref{GenOpphi}. 
It will be shown below we will see that it is possible to achieve gauge invariance to order $\mc{O}(h)$ without such terms. However, they turn out to be necessary to close the gauge algebra, hence we will revisit them shortly.

\subsubsection{Noether procedure to order $\mc{O}(h)$}

From $\varphi$ and its conjugate $\bar{\varphi}$ one may construct the rank $s\geq 0$ descendants
\begin{align}
	j_{\a(s)\ad(s)}=\ri^s\sum_{k=0}^{s}(-1)^k\binom{s}{k}^2(\nabla_{\a\ad})^k\varphi(\nabla_{\a\ad})^{s-k}\bar{\varphi}~  \label{CScurrent}
\end{align}
It may be shown that $j_{\a(s)\ad(s)}$ possesses the following crucial features:
\begin{enumerate}[label=(\roman*)]
	\item Reality
	\begin{subequations}
		\begin{align}
			j_{\a(s)\ad(s)}=\bar{j}_{\a(s)\ad(s)}~;
		\end{align}
		\item Conformal covariance on arbitrary gravitational backgrounds
		\begin{align}
			K^\bb j_{\a(s)\ad(s)} = 0~, \qquad \mb{D}j_{\a(s)\ad(s)}=(s+2)j_{\a(s)\ad(s)}~; \label{CScurrentprop}
		\end{align}
		\item Transverse on-shell when restricted to conformally flat backgrounds (for $s>0$)
		\begin{align}
			C_{abcd}=0\qquad \implies \qquad \nabla^{\b \bd}j_{\b\a(s-1) \bd \ad(s-1)}\approx 0~, \label{CSTOS}
		\end{align}
		where we have made use of the symbol $\approx$ to denote equality on-shell.
	\end{subequations}
\end{enumerate}
The conformal currents \eqref{CScurrent} in Minkowski space were introduced for the first time by Craigie, Dobrev and Todorov \cite{CDT}, although the method to derive these currents had been described earlier by Makeenko 
\cite{Makeenko}.

The currents \eqref{CScurrent} with $s=1$ and $s=2$ are actually conserved on-shell in arbitrary gravitational backgrounds. However, it was shown in \cite{BT}  (building on the analysis 
in \cite{GrigorievT})
that the Weyl tensor $C_{abcd}$ is the obstruction for this property to hold in the case $s=3$. This is also true for spin $s> 3$, hence for the remainder of this section we restrict ourselves to conformally flat backgrounds. 

The conformal currents $j_{\a(s)\ad(s)}$ naturally couple to the integer spin-$s$ conformal gauge fields $h_{\a(s)\ad(s)}$ via the cubic Noether coupling
\begin{subequations}
	\label{CSNoether}
	\begin{align}	
		\mc{S}_{\rm{N.C.}}[\varphi,\bar{\vf},h]&=\sum_{s=0}^{\infty}c_s\,\mc{S}^{(s)}_{\rm{N.C.}}[\varphi,\bar{\vf},h_{(s)}]~,\\ \mc{S}^{(s)}_{\rm{N.C.}}[\varphi,\bar{\vf},h_{(s)}]&=\int \text{d}^{4}x \, e \, h_{\a(s)\ad(s)}j^{\a(s)\ad(s)}~,
	\end{align}
\end{subequations}
for undetermined coefficients $c_s$. On account of \eqref{CSTOS}, the action \eqref{CSNoether} is invariant under the gauge transformations
\begin{align}
	\label{4.132}
	\d_\ell h_{\a(s)\ad(s)} = \nabla_\aa \ell_{\a(s-1) \ad(s-1)}~, \qquad s \geq 1~,
\end{align}
provided $\varphi$ is on-shell, $\delta_{\ell}\mc{S}_{\rm{N.C.}}[\vf,\bar{\vf},h]\approx 0$. Further, owing to the conformal properties \eqref{CHSprimary}, the cubic action (see \eqref{CSaction} for the kinetic action for $\vf$)
\begin{align}
	\mc{S}_{\text{Cubic}}[\varphi,\bar{\vf},h]=\mc{S}[\varphi,\bar{\vf}]+\mc{S}_{\rm{N.C.}}[\varphi,\bar{\varphi},h] \label{CSCubicact}
\end{align}
is clearly conformal.\footnote{It follows from \eqref{CScurrentprop} that \eqref{CSCubicact} is actually conformal on arbitrary backgrounds.}

In order to elevate \eqref{4.132} to an off-shell symmetry, it is necessary to endow $\varphi$ with its own transformation rule. For this we use the rule \eqref{4.127} with $\cU^{(s)}$ given by \eqref{CSOp} and impose the reality condition $\bar{\z}^{\a(s) \ad(s)} = - \z^{\a(s) \ad(s)}$. Then, the properties \eqref{4.129} allow us to make the field identification  $\z^{\a(s)\ad(s)}\equiv \ri \ell^{\a(s)\ad(s)}$, where $\ell_{\a(s)\ad(s)}$ is the gauge parameter associated with $h_{\a(s+1) \ad(s+1)}$.
Now we may fix the coefficients $c_s$ in \eqref{CSNoether} by requiring the cubic action \eqref{CSCubicact} to be invariant up to terms linear in gauge fields,
\begin{align}
	c_s=\frac{(-1)^{s}}{2} \binom{2s}{s}^{-1} \quad \implies \quad	\delta_{\ell}\mc{S}_{\text{Cubic}}[\varphi,\bar{\vf},h] = \mc{O}\big(h\big)~. \label{CSCubic}
\end{align} 
It should be emphasised that this statement holds off-shell.

\subsubsection{Closing the gauge algebra}\label{CSsec2}

Fundamental to our analysis above was the family of linear operators \eqref{4.127}, which, together with \eqref{4.132}, constitute gauge transformations for the action \eqref{CSNoether}. It may be verified, however, that they do not form a closed algebra (modulo trivial symmetries\footnote{Given a functional $\mc{S}[\U^I]$ of some fields $\U^I$, a transformation $\delta \U^I$ is said to be a trivial symmetry if it may be expressed in the form $\delta \U^I=A^{IJ}\frac{\delta \mc{S}}{\delta \U^J}$ (here we are using DeWitt's condensed notation). This is possible if and only if $\delta\mc{S}=0$ and $\delta\U^I\approx 0$ (see e.g. \cite{HenneauxT}). Here $A^{IJ}$ is some differential operator satisfying $A^{IJ}=-A^{JI}$.
}), as their commutator involves terms proportional to $\square$. Hence, the latter cannot be expressed in the form \eqref{4.127}. Consequently, the cubic coupling of $\vf$ to the tower of CHS fields $h_{\a(s)\ad(s)}$ is already inconsistent at order $\mathcal{O}\big(h\big)$.

Therefore, it is necessary to consider more general matter transformations than \eqref{4.127} to ensure that we obtain a closed algebra. Such transformations are non-trivially constrained by the requirement that they must be consistent with the conformal properties of $\vf$. Naively, one may consider the ansatz
\begin{align}
	\d \vf = - \sum_{s=0}^{\infty}\sum_{l=0}^{\lfloor s/2 \rfloor} \sum_{k=0}^{s-2l} \z_{(s,l)}^{\a(k) \ad(k)} (\nabla_\aa)^k \square^l \vf~,
\end{align}
and require that it preserves the kinematic properties of $\vf$. Such an approach proves to be computationally difficult, hence we will not pursue it further. Instead, we will make use of a nowhere-vanishing real scalar field $\eta$ with the conformal properties
\begin{align}
	\mb{D}\eta=-2\eta~, \qquad K^{\a\ad}\eta=0~. \label{CScompprop}
\end{align}
It is clear to see that the operator $\hat{\square} :=\eta \square$ takes a primary scalar of unit dimension to a primary scalar of unit dimension. 

Employing this observation, we postulate the following matter transformations for $\vf$
\begin{subequations}
	\label{4.138}
	\begin{align}
		\delta \vf = -{\mc{V}}\vf= \sum_{s=0}^{\infty}{\mc{V}}^{(s)}\vf~, \qquad {\mc{V}}^{(s)} =\sum_{l=0}^{\lfloor s/2 \rfloor}{\mc{U}}^{(s-2l)}_{[s,l]}\hat{\square}^l ~.\label{genvarCS}
	\end{align}
	where the operator $\cU_{[s,l]}^{(s-2l)}$ takes the same functional form as \eqref{CSOp}
	\begin{align}
		{\mc{U}}^{(s-2l)}_{[s,l]}= \ri^{s-2l} \sum_{k=0}^{s-2l} \binom{s-2l}{k}^2\binom{2s-4l+2}{s-2l-k}^{-1} (\nabla_{\aa})^{s-2l-k} \z_{[s,l]}^{\a(s-2l) \ad(s-2l)} (\nabla_\aa)^k~. \label{newGPop}
	\end{align}
\end{subequations}
Requiring the transformation \eqref{4.138} to preserve the conformal properties of $\vf$, we find that the highest rank parameter of \eqref{newGPop} obeys
\begin{align}
	K^{\b\bd}\z_{[s,l]}^{\a(s-2l) \ad(s-2l)}=0~,\qquad \mb{D}\z_{[s,l]}^{\a(s-2l) \ad(s-2l)}=-(s-2l)\z_{[s,l]}^{\a(s-2l) \ad(s-2l)}~. \label{ConfpropCSgp}
\end{align}
In particular, the gauge parameters of \eqref{CSOp} correspond to the $l=0$ case, that is, $\z_{[s,0]}^{\a(s)\ad(s)} = \z^{\a(s)\ad(s)}$. In closing, we emphasise that the matter transformations \eqref{4.138} do indeed form a closed algebra.

\subsubsection{Inclusion of auxiliary gauge fields} \label{CSsec3}

In the previous subsection we extended the gauge group such that the matter transformations \eqref{4.138} form a closed algebra. They were shown to be parametrised by the primary fields $\z_{[s,l]}^{\a(s-2l)\ad(s-2l)}$, where $\z_{[s,0]}^{\a(s)\ad(s)} = \z^{\a(s) \ad(s)}$. In order to gauge this symmetry, we introduce new gauge fields\footnote{For $l > 0$, they correspond to the traceful component of the CHS fields of \cite{Segal, BJM}.} $h_{[s,l]}^{\a(s-2l)\ad(s-2l)}$ associated with these parameters. Their conformal properties are as follows:
\begin{align}
	K^{\b\bd}h_{[s,l]}^{\a(s-2l)\ad(s-2l)}=0~,\qquad \mb{D}h_{[s,l]}^{\a(s-2l)\ad(s-2l)}=(2-s+2l)h_{[s,l]}^{\a(s-2l)\ad(s-2l)}~.
\end{align}
It should be emphasised that the original CHS gauge fields correspond to those with $l=0$, $h_{[s,0]}^{\a(s)\ad(s)} \equiv  h^{\a(s)\ad(s)}$.

Omitting the techincal details, it may be shown that the cubic couplings between $\vf$ and $h_{[s,l]}^{\a(s-2l)\ad(s-2l)}$ are described by the action:
\begin{subequations}
	\label{4.141}
	\begin{align}
		\mathcal{S}_{\text {Cubic}}[\vf,\bar{\vf},h]&=\int \text{d}^{4} x \, e \, \Big\{ \bar{\vf}\square \vf + \sum_{s=0}^{\infty}\sum_{l=0}^{\lfloor s/2 \rfloor} h_{[s,l]}^{\a(s-2l)\ad(s-2l)}j_{\a(s-2l)\ad(s-2l)}^{[s,l]} \Big\}~,\\
		j_{\a(s-2l)\ad(s-2l)}^{[s,l]}&= \frac{\ri^s}{2}\sum_{k=0}^{s-2l}(-1)^k\binom{s-2l}{k}^2\Big((\nabla_{\a\ad})^k\hat{\square}^l\vf (\nabla_{\a\ad})^{s-2l-k}\bar{\vf} \non \\
		& \qquad \qquad \qquad \qquad \qquad \qquad + (\nabla_{\a\ad})^{k}\vf (\nabla_{\a\ad})^{s-2l-k}\hat{\square}^{l}\bar{\vf}\Big)~,
	\end{align}
\end{subequations}
By construction $j_{\a(s-2l)\ad(s-2l)}^{[s,l]}$ is a real primary field of dimension $(2+s-2l)$. Additionally, for $l=0$ it coincides with \eqref{CScurrent}. At the same time, it is easily seen that for $l > 0$ the current vanishes on-shell. 

We now spell out the gauge transformations of the fields $h_{[s,l]}^{\a(s-2l)\ad(s-2l)}$ which, together with the matter transformations \eqref{4.138}, ensure off-shell gauge invariance of \eqref{4.141}. First, the CHS fields transform as follows:\footnote{The additional factor of $c_s$ in \eqref{CHSdeformedGTb} occurs as we have absorbed this factor in the definition of $h_{\a(s)\ad(s)}$, c.f. \eqref{CSCubic}.}
\begin{subequations}\label{CHSdeformedGT}
	\begin{align}
		\delta h_{\a(s)\ad(s)}&= \nabla_{\a\ad}\ell_{\a(s-1)\ad(s-1)}+\mc{O}(h)~,\label{CHSdeformedGTa}\\
		\ell_{\a(s-1)\ad(s-1)}&\equiv c_s\ri \big(\z_{\a(s-1)\ad(s-1)}-\bar{\z}_{\a(s-1)\ad(s-1)}\big)~,\label{CHSdeformedGTb}
	\end{align}
\end{subequations}
In particular, we note that \eqref{CHSdeformedGTa} contains a non-abelian component (whose explicit form we do not give here). On the other hand, the gauge fields $h_{[s,l]}^{\a(s-2l)\ad(s-2l)}$ with $l>0$ begin their transformations algebraically:
\begin{align}
	\delta h_{[s,l]}^{\a(s-2l)\ad(s-2l)} \propto \eta^{-1}\big(\z_{[s-2,l-1]}^{\a(s-2l)\ad(s-2l)}+\bar{\z}_{[s-2,l-1]}^{\a(s-2l)\ad(s-2l)}\big)+\cdots~,\qquad 1\leq l \leq \lfloor s/2 \rfloor~. \label{CSalgebraic}
\end{align}
Hence, the real component of each $\z_{[s-2,l-1]}^{\a(s-2l)\ad(s-2l)}$ may be utilised to fix the gauge
\begin{align}
	h_{[s,l]}^{\a(s-2l)\ad(s-2l)}=0~, \qquad s\geq 0 ~\text{ and }~ 1\leq l \leq \lfloor s/2 \rfloor~.\label{CSGfix}
\end{align}
Consequently, the only remaining gauge parameters are the imaginary parts of $\z^{\a(s-1)\ad(s-1)}$.\footnote{For $l>0$, the imaginary components of $\z_{[s,l]}^{\a(s-2l)\ad(s-2l)}$ do not appear in \eqref{CHSdeformedGT}.} Thus, for the variations of both the matter and gauge fields to preserve this gauge \eqref{CSGfix}, the former should be supplemented with a compensating $h^{\a(s)\ad(s)}$ and $\ell^{\a(s-1)\ad(s-1)}$ dependent transformation. However, due to its complexity, we omit this analysis.

\subsubsection{Rigid symmetries of free Klein-Gordon action} \label{secCSrigid}

As is well known, the reducibility parameters of $h_{\a(s)\ad(s)}$ define rigid symmetries of the matter action $\cS[\vf,\bar{\vf}]$, see e.g. \cite{Bekaert}. We recall that the former are those gauge parameters $\ell^{(0)}_{\a(s-1)\ad(s-1)}$ leaving $h_{\a(s) \ad(s)}$ unchanged
\begin{align}
	\label{4.152}
	\d_{\ell^{(0)}} h_{\a(s)\ad(s)} = \nabla_\aa \ell^{(0)}_{\a(s-1) \ad(s-1)} = 0~.
\end{align}
In particular, this implies that $\ell^{(0)}_{\a(s-1)\ad(s-1)}$ constitutes a conformal Killing tensor field \eqref{N=2CKT}. It trivially follows from eq. \eqref{4.152} that
\begin{align}
	\delta_{\ell^{(0)}}\mc{S}[\vf,\bar{\vf}]=-\delta_{\ell^{(0)}}\mc{S}_{\text{N.C.}}[\vf,\bar{\vf},h]\big|_{h=0}=0
\end{align}
Hence, the matter action $\cS[\vf,\bar{\vf}]$ is invariant under transformations \eqref{CSOp} upon requiring that $\z_{\a(s-1) \ad(s-1)} = \ri \ell_{\a(s-1) \ad(s-1)}^{(0)}$. We note that such matter transformations may be obtained from the higher symmetries of the conformal d'Alembertian, eq. \eqref{HSCD}, by fixing $\xi_{\a(s-1)\ad(s-1)} = \ri^{s+1} \ell^{(0)}_{\a(s-1)\ad(s-1)}$.

\subsection{Chiral multiplet coupled to superspin-($s+\frac12$) SCHS fields } \label{masslesschiral}

The results of this subsection, excluding part of the calculation to obtain \eqref{4141b} and those of ``Rigid symmetries of the free massless Wess-Zumino action'' were obtained by Michael Ponds.

As an extension of the non-supersymmetric analysis above, this subsection is devoted to the construction of a superspace action describing the interactions between a massless chiral multiplet $\Phi$ and the superspin-($s+\frac12$) SCHS prepotentials $H_{\a(s)\ad(s)}$ in conformally flat superspace backgrounds. 

\subsubsection{Differential operators on the space of primary chiral superfields}

As in the non-supersymmetric case, we begin our analysis by constructing the permissible higher-derivative transformation laws for the chiral scalar $\Phi$ consistent with its kinematics, specifically its chirality and superconformal properties. We propose the transformation
\begin{subequations}
	\label{4.154}
	\begin{align}
		\delta\Phi=-{\mc{U}}\Phi~,\qquad {\mc{U}}=\sum_{s=0}^{\infty}{\mc{U}}^{(s)}, \label{Mattergt}
	\end{align}
	where ${\mc{U}}^{(s)}$ is a linear differential operator of order $s$
	\begin{align}
		{\mc{U}}^{(s)}=\sum_{k=0}^{s}\O_{(s)}^{\a(k)\ad(k)}(\nabla_\aa)^k+\sum_{k=1}^{s}\O_{(s)}^{\a(k)\ad(k-1)}(\nabla_{\a\ad})^{k-1}\nabla_{\a} ~. \label{GenOpPhi}
	\end{align}
\end{subequations}
It is clear that each $\cU^{(s)}$ must preserve the kinematic properties of $\Phi$. These conditions uniquely determine all coefficients of \eqref{GenOpPhi} (up to overall normalisation) in terms of a single superfield, which will be shown to be proportional to the gauge parameter of $H_{\a(s)\ad(s)}$.

We begin by imposing the chirality condition $\bar{\nabla}_\ad \mc{U}^{(s)} \Phi = 0$, which yields the constraints:
\begin{subequations}
	\begin{align}
		\bar{\nabla}^\ad \O^{\a(k) \ad(k)}_{(s)} &= 0 ~, \qquad &0 \leq k \leq s~, \label{4.155a} \\
		\O^{\a(k) \ad(k-1)}_{(s)} &= \frac{\ri k}{2(k+1)} \bar{\nabla}_\bd \O^{\a(k) \ad(k-1)\bd}_{(s)}~, \qquad &~1 \leq k \leq s~.
	\end{align}
\end{subequations}
In particular, \eqref{4.155a} implies that $\O_{(s)}^{\a(k)\ad(k)}$ is longitudinal linear for $s>0$ and chiral for $s=0$. The former condition may be used to solve for $\O_{(s)}^{\a(s) \ad(s)}$ in terms an unconstrained parameter $\o$ by making the definition
\begin{subequations}
	\begin{align}
		s>0:& \qquad \O_{(s)}^{\a(s) \ad(s)} = \ri^s \bar{\nabla}^{\ad} \o^{\a(s)\ad(s-1)}~, \\
		s=0:& \qquad \O_{(0)} = \bar{\nabla}^{2} \o ~.
	\end{align}
\end{subequations}
Requiring the transformed field to be primary, $K^{B}{\mc{U}}^{(s)}\Phi=0$, we find that this parameter is a primary superfield
\begin{subequations}
	\label{gtpropid}
	\begin{align}
		s>0:& \qquad K^{B}\o^{\a(s)\ad(s-1)}=0~,\qquad \mathbb{D}\o^{\a(s)\ad(s-1)}=-\big(s+1/2\big)\o^{\a(s)\ad(s-1)} ~,  \\
		s=0:& \qquad K^B \o = 0 ~, \qquad \mathbb{D} \o = -1~.
	\end{align}
\end{subequations}
and that the remaining parameters are uniquely determined as its descendants. Specifically, it may be shown that \eqref{GenOpPhi} takes the form 
\begin{subequations}
	\label{ChiralPrimalOp}
	\begin{align}
		s > 0:& \qquad \mc{U}^{(s)} = \sum_{k=1}^{s} \ri^s\binom{s}{k}\binom{s-1}{k-1}\binom{2s+1}{s-k}^{-1}\Big\{ \non \\ &\qquad ~ (\nabla_{\b\bd})^{s-k}\bar{\nabla}^{\ad}\o^{\a(k)\b(s-k)\bd(s-k)\ad(k-1)}(\nabla_{\a\ad})^k \notag \\
		&\qquad +\frac{\ri}{4}\bar{\nabla}^2(\nabla_{\b\bd})^{s-k}\o^{\a(k)\b(s-k)\bd(s-k)\ad(k-1)}(\nabla_{\a\ad})^{k-1}\nabla_{\a}\notag\\
		&\qquad -\frac{\ri}{4}\frac{k}{s+k+1}\bar{\nabla}^2\nabla_{\b}(\nabla_{\b\bd})^{s-k}\o^{\a(k-1)\b(s-k+1)\bd(s-k)\ad(k-1)}(\nabla_{\a\ad})^{k-1} \Big\}~, \\
		s = 0:& \qquad \mc{U}^{(0)} = \bar{\nabla}^2 \o~.
	\end{align}
\end{subequations}

It should be noted that did not include contributions to $\d \F$ which vanish on-shell. As will be shown below, it is possible for the action for $\Phi$ coupled to the infinite tower of SCHS fields $H_{\a(s)\ad(s)}$, $s \geq 0$, to be gauge-invariant off-shell at order $\mathcal{O}(H)$ in their absence. However, such on-shell vanishing terms prove to be necessary for the algebra of matter transformations to close, hence they will be revisited below.

\subsubsection{Noether procedure to order $\mc{O}(H)$}

From $\Phi$ and $\bar{\Phi}$ we construct the rank-$s\geq0$ descendants
\begin{align}
	J_{\a(s)\ad(s)}&=\ri^s\sum_{k=0}^{s}(-1)^k\binom{s}{k}^2\Big((\nabla_{\a\ad})^k\Phi(\nabla_{\a\ad})^{s-k}\bar{\Phi} \non \\
	& \qquad \qquad \qquad \qquad \qquad \qquad  -\frac{\text{i}}{2}\frac{s-k}{k+1}(\nabla_{\a\ad})^{k}\nabla_{\a}\Phi(\nabla_{\a\ad})^{s-k-1}\bar{\nabla}_{\ad}\bar{\Phi}\Big)~, \label{supercur}
\end{align}
It may be shown that $J_{\a(s)\ad(s)}$ possesses the following crucial features:
\begin{enumerate}[label=(\roman*)]
	\item Reality
	\begin{subequations}\label{3.15}
		\begin{align}
			J_{\a(s)\ad(s)}=\bar{J}_{\a(s)\ad(s)}~;
		\end{align}
		\item Superconformal covariance in an arbitrary supergravity background
		\begin{align}
			K^B J_{\a(s)\ad(s)} = 0~, \qquad \mb{D}J_{\a(s)\ad(s)}=(s+2)J_{\a(s)\ad(s)}~; \label{currentprop}
		\end{align}
		\item When restricted to a conformally flat background, $W_{\a\b\g}=0$,
		$J_{\a(s)\ad(s)}$ constitutes a conserved supercurrent if $\F$ is on-shell\footnote{The current multiplets  \eqref{CScurrent} with $s=0$ and $s=1$ are actually conserved on-shell on arbitrary backgrounds. However, the super-Weyl tensor $W_{\a\b\g}$ proves to be the obstruction for this property to hold for case $s\geq 2$. }
		\begin{align}
			s \geq 1:& \qquad \nabla_\b J^{\a(s-1)\b \ad(s)} = 0 ~, \quad \bar{\nabla}_\bd J^{\a(s) \ad(s-1)\bd} = 0 ~, \\
			s = 0:& \qquad \nabla^2 J = 0 ~, \quad \bar{\nabla}^2 J = 0~.
		\end{align}
	\end{subequations}
\end{enumerate}
For $s=0$ and $s=1$, \eqref{supercur} coincides with the flavour current multiplet \cite{FWZ}
and the Ferrara-Zumino supercurrent \cite{FZ}, respectively. In the higher-spin case, $s \geq 2$, the corresponding supercurrents were introduced in \cite{KMT} in Minkowksi superspace, and shortly afterwards they were extended to AdS superspace \cite{BHK18}.

The above supercurrents naturally couple to the half-integer superspin-$(s+\hf)$ conformal prepotentials $H_{\a(s)\ad(s)}$ via the cubic Noether coupling
\begin{subequations}
	\label{4132}
	\begin{align}	
		\mc{S}_{\text{N.C.}}[\Phi,\bar{\Phi},H]&=\sum_{s=0}^{\infty}c_s\mc{S}^{(s)}_{\text{N.C.}}[\Phi,\bar{\Phi},H]~,\\ \mc{S}^{(s)}_{\text{N.C.}}[\Phi,\bar{\Phi},H]&=\int \text{d}^{4|4}z \, E \, H^{\a(s)\ad(s)}J_{\a(s)\ad(s)}~,
	\end{align}
\end{subequations}
for undetermined coefficients $c_s \in \mathbb{R}$. It is clear to see that \eqref{4132} is invariant under the gauge transformations
\begin{subequations}
	\label{4.155}
	\begin{align}
		s\geq 1:&\qquad \qquad \qquad \delta_{\z} H_{\a(s)\ad(s)}=\bar{\nabla}_{\ad}\z_{\a(s)\ad(s-1)}- \nabla_{\a}\bar{\z}_{\a(s-1)\ad(s)}~, \qquad \qquad \label{SCHSgt.a} \\[5pt]
		s= 0:&\qquad \qquad \qquad \qquad ~~ \delta_{\z} H = \bar{\nabla}^2\z+\nabla^2\bar{\z} ~, \qquad \qquad \label{3.19b}
	\end{align}
\end{subequations}
provided $\Phi$ is on-shell, $\delta_{\z}\mc{S}_{\text{N.C.}}[\F,\bar{\F},H]\approx 0$. Further, the cubic action
\begin{align}
	\mc{S}_{\text{Cubic}}[\Phi,\bar{\Phi},H]=\mc{S}[\Phi,\bar{\Phi}]+\mc{S}_{\text{N.C.}}[\Phi,\bar{\Phi},H] \label{Cubicact}
\end{align}
is manifestly superconformal.\footnote{Action \eqref{Cubicact} is actually invariant under the superconformal group on an arbitrary backgrounds.}

In order to elevate \eqref{4.155} to an off-shell symmetry, it is necessary to endow $\Phi$ with its own gauge transformation rule. For this we employ the operators \eqref{ChiralPrimalOp} derived above with the identifications $\z^{\a(s)\ad(s-1)} \equiv  \o^{\a(s)\ad(s-1)} $ and $\z \equiv \o$. Now, requiring \eqref{4132} to be invariant to order $\mathcal{O}(H)$, we determine the coefficients $c_s$
\begin{align}
	c_s = (-1)^{s}\binom{2s+1}{s}^{-1}~ \quad \implies \quad \delta_{\z}\mc{S}_{\text{Cubic}}[\Phi,\bar{\Phi},H] = \mc{O}\big(H\big)~. \label{CubicGI}
\end{align}
Thus, we have shown a correspondence between the matter transformations\footnote{As will be shown at the end of this subsection, the former yield rigid symmetries of the free massless Wess-Zumino model.} \eqref{ChiralPrimalOp} and the conformal supercurrents \eqref{supercur}.  This superspace Noether procedure may be understood as a generalisation of the one described in \cite{MSW} to the superconformal higher-spin case.

It should be noted that, in the rigid supersymmetric case, cubic vertices \eqref{4132} were introduced in \cite{KMT}. However, they took $\F$ to be on-shell, hence the coefficients \eqref{CubicGI} were not fixed.

\subsubsection{Consistency to all orders} \label{secChiralBack}

As was the case in the non-supersymmetric model, the gauge transformations \eqref{4.154} with $\mc{U}^{(s)}$ given by \eqref{ChiralPrimalOp}  do not form a closed algebra. This is because we have restricted our attention to transformation operators which, when acting on $\Phi$, do not vanish on the free equations of motion. This will be done by utilising a superconformal extension of the operator $\hat{\square}$ introduced in the previous subsection.

To this end, we introduce a nowhere-vanishing chiral scalar $\U$
\begin{subequations}
	\begin{align}
		\bar{\nabla}_{\ad} \U =0~,
	\end{align}
	with the superconformal properties
	\begin{align}
		\mathbb{D} \U = - \U~,\qquad K^{B}\U =0 \qquad \implies \qquad \mathbb{Y} \U = \frac{2}{3}\U~.
	\end{align}
\end{subequations}
From $\U$ we construct the following two operators
\begin{align}
	\Delta:= \U \bar{\nabla}^2~,\qquad \bar{\Delta} := \bar{\U}\nabla^2~.
\end{align}
It then follows that their product, $\bm{\Delta}:=\Delta \bar{\Delta}$, preserves all kinematic properties of $\Phi$:
\begin{align}
	\mathbb{D}\bm{\Delta} \F =  \bm{\Delta}\F~,\qquad &\mathbb{Y}\bm{\Delta}\F = -\frac{2}{3}\bm{\Delta}\F~,\qquad K^{B}\bm{\Delta}\F =0~,\qquad \bar{\nabla}_{\ad}\bm{\Delta}\Phi = 0~.
\end{align}
Thus, $\bm{\Delta}$ is a superconformal extension of $\hat{\square}$.

It follows that we may generate higher-order primary operators by combining powers of $\bm{\Delta}$ with \eqref{ChiralPrimalOp}. In particular, the most general transformation rule for $\Phi$ preserving all its kinematic properties is
\begin{subequations}\label{genvarChiral}
	\begin{align}
		\delta \F = -{\mc{V}}\F~, \qquad {\mc{V}}= \sum_{s=0}^{\infty} {\mc{V}}^{(s)}\F~.
	\end{align}
	where we have introduced the order-$s$ operator 
	\begin{align}
		{\mc{V}}^{(s)} = \sum_{l=0}^{\lfloor s/2 \rfloor} {\mc{U}}^{(s-2l)}_{[s,l]}\bm{\Delta}^l
		\label{genvarChiral.b}
	\end{align}
	and the notation
	\begin{align}
		{\mc{U}}^{(s-2l)}_{[s,l]} &= \sum_{k=1}^{s-2l} \ri^s\binom{s}{k}\binom{s-1}{k-1}\binom{2s+1}{s-k}^{-1} \Big\{ \bar{\nabla}^{\ad}\nabla^{s-2l-k}_{\b\bd}\o_{[s,l]}^{\a(k)\b(s-2l-k)\bd(s-2l-k)\ad(k-1)}\nabla^k_{\a\ad}\notag \\
		&\phantom{=}+\frac{\ri}{4}\bar{\nabla}^2\nabla^{s-2l-k}_{\b\bd}\o_{[s,l]}^{\a(k)\b(s-2l-k)\bd(s-2l-k)\ad(k-1)}\nabla^{k-1}_{\a\ad}\nabla_{\a}\notag\\
		&\phantom{=}-\frac{\ri}{4}\frac{k}{s-2l+k+1}\bar{\nabla}^2\nabla_{\b}\nabla^{s-2l-k}_{\b\bd}\o_{[s,l]}^{\a(k-1)\b(s-2l-k+1)\bd(s-2l-k)\ad(k-1)}\nabla^{k-1}_{\a\ad}\Big\}~.
	\end{align}
\end{subequations}
The gauge parameters introduced above are primary with weights and charges given by
\begin{subequations}
	\begin{align}
		\mathbb{D}\o_{[s,l]}^{\a(s-2l)\ad(s-2l-1)}&=-\big(s-2l+1/2\big)\o_{[s,l]}^{\a(s-2l)\ad(s-2l-1)}~,\\ \mathbb{Y}\o_{[s,l]}^{\a(s-2l)\ad(s-2l-1)}&=\o_{[s,l]}^{\a(s-2l)\ad(s-2l-1)}~.
	\end{align}
\end{subequations}
It is clear that the algebra of gauge transformations \eqref{genvarChiral} now closes. 

Omitting the technical details, it may be shown that the cubic couplings between $\F$ and $H_{[s,l]}^{\a(s-2l)\ad(s-2l)}$ are described by the action
\begin{subequations}
	\label{4141}
	\begin{align}
		\mathcal{S}_{\text{Cubic}}[\Phi,\bar{\Phi},H]&=\int \text{d}^{4|4} z \, E \, \Big\{ \bar{\Phi} \Phi + \sum_{s=0}^{\infty}\sum_{l=0}^{\lfloor s/2 \rfloor} H_{[s,l]}^{\a(s-2l)\ad(s-2l)}J_{\a(s-2l)\ad(s-2l)}^{[s,l]} \Big\}~,\\
		J^{[s,l]}_{\a(s-2l)\ad(s-2l)}&=\frac{\ri^s}{2}\sum_{k=0}^{s-2l} (-1)^k \binom{s-2l}{k}^2 \Big \{ \nabla_{\a\ad}^k\Phi\nabla_{\a\ad}^{s-k-2l} \bm{\Delta}^l \bar{\Phi} + \nabla_{\a\ad}^k \bm{\Delta}^l \Phi\nabla_{\a\ad}^{s-k-2l}\bar{\Phi} \non \\  -\frac{\text{i}}{2}&\frac{s-2l-k}{k+1} \Big (\nabla_{\a\ad}^{k}\nabla_{\a}\Phi\nabla_{\a\ad}^{s-2l-k-1}\bar{\nabla}_{\ad} \bm{\Delta}^l \bar{\Phi} + 
		\nabla_{\a\ad}^{k}\nabla_{\a} \bm{\Delta}^l \Phi\nabla_{\a\ad}^{s-2l-k-1}\bar{\nabla}_{\ad}\bar{\Phi} \Big ) \Big\}~. 
		\label{4141b}
	\end{align}
\end{subequations}
For $l=0$, the supercurrents $J_{\a(s-2l)\ad(s-2l)}^{[s,l]}$ coincide with \eqref{supercur}, while for $l>0$ they vanish on-shell. The gauge transformations of SCHS prepotentials $H^{\a(s)\ad(s)}\equiv H^{\a(s)\ad(s)}_{[s,0]}$ may be shown to take the form
\begin{subequations}\label{SCHSdeformedGT}
	\begin{align}
		\delta H_{\a(s)\ad(s)}&= \bar{\nabla}_{\ad}\z_{\a(s)\ad(s-1)}-\nabla_{\a}\bar{\z}_{\a(s-1)\ad(s)}+\mc{O}(H)~,\label{SCHSdeformedGTa}\\
		\z_{\a(s)\ad(s-1)}&\equiv c_s \o^{[s,0]}_{\a(s)\ad(s-1)}~,\label{SCHSdeformedGTb}
	\end{align}
\end{subequations} 
where $c_s$ are defined in \eqref{CubicGI}. In particular, it should be noted that \eqref{SCHSdeformedGTa} contains a non-abelian component (whose explicit form we do not determine here). We expect that the `auxiliary' superfields $H_{[s,l]}^{\a(s-2l)\ad(s-2l)}$ with $l>0$ begin their gauge transformations algebraically, allowing them to be gauged away (such a fixing will lead to new non-trivial contributions to the gauge transformation law of $H_{\a(s)\ad(s)}$).

\subsubsection{Rigid symmetries of the free massless Wess-Zumino model}

It remains to provide some comments on the rigid symmetries of the free matter action $\cS[\F,\bar{\F}]$. 
From \eqref{CubicGI} one may deduce that
$\delta_{\z}\mc{S}[\F,\bar{\F}]=-\delta_{\z}\mc{S}_{\text{N.C.}}[\F,\bar{\F},H]\big|_{H=0}$,
from which it follows that 
\begin{align}
	\delta_{\z}H_{\a(s)\ad(s)}=0 \qquad \implies \qquad \delta_{\z}\mc{S}[\F,\bar{\F}]=0~. \label{RigidSym}
\end{align}
Hence, the reducibility parameters of $H_{\a(s)\ad(s)}$ define rigid symmetries of $\mc{S}[\F,\bar{\F}]$. The former are those gauge parameters $\z^{(0)}_{\a(s) \ad(s-1)}$ satisfying the constraint
\begin{align}
	\label{4.152}
	\d_{\z^{(0)}} H_{\a(s)\ad(s)} = \bar{\nabla}_\ad \z^{(0)}_{\a(s) \ad(s-1)} - {\nabla}_\a \bar{\z}^{(0)}_{\a(s-1) \ad(s)} = 0~.
\end{align}
The above constraint implies that $\chi_{\a(s)\ad(s)} = \ri \bar{\nabla}_{\ad}\z^{(0)}_{\a(s)\ad(s-1)} = \bar{\chi}_{\a(s)\ad(s)}$ constitutes a real conformal Killing tensor superfield of the background \eqref{N=2CKT}
\begin{align}
	\bar{\nabla}_{\ad}\chi_{\a(s)\ad(s)}=0\qquad \Longleftrightarrow \qquad \nabla_{\a}\chi_{\a(s)\ad(s)}=0~. \label{4.167}
\end{align}

It should be noted that, when \eqref{4.167} is imposed, the operators $\cU^{(s)}$, which define the gauge transformations for the matter multiplet, may be written solely in terms of $\chi_{\a(s)\ad(s)}$
\bea
\label{4.169}
\cU^{(s)} &=& - \ri^s \sum_{k=0}^n  {n \choose k}^2  {2n+1 \choose n-k}^{-1} \Big \{ (\nabla_\bb)^{n-k} - \frac{\ri (n-k)}{2(n+1)} (\nabla_\bb)^{n-k-1} \nabla_\b \bar{\nabla}_\bd \Big \} \chi^{\a(k) \b(n-k) \ad(k) \bd(n-k)} \non \\
&\phantom{=}& \times (\nabla_\aa)^k - \ri^{s+1} \sum_{k=1}^n \frac{ n}{2(n+1)}  {n \choose k} {n-1 \choose k-1}{{2n+1 \choose n-k}}^{-1} (\nabla_\bb)^{n-k} \bar{\nabla}_\bd \chi^{\a(k) \b(n-k) \ad(k-1) \bd(n-k+1)} \non \\
&& \times   (\nabla_\aa)^{k-1} \nabla_\a ~,
\eea
We emphasise that the transformations they induce constitute rigid symmetries of the free action, and consequently of its equation of motion. Further, we note that such transformations may be obtained from the higher symmetries of the massless Wess-Zumino operator, eq. \eqref{WZHSsoln}, by fixing $\xi_{\a(s)\ad(s)} = \ri^{s} \chi_{\a(s)\ad(s)}$.

\section{Summary of results} \label{Chapter4.5}

This chapter was devoted to the study of models for (super)conformal higher-spin gauge multiplets in curved backgrounds. Their associated gauge (super)fields may be readily deduced by employing the method of supercurrent multiplets. To this end, in section \ref{Chapter4.1} we classified the conformal higher-spin (super)current multiplets. Following this, in section \ref{Chapter4.2}, by virtue of the Noether coupling \eqref{NoetherCoupling} we obtained the (S)CHS gauge multiplets and their corresponding linearised gauge transformations. Utilising this result, we obtained the (super)conformal and gauge-invariant free  action for each gauge (super)field in conformally flat backgrounds. Additionally, in appendix \ref{Appendix4A} we performed such an analysis for the $\cN=2$ superconformal gravitino multiplet and identified the gauge prepotential describing its dynamics in addition to its associated gauge-invariant action. These results greatly extend the existing literature\footnote{For the appropriate references, we refer the reader to the main body of this chapter.} of superconformal higher-spin gauge multiplets in the $\cN \geq 2$ case.  

Building on the above result, we then proceeded to study $\sU(1)$ duality-invariant models for the (S)CHS fields described above in section \ref{Chapter4.3}. Specifically, we deduced the self-duality equation for real gauge (super)fields.\footnote{In the non-supersymmetric case the extension of this self-duality equation for general CHS fields was obtained in appendix \ref{Appendix4A}.} It was also shown that such models are self-dual under Legendre transformations. Next, we generalised the Ivanov-Zupnik auxiliary variable formulation \cite{IZ_N3,IZ1,IZ2} and its $\cN=1$ supersymmetric extension \cite{K13,ILZ} to the case of higher-spins with $\cN$-extended superconformal symmetry. This formulation allowed us to derive several interesting self-dual models, including ($\cN=1$ supersymmetric) higher-spin generalisations of ModMax electrodynamics \cite{BLST,BLST2,K21}.

To conclude this chapter, we constructed manifestly (super)conformal gauge-invariant models for conformal matter (super)fields interacting with an infinite tower of background (S)CHS gauge multiplets in section \ref{Chapter4.4}. Specifically, we coupled a massless complex scalar to the spin-$s$ gauge fields $h_{\a(s) \ad(s)}$, $s \geq 1$, and also considered interactions between a massless chiral multiplet and the superspin-$(s+\hf)$ gauge multiplets $H_{\a(s) \ad(s)}$, $s \geq 0$. To ensure off-shell gauge-invariance of the corresponding actions, it was necessary for each matter multiplet to transform under gauge transformations of the (S)CHS fields. Further, upon considering the class of gauge transformations preserving the latter, we obtained the rigid symmetries for the free matter theory. The latter are shown to be intimately connected to the higher symmetries described in section \ref{Chapter3.3}.

\begin{subappendices}

\section{The $\cN=2$ superconformal vector and gravitino multiplets} \label{Appendix4A}

In sections \ref{Chapter4.1}, we derived the $\cN$-extended (S)CHS gauge (super)fields by first identifying all possible conformal (super)current multiplets and then utilising an appropriate Noether coupling to deduce the gauge transformations of the former. Since we restricted our analysis in section \ref{Chapter4.1} to (super)currents inert under $\sSU(\cN)$ rotations, we have ignored the study of several important gauge multiplets. This appendix serves to bridge this gap, in the $\cN=2$ case. 

\subsection{The $\cN=2$ vector multiplet}

In rigid supersymmetry, the off-shell $\cN=2$ vector multiplet was formulated by Grimm, Sohnius and Wess \cite{GSW}.
The multiplet of currents associated with its prepotential is the so-called linear multiplet $J^{ij} = J^{ji}$ \cite{BS1,SSW}. It is a primary, real iso-triplet of dimension $2$  subject to the conservation equations:
\begin{align}
	\nabla_\a^{(i} J^{jk)} = 0 ~, \qquad \bar{\nabla}_\ad^{(i} J^{jk)} = 0 ~.
\end{align}
Hence, by making use of the Noether coupling
\begin{align}
	\mathcal{S}_{\rm N.C.} = \int \text{d}^{4|8}z \, E \, H_{ij} J^{ij} ~,
\end{align}
we see that the gauge superfield $H_{ij}$ is real, primary and of dimension $-2$
\begin{align}
	K^C H_{ij} = 0 ~, \qquad \mathbb{D} H_{ij} = -2 H_{ij}~,
\end{align}
and is defined only up to gauge transformations of the form
\begin{align}
	\delta_{\L} H_{ij} &= \nabla^{\alpha k} \Lambda_\alpha{}_{kij}
	+ \text{c.c.} ~, \qquad
	\Lambda_\alpha{}_{kij} = \Lambda_\alpha{}_{(kij)}~,
	\label{pre-gauge1}
\end{align}
where $ \Lambda_{\alpha kij} $ is arbitrary modulo the algebraic condition given. The superfield $H_{ij}$ is a curved superspace extension 
of Mezincescu's prepotential \cite{Mezincescu}.

From $H_{ij}$, we construct the field strength
\begin{align}
	\mathfrak{W}(H) = \frac{1}{48} \bar \nabla^4 \nabla^{ij} H_{ij}~,
\end{align}
which is a primary chiral superfield of unit dimension
\begin{align}
	K^A \mathfrak{W}(H) = 0 ~, \quad \bar{\nabla}^\ad_i \mathfrak{W}(H) = 0 ~, \quad \mathbb{D} \mathfrak{W}(H) = \mathfrak{W}(H)~,
\end{align}
and is gauge-invariant \eqref{pre-gauge1} on arbitrary curved backgrounds; $\d_{\L} \mathfrak{W}(H) = 0$. Further, owing to the Bianchi identity
\begin{subequations}
	\bea
	\nabla^{ij}\mathfrak{W}(H)&=&\bar \nabla^{ij} \bar{\mathfrak{W}}(H) ~,
	\eea
	the following is a total derivative
	\begin{align}
		\ri \int \rd^4x \rd^{2\N} \q \, \cE\, \mathfrak{W}(H)^2 +\text{c.c.} = 0~.
	\end{align}
\end{subequations}
Hence, the free action for $H_{ij}$ is simply 
\begin{align}
	\label{4.176}
	\cS_{\text{VM}}[H] = \int \rd^4x \rd^{2\N} \q \, \cE\, \mathfrak{W}(H)^2 ~,
\end{align}
where it should be emphasised that \eqref{4.176} is gauge-invariant and superconformal on arbitrary curved backgrounds.

\subsection{The $\cN=2$ superconformal gravitino multiplet}
\label{Appendix4A2}

The supercurrent multiplet which couples to the $\cN=2$ superconformal gravitino multiplet takes the form of a primary, complex isospinor ${J}^i$. It is subject to the conservation equations:
\begin{align}
	\label{2.4}
	{\nabla}_\a^{(i} {J}^{j)} = 0 ~, \qquad \bar{\nabla}^{(ij} {J}^{k)} = 0 ~,
\end{align}
and characterised by the superconformal properties
\begin{align}
	K^M J^i = 0 ~, \qquad \mathbb{D} J^i = \mathbb{Y} J^i = 2 J^i~.
\end{align}
The present multiplet naturally arises via an $\cN = 3 \rightarrow 2$ superspace reduction of the $\cN=3$ conformal supercurrent multiplet \cite{HST}.

We now utilise a Noether coupling to determine the gauge prepotential describing the superconformal gravitino multiplet, which we denote $H_i$. Specifically, the latter may be determined by requiring that
\begin{align}
	\label{2.5}
	\mathcal{S}_{\text{N.C.}} = \int \rd^{4|8}z \, E\, H_i \, {J}^i + \text{c.c.} ~,
\end{align}
is locally superconformal and gauge-invariant. It is clear to see that $H_i$ is defined modulo the gauge freedom
\begin{align}
	\label{2.6}
	\d H_i = \nabla^{\a j} \z_{\a ij} + \bar{\nabla}^{jk} \o_{ijk}~,
\end{align}
and its superconformal transformation law is characterised by the properties:
\begin{align}
	\label{2.7}
	K^B H_i = 0 ~, \qquad \mathbb{D} H_i = \mathbb{Y} H_i = -2 H_i~.
\end{align}
It should be emphasised that the superconformal properties \eqref{2.7} imply that the gauge transformations \eqref{2.6} are superconformal.

Having determined the new gauge prepotential $H_i$ above, it remains to determine its kinetic action. For simplicity, we perform this analysis in conformally flat superspace backgrounds. In particular, one may show that the following chiral descendants
\begin{align}
	\label{2.8}
	\hat{\mathfrak{W}}_\a(H) = \frac{1}{48} \bar{\nabla}^{4} \nabla^{ij} \nabla_{\a i} H_j ~, \qquad \check{\mathfrak{W}}_\a(H) = \frac{1}{48} \bar{\nabla}^{4} \nabla_\a^i \bar{H}_i ~,
\end{align}
are gauge-invariant in conformally flat superspaces and possess the superconformal properties:
\begin{subequations}
	\label{2.9}
	\begin{align}
		K^M \hat{\mathfrak{W}}_\a (H) &= 0 ~, \qquad \mathbb{D} \hat{\mathfrak{W}}_\a (H) = \frac{3}{2} \hat{\mathfrak{W}}_\a (H) ~, \\
		K^M \check{\mathfrak{W}}_\a (H) &= 0 ~, \qquad \mathbb{D} \check{\mathfrak{W}}_\a (H) = \frac{1}{2} \check{\mathfrak{W}}_\a (H) ~,
	\end{align}
\end{subequations}
which are valid in generic backgrounds. It then follows that
\begin{align}
	\label{2.10}
	\mathcal{S}_{\text{GM}}[H,\bar{H}] = \int \rd^4x \rd^4 \q \, \cE\, \hat{\mathfrak{W}}^\a (H) \check{\mathfrak{W}}_\a (H) + \text{c.c.} ~,
\end{align}
is the locally superconformal and gauge-invariant action for the $\cN=2$ superconformal gravitino multiplet in conformally flat backgrounds. The overall coefficient in \eqref{2.10} has been chosen due to the identity
\begin{align}
	\label{2.11}
	\ri \int \rd^4x \rd^4 \q \, \cE\, \hat{\mathfrak{W}}^\a (H) \check{\mathfrak{W}}_\a (H) + \text{c.c.} = 0~,
\end{align}
which holds up to a total derivative on conformally flat backgrounds.

\section{$\sU(1)$ duality rotations for complex CHS fields} \label{Appendix4B}

In section \ref{section4.3.1} we restricted our attention to the analysis of duality-invariant models described by real gauge fields. However, our construction readily extends itself to the complex case. In particular, one may consider a complex CHS gauge field $h_{\a(m)\ad(n)}$, $m \neq n$ and $m,n\geq1$. The action functional describing its dynamics is required to depend only on the field strengths \eqref{4.20}, that is $\cS^{(m,n)}[\hat{\mathfrak{C}}, \check{\mathfrak{C}},\bar{\hat{\mathfrak{C}}},\bar{\check{\mathfrak{C}}}]$, where we have made the definitions.
It is important to note that the field strengths \eqref{4.20} also obey the Bianchi identity
\be
\label{ComplexBI}
(\nabla^\b{}_\ad)^m \hat{\mathfrak{C}}_{\a(n) \b(m)} 
= (\nabla_\a{}^{\bd})^n \bar{\check{\mathfrak{C}}}_{\ad(m) \bd(n)} ~.
\ee

Now, considering $\cS^{(m,n)}[\hat{\mathfrak{C}},\check{\mathfrak{C}},\bar{\hat{\mathfrak{C}}},\bar{\check{\mathfrak{C}}}]$ as a functional of the unconstrained fields $\hat{\mathfrak{C}}_{\a(m+n)}$, $\check{\mathfrak{C}}_{\a(m+n)}$ and their conjugates, we may introduce the primary fields
\be
\ri^{m+n+1} \hat{\mathfrak M}_{\a(m+n)} :=  \frac{\d \cS^{(m,n)}[\hat{\mathfrak{C}},\check{\mathfrak{C}},\bar{\hat{\mathfrak{C}}},\bar{\check{\mathfrak{C}}}]}{\d \check{\mathfrak{C}}^{\a(m+n)}} ~, 
\quad \ri^{m+n+1} \check{\mathfrak M}_{\a(m+n)} :=  \frac{\d \cS^{(m,n)}[\hat{\mathfrak{C}},\check{\mathfrak{C}}~,\bar{\hat{\mathfrak{C}}},\bar{\check{\mathfrak{C}}}]}{\d \hat{\mathfrak{C}}^{\a(m+n)}}~,
\label{5.15}
\ee
where we have made the definition
\begin{align}
	\d \cS^{(m,n)}[\hat{\mathfrak{C}},\check{\mathfrak{C}},\bar{\hat{\mathfrak{C}}},\bar{\check{\mathfrak{C}}}] &= \int \rd^4x\, e \, 
	\Big \{ \d \hat{\mathfrak{C}}^{\a(m+n)} \frac{\d \cS^{(m,n)}[\hat{\mathfrak{C}},\check{\mathfrak{C}},\bar{\hat{\mathfrak{C}}},\bar{\check{\mathfrak{C}}}]}{\d \hat{\mathfrak{C}}^{\a(m+n)}} \non \\
	& \qquad \qquad \qquad \qquad + \d \check{\mathfrak{C}}^{\a(m+n)} \frac{\d \cS^{(m,n)}[\hat{\mathfrak{C}},\check{\mathfrak{C}},\bar{\hat{\mathfrak{C}}},\bar{\check{\mathfrak{C}}}]}{\d \check{\mathfrak{C}}^{\a(m+n)}} \Big \} + \text{c.c.}
\end{align}
The conformal properties of the dual fields \eqref{5.15} are: 
\begin{subequations} 
	\bea 
	K^\bb \hat{\mathfrak M}_{\a(m+n)} &=&0~, \qquad 
	\mathbb{D} \hat{\mathfrak M}_{\a(m+n)} = \Big(2 + \hf(n-m)\Big) \hat{\mathfrak M}_{\a(m+n)}~,\\
	K^{\bb}\check{{\mathfrak M}}_{\a(m+n)}&=&0~,\qquad \mathbb{D}\check{{\mathfrak M}}_{\a(m+n)}
	=\Big(2+\frac{1}{2}(m-n)\Big)\check{{\mathfrak M}}_{\a(m+n)}~.
	\eea
\end{subequations} 
Varying $\cS^{(m,n)}[\hat{\mathfrak{C}},\check{\mathfrak{C}},\bar{\hat{\mathfrak{C}}},\bar{\check{\mathfrak{C}}}]$ with respect to $h_{\a(m) \ad(n)}$ yields the equation of motion
\bea
\label{ComplexEoM}
(\nabla^\b{}_\ad)^m \hat{\mathfrak M}_{\a(n) \b(m)} 
= (\nabla_\a{}^{\bd})^n \bar{\check{\mathfrak M}}_{\ad(m) \bd(n)} ~.
\eea

It is clear from the discussion above that the system of equations \eqref{ComplexBI} and \eqref{ComplexEoM} is invariant under the $\sU(1)$ duality rotations
\begin{subequations}
	\begin{align}
		\d_\l \hat{\mathfrak{C}}_{\a(m+n)} = \l \hat{\mathfrak M}_{\a(m+n)} ~, \quad \d_\l \check{\mathfrak{C}}_{\a(m+n)} = \l \check{\mathfrak M}_{\a(m+n)} ~, \\
		\d_\l \hat{\mathfrak M}_{\a(m+n)} = - \l \hat{\mathfrak{C}}_{\a(m+n)} ~, \quad \d_\l \check{\mathfrak M}_{\a(m+n)} = - \l \check{\mathfrak{C}}_{\a(m+n)} ~.
	\end{align}
\end{subequations}
One may then perform similar analyses to those undertaken in section \ref{section4.3.1} and construct $\sU(1)$ duality-invariant nonlinear models for such fields.\footnote{We mention in passing that such a construction can also be uplifted to the case of a SCHS theory described by a complex prepotential.} 
They satisfy  the self-duality equation
\bea
\ri^{m+n+1} \int \rd^4x \, e \, \Big \{ \hat{\mathfrak{C}}^{\a(m+n)}  \check{\mathfrak{C}}_{\a(m+n)}
+ \hat{\mathfrak M}^{\a(m+n)} \check{\mathfrak M}_{\a(m+n)} \Big \} + \text{c.c.}  = 0 ~,
\eea
which must hold for unconstrained fields $\hat{\mathfrak{C}}_{\a(m+n)}$ and $\check{\mathfrak{C}}_{\a(m+n)}$.
The simplest solution of this equation is the free CHS action \eqref{CHSaction}.
	
\end{subappendices}

\chapter{Conformal supergeometries in two dimensions} \label{Chapter5}

In the preceeding chapters, a central role was played by the framework of conformal superspace. In particular, it proved to be a natural formalism to study (conformal) isometries of a given background, higher derivative symmetries of matter multiplets and various models for (super)conformal higher-spin fields. It should be emphasised that, while our attention was restricted to the four-dimensional case, conformal superspace formulations have previously been constructed for $3 \leq d \leq 6$ spacetime dimensions \cite{ButterN=1, ButterN=2, BKNT-M1, BKNT-M3, BKNT} and have lead to a plethora of applications, see chapter \ref{Chapter1} for more details. More recently, the $d=2$ case was constructed in \cite{KR22} and is the subject of this chapter.

This chapter is organised as follows. Section \ref{Chapter5.1} is devoted to a derivation of the infinite-dimensional superconformal algebra of Minkowski superspace 
$\mathbb{M}^{(2|p,q)}$ by employing its conformal Killing supervector fields. 
We then describe the additional constraints on the conformal Killing supervector fields 
which single out the finite-dimensional superconformal algebra $\mathfrak{osp}(p|2;\mathbb{R}) \oplus \mathfrak{osp}(q|2;\mathbb{R})$.
In section \ref{Chapter5.2} we review the formulation of conformal gravity in two dimensions as the gauge theory of 
the conformal group
$\mathsf{SL} (2 ,{\mathbb R} ) \times \mathsf{SL} (2, {\mathbb R} ) $.
Building on the construction of conformal gravity,  in section \ref{Chapter5.3} 
we formulate conformal $(p,q)$ supergravity as a gauge theory of the superconformal group 
$\mathsf{OSp}_0 (p|2; {\mathbb R} ) \times  \mathsf{OSp}_0 (q|2; {\mathbb R} )$ with a flat connection.
The procedure of `degauging' from this superconformal formulation to the $\sSO(p) \times \sSO(q)$ superspaces is described in section \ref{Chapter5.4}. 
Section \ref{Chapter5.5} is mostly devoted to generalisations and applications of the results derived in
section \ref{Chapter5.4}. 
The main body of this chapter is accompanied by two technical appendices. 
In appendix \ref{Appendix5A} we construct $\mathcal{N}=(1,0)$ conformal superspace with non-vanishing curvature.
Appendix \ref{Appendix5B} is devoted to the supertwistor realisation of 
compactified Minkowski superspace $\overline{\mathbb M}^{(2|2n,q)}$
as a homogeneous space of the superconformal group
$ \mathsf{SU} (1,1|n )
\times  {\sOSp}_0 (q|2; {\mathbb R} )$.

Throughout this chapter we will make use of two types of notation, $(p,q)$ and $\cN=(p,q)$, to denote superspaces with $p$ left and $q$ right real spinor coordinates. Additionally, the special case $p=q$ is also referred to as  $\cN = p$.

\section{Rigid superconformal transformations of $\mathbb{M}^{(2|p,q)}$} \label{Chapter5.1}

This section is devoted the description of the conformal Killing supervector fields of 
$\mathbb{M}^{(2|p,q)}$, the $(p,q)$ Minkowski superspace in two dimensions \cite{Hull:1985jv}.
Our presentation is similar to that of section \ref{Chapter2.1}, though there are some features specific to the two-dimensional case, which will be emphasised.

\subsection{The conformal Killing supervector fields of $\mathbb{M}^{(2|p,q)}$}

Minkowski superspace $\mathbb{M}^{(2|p,q)}$
is  parametrised by the real coordinates $z^{A} = (x^{a},\q^{+ \Io},\q^{- \Iu})$, where $x^{a} = (x^{++},x^{--})= \frac{1}{\sqrt 2}(x^0 + x^1, x^0 - x^1)$, $\Io = \overline{1}, \dots , \overline{p}$ and $\Iu = \1, \dots , \underline{q}$. 
Its covariant derivatives $D_A = (\partial_{a}, D_+^{\Io}, D_-^{\Iu})$ take the form
\bea
\pa_{a}:=\frac{\pa}{\pa x^{a}}= ( \pa_{++} , \pa_{--} )~, \quad
D_{+}^{\Io}
:=
\frac{\pa}{\pa\q^{+ \Io}}
+ \ri \q^{+ \Io} \pa_{++}
~,\quad
D_{-}^{\Iu}
:=
\frac{\pa}{\pa\q^{- \Iu}}
+ \ri \q^{- \Iu}\pa_{--}
~,
\eea
and satisfy the algebra:
\bea
\{ D_{+}^{\Io} , D_{+}^{\Jo} \} =  2 \ri \d^{\Io \Jo} \partial_{++} ~, \qquad  
\{ D_{-}^{\Iu} , D_{-}^{\Ju} \} =  2 \ri \d^{\Iu \Ju} \partial_{--} ~.
\eea
We emphasise that for $p=0$ the left spinor covariant derivative $D_+^{\OI}$ does not appear, similarly $D_-^\UI$ is not present for $q=0$.

The conformal Killing supervector fields of $\mathbb{M}^{(2|p,q)}$,
\be 
\label{5.3}
\xi = \xi^{a} \partial_{a} + \xi^{+ \Io} D_+^{\Io} + \xi^{- \Iu} D_-^{\Iu} = \xi^{++} \partial_{++} + \xi^{--} \partial_{--} + \xi^{+ \Io} D_+^{\Io} + \xi^{- \Iu} D_-^{\Iu} = \bar{\xi} ~,
\ee
may be defined to satisfy the constraints
\be 
\label{5.4}
[\xi , D_+^{\Io} ] = - (D_+^{\Io} \xi^{+ \Jo}) D_+^{\Jo} ~, \qquad
[\xi , D_-^{\Iu} ] = - (D_-^{\Iu} \xi^{- \Ju}) D_-^{\Ju} ~.
\ee
We note that for vanishing $p$ ($q$), the spinor $\xi^{+ \OI}$ ($\xi^{- \UI}$) must be turned off.
From $\eqref{5.4}$ we obtain the fundamental equations
\begin{subequations}\label{5.5}
	\bea
	D_+^{\Io} \xi^{--} &=& 0 \quad \implies \quad \pa_{++} \x^{--}=0~,
	\label{5.5a}\\
	D_-^{\Iu} \xi^{++} &=& 0  \quad \implies \quad \pa_{--} \x^{++}=0~, \label{5.5b}
	\eea
\end{subequations}
and expressions for the spinor parameters
\begin{align}
	\label{5.6}
	\xi^{+ \Io} = - \frac{\ri}{2} D_+^{\Io} \xi^{++} ~, \qquad \xi^{- \Iu} = - \frac{\ri}{2} D_-^{\Iu} \xi^{--}~.
\end{align}
Hence, we see that every conformal Killing supervector field \eqref{5.3} is completely determined by its vector parameter $\xi^{a}$. Additionally, equations \eqref{5.5} tell us  that
\begin{align}
	\xi^{++} =\xi^{++}(x^{++}, \q^+ )	
	\equiv  \xi^{++}(\z_L)  ~, \qquad 
	\xi^{--} =  \xi^{--}(x^{--}, \q^- )   \equiv \xi^{--}(\z_R) ~.
\end{align}
Here $\xi^{++}(\z_L)$ and $\xi^{--}(\z_R)$ are arbitrary functions of $\z_L$ and $\z_R$, respectively.

Taking  \eqref{5.5} into account, the equations \eqref{5.4} 
can be rewritten in the form
\begin{subequations}
	\begin{align}
		[\xi , D_+^{\Io}] &= - \hf (\s[\xi] + K[\xi]) D_+^{\Io} - \r[\xi]^{\Io \Jo} D_{+}^{\Jo} ~, \\
		[\xi , D_-^{\Iu}] &= - \hf (\s[\xi] - K[\xi]) D_-^{\Iu} - \r[\xi]^{\Iu \Ju} D_{-}^{\Ju} ~,
	\end{align}
\end{subequations}
where we have defined the following parameters:
\begin{subequations}
	\label{5.8}
	\begin{align}
		\s[\xi] &:= \hf \big ( \partial_{++} \xi^{++} + \partial_{--} \xi^{--} \big)~, \label{5.8a}\\
		K[\xi] &:= \hf \big ( \partial_{++} \xi^{++} - \partial_{--} \xi^{--} \big)~, \label{5.8b}\\
		\r[\xi]^{\Io \Jo} &:= - \frac{\ri}{4} \big[ D_+^{\Io} , D_+^{\Jo} \big ] \xi^{++} ~, \\
		\r[\xi]^{\Iu \Ju} &:= - \frac{\ri}{4} \big[ D_-^{\Iu} , D_-^{\Ju} \big ] \xi^{--} ~.
	\end{align}
\end{subequations}
Their $z$-independent components generate scale, Lorentz, $\mathfrak{so}(p)$  
and $\mathfrak{so}(q)$ transformations, respectively.
We point out that the $\mathfrak{so}(p)$ parameter $\r[\xi]^{\Io \Jo} $ is a function of the left variables $\z_L =(x^{++},\q^+)$, while the $\mathfrak{so}(q)$ parameter $\r[\xi]^{\Iu \Ju} $ is a function of the right variables $\z_R =(x^{--},\q^-)$. Further, the former (latter) identically vanishes when $p<2$ ($q<2$).

Given two conformal Killing supervectors $\xi_1$ and $\xi_2$, their commutator is another conformal Killing supervector $\xi_3$
\begin{subequations}
	\label{5.10}
	\begin{align} 
		[\xi_1 , \xi_2] = \xi_3^{A} D_A = \xi_3~,
	\end{align}
	with the definitions
	\begin{align}
		\xi_3^{++} &= \xi_1^{++} \pa_{++} \xi_2^{++} - \xi_2^{++} \pa_{++} \xi_1^{++} + 2 \ri \xi_1^{+ \OI} \xi_2^{+ \OI}~ \implies~ D_{-}^{\UI} \xi_3^{++} = 0~, \\
		\xi_3^{--} &= \xi_1^{--} \pa_{--} \xi_2^{--} - \xi_2^{--} \pa_{--} \xi_1^{--} + 2 \ri \xi_1^{- \UI} \xi_2^{- \UI} ~ \implies ~ D_{+}^{\OI} \xi_3^{--} = 0~, 
	\end{align}
\end{subequations}
and the spinor parameters are determined by eq. \eqref{5.6}. 
Equations \eqref{5.5} imply that the algebra of conformal Killing supervector fields of $\mathbb{M}^{(2|p,q)}$ is infinite dimensional.
It
may be referred to as 
a $(p,q)$ super Virasoro algebra. For $p=q$, such superalgebras
were studied in \cite{Ademollo:1975an, Ramond:1976qq, Gastmans:1987up}.

The superconformal transformation law of a primary tensor superfield $U$ (with suppressed Lorentz, $\sSO(p)$ and $ \sSO(q)$ indices) is
\bea
\d_\x U &=&\Big\{
\x + \l_U K[\x] 
+\D_U\s[\x]  + \hf \r[\xi]^{\Io \Jo} \mathfrak{L}^{\OI \OJ} 
+ \hf \r[\xi]^{\Iu \Ju}  \mathfrak{R}^{\UI \UJ} \Big\}U~,
\eea
where $\mathfrak{L}^{\OI \OJ} $ and $ \mathfrak{R}^{\UI \UJ} $ are the generators 
of the groups $\sSO(p)$ and $ \sSO(q)$, respectively.
The parameters $\l_U$ and $\D_U$ are called the Lorentz weight and the dimension (or Weyl weight) of $U$, respectively. These weights are related if $U$ depends only on $\z_L$ or $\z_R$,
\begin{subequations}
	\bea
	U&=& U(\z_L) \quad \implies \quad \l_U = \D_U~,\qquad  \mathfrak{R}^{\UI \UJ} U=0~, \\
	U&=& U(\z_R) \quad \implies \quad \l_U = -\D_U~, \qquad 
	\mathfrak{L}^{\OI \OJ} U=0~.
	\eea
\end{subequations}

As pointed out above, 
the algebra of conformal Killing supervector fields of $\mathbb{M}^{(2|p,q)}$ is infinite dimensional. It contains a finite dimensional subalgebra 
which is singled out by the  constraints
\begin{subequations}
	\label{5.13}
	\begin{align}
		p=0:& \quad \pa_{++} \pa_{++} \pa_{++} \xi^{++}=0~,
		\\
		p=1:& \quad \pa_{++} \pa_{++} D_{+} \xi^{++}= 0~,
		\\
		p=2:& \quad \pa_{++} D_{+}^{[\OI} D_{+}^{\OJ]} \xi^{++}= 0~, 
		\\
		p>2:& \quad D_{+}^{[\OI} D_{+}^{\OJ} D_+^{\OK]} \xi^{++}= 0~, 
	\end{align}
\end{subequations}
and their counterparts in the right sector.
Physically, these conditions mean that $\xi$ generates those infinitesimal superconformal transformations that belong to the superconformal algebra $\mathfrak{osp}(p|2;\mathbb{R}) \oplus \mathfrak{osp}(q|2;\mathbb{R})$.
This is the Lie algebra of the superconformal group ${\sOSp}_0 (p|2; {\mathbb R} ) \times  {\sOSp}_0 (q|2; {\mathbb R} )$,
which acts on the compactified Minkowski superspace \eqref{5.145}.
It is instructive to check that \eqref{5.10} preserves the conditions \eqref{5.13}.
We will assume \eqref{5.13} in what follows.

Employing \eqref{5.5}, it is possible to show that the parameters 
\eqref{5.8a} and \eqref{5.8b} 
satisfy the constraints
\begin{subequations}
	\label{5.9}
	\begin{align}
		D_+^{\Io} \s[\xi] &=  D_+^{\Io} K[\xi] \quad \implies \quad \pa_{++} \s[\xi] = \pa_{++} K[\xi]~, \\
		D_-^{\Iu} \s[\xi] &=  - D_-^{\Iu} K[\xi] \quad \implies \quad  \pa_{--} \s[\xi] = - \pa_{--} K[\xi]~.
	\end{align}
	Next, by using \eqref{5.5} in conjunction with \eqref{5.13}, one obtains the following constraints on the $R$-symmetry parameters:
	\begin{align}
		D_+^{\Io} \r[\xi]^{\Jo \OK} &= 2 \d^{\Io [\Jo} D_+^{\OK]} \s[\xi] \quad \implies \quad \pa_{++} \r^{\OI \OJ}[\xi] = 0 ~,\\
		D_-^{\Iu} \r[\xi]^{\Jo \OK} &= 0 \quad \implies \quad \pa_{--} \r^{\OI \OJ}[\xi] = 0~, \\
		D_+^{\OI} \r[\xi]^{\UJ \UK} &= 0  \quad \implies \quad \pa_{++} \r^{\UI \UJ}[\xi] = 0~, \\
		D_-^{\UI} \r[\xi]^{\UJ \UK} &= 2 \d^{\UI [\UJ} D_-^{\UK]} \s[\xi] \quad \implies \quad \pa_{--} \r^{\UI \UJ}[\xi] = 0~,
	\end{align}
	and on the scaling parameter:
	\begin{align}
		D_+^{\OI} D_+^{\OJ} \s[\xi] &= \frac{\ri}{p} \d^{\OI \OJ} \partial_{++} \s[\xi] \quad \implies \quad \partial_{++} D_+^\OI \s[\xi] = 0~, \\
		D_-^{\UI} D_-^{\UJ} \s[\xi] &= \frac{\ri}{q} \d^{\UI \UJ} \partial_{--} \s[\xi] \quad \implies \quad \partial_{--} D_-^\UI \s[\xi] = 0~.
	\end{align}
\end{subequations}

The above results mean that we may parametrise the conformal Killing supervector fields obeying \eqref{5.13} as
\be 
\xi \equiv \xi(\l(P)^{a},\l(Q)^{+\OI}, \l(Q)^{-\UI} ,  \l(M) , \l(\mathbb{D}) ,\l(\mathfrak{L})^{\OI\OJ} , \l(\mathfrak{R})^{\UI\UJ} , \l(K)_a , \l(S)_+^{\OI}, \l(S)_-^{\UI}) ~,
\ee
where we have defined the parameters
\bsubeq
\begin{align} \l(P)^a &:= \xi^{a} |_{z = 0} ~, \qquad \l(Q)^{+ \OI} := \xi^{+\OI}|_{z = 0} ~, \qquad \l(Q)^{- \UI} := \xi^{- \UI}|_{z = 0} ~, \\
	& \qquad \quad ~\, \l(M) := K[\xi]|_{z = 0} ~, \qquad \quad \, \l({\mathbb D}) := \s[\xi]|_{z = 0} ~, \\
	& \qquad \quad  \l(\mathfrak{L})^{\OI \OJ} := \r[\xi]^{\OI \OJ}|_{z = 0} ~, \;\,\, \quad \l(\mathfrak{R})^{\UI \UJ} := \r[\xi]^{\UI \UJ}|_{z = 0} \\
	\l(K)_a &:= \hf \pa_a \s[\xi] |_{z = 0} ~, \quad \l(S)_{+}^{\OI} := \hf D_+^\OI \s[\xi]|_{z = 0} ~, \quad \l(S)_{-}^{\UI} := \hf D_-^\UI \s[\xi]|_{z = 0} ~.
\end{align}
\esubeq
In particular, $\xi$ may be represented as
\be
\xi = \l(X)^{\tilde{A}} X_{\tilde{A}} ~,
\ee
where we have introduced a condensed notation for the superconformal parameters
\begin{subequations}
	\begin{align}
		\l(X)^{\tilde{A}} &= (\l(P)^{A}, \l(M) , \l(\mathbb{D}) ,\l(\mathfrak{L})^{\OI\OJ} , \l(\mathfrak{R})^{\UI\UJ} , \l(K)_{A})~, \\
		\l(P)^A &= (\l(P)^{a}, \l(Q)^{+\OI}, \l(Q)^{-\UI}) , \qquad \l(K)_A = (\l(K)_a, \l(S)_{+}^{\OI}, \l(S)_{-}^{\UI}) ~,
	\end{align}
\end{subequations}
and for generators of the superconformal algebra
\begin{subequations}
	\label{5.18}
	\begin{align}
		X_{\tilde{A}} &= (P_A, M , \mathbb{D} ,\mathfrak{L}^{\OI\OJ} , \mathfrak{R}^{\UI\UJ} , K^A)~, \\
		P_A &= (P_{a}, Q^{\OI}_+, Q^{\UI}_{-}) ~, \qquad K^A = (K^{a}, S^{+\OI}, S^{-\UI})~.
	\end{align}
\end{subequations}

Making use of the above results allows us to derive the graded commutation relations for the superconformal algebra.
This may be achieved by utilising the relation
\bea
[\xi_1,\xi_2] = - \l(X)_2^{\tilde{B}} \l(X)_1^{\tilde{A}} \big[ X_{\tilde{A}}, X_{\tilde{B}} \big \}~.
\eea  
The commutation relations for the conformal algebra are as follows:\footnote{These relations can be compared with \eqref{2.17}.}
\begin{subequations}
	\label{5.20}
	\begin{align}
		[M,P_{\pm\pm}]&= \pm P_{\pm\pm}~, \qquad ~\, [\mathbb{D},P_{\pm\pm}]=P_{\pm \pm}~, \label{5.20a}\\ 
		[M,K^{\pm\pm}]&= \mp K^{\pm\pm}~, \qquad [\mathbb{D},K^{\pm\pm}]=-K^{\pm\pm}~,\\
		&\; [K^{\pm \pm},P_{\pm \pm}]= 2 (\mathbb{D} \pm  M) ~. \label{5.20c}
	\end{align}
\end{subequations}
The $R$-symmetry generators $\mathfrak{L}^{\OI \OJ}$ and $\mathfrak{R}^{\UI \UJ}$ commute with all the generators of the conformal group. Amongst themselves, they obey the algebra
\begin{subequations}
	\begin{align}
		[\mathfrak{L}^{\OI \OJ},\mathfrak{L}^{\OK \OL}]&= 2 \d^{\OK [\OI} \mathfrak{L}^{\OJ] \OL} - 2 \d^{\OL [\OI} \mathfrak{L}^{\OJ] \OK} ~,\\
		[\mathfrak{R}^{\UI \UJ},\mathfrak{R}^{\UK \UL}]&= 2 \d^{\UK [\UI} \mathfrak{R}^{\UJ] \UL} - 2 \d^{\UL [\UI} \mathfrak{R}^{\UJ] \UK} ~.
	\end{align}
\end{subequations}
The superconformal algebra is then obtained by extending the translation generator $P_{a}$ to $P_A$ and the special conformal generator $K^{a}$ to $K^A$. The commutation relations involving the $Q$-supersymmetry generators with the bosonic ones are:
\begin{subequations}
	\begin{align}
		\big[M, Q_+^\OI \big] &= \hf Q_+^\OI ~,\qquad \qquad \quad 
		\big[M, Q_-^\UI \big] = - \hf Q_-^\UI~,\\
		\big[\mathbb{D}, Q_+^\OI \big] &= \hf Q_+^\OI ~, \qquad \qquad \quad \,\,
		\big[\mathbb{D}, Q^\UI_- \big] = \hf Q^\UI_- ~, \\
		\big[\mathfrak{L}^{\OI \OJ}, Q_+^\OK \big] &=  2 \d^{\OK[\OI}Q_+^{\OJ]} ~, \qquad \; \;
		\big[\mathfrak{R}^{\UI \UJ}, Q_-^\UK \big] = 2 \d^{\UK[\UI}Q_-^{\UJ]} ~,  \\
		\big[K^{++}, Q_+^\OI \big] &= \ri S^{+ \OI} ~, \qquad \quad \quad \,
		\big[K^{--}, Q_-^\UI \big] = \ri S^{- \UI} ~.
	\end{align}
\end{subequations}
The commutation relations involving the $S$-supersymmetry generators 
with the bosonic operators are: 
\begin{subequations}
	\begin{align}
		\big[M, S^{+\OI} \big] &= - \hf S^{+\OI} ~,\qquad \qquad \quad 
		\big[M, S^{-\UI} \big] = \hf S^{-\UI}~,\\
		\big[\mathbb{D}, S^{+\OI} \big] &= - \hf S^{+\OI} ~, \qquad \qquad \quad \,\,
		\big[\mathbb{D}, S^{-\UI} \big] = - \hf S^{-\UI} ~, \\
		\big[\mathfrak{L}^{\OI \OJ}, S^{+\OK} \big] &=  2 \d^{\OK[\OI} S^{+\OJ]} ~, \qquad \;\;\;
		\big[\mathfrak{R}^{\UI \UJ}, S^{-\UK} \big] = 2 \d^{\UK[\UI} S^{-\UJ]} ~,  \\
		\big[S^{+\OI}, P_{++} \big] &= -2\ri Q_{+}^{\OI} ~, \qquad \quad \quad \;
		\big[S^{-\UI}, P_{--} \big] = -2\ri Q_{-}^{\UI} ~.
	\end{align}
\end{subequations}
Finally, the anti-commutation relations of the fermionic generators are: 
\begin{subequations}
	\label{5.25}
	\begin{align}
		\{Q_+^{\OI} , Q_+^{\OJ} \} &= 2 \ri \d^{\OI \OJ} P_{++}~, \qquad \quad\;\;\, \{Q_-^{\UI} , Q_-^{\UJ} \} = 2 \ri \d^{\UI \UJ} P_{--} ~, \\
		\{ S^{+\OI} , S^{+\OJ} \} &= - 4 \ri \d^{\OI \OJ} K^{++}~, \qquad \{ S^{-\UI} , S^{-\UJ} \} = - 4 \ri \d^{\UI \UJ} K^{--}~,\\
		& \;\; \{ S^{+\OI} , Q_+^{\OJ} \} = 2 \d^{\OI \OJ} (\mathbb{D} + M) - 2 \mathfrak{L}^{\OI \OJ} ~, \\
		& \;\; \{ S^{-\UI} , Q_-^{\UJ} \} = 2 \d^{\UI \UJ} (\mathbb{D} - M) - 2 \mathfrak{R}^{\UI \UJ}  ~.
	\end{align}
\end{subequations}
Note that all remaining (anti-)commutators not listed above vanish identically.

\section{Conformal geometry in two dimensions} \label{Chapter5.2}

Before turning to the superconformal case, it is instructive to first consider conformal gravity as the gauge theory of the $d=2$ conformal group
$\mathsf{SL} (2 ,{\mathbb R} ) \times \mathsf{SL} (2, {\mathbb R} ) $.
Such a formulation can be extracted from those for the $(1,0)$, $\cN=1$ and $\cN=2$  conformal supergravity theories
\cite{vanNieuwenhuizen:1985an, Uematsu:1984zy, Uematsu:1986de, Hayashi:1986ev, Uematsu:1986aa, Bergshoeff:1985qr, Bergshoeff:1985gc, Schoutens:1986kz}. However, our discussion below has some specific features not discussed in prior works. These will be important in our formulation of $(p,q)$ conformal supergravity in the next subchapter.
We will also emphasise those aspects of conformal gravity are unique to two dimensions (as compared with the $d>2$ case reviewed in chapter \ref{Chapter2.2}).

We recall from the previous subsection that the conformal algebra of $\mathbb{M}^{2}$, whose generators we collectively denote $X_{\tilde{a}}$, is spanned by 
the translation ($P_{a}$), Lorentz ($M$), dilatation ($\mathbb{D}$) and special conformal generators ($K^{a}$), which can be grouped into the two disjoint subalgebras spanned by $P_a$ and $X_{\underline a}$:
\begin{align}
	X_{\tilde{a}} = (P_{a}, X_{\underline{a}})~, \qquad X_{\underline{a}} = (M, \mathbb{D}, K^{a})~.
\end{align}
Then, the commutation relations \eqref{5.20} may be rewritten as follows
\bsubeq
\begin{align}
	[X_{\underline{a}} , X_{\underline{b}} ] &= -f_{\underline{a} \underline{b}}{}^{\underline{c}} X_{\underline{c}} \ , \\
	[X_{\underline{a}} , P_{{b}} ] &= -f_{\underline{a} { {b}}}{}^{\underline{c}} X_{\underline{c}}
	- f_{\underline{a} { {b}}}{}^{ {c}} P_{ {c}} \label{nonsusymixing}
	~,
\end{align}
\esubeq
where $f_{\underline{a} \underline{b}}{}^{\underline{c}}$, $f_{\underline{a} { {b}}}{}^{\underline{c}}$ and $f_{\underline{a} { {b}}}{}^{ {c}}$ denote the structure coefficients of the conformal algebra.

\subsection{Gauging the conformal algebra}

Let $\mathcal{M}^{2}$ be a two-dimensional curved spacetime parametrised by local coordinates $x^m$.
To gauge the conformal algebra we associate each non-translational generator $X_{\underline{a}}$ with a connection one-form $\o^{\underline{a}} = (\o,b,\mathfrak{f}_a)=\rd x^{m} \o_m{}^{\underline{a}}$ and with $P_a$ a vielbein one-form $e^a = \rd x^m e_m{}^a$, where it is assumed that $e:={\rm det}(e_m{}^a) \neq 0$, hence there exists a unique inverse vielbein $e_a{}^m$
\begin{align}
	e_a{}^m e_m{}^b = \d_a{}^b~, \qquad e_m{}^a e_a{}^n=\d_m{}^n~.
\end{align}
The latter may be used to construct
the vector fields $e_a = e_a{}^m \pa_m $, with 
$\pa_m = \pa /\pa x^m$, which constitute a basis for the tangent space at each
point of $\mathcal{M}^{2}$. It may then be used to express the connection in the vielbein basis
as $\omega^{\underline{a}} =e^b\omega_b{}^{\underline{a}}$, 
where $\omega_b{}^{\underline{a}}=e_b{}^m\omega_m{}^{\underline{a}}$. 
The covariant derivatives $\nabla_a$ then take the form 
\bea
\label{CGNabla}
\nabla_a
&=& e_a  - \o_a{}^{\underline b} X_{\underline b}=
e_a -  \o_a M - b_a \mathbb{D} - \mathfrak{f}_{ab} K^b
~.
\eea
It should be noted that the translation generators $P_a$ do not appear in the above expression.
Instead, we assume that they are replaced by the covariant derivatives $\nabla_a$ and obey the graded commutation relations (c.f. \eqref{nonsusymixing})
\be
[ X_{\underline{a}} , \nabla_b ] = - f_{\underline{a} b}{}^{\underline{c}} X_{\underline{c}} -f_{\underline{a} b}{}^c \nabla_c ~.
\ee

By definition, the gauge group of conformal gravity is generated by local transformations of the form
\begin{subequations}\label{5.31}
	\bea
	\delta_{\mathscr K} \nabla_a &=& [\mathscr{K},\nabla_a] \ , \\
	\mathscr{K} &=& \xi^b \nabla_b +  \L^{\underline{b}} X_{\underline{b}}
	=  \xi^b \nabla_b + K M + \s \mathbb{D} + \L_b K^b ~,
	\eea
\end{subequations}
where  the gauge parameters satisfy natural reality conditions. These gauge transformations act
on a conformal tensor field $U$ (with its indices suppressed) as 
\bea 
\label{5.32}
\d_{\mathscr K} U = {\mathscr K} U ~.
\eea
Further, we will say that $U$ is primary if (i) it is annihilated by the special
conformal generators, $K^a U = 0$; and (ii) it is an eigenvector of $\mathbb D$.  It  will be said to have 
dimension $\D$ and Lorentz weight $\l$ if 
\begin{align}
	\mathbb D U = \D U~, \qquad M U = \l \mathcal{U}~. 
\end{align}

The covariant derivatives \eqref{CGNabla} obey the commutation relations
\begin{align}
	\big[\nabla_{++} , \nabla_{--}\big] &= -\mathcal{T}^{a} \nabla_a - \mathcal{R}(X)^{\underline{a}} X_{\underline{a}} = - \cT^{++} \nabla_{++} - \cT^{--} \nabla_{--} - \mathcal{R}(X)^{\underline{a}} X_{\underline{a}} ~,
\end{align}
where the torsion and curvatures take the form
\begin{subequations}
	\begin{align}
		\mathcal{T}^{++} &= - \mathscr{C}^{++} + \o^{++} + b^{++}~,\\
		\mathcal{T}^{--} &= - \mathscr{C}^{--} + \o^{--} - b^{--} ~,\\
		\mathcal{R}(M) &= - \hf R - 2 (\mathfrak{f}_{++,--} + \mathfrak{f}_{--,++})~,\\
		\mathcal{R}(\mathbb{D}) &= - \mathscr{C}^a b_a + e_{++} b_{--} - e_{--} b_{++} + 2(\mathfrak{f}_{++,--} - \mathfrak{f}_{--,++}) ~,\\
		\mathcal{R}(K)_{++} &= - \mathscr{C}^a \mathfrak{f}_{a,++} + e_{++} \mathfrak{f}_{--,++} - e_{--} \mathfrak{f}_{++,++} - \o_{++} \mathfrak{f}_{--,++} \non \\
		&\phantom{=}~ - b_{++} \mathfrak{f}_{--,++} + \o_{--} \mathfrak{f}_{++,++} + b_{--} \mathfrak{f}_{++,++}~,\\
		\mathcal{R}(K)_{--} &= - \mathscr{C}^a \mathfrak{f}_{a,--} + e_{++} \mathfrak{f}_{--,--} - e_{--} \mathfrak{f}_{++,--} \non \\
		&\phantom{=}~ - b_{++} \mathfrak{f}_{--,--} - \o_{--} \mathfrak{f}_{++,--} + b_{--} \mathfrak{f}_{++,--} + \o_{++} \mathfrak{f}_{--,--} ~, \\
		R &= 2 \mathscr{C}^{a} \o_a -2 e_{++} \o_{--} +  e_{--} \o_{++} ~.
	\end{align}
\end{subequations}
Here $R$ is the scalar curvature constructed from the Lorentz connection $\o_{a}$ and we have introduced the anholonomy coefficients $\mathscr{C}^a$
\begin{align}
	[e_{++},e_{--}] = \mathscr{C}^a e_a = \mathscr{C}^{++} e_{++} + \mathscr{C}^{--} e_{--}~.
\end{align}

In order for this geometry to describe conformal gravity, it is necessary to impose certain covariant constraints. Specifically, we require that the torsion and both Lorentz and dilatation curvatures vanish
\begin{align}
	\label{5.37}
	\cT^{a} = 0 ~, \qquad \cR(M) = 0~, \qquad \cR(\mathbb{D}) = 0~.
\end{align}
The constraint $\cT^{a} = 0$ determines the Lorentz connection in terms of the vielbein and dilatation connection
\begin{align}
	\o_{\pm\pm} = \mathscr{C}_{\pm\pm} \pm b_{\pm\pm}~,
\end{align}
while $\cR(M) = \cR(\mathbb{D}) = 0$ fixes several components of the special conformal connection
\begin{subequations}
	\label{5.39}
	\begin{align}
		\mathfrak{f}_{++,--} &= -\frac{1}{8} \big ( \mathcal{R} -2 \mathscr{C}^a b_a +2 e_{++} b_{--} -2 e_{--} b_{++} \big) ~, \\
		\mathfrak{f}_{--,++} &= -\frac{1}{8} \big ( \mathcal{R} +2 \mathscr{C}^a b_a -2 e_{++} b_{--} +2 e_{--} b_{++} \big) ~.
	\end{align}
\end{subequations}
As a result, the algebra of conformal covariant derivatives takes the form\footnote{The $(1,0)$ superconformal extension of this geometry is described in appendix \ref{Appendix5A}.}
\begin{align}
	\label{5.40}
	\big[\nabla_{++} , \nabla_{--}\big] &= W_{++} K^{++} + W_{--} K^{--} ~, \qquad W_{a} \equiv - \mathcal{R}(K)_{a}~,
\end{align}
where $W_{++}$ and $W_{--}$ have the following conformal properties 
\begin{subequations}
	\begin{align}
		K^{a} W_{++} = 0 ~, \qquad \mathbb{D} W_{++} = 3 W_{++} ~, \\
		K^{a} W_{--} = 0 ~, \qquad \mathbb{D} W_{--} = 3 W_{--} ~.
	\end{align}
\end{subequations}

Strictly speaking, it is necessary to impose the constraint $W_a = 0$ for this geometry to describe conformal gravity.\footnote{This is in contrast to the situation in $d>3$ dimensions, see chapter \ref{Chapter2.1}.} Specifically, in the absence of this constraint, there are extra degrees of freedom, in addition to the vielbein, which correspond to the special conformal connections $\mathfrak{f}_{++,++}$ and $\mathfrak{f}_{--,--}$. This will be addressed in further detail below.

\subsection{Degauging to Lorentzian geometry}

According to \eqref{5.31}, under an infinitesimal special superconformal gauge transformation $\mathscr{K} = \Lambda_{b} K^{b}$, the dilatation connection transforms algebraically
\bea
\d_{\mathscr{K}} b_{a} = - 2 \L_{a} ~.
\eea
As a result, we may enforce the gauge $b_{a} = 0$, which completely fixes 
the freedom to perform special superconformal transformations with unconstrained $\L_b$. Hence, the connection $\mathfrak{f}_{ab}$ is not required for the covariance of $\nabla_a$ and it may be separated
\bea
\nabla_{a} &=& \cD_{a} - \mathfrak{f}_{ab} K^{b} ~,
\eea
where the operator $\cD_a$ takes the form
\begin{align}
	\cD_a = e_a -  \o_a M ~.
\end{align}

An important feature of this gauge, which follows from \eqref{5.39}, is 
\bea
\mathfrak{f}_{++,--} = \mathfrak{f}_{--,++} = - \frac{1}{8} \mathcal{R} ~,
\eea
which, keeping in mind the relation
\bea
\label{5.46}
[ \cD_{++} , \cD_{--} ] &=& [ \nabla_{++} , \nabla_{--} ] + \big(\cD_{++} \mathfrak{f}_{--,a} - \cD_{--} \mathfrak{f}_{++,a} \big) K^a \non \\
&\phantom{=}& + \mathfrak{f}_{++,a} [ K^{a} , \nabla_{--}] - \mathfrak{f}_{--,a} [ K^{a} , \nabla_{++}]  ~,
\eea
allows one to determine $[ \cD_{++} , \cD_{--} ]$ by a routine computation. One finds
\begin{align}
	[ \cD_{++}, \cD_{--}] = \hf \cR M~.
\end{align}
Additionally, by analysing the special conformal sector of \eqref{5.46}, we obtain the relations
\begin{align}
	W_{++} = -\frac{1}{8} \cD_{++} \cR -\cD_{--} \mathfrak{f}_{++,++} ~, \qquad
	W_{--} = \frac{1}{8} \cD_{--} \cR + \cD_{++} \mathfrak{f}_{--,--} ~.
	\label{5.48}
\end{align}
In particular, for vanishing $W_{++}$ ($W_{--}$), we see that $\mathfrak{f}_{++,++}$ ($\mathfrak{f}_{--,--}$) is a non-local function of the vielbein, which is in contrast to the situation in $d>2$ spacetime dimensions, see chapter \ref{Chapter2.1}.

If no constraint is imposed on the conformal curvature tensors $W_{++}$ and $W_{--}$, then the components $\mathfrak{f}_{++,++} $ and $ \mathfrak{f}_{--,--} $ of the 
special conformal connection remain independent fields in addition to the vielbein. 
Therefore, we are forced to impose the constraints 
\bea
W_{++}=0~, \qquad W_{--}=0~,
\eea
in order for the  vielbein to be the only independent field. Then it follows from 
\eqref{5.48} that $\mathfrak{f}_{++,++} $ and $ \mathfrak{f}_{--,--} $ become non-local functions of the gravitational field. 

Next, it is important to describe the gauge freedom of this geometry, which corresponds to the residual gauge transformations of \eqref{5.31} in the gauge $b_a = 0$. These include local $\mathcal{K}$-transformations of the form
\begin{subequations}\label{CGtransformations}
	\bea
	\delta_{\mathcal K} \cD_A = [\mathcal{K},\cD_A] ~ , \qquad
	\mathcal{K} = \xi^b \cD_b + K M ~,
	\eea
	which act on tensor fields $U$ (with their indices suppressed) as
	\begin{align}
		\d_\cK U = \cK U ~.
	\end{align}
\end{subequations}

The gauge transformations \eqref{CGtransformations} are not the most general conformal gravity gauge transformations preserving the gauge $b_a=0$.
Specifically, it may be shown that the following transformation also enjoys this property
\begin{align}
	\mathscr{K}(\s) = \s \mathbb{D} + \frac{1}{2} \nabla_b \s K^b \quad \Longrightarrow \quad \d_{\mathscr{K}(\s)} b_a = 0~,
\end{align}
where $\s$ is real but otherwise unconstrained. As a result, it is necessary to consider how this transformation manifests in the degauged geometry
\begin{align}
	\d_{\mathscr{K}(\s)} \nabla_A \equiv \d_\s \nabla_a = \d_\s \cD_a - \d_\s \mathfrak{f}_{ab} K^b~.
\end{align}
Employing this relation, we arrive at the transformation laws
\begin{subequations}
	\begin{align}
		\d_\s \cD_{++} &= \s \cD_{++} + \cD_{++} \s M ~, \\
		\d_\s \cD_{--} &= \s \cD_{--} - \cD_{--} \s M ~, \\
		\d_\s \cR &= 2 \s \cR - 4 \cD_{++} \cD_{--} \s ~,
	\end{align}
\end{subequations}
which are exactly the Weyl transformations of spacetime.

\section{Conformal $(p,q)$ superspace} \label{Chapter5.3}

In chapter \ref{Chapter5.2}, we realised the superconformal algebra 
$\mathfrak{osp}(p|2;\mathbb{R}) \oplus \mathfrak{osp}(q|2;\mathbb{R})$
as the maximal finite-dimensional subalgebra of the $(p,q)$ super Virasoro algebra.
Now we turn to formulating the corresponding gauge theory. This is known as conformal superspace and is identified with a pair $(\cM^{(2|p,q)}, \nabla)$, where $\mathcal{M}^{(2|p,q)}$ denotes a supermanifold parametrised by local  coordinates $z^M$,
and $\nabla$ is a covariant derivative associated with the superconformal algebra. The latter is generated by the operators $X_{\tilde A} $, eq.  \eqref{5.18}, which can be grouped into the two disjoint 
subsets
$P_A$ and $X_{\underline A}$,
\bea 
\label{5.55}
X_{\tilde A} = (P_A, X_{\underline{A}} )~, \qquad
X_{\underline{A}} =
( M ,{\mathbb D}, \mathfrak{L}^{\OI\OJ}, \mathfrak{R}^{\UI\UJ} , K^A)~,
\eea 
each of which constitutes a superalgebra:
\bsubeq
\begin{align}
	[P_{ {A}} , P_{ {B}} \} &= -f_{{ {A}} { {B}}}{}^{{ {C}}} P_{ {C}}
	\ , \\
	[X_{\underline{A}} , X_{\underline{B}} \} &= -f_{\underline{A} \underline{B}}{}^{\underline{C}} X_{\underline{C}} \ , \\
	[X_{\underline{A}} , P_{{B}} \} &= -f_{\underline{A} { {B}}}{}^{\underline{C}} X_{\underline{C}}
	- f_{\underline{A} { {B}}}{}^{ {C}} P_{ {C}}
	\ . \label{mixing}
\end{align}
\esubeq

In order to define covariant derivatives, it is necessary to associate with each non-translational generator $X_{\underline A}$
a connection one-form 
$\Omega^{\underline{A}} = (\O,B,\Phi^{\OI\OJ},\F^{\UI\UJ},\frak{F}_{A})= \rd z^M \Omega_M{}^{\underline{A}}$,  
and with $P_{ {A}}$ a supervielbein one-form
$E^{ {A}} = (E^{a}, E^{+ \OI},E^{-\UI}) = \rd z^{ {M}} E_M{}^A$. 
It is assumed that the supermatrix $E_M{}^A$ is nonsingular, $E:= {\rm Ber} (E_M{}^A) \neq 0$, 
hence there exists a unique inverse supervielbein $E_A{}^M$
\begin{align}
	E_A{}^ME_M{}^B=\d_A{}^B~, \qquad E_M{}^AE_A{}^N=\d_M{}^N~.
\end{align}
The latter may be used to construct
the supervector fields $E_A = E_A{}^M \pa_M $, with 
$\pa_M = \pa /\pa z^M$, which constitute a basis for the tangent space at each
point of $\mathcal{M}^{(2|p,q)}$. The connection may then be expressed in 
the supervielbein basis as $\Omega^{\underline{A}} =E^B\Omega_B{}^{\underline{A}}$, 
where $\Omega_B{}^{\underline{A}}=E_B{}^M\Omega_M{}^{\underline{A}}$. 
The covariant derivatives $\nabla_A= (\nabla_{a}, \nabla_{+}^{\OI}, \nabla_{-}^{\UI})$ then take the form
\bea
\label{5.58}
\nabla_A 
&=& E_A  - \O_A{}^{\underline B} X_{\underline B}=
E_A -  \Omega_A M - B_A \mathbb{D} - \hf \Phi_A^{\OI \OJ} \mathfrak{L}^{\OI \OJ} - \hf \Phi_A^{\UI \UJ} \mathfrak{R}^{\UI \UJ} - \mathfrak{F}_{AB} K^B
~.~~~
\eea
It should be noted that the translation generators $P_A$ do not appear in the above expression.
Instead, we assume that they are replaced by the covariant derivatives $\nabla_A$ and obey the graded commutation relations
\be
[ X_{\underline{B}} , \nabla_A \} = -f_{\underline{B} A}{}^C \nabla_C
- f_{\underline{B} A}{}^{\underline{C}} X_{\underline{C}} ~,
\ee
where the relevant structure constants were defined in equation \eqref{mixing}.

By definition, the gauge group of conformal supergravity  is generated by local transformations of the form
\begin{subequations}\label{5.60}
	\bea
	\delta_{\mathscr K} \nabla_A &=& [\mathscr{K},\nabla_A] \ , \\
	\mathscr{K} &=& \xi^B \nabla_B +  \L^{\underline{B}} X_{\underline{B}}
	=  \xi^B \nabla_B + K M + \s \mathbb{D} + \hf \r^{\OI \OJ} \mathfrak{L}^{\OI \OJ}
	+ \hf \r^{\UI \UJ} \mathfrak{R}^{\UI \UJ}	+ \L_B K^B ~,~~~~~~~~
	\eea
\end{subequations}
where  the gauge parameters satisfy natural reality conditions. These gauge transformations act
on a conformal tensor superfield ${U}$ (with its indices suppressed) as 
\bea 
\label{5.61}
\d_{\mathscr K} {U} = {\mathscr K} {U} ~.
\eea
Further, we will say that ${U}$ is primary if (i) it is annihilated by the special
conformal generators, $K^A {U} = 0$; and (ii) it is an eigenvector of $\mathbb D$. The superfield is said to have 
dimension $\D$ and Lorentz weight $\l$ if 
$\mathbb D U = \D U$ and  $ M {U} = \l {U}$. 

The covariant derivatives \eqref{5.58} obey the graded commutation relations
\begin{align}
	\label{5.62}
	\big[\nabla_{A} , \nabla_B\big\} = -\mathcal{T}_{AB}{}^{C} \nabla_C - \mathcal{R}(X)_{AB}{}^{\underline{C}} X_{\underline{C}}~.
\end{align}
In conformal superspace, we impose the requirement that torsion $\mathcal{T}_{AB}{}^{C}$ and curvature tensors $\mathcal{R}(X)_{AB}{}^{\underline{C}}$ differ from their flat counterparts only by terms proportional 
to the conformal curvatures of $\cM^{(2|p,q)}$. Here we will assume that all such superfields vanish\footnote{See appendix \ref{Appendix5A} for a construction of conformal $(1,0)$ superspace with non-vanishing curvature.} and thus the only non-vanishing graded commutators are:
\begin{align}
	\label{5.63}
	\{ \nabla_{+}^{\Io} , \nabla_{+}^{\Jo} \} =  2 \ri \d^{\Io \Jo} \nabla_{++} ~, \qquad  
	\{ \nabla_{-}^{\Iu} , \nabla_{-}^{\Ju} \} =  2 \ri \d^{\Iu \Ju} \nabla_{--} ~.
\end{align}

\section{The superspace geometry of $(p,q)$ supergravity} \label{Chapter5.4}

According to \eqref{5.60}, under an infinitesimal special superconformal gauge transformation $\mathscr{K} = \Lambda_{B} K^{B}$, the dilatation connection transforms algebraically
\bea
\label{563}
\d_{\mathscr{K}} B_{A} = - 2 \Lambda_{A} ~.
\eea
Hence, we may enforce the gauge $B_{A} = 0$, which completely fixes 
the freedom to perform special superconformal transformations with unconstrained $\L_B$. As a result, the corresponding connection $\mathfrak{F}_{AB}$ is not required for the covariance of $\nabla_A$, and it may be separated
\bea
\nabla_{A} &=& \cD_{A} - \mathfrak{F}_{AB} K^{B} ~. \label{5.65}
\eea
Here the degauged covariant derivative $\cD_A$ involves only the Lorentz and $R$-symmetry connections (depending on the choice of $p$ and $q$).
Additionally, the special superconformal connection
$\mathfrak{F}_{AB}$ may be related to the torsion and curvatures of the degauged geometry by analysing the relation
\bea
\label{5.66}
[ \cD_A , \cD_B \}  &=&[ \nabla_{A} , \nabla_{B} \}+ \big(\cD_A \mathfrak{F}_{BC} - (-1)^{\ve_A \ve_B} \cD_B \mathfrak{F}_{AC} \big) K^C + \mathfrak{F}_{AC} [ K^{C} , \nabla_B \} \non \\
&& - (-1)^{\ve_A \ve_B} \mathfrak{F}_{BC} [ K^{C} , \nabla_A \} - (-1)^{\ve_B \ve_C} \mathfrak{F}_{AC} \mathfrak{F}_{BD} [K^D , K^C \} ~.
\eea

We will refer to the superspace geometry described by the covariant derivatives $\cD_A$ as
curved $\sSO(p) \times \sSO(q)$ superspace.

\subsection{$p,q > 1$ case}

First, we consider the case where $p,q >1$. By a routine calculation, one finds that the degauged connection $\mathfrak{F}_{AB}$ takes the form
\begin{subequations}
	\label{5.67}
	\bea
	\mathfrak{F}_{+,-}^{\OI \phantom{,,} \UJ} & = & - \mathfrak{F}_{-,+}^{\UJ \phantom{,,} \OI}  = S^{\OI \UJ} ~, \quad
	\mathfrak{F}_{+,+}^{\OI \phantom{,,} \OJ} =  - \mathfrak{F}_{+,+}^{\OJ \phantom{,,} \OI}  = X_{++}^{\OI \OJ}~, \quad
	\mathfrak{F}_{-,-}^{\UI \phantom{,,} \UJ} =  - \mathfrak{F}_{-,-}^{\UJ \phantom{,,} \UI}  = X_{--}^{\UI \UJ}~, \quad
	\\
	\mathfrak{F}_{+, --}^{\OI} &=& \mathfrak{F}_{--, +}^{~~~~\phantom{,}\OI}  = \frac{\ri}{q} \cD_-^{\UJ} S^{\OI \UJ} ~, \qquad \qquad \;\;\,\,
	\mathfrak{F}_{-, ++}^{\UI} = \mathfrak{F}_{++, -}^{~~~~\phantom{,}\UI}  = - \frac{\ri}{p} \cD_+^{\OJ} S^{\OJ \UI} ~, 
	\\
	\mathfrak{F}_{+, ++}^{\OI} &=& \mathfrak{F}_{++, +}^{~~~~\phantom{,}\OI}  = - \frac{\ri}{p-1} \cD_+^{\OJ} X_{++}^{\OJ \OI} ~, \qquad
	\mathfrak{F}_{-, --}^{\UI} = \mathfrak{F}_{--, -}^{~~~~\phantom{,}\UI}  = - \frac{\ri}{q-1} \cD_-^{\UJ} X_{--}^{\UJ \UI} ~, 
	\\
	&& \mathfrak{F}_{++, --} = \mathfrak{F}_{--,++} =  \frac{1}{2pq} \big [ \cD_+^\OI , \cD_-^\UJ \big ] S^{\OI \UJ} - \frac{p+q}{pq} S^{\OI \UJ} S^{\OI \UJ}~,
	\\
	&& \qquad \mathfrak{F}_{++, ++} = \frac{1}{p(p-1)} \cD_+^\OI \cD_+^\OJ X_{++}^{\OI \OJ} - \frac{2}{p} X_{++}^{\OI\OJ} X_{++}^{\OI\OJ} ~,
	\\
	&& \qquad \mathfrak{F}_{--, --} = \frac{1}{q(q-1)} \cD_-^\UI \cD_-^\UJ X_{--}^{\UI \UJ} - \frac{2}{q} X_{--}^{\UI\UJ} X_{--}^{\UI\UJ} ~,
	\eea
\end{subequations}
where we have introduced the {\it imaginary} dimension-$1$ torsion tensors $S^{\OI \UJ}$, $X_{++}^{\OI \OJ}$ and $X_{--}^{\UI \UJ}$.
It should be emphasised that, in contrast to the non-supersymmetric case, all components of ${\mathfrak F}_{AB}$ are determined in terms of the supergravity multiplet.

The torsion tensors obey the Bianchi identities
\begin{subequations} \label{5.68}
	\begin{align}
		\cD_+^{\OI} S^{\OJ \UK} &= \frac 1 p \d^{\OI \OJ} \cD_+^{\OL} S^{\OL \UK} + \cD_-^\OK X_{++}^{\OI \OJ} ~, \quad 
		\cD_-^{\UI} S^{\OJ \UK} = \frac 1 q \d^{\UI \UK} \cD_-^{\UL} S^{\OJ \UL} - \cD_+^\OJ X_{--}^{\UI \UK} ~, \\
		\cD_+^{\OI} X_{++}^{\OJ \OK} &= \frac{2}{p-1} \d^{\OI [\OJ} \cD_+^{|\OL} X_{++}^{\OL| \OK]} ~, \qquad \quad 
		\cD_-^{\UI} X_{--}^{\UJ \UK} = \frac{2}{q-1} \d^{\UI [\UJ} \cD_-^{|\UL} X_{--}^{\UL| \UK]} ~.
	\end{align}
\end{subequations}
Additionally, the algebra obeyed by $\cD_A$ takes the form
\begin{subequations} \label{5.69}
	\bea
	\{ \cD_{+}^{\OI}, \cD_{+}^{\OJ} \} &=& 2 \ri \d^{\OI \OJ} \cD_{++} - 4 X_{++}^{\OK (\OI} \mathfrak{L}^{\OJ) \OK} ~, \\
	\{ \cD_{+}^{\OI}, \cD_{-}^{\UJ} \} &=& - 4 S^{\OI \UJ} M + 2 S^{\OK \UJ} \mathfrak{L}^{\OK \OI} - 2 S^{\OI \UK} \mathfrak{R}^{\UK \UJ} ~, \label{5.69b}\\
	\{ \cD_{-}^{\UI}, \cD_{-}^{\UJ} \} &=& 2 \ri \d^{\UI \UJ} \cD_{--} - 4 X_{--}^{\UK (\UI} \mathfrak{R}^{\UJ) \UK} ~, \\
	\big[ \cD_{+}^{\OI} , \cD_{--} \big]
	& = & - 2 \ri S^{\OI \UJ} \cD_{-}^{\UJ} - \frac{4 \ri}{q} \cD_{-}^{\UJ} S^{\OI \UJ} M + \frac{2 \ri}{q} \cD_{-}^{\UK} S^{\OJ \UK} \mathfrak{L}^{\OJ \OI}
	~, \\
	\big[ \cD_{-}^{\UI} , \cD_{++} \big]
	& = & 2 \ri S^{\OJ \UI} \cD_{+}^{\OJ} - \frac{4 \ri}{p} \cD_{+}^{\OJ} S^{\OJ \UI} M - \frac{2 \ri}{p} \cD_{+}^{\OK} S^{\OK \UJ} \mathfrak{R}^{\UJ \UI} ~,
	\\
	\big[ \cD_{+}^{\OI} , \cD_{++} \big]
	& = & - 2 \ri X_{++}^{\OI \OJ} \cD_+^{\OJ} - \frac{2 \ri}{p-1} \cD_+^\OJ X_{++}^{\OJ \OK} \mathfrak{L}^{\OK \OI} ~,
	\\
	\big[ \cD_{-}^{\UI} , \cD_{--} \big]
	& = & - 2 \ri X_{--}^{\UI \UJ} \cD_-^{\UJ} - \frac{2 \ri}{q-1} \cD_-^\UJ X_{--}^{\UJ \UK} \mathfrak{R}^{\UK \UI}  ~,
	\\
	\big[ \cD_{++} , \cD_{--} \big]
	& = & -\frac{2}{p} \cD_+^{\OJ} S^{\OJ \UI} \cD_-^{\UI} - \frac{2}{q} \cD_-^{\UJ} S^{\OI \UJ} \cD_+^{\OI} \non \\
	&& - \frac{2}{pq} \big( \big[\cD_+^{\OI}, \cD_-^{\UJ}\big] - 2(p+q) S^{\OI \UJ} \big) S^{\OI \UJ} M
	~.
	\eea
\end{subequations}

Next, it is important to describe the supergravity gauge freedom of this geometry, which corresponds to the residual gauge transformations of \eqref{5.60} in the gauge $B_A = 0$. This freedom includes local $\mathcal{K}$-transformations of the form
\begin{subequations}\label{5.70}
	\bea
	\delta_{\mathcal K} \cD_A &=& [\mathcal{K},\cD_A] \ , \\
	\mathcal{K} &=& \xi^B \cD_B + K M +\hf \r^{\OI \OJ} \mathfrak{L}^{\OI \OJ}
	+ \hf \r^{\UI \UJ} \mathfrak{R}^{\UI \UJ} ~,
	\eea
	which act on tensor superfields ${U}$ (with their indices suppressed) as
	\begin{align}
		\d_\cK U = \cK U ~.
	\end{align}
\end{subequations}
The gauge transformations \eqref{5.70} prove to not be the most general conformal supergravity gauge transformations preserving the gauge $B_A=0$.
Specifically, it may be shown that the following transformation also enjoys this property
\begin{align}
	\mathscr{K}(\s) = \s \mathbb{D} + \frac{1}{2} \nabla_B \s K^B \quad \Longrightarrow \quad \d_{\mathscr{K}(\s)} B_A = 0~,
\end{align}
where $\s$ is real but otherwise unconstrained. 

As a result, it is necessary to consider how this transformation manifests in the degauged geometry
\begin{align}
	\d_{\mathscr{K}(\s)} \nabla_A \equiv \d_\s \nabla_A = \d_\s \cD_A - \d_\s \mathfrak{F}_{AB} K^B~.
\end{align}
Employing this relation, we arrive at the transformation laws for $\cD_A$
\begin{subequations}
	\begin{align}
		\d_\s \cD_+^\OI &= \hf \s \cD_+^\OI + \cD_+^\OI \s M - \cD_+^\OJ \s \mathfrak{L}^{\OJ \OI} ~, \\
		\d_\s \cD_-^\UI &= \hf \s \cD_-^\UI - \cD_-^\UI \s M - \cD_-^\UJ \s \mathfrak{R}^{\UJ \UI} ~, \\
		\d_\s \cD_{++} &= \s \cD_{++} - \ri \cD_+^\OI \s \cD_+^\OI + \cD_{++} \s M ~, \\
		\d_\s \cD_{--} &= \s \cD_{--} - \ri \cD_-^\UI \s \cD_-^\UI - \cD_{--} \s M ~,
	\end{align}
	and, by making use of \eqref{5.67}, it may be shown that the torsions transform as follows
	\begin{align}
		\d_\s S^{\OI \UJ} &= \s S^{\OI \UJ} + \frac 1 2 \cD_+^\OI \cD_-^\UJ \s ~, \\
		\d_\s X_{++}^{\OI \OJ} &= \s X_{++}^{\OI \OJ} + \frac 1 4 \big[\cD_+^\OI , \cD_+^\OJ \big]  \s ~, \\
		\d_\s X_{--}^{\UI \UJ} &= \s X_{--}^{\UI \UJ} + \frac 1 4 \big[\cD_-^\UI , \cD_-^\UJ \big]  \s ~.
	\end{align}
\end{subequations}
These are the super-Weyl transformations of the degauged geometry.

It should be mentioned that, for the special case $\cN= (2,2)$, an equivalent superspace geometry was formulated in the works \cite{Howe:1987ba, Grisaru:1994dm,Grisaru:1995dr, Gates:1995du}.\footnote{This follows from the fact that the superconformal groups $\mathsf{OSp}_0 (2|2; {\mathbb R} ) \times  \mathsf{OSp}_0 (2|2; {\mathbb R} )$ and  
	$\sSU(1,1|1) \times  \sSU(1,1|1) $ are isomorphic.} To see this, we first eliminate the superfields $X_{++}^{\overline{I} \overline{J}}$ and $X_{--}^{\underline{I} \underline{J}}$ in \eqref{5.69} by redefining the vector covariant derivatives
\begin{align}
	\hat{\cD}_{++} = \cD_{++} -  \ri X_{++}^{\overline{I} \overline{J}} \mathfrak{L}^{\overline{I} \overline{J}} ~, \qquad
	\hat{\cD}_{--} = \cD_{--} -  \ri X_{--}^{\underline{I} \underline{J}} \mathfrak{R}^{\underline{I} \underline{J}} ~.
\end{align}
The resulting algebra is as follows
\begin{subequations} \label{5.75}
	\bea
	\{ \cD_{+}^{\OI}, \cD_{+}^{\OJ} \} &=& 2 \ri \d^{\OI \OJ} \hat{\cD}_{++} ~,  \qquad
	\{ \cD_{-}^{\UI}, \cD_{-}^{\UJ} \} = 2 \ri \d^{\UI \UJ} \hat{\cD}_{--} ~, 
	\\
	\{ \cD_{+}^{\OI}, \cD_{-}^{\UJ} \} &=& - 4 S^{\OI \UJ} M + 2 S^{\OK \UJ} \mathfrak{L}^{\OK \OI} - 2 S^{\OI \UK} \mathfrak{R}^{\UK \UJ} ~,\\
	\big[ \cD_{+}^{\OI} , \hat{\cD}_{++} \big]
	& = & 0 ~, \qquad
	\big[ \cD_{-}^{\UI} , \hat{\cD}_{--} \big]
	= 0  ~,
	\\
	\big[ \cD_{+}^{\OI} , \hat{\cD}_{--} \big]
	& = & - 2 \ri S^{\OI \UJ} \cD_{-}^{\UJ} - 2 \ri \cD_{-}^{\UJ} S^{\OI \UJ} M + \ri \cD_{-}^{\UK} S^{\OJ \UK} \mathfrak{L}^{\OJ \OI} + \ri \cD_-^\UJ S^{\OI \UK} \mathfrak{R}^{\UJ \UK}
	~, \\
	\big[ \cD_{-}^{\UI} , \hat{\cD}_{++} \big]
	& = & 2 \ri S^{\OJ \UI} \cD_{+}^{\OJ} - 2 \ri \cD_{+}^{\OJ} S^{\OJ \UI} M - \ri \cD_+^\OJ S^{\OK \UI} \mathfrak{L}^{\OJ \OK} - \ri \cD_{+}^{\OK} S^{\OK \UJ} \mathfrak{R}^{\UJ \UI} ~,
	\\
	\big[ \hat{\cD}_{++} , \hat{\cD}_{--} \big]
	& = & - \cD_+^{\OJ} S^{\OJ \UI} \cD_-^{\UI} - \cD_-^{\UJ} S^{\OI \UJ} \cD_+^{\OI} - \frac{1}{2} \big( \big[\cD_+^{\OI}, \cD_-^{\UJ}\big] - 8 S^{\OI \UJ} \big) S^{\OI \UJ} M
	~, \non \\
	&& - \frac 1 4 \cD_-^\UK \cD_+^\OI S^{\OJ \UK} \mathfrak{L}^{\OI \OJ} - \frac 1 4 \cD_+^\OK \cD_-^\UI S^{\OK \UJ} \mathfrak{R}^{\UI \UJ}~.
	\eea
\end{subequations}
It should be emphasised that the resulting geometry is described solely in terms of $S^{\OI \UJ}$.
Then, to relate the geometry of \cite{Howe:1987ba, Grisaru:1994dm, Grisaru:1995dr, Gates:1995du} to ours it is necessary to express \eqref{5.75} in a complex basis of spinor covariant derivatives
\begin{subequations}
	\label{5.76}
	\begin{align}
		\cD_+ := \frac{1}{\sqrt 2} (\cD_+^{\overline 1} - \ri \cD_+^{\overline 2}) ~, \qquad \bar{\cD}_+ = -\frac{1}{\sqrt 2} (\cD_+^{\overline 1} + \ri \cD_+^{\overline 2}) ~, \\
		\cD_- := \frac{1}{\sqrt 2} (\cD_-^{\underline 1} - \ri \cD_-^{\underline 2}) ~, \qquad \bar{\cD}_- = -\frac{1}{\sqrt 2} (\cD_-^{\underline 1} + \ri \cD_-^{\underline 2})
		~.
	\end{align}
\end{subequations}
We omit further technical details regarding this procedure, which will appear in a future work. 

To the best of our knowledge, for $p,q> 2$ our superspace geometry described by 
the equations \eqref{5.68} and \eqref{5.69} has not appeared in the literature. In particular, in the  $\cN=(4,4)$ case, the above $\sSO(4) \times \sSO(4)$ superspace geometry differs from the one proposed in \cite{TM} by the choice of structure group, see section \ref{Chapter5.5} for the discussion of this formulation. It should also be pointed out that the formulation for $\cN=(4,4)$ matter-coupled supergravity in $\sSU(2) \times \sSU(2)$ harmonic superspace was constructed in \cite{BeIv}.\footnote{I am grateful to Evgeny Ivanov for bringing this reference to my attention.}

It follows from eq. \eqref{5.67} that this geometry is ill-defined if at least one of the parameters $p$ and $q$ takes values $0$ or $1$. Each of these cases should be studied separately, 
which is done in the remainder of this section.

\subsection{$p > 1,~ q = 1$ case}

Next, let us consider the case where $p>1,~q=1$. For convenience, we will unambiguously remove bars over left isovector indices, e.g. $\OI \equiv I$. By a routine calculation, one obtains the following components of the degauged connection $\mathfrak{F}_{AB}$
\begin{subequations}
	\label{5.77}
	\bea
	\mathfrak{F}_{+,-}^{I \phantom{,,}} & = & - \mathfrak{F}_{-,+}^{\phantom{,,} I}  = S^{I} ~, \quad
	\mathfrak{F}_{+,+}^{I \phantom{,,} J} =  - \mathfrak{F}_{+,+}^{J \phantom{,,} I}  = X_{++}^{I J}~, \quad
	\mathfrak{F}_{-,-} = 0~, \quad
	\\
	\mathfrak{F}_{+, --}^{I} &=& \mathfrak{F}_{--, +}^{~~~~\phantom{,}I}  = \ri \cD_- S^{I} ~, \qquad 
	\mathfrak{F}_{-, ++} = \mathfrak{F}_{++, -}  = - \frac{\ri}{p} \cD_+^{I} S^{I} ~, 
	\\
	\mathfrak{F}_{+, ++}^{I} &=& \mathfrak{F}_{++, +}^{~~~~\phantom{,}I}  = - \frac{\ri}{p-1} \cD_+^{J} X_{++}^{J I} ~, 
	\\
	\mathfrak{F}_{++, --} &=& \mathfrak{F}_{--,++} =  \frac{1}{2p} \big [ \cD_+^I , \cD_- \big ] S^{I} - \frac{p+1}{p} S^{I} S^{I}~,
	\\
	\mathfrak{F}_{++, ++} &=& \frac{1}{p(p-1)} \cD_+^I \cD_+^J X_{++}^{I J} - \frac{2}{p} X_{++}^{IJ} X_{++}^{IJ} ~,
	\eea
	where we have introduced the dimension-$1$ torsions $S^{I}$ and $X_{++}^{I J}$.
	The remaining components of $\mathfrak{F}_{AB}$ do not play a role in the degauged geometry, though they satisfy
	\begin{align}
		\label{5.76f}
		\mathfrak{F}_{-, --} = \mathfrak{F}_{--, -}~,
		\quad
		\cD_+^I \mathfrak{F}_{-,--} =\cD_{--} S^I~,
		\quad
		\mathfrak{F}_{--, --} = - \ri \cD_{-} \mathfrak{F}_{-,--}~.
	\end{align}
\end{subequations}
Equations \eqref{5.76f} imply that $\mathfrak{F}_{-,--}$ is a non-local function of the supergravity multiplet.

The superfields $S^I$ and $X_{++}^{IJ}$ obey the Bianchi identities
\begin{align}
	\cD_+^{I} S^{J} = \frac 1 p \d^{I J} \cD_+^{K} S^{K} + \cD_- X_{++}^{I J} ~, \qquad 
	\cD_+^{I} X_{++}^{J K} = \frac{2}{p-1} \d^{I [J} \cD_+^{|L} X_{++}^{L| K]} ~.
\end{align}
Additionally, the algebra obeyed by $\cD_A$ takes the form
\begin{subequations} \label{5.79}
	\bea
	\{ \cD_{+}^{I}, \cD_{+}^{J} \} &=& 2 \ri \d^{I J} \cD_{++} - 4 X_{++}^{K (I} \mathfrak{L}^{J) K} ~, \\
	\{ \cD_{+}^{I}, \cD_{-} \} &=& - 4 S^{I } M + 2 S^{J} \mathfrak{L}^{J I} ~, \\
	\{ \cD_{-}, \cD_{-} \} &=& 2 \ri  \cD_{--} ~, \\
	\big[ \cD_{+}^{I} , \cD_{--} \big]
	& = & - 2 \ri S^{I} \cD_{-} - 4 \ri \cD_{-} S^{I} M + 2 \ri \cD_{-} S^{J} \mathfrak{L}^{J I}
	~, \\
	\big[ \cD_{-} , \cD_{++} \big]
	& = & 2 \ri S^{I} \cD_{+}^{I} - \frac{4 \ri}{p} \cD_{+}^{I} S^{I} M ~,
	\\
	\big[ \cD_{+}^{I} , \cD_{++} \big]
	& = & - 2 \ri X_{++}^{I J} \cD_+^{J} - \frac{2 \ri}{p-1} \cD_+^J X_{++}^{J K} \mathfrak{L}^{K I} ~,
	\\
	\big[ \cD_{-} , \cD_{--} \big] &=& 0 ~, \\
	\big[ \cD_{++} , \cD_{--} \big]
	& = & -\frac{2}{p} \cD_+^{I} S^{I} \cD_- - 2 \cD_- S^{I} \cD_+^{I} \non \\
	&& - \frac{2}{p} \big( \big[\cD_+^{I}, \cD_-\big] - 2(p+1) S^{I} \big) S^{I} M
	~.
	\eea
\end{subequations}

The supergravity gauge freedom of this geometry corresponds to the residual gauge transformations of \eqref{5.60} in the gauge $B_A = 0$. In particular, this includes local $\mathcal{K}$-transformations of the form
\begin{subequations}\label{5.80}
	\bea
	\delta_{\mathcal K} \cD_A &=& [\mathcal{K},\cD_A] \ , \\
	\mathcal{K} &=& \xi^B \cD_B + K M + \hf \r^{I J} \mathfrak{L}^{I J}~,
	\eea
	which act on tensor superfields ${U}$ (with their indices suppressed) as $\d_\cK U = \cK U$.
\end{subequations}
Transformations \eqref{5.80} prove to not be the most general conformal supergravity gauge transformations preserving the gauge $B_A=0$.
Specifically, the following transformation also enjoys this property
\begin{align}
	\label{5.81}
	\mathscr{K}(\s) = \s \mathbb{D} + \frac{1}{2} \nabla_B \s K^B \quad \Longrightarrow \quad \d_{\mathscr{K}(\s)} B_A = 0~,
\end{align}
where $\s$ is real but otherwise unconstrained. 

As a result, it is necessary to consider how transformation \eqref{5.81} manifests in the degauged geometry
\begin{align}
	\d_{\mathscr{K}(\s)} \nabla_A \equiv \d_\s \nabla_A = \d_\s \cD_A - \d_\s \mathfrak{F}_{AB} K^B~.
\end{align}
Employing this relation, we arrive at the transformation laws for $\cD_A$
\begin{subequations}
	\begin{align}
		\d_\s \cD_+^I &= \hf \s \cD_+^I + \cD_+^I \s M - \cD_+^J \s \mathfrak{L}^{J I} ~, \\
		\d_\s \cD_- &= \hf \s \cD_- - \cD_-\s M~, \\
		\d_\s \cD_{++} &= \s \cD_{++} - \ri \cD_+^I \s \cD_+^I + \cD_{++} \s M ~, \\
		\d_\s \cD_{--} &= \s \cD_{--} - \ri \cD_- \s \cD_- - \cD_{--} \s M ~,
	\end{align}
	and it may be shown that the torsions transform as follows
	\begin{align}
		\d_\s S^{I} &= \s S^{I} + \frac 1 2 \cD_+^I \cD_- \s ~, \\
		\d_\s X_{++}^{I J} &= \s X_{++}^{I J} + \frac 1 4 \big [ \cD_+^I , \cD_+^J \big ]  \s ~.
	\end{align}
\end{subequations}
These are the super-Weyl transformations of the degauged geometry.

It may be shown that, if $p=2$, the superfield $X_{++}^{IJ}$ can be eliminated by performing the redefinition
\begin{align}
	\label{5.84}
	\hat{\cD}_{++} = \cD_{++} -  \ri X_{++}^{\overline{I} \overline{J}} \mathfrak{L}^{\overline{I} \overline{J}} ~.
\end{align}
The resulting algebra then takes the form
\begin{subequations} \label{(2,1)algebra}
	\bea
	\{ \cD_{+}^{I}, \cD_{+}^{J} \} &=& 2 \ri \d^{I J} \hat{\cD}_{++} ~, \qquad \{ \cD_{-}, \cD_{-} \} = 2 \ri  \cD_{--} ~, \\
	\{ \cD_{+}^{I}, \cD_{-} \} &=& - 4 S^{I } M + 2 S^{J} \mathfrak{L}^{J I} ~, \\
	\big[ \cD_{+}^{I} , \cD_{--} \big]
	& = & - 2 \ri S^{I} \cD_{-} - 4 \ri \cD_{-} S^{I} M + 2 \ri \cD_{-} S^{J} \mathfrak{L}^{J I}
	~, \\
	\big[ \cD_{-} , \hat{\cD}_{++} \big]
	& = & 2 \ri S^{I} \cD_{+}^{I} - 2 \ri \cD_{+}^{I} S^{I} M ~,
	\\
	\big[ \cD_{+}^{I} , \hat{\cD}_{++} \big]
	& = & 0 ~, \qquad \big[ \cD_{-} , \cD_{--} \big]
	=  0 ~,
	\\
	\big[ \hat{\cD}_{++} , \cD_{--} \big]
	& = & - \cD_+^{I} S^{I} \cD_- - 2 \cD_- S^{I} \cD_+^{I} - \big( \big[\cD_+^{I}, \cD_-\big] - 6 S^{I} \big) S^{I} M \non \\
	&&- \frac 1 4 \cD_- \cD_+^I S^{J} \mathfrak{L}^{IJ}
	~.
	\eea
\end{subequations}
Hence, the resulting geometry is described solely in terms of $S^{I}$.

\subsection{$p > 1,~ q = 0$ case}
Now, we consider the case $p >1,~ q=0$. As in the previous subsection, we remove bars over left isovector indices; $\OI \equiv I$. By a routine calculation, we readily obtain the following components of the degauged connection $\mathfrak{F}_{AB}$
\begin{subequations}
	\bea
	\mathfrak{F}_{+,+}^{I \phantom{,,} J} &=& - \mathfrak{F}_{+,+}^{J \phantom{,,} I}  = X_{++}^{I J}~,  \\
	\mathfrak{F}_{+, --}^{I} &=& \mathfrak{F}_{--, +}^{~~~~\phantom{,}I}  = G_{-}^{I} ~, \qquad \qquad \phantom{-} \,
	\mathfrak{F}_{+, ++}^{I} = \mathfrak{F}_{++, +}^{~~~~\phantom{,}I}  = - \frac{\ri}{p-1} \cD_+^{J} X_{++}^{J I} ~, \\
	\mathfrak{F}_{++, --} &=& \mathfrak{F}_{--,++} =  -\frac \ri p \cD_+^I G_-^I~, \quad 
	\mathfrak{F}_{++, ++} = \frac{1}{p(p-1)} \cD_+^I \cD_+^J X_{++}^{I J} - \frac{2}{p} X_{++}^{I J} X_{++}^{I J} ~, \qquad
	\eea
	while $\mathfrak{F}_{--,--}$ only appears through the differential constraint
	\begin{align}
		\label{5.86d}
		\cD_+^I \mathfrak{F}_{--,--} = \cD_{--} G_{-}^{I}~.
	\end{align}
\end{subequations}
In the above equations we have introduced the torsions $X_{++}^{I J}$ and $G_{-}^I$, which obey the Bianchi identities
\begin{align}
	\label{5.87}
	\cD_+^{I} X_{++}^{J K} = \frac{2}{p-1} \d^{I [J} \cD_+^{|L} X_{++}^{L| K]} ~, \qquad
	\cD_+^{I} G_-^{J} = \frac{1}{p} \d^{I J} \cD_+^{K} G_-^{K} + \cD_{--} X_{++}^{I J} ~.
\end{align}

Making use of these results, we find that the algebra obeyed by $\cD_A$ takes the form
\begin{subequations} 
	\label{5.89}
	\bea
	\{ \cD_{+}^{I}, \cD_{+}^{J} \} &=& 2 \ri \d^{I J} \cD_{++} - 4 X_{++}^{K (I} \mathfrak{L}^{J) K} ~, \\
	\big[ \cD_{+}^{I} , \cD_{--} \big]
	& = & -4 G_-^I M + 2 G_-^J \mathfrak{L}^{J I}
	~, \\
	\big[ \cD_{+}^{I} , \cD_{++} \big]
	& = & - 2 \ri X_{++}^{I J} \cD_+^{J} - \frac{2 \ri}{p-1} \cD_+^J X_{++}^{J K} \mathfrak{L}^{K I} ~,
	\\
	\big[ \cD_{++} , \cD_{--} \big]
	& = & 2 \ri G_-^I \cD_+^I - \frac{4 \ri}{p} \cD_{+}^I G_{-}^I M ~.
	\eea
\end{subequations}
It should be noted that $(p,0)$ superspace geometries, with $p \geq 2$, were discussed in \cite{EO}, where the structure was chosen to be the Lorentz group,
$\sSO(1,1)$,
though the $p=2$ case appeared earlier \cite{BMG}.\footnote{In \cite{Govindarajan:1991sx} the structure group 
	for $(2,0) $ supergravity was enlarged from $\sSO(1,1)$ to 
	$\sSO(1,1) \times \sU(1)$.}
The emergence of the $(2,0)$ geometry of \cite{BMG} in our framework will be discussed below.

The supergravity gauge transformations of the $(p,0)$ geometry \eqref{5.89} may be obtained from \eqref{5.60} after imposing $B_A = 0$. They include the local $\mathcal{K}$-transformations
\begin{subequations}\label{5.90}
	\bea
	\delta_{\mathcal K} \cD_A &=& [\mathcal{K},\cD_A] \ , \\
	\mathcal{K} &=& \xi^B \cD_B + K M + \hf \r^{I J} \mathfrak{L}^{I J} ~,
	\eea
	which act on tensor superfields ${U}$ (with indices suppressed) as $\d_\cK U = \cK U$.
\end{subequations}
In addition to \eqref{5.90}, the following transformation also preserves the gauge $B_A=0$
\begin{align}
	\mathscr{K}(\s) = \s \mathbb{D} + \frac{1}{2} \nabla_B \s K^B \quad \Longrightarrow \quad \d_{\mathscr{K}(\s)} B_A = 0~,
\end{align}
where $\s$ is real but otherwise unconstrained. 

It is then necessary to consider how this transformation manifests in the degauged geometry
\begin{align}
	\d_{\mathscr{K}(\s)} \nabla_A \equiv \d_\s \nabla_A = \d_\s \cD_A - \d_\s \mathfrak{F}_{AB} K^B ~.
\end{align}
Employing this relation, we arrive at the transformation laws for $\cD_A$
\begin{subequations}
	\label{5.93}
	\begin{align}
		\d_\s \cD_+^I &= \hf \s \cD_+^I + \cD_+^I \s M - \cD_+^J \s \mathfrak{L}^{J I} ~, \\
		\d_\s \cD_{++} &= \s \cD_{++} - \ri \cD_+^I \s \cD_+^I + \cD_{++} \s M ~, \\
		\d_\s \cD_{--} &= \s \cD_{--} - \cD_{--} \s M ~,
	\end{align}
	and it may be shown that the torsions transform as follows
	\begin{align}
		\d_\s X_{++}^{I J} &= \s X_{++}^{I J} + \frac 1 4 \big [ \cD_+^I , \cD_+^J \big ]  \s ~, \\
		\d_\s G_{-}^I &= \frac{3}2 \s G_{-}^I + \frac 1 2 \cD_+^I \cD_{--} \s ~.
	\end{align}
\end{subequations}
These are exactly the super-Weyl transformations of the degauged geometry.

We emphasise that in special case $p=2$, the superfield $X_{++}^{IJ}$ can be eliminated by performing the redefinition \eqref{5.84}. Additionally, it is useful to work in a complex basis of spinor covariant derivatives, where $\cD_+^I$ is replaced by $\cD_+ = \frac{1}{\sqrt 2} (\cD_+^{1} - \ri \cD_+^{2})$ and its conjugate $\bar \cD_+$.
The resulting algebra of covariant derivatives takes the form
\begin{subequations} \label{5.94}
	\bea
	\{ \cD_{+}, \cD_{+} \} &=& 0 ~, \qquad \{ \cD_{+}, \bar{\cD}_{+} \} = 2 \ri \hat{\cD}_{++} ~, \\
	\big[ \cD_{+} , \cD_{--} \big]
	& = & -4 \bar{G}_- M - 2 \ri \bar{G}_- \mathfrak{L}^{12}
	~, \\
	\big[ \cD_{+}, \hat{\cD}_{++} \big]
	& = &0 ~,
	\\
	\big[ \hat{\cD}_{++} , \cD_{--} \big]
	& = & 2 \ri G_- \cD_+ + 2 \ri \bar{G}_- \bar{\cD}_+ + 2 \ri (\cD_+ {G}_- + \bar{\cD}_+ \bar{G}_-) M \non \\
	&& + (\cD_+ {G}_- - \bar{\cD}_+ \bar{G}_-) \mathfrak{L}^{12}~.
	\eea
\end{subequations}
This supergeometry is described solely in terms of the complex superfield 
\begin{align}
	\label{5.95}
	G_{-} := -\frac{1}{\sqrt 2} (G_-^{1} + \ri G_-^{2}) ~.
\end{align}
It follows from \eqref{5.87} that it satisfies the chirality constraint
\begin{align}
	\bar{\cD}_{+} G_- = 0~.
\end{align}

The $\sSO(2)$ gauge freedom enjoyed by the geometry \eqref{5.94} may be 
used to gauge away its corresponding spinor $\sSO(2)$ connection $\Phi_+^{I,\,JK } = -\Phi_+^{I,\,KJ } $. This can be seen   
by coupling the conformal supergravity multiplet to a conformally primary compensator $\phi$ obeying the chirality condition $\bar{\cD}_+ \phi = 0$. Choosing its weight to be $\D_\f =1$,  it transforms under super-Weyl \eqref{5.93} and local $\sSO(2)$ transformations \eqref{5.90} as follows:
\begin{align}
	\d \phi = (\s - \ri \r^{12} ) \phi ~,
\end{align}
hence this freedom may be used to impose the gauge $\phi = 1$. Associated with this gauge are the consistency conditions
\begin{align}
	\Phi_+^{I,12} = 0 ~, \qquad 
	\ri \bar{\cD}_+ \G_{--} = G_- ~, \qquad
	\G_{--} = \bar{\G}_{--}:= \hf \Phi_{--}^{\phantom{--}12} ~.
\end{align}
As the $\sSO(2)$ connection is now auxiliary, it should be separated by performing the redefinition
\begin{align}
	\hat{\cD}_{--} = \cD_{--} + 2 \G_{--} \mathfrak{L}^{12}~.
\end{align}
The resulting algebra of covariant derivatives is
\begin{subequations}
	\bea
	\{ \cD_{+}, \cD_{+} \} &=& 0 ~, \qquad \{ \cD_{+}, \bar{\cD}_{+} \} = 2 \ri \hat{\cD}_{++} ~, \\
	\big[ \cD_{+} , \hat{\cD}_{--} \big]
	& = & -2\ri\G_{--} \cD_+ -4 \ri \bar{\cD}_+ \G_{--} M ~, \\
	\big[ \cD_{+}, \hat{\cD}_{++} \big]
	& = &0 ~,
	\\
	\big[ \hat{\cD}_{++} , \hat{\cD}_{--} \big]
	& = & -2 \bar{\cD}_+ \G_{--} \cD_+ +2 {\cD}_+ \G_{--} \bar{\cD}_+ - 2 [\cD_+, \bar{\cD}_+] \G_{--} M ~,
	\eea
\end{subequations}
which coincides with the one appearing in \cite{BMG}. 

\subsection{$p = q = 1$ case}

Next, we fix $p=q=1$. By a routine calculation, we obtain the following components of the degauged special conformal connection
\begin{subequations}
	\bea
	\mathfrak{F}_{+,+} &=& 0 ~, \qquad \mathfrak{F}_{-,-} = 0 ~, \qquad \mathfrak{F}_{+,-} = - \mathfrak{F}_{-,+} = S  \\
	\mathfrak{F}_{+, --} &=& \mathfrak{F}_{--, +}  = \ri \cD_- S ~, \qquad
	\mathfrak{F}_{-, ++} = \mathfrak{F}_{++, -}  = - \ri \cD_+ S ~, \\
	\mathfrak{F}_{++,++} &=& - \ri \cD_+ \mathfrak{F}_{+,++} ~, \qquad \mathfrak{F}_{--,--} = -\ri \cD_- \mathfrak{F}_{-,--}~, \\
	\mathfrak{F}_{++, --} &=& \mathfrak{F}_{--,++} =  \hf [\cD_+, \cD_-]S - 2 S^2 ~,
	\eea
	where we have introduced the imaginary scalar $S$. The remaining components of $\mathfrak{F}_{AB}$ do not play a role in the degauged geometry, though they satisfy the constraints
	\begin{align}
		\mathfrak{F}_{+,++} &= \mathfrak{F}_{++,+} ~, \qquad \mathfrak{F}_{-,--} = \mathfrak{F}_{--,-}~, \\
		\cD_- \mathfrak{F}_{+,++} &= - \cD_{++} S ~, \qquad \cD_+ \mathfrak{F}_{-,--} =\cD_{--} S ~.
	\end{align}
\end{subequations}
It follows that $\mathfrak{F}_{+,++}$ and $\mathfrak{F}_{-,--}$ are non-local functions of the supergravity multiplet.

It follows from the above results that the algebra obeyed by $\cD_A$ takes the form
\begin{subequations}
	\label{5.102}
	\bea
	\{ \cD_{+}, \cD_{+} \} &=& 2 \ri \cD_{++}~, \qquad \{ \cD_{-}, \cD_{-} \} = 2 \ri \cD_{--} ~, \\
	&& \{ \cD_{+}, \cD_{-} \} = - 4 S M~, \\
	\big[ \cD_{+} , \cD_{--} \big]
	& = & - 2 \ri S \cD_- - 4 \ri (\cD_- S) M ~, \\
	\big[ \cD_{-} , \cD_{++} \big]
	& = &  2 \ri S \cD_+ - 4 \ri (\cD_+ S) M ~,
	\\
	\big[ \cD_{++} , \cD_{--} \big]
	& = & -2(\cD_+ S) \cD_- - 2 (\cD_- S) \cD_+ \non \\
	& \phantom{=} &- 2 \big( [\cD_+ , \cD_-] - 4 S \big)S M  ~.
	\eea
\end{subequations}

The supergravity gauge freedom of this geometry may be obtained from the conformal superspace transformations \eqref{5.60} after restricting to the gauge $B_A = 0$. These include local $\mathcal{K}$-transformations of the form
\bea
\label{5.103}
\delta_{\mathcal K} \cD_A &=& [\mathcal{K},\cD_A] ~, \qquad
\mathcal{K} = \xi^B \cD_B + K M~,
\eea
which act on tensor superfields $\mathcal{U}$ (with indices suppressed) as $\d_\cK \cU = \cK \cU$.
In addition to \eqref{5.103}, it may be shown that the following also preserves the gauge $B_A=0$
\begin{align}
	\mathscr{K}(\s) = \s \mathbb{D} + \frac{1}{2} \nabla_B \s K^B \quad \Longrightarrow \quad \d_{\mathscr{K}(\s)} B_A = 0~,
\end{align}
where $\s$ is real but otherwise unconstrained. 

As a result, it is necessary to consider how this transformation manifests in the degauged geometry
\begin{align}
	\d_{\mathscr{K}(\s)} \nabla_A \equiv \d_\s \nabla_A = \d_\s \cD_A - \d_\s (\mathfrak{F}_{AB} K^B)~.
\end{align}
Employing this relation, we arrive at the transformation laws for $\cD_A$ and $S$
\begin{subequations}
	\begin{align}
		\d_\s \cD_+ &= \hf \s \cD_+ + \cD_+\s M  ~, \\
		\d_\s \cD_- &= \hf \s \cD_- - \cD_-\s M  ~, \\
		\d_\s \cD_{++} &= \s \cD_{++}  - \ri \cD_+ \s \cD_+ + \cD_{++} \s M ~, \\
		\d_\s \cD_{--} &= \s \cD_{--} - \ri \cD_- \s \cD_- - \cD_{--} \s M ~, \\
		\d_\s S &= \s S + \frac 1 2 \cD_+  \cD_- \s~.
	\end{align}
\end{subequations}
These are exactly the super-Weyl transformations of the degauged geometry.

The superspace geometry of $\cN=1$ supergravity described above was originally constructed in \cite{Howe1979,BrGa}, see also \cite{Martinec,GN,RvanNZ}.

\subsection{$p = 1,~ q = 0$ case}

Finally, we consider the case $p=1,~q=0$. 
A routine computation leads to the degauged special conformal connections
\begin{subequations}
	\bea
	\label{5.107a}
	\mathfrak{F}_{+,+} &=& 0 ~, \qquad \mathfrak{F}_{+,--} = \mathfrak{F}_{--,+} = G_{-} ~, \qquad \mathfrak{F}_{++,--} = \mathfrak{F}_{--,++} = - \ri \cD_+ G_-~.~~~
	\eea
	where we have introduced the spinor $G_-$. The remaining components of $\mathfrak{F}_{AB}$ do not play a role in the degauged geometry, though they satisfy the constraints
	\begin{align}
		\label{5.107b}
		\mathfrak{F}_{+,++} &= \mathfrak{F}_{++,+} ~, \qquad \mathfrak{F}_{++,++} = - \ri \cD_+ \mathfrak{F}_{+,++}~, \\
		\label{5.107c}
		\cD_{++} G_- &= \cD_{--} \mathfrak{F}_{+,++} ~, \qquad \cD_{--} G_{-} = \cD_+ \mathfrak{F}_{--,--}~.
	\end{align}
\end{subequations}
It is clear that $\mathfrak{F}_{+,++}$ and $\mathfrak{F}_{--,--}$ are non-local functions of the supergravity multiplet.

It immediately follows that the algebra obeyed by $\cD_A$ takes the form
\begin{subequations}
	\label{5.108}
	\bea
	\{ \cD_{+}, \cD_{+} \} &=& 2 \ri \cD_{++}~,\\
	\big[ \cD_+, \cD_{--} \big ] &=& - 4 G_- M ~, \quad \big[\cD_{+}, \cD_{++} \big] = 0 \\
	\big[ \cD_{++} , \cD_{--} \big] &=& 2 \ri G_{-} \cD_{+} + 4 \ri (\cD_+ G_-) M  ~.
	\eea
\end{subequations}

As described in the previous subsections, the supergravity gauge transformations of this geometry correspond to \eqref{5.60} in the gauge $B_A = 0$. They include local $\mathcal{K}$-transformations of the form
\bea
\label{5.109}
\delta_{\mathcal K} \cD_A &=& [\mathcal{K},\cD_A] ~, \qquad
\mathcal{K} = \xi^B \cD_B + \o M~,
\eea
which acts on tensor superfields ${U}$ (with indices suppressed) as	$\d_\cK U = \cK U$.
One may also show that the following transformation also preserves the $B_A=0$ gauge
\begin{align}
	\mathscr{K}(\s) = \s \mathbb{D} + \frac{1}{2} \nabla_B \s K^B \quad \Longrightarrow \quad \d_{\mathscr{K}(\s)} B_A = 0~,
\end{align}
where $\s$ is real but otherwise unconstrained. 

As a result, it is necessary to consider how this transformation manifests in the degauged geometry
\begin{align}
	\d_{\mathscr{K}(\s)} \nabla_A \equiv \d_\s \nabla_A = \d_\s \cD_A - \d_\s \mathfrak{F}_{AB} K^B~.
\end{align}
Employing this relation, we arrive at the transformation laws for $\cD_A$ and $G_-$
\begin{subequations}
	\begin{align}
		\d_\s \cD_+ &= \hf \s \cD_+ + \cD_+\s M  ~, \\
		\d_\s \cD_{++} &= \s \cD_{++} - \ri \cD_+ \s \cD_+ + \cD_{++} \s M ~, \\
		\d_\s \cD_{--} &= \s \cD_{--} - \cD_{--} \s M ~, \\
		\d_\s G_- &= \frac 3 2 \s G_- + \frac 1 2 \cD_+ \cD_{--} \s~.
	\end{align}
\end{subequations}
These are exactly the super-Weyl transformations of the degauged geometry.

It should be noted that the above curved $ (1,0)$ superspace geometry was originally constructed in \cite{BMG,Gates:1986ez,EO}.

\section{Applications and generalisations} \label{Chapter5.5}

So far, we have described a universal superspace formulation for conformal $(p,q)$ supergravity, the $\sSO(p) \times \sSO(q)$ superspace. This subsection is, in part, devoted to applications of this formalism. In particular, we describe the $\cN$-extended AdS superspace and also study supersymmetric extensions of the Gauss-Bonnet topological invariant and Fradkin-Tseytlin term in subsections \ref{section5.5.1} and \ref{section5.5.2}, respectively.

We also recall that the $\sSO(p) \times \sSO(q)$ superspace geometry was obtained by imposing an appropriate gauge within the conformal $(p,q)$ superspace described in section \ref{Chapter5.3}, where the latter is a gauge theory of the superconformal group ${\sOSp}_0 (p|2; {\mathbb R} ) \times  {\sOSp}_0 (q|2; {\mathbb R} )$. As is well known, alternative two-dimensional superconformal groups, and thus conformal superspace formulations, exist. This is discussed further in subsection \ref{section5.5.3}.

\subsection{$\cN$-extended AdS superspace}
\label{section5.5.1}

Supersymmetric theories in AdS$_2$ have recently attracted much interest, see e.g.
\cite{Beccaria:2019dju} and references therein. Using our construction of $\mathsf{SO}(p) \times \mathsf{SO}(q)$ superspace, it is possible to work out the structure of AdS superspaces in two dimensions. This can be achieved by analogy with the derivation of $(p,q)$ superspace in three dimensions \cite{KLT-M12}. Specifically, two-dimensional AdS superspaces 
correspond to those supergravity backgrounds which 
satisfy the following requirements:

(i) the torsion and curvature tensors are Lorentz invariant;

(ii) the torsion and curvature tensors are covariantly constant.\\
Condition (i) means that 
\begin{align}
	X_{++}^{\OI \OJ} = 0~, \qquad X_{--}^{\UI \UJ} = 0~.
\end{align}
Condition (ii) is equivalent to the requirement
\begin{align}
	\cD_A S^{\OI \UJ} = 0~,
\end{align}
which has nontrivial implications. This requirement and 
the relation \eqref{5.69b} give 
\bea
0= \{ \cD_{+}^{\OI}, \cD_{-}^{\UJ} \} S^{\OK \UL} = \d^{\OI \OK} S^{\OM \UJ} S^{\OM \UL} - \d^{\UL \UJ} S^{\OI \UM} S^{\OK \UM}~.
\eea 
The simplest solution for $p=q \equiv \cN$  is
\begin{align}
	S^{\OI \UJ} = S \d^{\OI \UJ}~, \qquad \cD_A S = 0~.
\end{align}
It corresponds to a special frame (or a special gauge condition) in which  
the left and right $R$-symmetry connections coincide 
\begin{align}
	\Phi_A^{\OI \OK} \d^{\OK \UJ} = \d^{\OI \UK} \Phi_A^{\UK \UJ} 
	~.
\end{align}
This means that the two types of $R$-symmetry indices turn into a single type, $\OI = \UI \equiv I$, and we stay with the diagonal subgroup of the $R$-symmetry group $\mathsf{SO}(\cN) \times \mathsf{SO}(\cN)$. The latter is generated by $J^{I J} = - J^{J I} = \mathfrak{L}^{IJ} + \mathfrak{R}^{IJ}$, which acts on isovectors as follows
\begin{align}
	J^{IJ} \chi^{K} = 2 \d^{K[I} \chi^{J]}~.
\end{align}
To summarise, the algebra of covariant derivatives for $\cN$-extended anti-de-Sitter superspace is
\begin{subequations} \label{AdSalgebra}
	\bea
	\{ \cD_{+}^{I}, \cD_{+}^{J} \} &=& 2 \ri \d^{I J} \cD_{++} ~, \quad \{ \cD_{-}^{I}, \cD_{-}^{J} \} = 2 \ri \d^{I J} \cD_{--} ~,  \\
	\{ \cD_{+}^{I}, \cD_{-}^{J} \} &=& - 4 \d^{I J} S M - 2 S J^{I J} ~, \\
	\big[ \cD_{+}^{I} , \cD_{--} \big]
	& = & - 2 \ri S \cD_{-}^{I}
	~, \quad 
	\big[ \cD_{-}^{I} , \cD_{++} \big]
	= 2 \ri S \cD_{+}^{I} ~,
	\\
	\big[ \cD_{++} , \cD_{--} \big]
	& = & 8 S^2 M
	~,
	\eea
\end{subequations}
where it should be understood that $J^{IJ}$ is not present for $\cN=1$. The AdS curvature $S$ is related to the scalar curvature by $\mathcal{R} = 16 S^2 < 0$.
The isometry supergroup of this $\cN$-extended AdS superspace is $\sOSp(\cN|2; {\mathbb R})$, see  \cite{GST} for the complete list of AdS$_2$ supergroups.

\subsection{The Gauss-Bonnet invariant and Fradkin-Tseytlin term}
\label{section5.5.2}

The formalism developed in this chapter may also be used to construct supersymmetric extensions of the Gauss-Bonnet invariant by a generalisation of four-dimensional logarithm construction of \cite{BdeWKL}. To this end, we consider a nowhere vanishing primary scalar (super)field $\varphi$ of non-zero dimension $\D$. From $\varphi$, one may construct the following primary descendants:
\begin{subequations} \label{5.131}
	\begin{align}
		\cN = (0,0):& \qquad \nabla_{++} \nabla_{--} \, \text{ln} \, \varphi ~, \label{5.131a}\\
		\cN = (1,0):& \qquad \nabla_{--} \nabla_{+} \, \text{ln} \, \varphi ~, \\
		\cN = (1,1):& \qquad \hf [ \nabla_{+} , \nabla_{-} ] \, \text{ln} \, \varphi ~.
		\label{5.131c}
	\end{align}
\end{subequations}
They can be used to define the (super)conformal functionals:
\begin{subequations}
	\label{5.132}
	\begin{align}
		\cN = (0,0):& \qquad \cS_{(0,0)} = - \frac{1}{2\D} \int {\rm d}^2x \, e \, \nabla_{++} \nabla_{--} \, \text{ln} \, \varphi = - \frac{1}{8} \int {\rm d}^2x \, e \, \cR  ~, \label{5.132a}\\
		\cN = (1,0):& \qquad \cS_{(1,0)} = - \frac{1}{2\D} \int {\rm d}^{(2|1,0)}z^{-} \, E \, \nabla_{--} \nabla_{+} \, \text{ln} \, \varphi = \int {\rm d}^{(2|1,0)}z^{-} \, E \, G_{-} ~, \\
		\cN = (1,1):& \qquad \cS_{(1,1)} = - \frac{1}{4\D} \int {\rm d}^{(2|1,1)}z \, E \, [ \nabla_{+} , \nabla_{-} ] \, \text{ln} \, \varphi = \int {\rm d}^{(2|1,1)}z \, E \, S ~.
		\label{5.132c}
	\end{align}
\end{subequations}
Where in the latter expressions we have degauged the Lagrangian and then ignored all $\vf$-dependent surface terms. Remarkably, these functionals have proven to be independent of $\varphi$. Additionally, the first is simply the Gauss-Bonnet invariant, while the latter two are its simplest supersymmetric extensions. 

Making use of the primary (super)fields \eqref{5.131}, we introduce manifestly (super)conformal generalisations of the Fradkin-Tseytlin term in string theory \cite{Fradkin:1985ys}
\bea
S_{\rm FT}=  \frac{1}{4\p} \int {\rm d}^2x \, e \, \cR {\bm \F}~,
\eea
where $\bm \F$ denotes the dilaton field in the curved spacetime in which the string propagates. If we symbolically denote by $\O(\vf)$ any of the primaries in \eqref{5.131} and by $\int \rd \m $ the integration measures in 
\eqref{5.132}, then $I:= \int \rd \m \, \O(\vf)  {\bm \F}$ is invariant under the gauge group of conformal (super)gravity for any dimensionless primary scalar $\bm \F$. Here $\vf$ plays the role of a conformal compensator.  Choosing a (super-)Weyl gauge $\vf =1$ leads to standard expressions for the Fradkin-Tseytlin term and its supersymmetric extensions, modulo an overall numerical coefficient. 

The analysis above may be extended to the $\cN=(2,2)$ case.
To this end, it is necessary to work in the complex basis of spinor covariant derivatives, which is obtained from \eqref{5.76} upon the replacement $\cD \rightarrow \nabla$. This allows us to define two types of constrained superfields, namely chiral superfields $\Phi$
\begin{subequations}
	\label{5.134}
	\begin{align}
		\bar{\nabla}_+ \Phi = 0 ~, \qquad \bar{\nabla}_- \Phi = 0~,
	\end{align}
	and twisted chiral superfields $\chi$
	\begin{align}
		\bar{\nabla}_+ \chi = 0 ~, \qquad \nabla_- \chi = 0~,
	\end{align}
\end{subequations}
where the Lorentz weights of $\Phi$ and $\chi$ are not indicated. Assuming that they are primary, their superconformal properties are related as follows:
\begin{subequations}
	\begin{align}
		q^L_\Phi &= \D_\Phi + \l_\Phi ~, \qquad q^R_\Phi = \D_\Phi - \l_\Phi~, \\
		q^L_\chi &= \D_\chi + \l_\chi  ~, \qquad q^R_\chi = \l_\chi - \D_\chi  ~.
	\end{align}
\end{subequations}
Here $q^L_\Phi$ is defined by $\ri \mathfrak{L}^{\overline{1} \overline{2}} \Phi = q^L_\Phi \Phi$ and similarly for $q^R_\Phi$. 

We now specify to primary Lorentz scalars $\Phi$ and $\chi$ of non-zero dimensions. From these superfields we may construct the primary descendants 
\bea
\cN = (2,2): \qquad 
\bar{\nabla}_+ \bar{\nabla}_- \, \text{ln} \, \bar{\Phi}~,
\qquad \bar{\nabla}_+ {\nabla}_- \, \text{ln} \, \bar{\chi}~, 
\eea
which are chiral and twisted chiral, respectively. They may be used to define the superconformal functionals
\begin{subequations}
	\label{5.137}
	\begin{align}
		\mathcal{S}^{\rm C}_{(2,2)} &= - \frac{1}{\D_\Phi} \int {\rm d}^{2}x \rd^2 \q \, \cE \, \bar{\nabla}_+ \bar{\nabla}_- \, \text{ln} \, \bar{\Phi} = \int {\rm d}^{2}x \rd^2 \q \, \cE \, \Xi^{\rm C} +\dots ~, \label{5.137a} \\
		\mathcal{S}^{\rm TC}_{(2,2)} &= \frac{1}{\D_\chi} \int {\rm d}^{2}x \rd {\q}^+ \rd \bar{\q}^- \, \mathfrak{E} \, \bar{\nabla}_+ {\nabla}_- \, \text{ln} \, \bar{\chi} = \int {\rm d}^{2}x \rd {\q}^+ \rd \bar{\q}^- \, \mathfrak{E} \, \Xi^{\rm TC}  +\dots ~, \label{5.137b}
	\end{align}
\end{subequations}
where $\cE$ and $\mathfrak{E}$ are appropriately defined measures for the chiral and twisted chiral subspaces, respectively, and we have made the definitions
\begin{subequations}
	\begin{align}
		\Xi^{\rm C} &:= S^{\overline{1} \underline{1}} + \ri S^{\overline{1} \underline{2}} + \ri S^{\overline{2} \underline{1}} - S^{\overline{2} \underline{2}}~, \quad \quad \quad \bar{\cD}_+ \Xi^{\rm C} = 0 ~, \quad \bar{\cD}_-\Xi^{\rm C} = 0~, \\
		\Xi^{\rm TC} &:= S^{\overline{1} \underline{1}} - \ri S^{\overline{1} \underline{2}} + \ri S^{\overline{2} \underline{1}} + S^{\overline{2} \underline{2}}~, \quad \quad \quad \bar{\cD}_+ \Xi^{\rm TC} = 0 ~, \quad {\cD}_-\Xi^{\rm TC} = 0~.
	\end{align}
\end{subequations}
It is not difficult to to show that the functionals \eqref{5.137} are independent of $\bar \F$ and $\bar \c$, respectively.  It may also be seen that these functionals are topological. Degauging the integrand in \eqref{5.137a} gives 
\bea
\bar{\nabla}_+ \bar{\nabla}_- \, \text{ln} \, \bar{\Phi}  = - \D_\F \X^{\rm C} 
+  \bar{\cD}_+ \bar{\cD}_- \, \text{ln} \, \bar{\Phi}  \equiv - \D_\F \X^{\rm C}  +\dots~,
\eea
where the ellipsis denotes a chiral superfield for which we do not yet have an explicit expression. The functionals \eqref{5.137} define $\cN=(2,2)$ extensions of the Gauss-Bonnet invariant, eq. \eqref{5.132a}. Component analyses of these invariants will be given elsewhere.

Given primary dimensionless chiral $\J$ and twisted chiral $\S$  scalars, the following functionals 
\bea
\int {\rm d}^{2}x \rd^2 \q \, \cE \, \J \bar{\nabla}_+ \bar{\nabla}_- \, \text{ln} \, \bar{\Phi} ~,
\qquad
\int {\rm d}^{2}x \rd {\q}^+ \rd \bar{\q}^- \, \mathfrak{E} \, \S \bar{\nabla}_+ {\nabla}_- \, \text{ln} \, \bar{\chi} 
\eea
are superconformal invariants. They may be viewed as $\cN=2$ supersymmetric extensions  of the 
Fradkin-Tseytlin term.

The above $\cN=(2,2)$  constructions may also be generalised to the $\cN=(2,0)$ and $\cN=(2,1)$ cases. In both cases it is necessary to consider a scalar superfield $\F$ of dimension $\D$ which is chiral with respect to the left coordinates, $\bar{\nabla}_+ \Phi = 0$. Using $\Phi$, we construct the primary left chiral descendants
\begin{subequations}
	\begin{align}
		\cN = (2,0):& \qquad \bar{\nabla}_{+} \nabla_{--} \, \text{ln} \, \bar{\Phi} ~, \label{5.141a} \\
		\cN = (2,1):& \qquad \bar{\nabla}_{+} \nabla_{-} \, \text{ln} \, \bar{\Phi} ~.
		\label{5.141b}
	\end{align}
\end{subequations}
They may be used to define the superconformal functionals
\begin{subequations}
	\begin{align}
		\cN = (2,0): \quad \mathcal{S}_{(2,0)} &= -\frac{1}{2\D} \int {\rm d}^{2}x \rd {\q}^+ \, \cE_L^{(2,0)} \, \bar{\nabla}_+ {\nabla}_{--} \, \text{ln} \, \bar{\Phi} \non \\
		\qquad &= 2\int {\rm d}^{2}x \rd {\q}^+ \, \cE_L^{(2,0)} \, G_{-} ~,  \\
		\cN = (2,1): \quad \mathcal{S}_{(2,1)} &= \frac{1}{\sqrt{2}\D} \int {\rm d}^{2}x \rd {\q}^+ \rd {\q}^- \, \cE_L^{(2,1)} \, \bar{\nabla}_+ {\nabla}_- \, \text{ln} \, \bar{\Phi} \non \\
		\qquad &= \int {\rm d}^{2}x \rd {\q}^+ \rd {\q}^- \, \cE_L^{(2,1)} \, \Xi^{\rm LC} +\dots ~, 
	\end{align}
\end{subequations}
where $\cE_L^{(2,0)}$ and $\cE_L^{(2,1)}$ are the appropriate integration measures, $G_-$ is defined in \eqref{5.95} and we have made the definition
\begin{align}
	\Xi^{\rm LC} &:= S^{\overline{1}} + \ri S^{\overline{2} } ~, \quad \quad \quad \bar{\cD}_+ \Xi^{\rm LC} = 0 ~.
\end{align}
Making use of the primary chiral descendants \eqref{5.141a} and \eqref{5.141b} allows us to define $(2,0)$ and $(2,1)$ supersymmetric analogues of the Fradkin-Tseytlin term. 
In both cases such invariants are associated with a primary dimensionless chiral scalar $\J$ and have the explicit form 
\begin{subequations}
	\begin{align}
		\cN = (2,0): \qquad & \int {\rm d}^{2}x \rd {\q}^+ \, \cE_L^{(2,0)} \,\J \bar{\nabla}_+ {\nabla}_{--} \, \text{ln} \, \bar{\Phi} \\
		\cN = (2,1): \qquad  & \int {\rm d}^{2}x \rd {\q}^+ \rd {\q}^- \, \cE_L^{(2,1)} \, \J\bar{\nabla}_+ {\nabla}_- \, \text{ln} \, \bar{\Phi} ~.
	\end{align}
\end{subequations}

\subsection{Alternative superconformal groups}
\label{section5.5.3}

In this chapter, a central role has been played by the $(p,q)$ superconformal group ${\sOSp}_0 (p|2; {\mathbb R} ) \times  {\sOSp}_0 (q|2; {\mathbb R} )$. Actually, alternative superconformal groups exist and were classified almost fourty years ago by 
G\"unaydin, Sierra and Townsend \cite{GST}. In particular, they were shown to have the structure 
\bea
\label{5.123}
G = G_L \times G_R~,
\eea
where $G_L$ and $G_R$ are simple supergroups. The supergroups  $G_L$ and $G_R$ can be any of the following: (i) $\sOSp_0 (m|2;{\mathbb R})$; 
(ii) $\sSU(1,1|m)$, for $m\neq 2$, or $\mathsf{PSU}(1,1|2)$;
(iii) $\sOSp(4^*|2m )$; (iv) $\mathsf{G}(3)$; (v) $\mathsf{F}(4)$; and 
(vi) $\mathsf{D}^1(2,1,\a)$. Below, by following a similar approach to the one utilised in section \ref{Chapter5.1}, we will realise the superconformal group $\sSU(1,1|n) \times \sOSp_0(q|2;{\mathbb R})$ (or $\sPSU(1,1|2) \times \sOSp_0(q|2;{\mathbb R})$ if $n=2$) in terms of conformal Killing supervector fields.\footnote{See appendix \ref{Appendix5B} for a supermatrix realisation of $\sSU(1,1|n) \times \sOSp_0(q|2;{\mathbb R})$.}

Our approach to rigid $(p,q)$ superconformal symmetry was based on the use of real coordinates $z^{A} = (x^{a},\q^{+ \Io},\q^{- \Iu})$  to parametrise Minkowski superspace $\mathbb{M}^{(2|p,q)}$. The conformal Killing supervector fields were defined to satisfy
the equation \eqref{5.4}, which implied 
that the algebra of conformal Killing supervector fields of $\mathbb{M}^{(2|p,q)}$ is infinite dimensional. 
Its maximal finite-dimensional subalgebra is singled out by 
the conditions \eqref{5.13}, and is isomorphic to  
$\mathfrak{osp}(p|2;\mathbb{R}) \oplus \mathfrak{osp}(q|2;\mathbb{R})$.
The latter is the Lie algebra of the supergroup
${\sOSp}_0 (p|2; {\mathbb R} ) \times  {\sOSp}_0 (q|2; {\mathbb R} )$,
which is the superconformal group of the compactified Minkowski superspace \eqref{5.145}.

In the case that $p$ is even, $p=2n$, the real Grassmann variables $\q^{+ \Io}$ can be replaced with complex ones, 
\bea
\q^{+ \Io} \to (\q^{+ i}~, \bar \q^+_i)~, \qquad \bar \q^+_i := \overline{\q^{+i}} ~, \quad 
i= 1, \dots , n~.
\eea   
At the same time, the real covariant derivatives $D_{+}^\OI$ should be replaced with complex ones
\bea
D^{\Io}_+ \to (D_{+ i}~, \bar D_+^i)~, \qquad \bar D_+^i := \overline{D_{+i}} ~,
\eea
which obey the algebra
\bea
\big \{ D_{+i} , \bar D_+^j \big \} = 2 \ri \d_i^j \partial_{++}~.
\eea

Then, equation \eqref{5.4} should be replaced with 
\begin{subequations}
	\bea 
	\label{5.116a}
	[\xi , D_{+i} ] = - (D_{+i} \xi^{+ j}) D_{+ j} ~, \qquad
	[\xi , D_-^{\Iu} ] = - (D_-^{\Iu} \xi^{- \Ju}) D_-^{\Ju} ~,
	\eea
	where $\xi$ takes the form
	\bea
	\xi = \xi^{a} \partial_{a} + \xi^{+ i} D_{+i} + \bar{\xi}^{+}_i \bar{D}_+^i + \xi^{- \UI} D_{-}^\UI = \bar{\xi}~.
	\eea
\end{subequations}
One may then perform a similar analysis to that of section \ref{Chapter5.1} to obtain the corresponding superconformal algebra as a subalgebra of the $(p,q)$ super Virasoro algebra. We will not perform a complete analysis here and instead sketch the 
salient points.

The defining relations \eqref{5.116a} imply the master equations
\begin{align}
	\label{5.117}
	D_{+i} \xi^{--} = 0 ~, \qquad D_{-}^{\UI} \xi^{++}=0 ~, \qquad D_{+i} D_{+j} \xi^{++} = 0~,
\end{align}
and yield the following expressions for the spinor parameters
\begin{align}
	\xi^{+ i} = - \frac \ri 2 \bar D_{+}^i \xi^{++} ~, \qquad \xi^{- \UI} = - \frac \ri 2 D_{-}^\UI \xi^{--}~.
\end{align}
It should be noted that the final relation of \eqref{5.117} has the following non-trivial implications, depending on the value of $n$
\begin{subequations}
	\begin{align}
		n = 2:& \quad \partial_{++} [D_{+i}, \bar{D}_{+}^i] \xi^{++} = 0 ~, \\
		n > 2:& \quad \partial_{++} [D_{+i}, \bar{D}_{+}^j] \xi^{++} = 0 ~.
	\end{align}
\end{subequations}
In particular, it follows from the latter constraint that
\begin{align}
	\label{5.120}
	\partial_{++} \partial_{++} D_{+i} \xi^{++} = 0~,
\end{align}
and thus, for $n>2$, the vector $\xi^{++}$ encodes finitely many parameters. This is in contrast to the situation in chapter \ref{Chapter5.1}, where it was necessary to impose the conditions \eqref{5.13}. This difference is a consequence of \eqref{5.116a} being more restrictive than \eqref{5.4}. For $n=1,2$ condition \eqref{5.120} must instead be imposed by hand.

Making use of \eqref{5.120}, the master equations \eqref{5.116a} may be written in the form
\begin{subequations}
	\begin{align}
		[\xi , D_{+i}] &= - \hf (\s[\xi] + K[\xi] + 2 \chi[\xi]) D_{+i} - \chi[\xi]_{i}{}^j D_{+j} ~, \\
		[\xi , D_-^{\Iu}] &= - \hf (\s[\xi] - K[\xi]) D_-^{\Iu} - \r[\xi]^{\Iu \Ju} D_{-}^{\Ju} ~,
	\end{align}
\end{subequations}
where $\s[\xi]$, $K[\xi]$ and $\r[\xi]^{\UI \UJ}$ were defined in \eqref{5.8} and we have made the definitions
\begin{subequations}
	\begin{align}
		\chi[\xi] &:= - \frac{\ri}{4n} [D_{+i}, \bar{D}_+^i] \xi^{++} ~, \\ \chi[\xi]_i{}^j &:= - \frac{\ri}{4n} \Big ( [D_{+i}, \bar{D}_+^j] - \frac 1 n \d_i^j [D_{+k}, \bar{D}_+^k] \Big ) \xi^{++} ~. 
	\end{align}
\end{subequations}
The former are constrained to satisfy \eqref{5.9}, while the new parameters obey
\begin{subequations}
	\begin{align}
		D_{+i} \chi[\xi] &= \frac{n-2}n D_{+i} \s[\xi] ~, \qquad  D_{-}^\UI \chi[\xi] = 0 ~, \label{5.143a} \\
		D_{+i} \chi[\xi]_j{}^k &= -2 \d_i^k D_{+j} \s[\xi] + \frac 2 n \d_j^k D_{+i} \s[\xi]~,
		\qquad D_{-}^\UI \chi[\xi]_j{}^k = 0~.
	\end{align}
\end{subequations}
One may then continue this analysis, keeping in mind the philosophy of section \ref{Chapter5.1}, to derive the corresponding superconformal algebra.

The resulting superconformal group for $n\neq2$ proves  to be 
$\mathsf{SU} (1,1|n) \times  \mathsf{OSp}_0 (q|2; {\mathbb R})$, with 
$\sU(n)  \times \mathsf{SO}(q)$ being  its $R$-symmetry subgroup. 
The $n=2$ case is special since, as follows from eq. \eqref{5.143a}, the diagonal $\sU(1)$ subgroup of 
$\mathsf{SU} (1,1|2 ) $
can be factored out, and the $R$-symmetry subgroup of the superconformal group becomes $\sSU(2) \times \mathsf{SO}(q)$. 
Now, the construction of conformal $(2n,q)$ superspace can be carried out by gauging 
the superconformal group $\mathsf{SU} (1,1|n ) \times  \mathsf{OSp}_0 (q|2; {\mathbb R} )$ for $n\neq 2$, or $\mathsf{PSU} (1,1|2 ) \times  \mathsf{OSp}_0 (q|2; {\mathbb R} )$ for $n=2$. The degauged version of this superspace may be denoted 
$\sU(n)  \times \mathsf{SO}(q)$ superspace, for $n\neq 2$, and $\sSU(2)  \times \mathsf{SO}(q)$ superspace, for $n=2$. Analogous considerations apply in the case that $q$ is even. In particular, for $p=q=4$, one can introduce conformal $(4,4)$ superspace as the gauge theory of $\mathsf{PSU} (1,1|2 ) \times \mathsf{PSU} (1,1|2 ) $. Its degauged version turns out to be the curved  $\sSU(2)  \times \sSU(2)$ superspace geometry
introduced in \cite{TM}.\footnote{As pointed out in \cite{TM}, 
	the super-Weyl and local $R$-symmetry transformations can be used to partially fix the gauge freedom such that the resulting supergravity multiplet turns into the one proposed 
	many years ago in  \cite{Gates:1988ey, Gates:1988tn}.}

\section{Summary of results} \label{Chapter5.6}

This chapter was devoted to the construction of two-dimensional conformal $(p,q)$ supergravity as the gauge theory of the superconformal group $\sOSp_0(p|2;\mathbb{R}) \times \sOSp_0(q|2;\mathbb{R})$ and applications thereof which appeared in the recent work \cite{KR22}. We begun by studying the conformal Killing supervectors of $\mathbb{M}^{(2|p,q)}$ in section \ref{Chapter5.1}, which were shown to define a $(p,q)$ supersymmetric extension of the Virasoro algebra. Upon appropriately truncating these supervectors via the constraints \eqref{5.13}, they were shown to generate the infinitesimal superconformal transformations associated with the superconformal algebra $\mathfrak{osp}(p|2;\mathbb{R}) \oplus \mathfrak{osp}(q|2;\mathbb{R})$. Employing this construction, we readily obtained the defining relations for the superconformal generators \eqref{5.18}, which are spelled out in equations \eqref{5.20} -- \eqref{5.25}.

Utilising the above results, it was shown that, as expected, $\mathfrak{osp}(p|2;\mathbb{R}) \oplus \mathfrak{osp}(q|2;\mathbb{R})$ contains the conformal algebra of Minkowksi space, $\mathfrak{sl}(2,\mathbb{R}) \oplus \mathfrak{sl}(2,\mathbb{R})$, as a subalgebra. In preparation for studying the supersymmetric case, we gauged the latter and studied the resulting geometry.\footnote{Such a geometry is encoded within the conformal supergravity theories of \cite{vanNieuwenhuizen:1985an, Uematsu:1984zy, Uematsu:1986de, Hayashi:1986ev, Uematsu:1986aa, Bergshoeff:1985qr, Bergshoeff:1985gc, Schoutens:1986kz}.} Keeping in mind $d\geq3$-dimensional construction described in section \ref{Chapter2.2}, we imposed the constraints \eqref{5.37} on the algebra of covariant derivatives. The resulting geometry was shown to be characterised by the primary dimension-$3$ superfields $W_{++}$ and $W_{--}$. In particular, these superfields were required to vanish for the geometry to describe conformal gravity. It was then shown that, upon degauging to Lorentzian geometry, this condition leads to a highly non-trivial outcome; the degauged special conformal connections $\mathfrak{f}_{++,++}$ and $\mathfrak{f}_{--,--}$ become non-local functions of the vielbein, which is in contrast to the $d\geq3$ case.\footnote{The $(1,0)$ supersymmetric extension of this geometry was described in appendix \ref{Appendix5A}.}

Building on the non-supersymmetric case described in section \ref{Chapter5.2}, we sketched the construction of conformal $(p,q)$ superspace with vanishing curvature in section \ref{Chapter5.3}. This was achieved by gauging the superconformal algebra $\mathfrak{osp}(p|2;\mathbb{R}) \oplus \mathfrak{osp}(q|2;\mathbb{R})$ described in section \ref{Chapter5.1} and imposing the constraints \eqref{5.63} on the covariant derivatives. Following this, in section \ref{Chapter5.4} we degauged this geometry to obtain curved $\sSO(p) \times \sSO(q)$ superspace, deriving both the algebra of covariant derivatives and super-Weyl transformations of the resulting geometry. Upon making the appropriate comparisons with the literature, it was shown that our formulation reproduces all known $(p,q)$ superspace geometries and, for $p,q>2$, yields previously unknown supergeometries.

To conclude, in section \ref{Chapter5.5} we studied various applications and generalisations of these supergeometries. In particular, we described the $\cN$-extended AdS superspaces as those supergravity backgrounds with Lorentz invariant and covariantly constant curvatures. A second application was the construction of $(p,q)$ supersymmetric extensions of the Gauss-Bonnet invariant for $p,q\leq2$. Such invariants were shown to yield manifestly (super)conformal generalisations of the Fradkin-Tseytlin term in string theory \cite{Fradkin:1985ys}. Additionally, with regard to extensions of our supergeometries, we briefly described formulations of conformal superspace based on the superconformal groups classified in \cite{GST}. An interesting example of such a superconformal group is $\sSU(1,1|n) \times \sOSp_0(q|2;\mathbb{R})$ (for $n=2$ the diagonal $\sU(1)$ subgroup of $\sSU(1,1|2)$ may be quotiented out, which leads to $\sPSU(1,1|2) \times \sOSp_0(q|2;\mathbb{R})$). To study this group further, we computed the `truncated' conformal Killing supervectors of $\mathbb{M}^{(2|2n,q)}$, which generate the infinitesimal superconformal transformations belonging to its Lie algebra. The latter were also obtained in appendix \ref{Appendix5B} via the supertwistor realisation for the compactified Minkowski superspace $\overline{\mathbb M}^{(2|2n,q)}$. 

\begin{subappendices}
	
\section{Conformal $(1,0)$ superspace with non-vanishing curvature} \label{Appendix5A}

In section \ref{Chapter5.3} we proposed conformal $(p,q)$ superspaces characterised by the relations \eqref{5.63}; all conformal curvatures were set to zero. This appendix is devoted to deriving conformal $(1,0)$ superspace with non-vanishing curvature as an extension of the non-supersymmetric geometry \eqref{5.40}. 

Guided by the construction in higher dimensions, see section \ref{Chapter2.3} and the important publications \cite{ButterN=1,ButterN=2,BKNT-M1,BKNT-M3,BKNT}, we require that the covariant derivatives $\nabla_A = (\nabla_+ , \nabla_{++} , \nabla_{--})$ obey  constraints that are similar to those of super Yang-Mills theory
\begin{align}
	\label{(1,0)CSSAlgebra}
	\big \{ \nabla_{+} , \nabla_{+} \big \} = 2 \ri \nabla_{++} ~, \qquad \big[ \nabla_{+} , \nabla_{--} \big] = \ri \mathscr{W}_{-} ~.
\end{align}
Here $\mathscr{W}_- = \mathscr{W}(X)_-{}^{\tilde A} X_{\tilde A}$, and $X_{\tilde A}$ denotes the generators of the superconformal algebra \eqref{5.18}. The Bianchi identities yield
\begin{align}
	\label{5.146}
	\big [ \nabla_{+} , \nabla_{++} \big ] = 0 ~, \qquad \big[ \nabla_{++} , \nabla_{--} \big] = \big \{ \nabla_{+} , \mathscr{W}_{-} \big \} ~.
\end{align}
Additionally, requiring consistency of \eqref{(1,0)CSSAlgebra} with the superconformal algebra leads to
\begin{align}
	\label{5.147}
	\big [K^{A}, \mathscr{W}_{-} \big \} = 0 ~, \qquad \big[\mathbb{D} , \mathscr{W}_- \big] = \frac 3 2 \mathscr{W}_- ~, \qquad \big[M , \mathscr{W}_- \big] = - \frac 1 2 \mathscr{W}_- ~.
\end{align}

Hence, we constrain $\mathscr{W}_-$ to be
\begin{align}
	\label{5.148}
	\mathscr{W}_{-} = \chi_{+} K^{++} + \psi_{---} K^{--}~,
\end{align} 
where $\chi_{+}$ and $\psi_{---}$ are primary superfields of dimension $5/2$. At the component level, they contain the conformal curvatures $W_{++}$ and $W_{--}$ \eqref{5.40}
\begin{align}
	\label{5.149}
	W_{++} = \nabla_{+} \chi_{+} |_{\q^+ = 0} ~, \qquad W_{--} = \nabla_{+} \psi_{---} |_{\q^+ = 0}~.
\end{align}
Finally, inserting \eqref{5.148} into \eqref{5.146}, we obtain the commutator of vector derivatives
\begin{align}
	\big[ \nabla_{++} , \nabla_{--} \big] = \ri \chi_+ S^+ + (\nabla_+ \chi_+) K^{++} + (\nabla_+ \psi_{---}) K^{--}~.
\end{align} 

We note that, according to \eqref{5.60}, the dilatation connection transforms algebraically under infinitesimal special superconformal gauge transformations \eqref{563}. This allows us to fix $B_A=0$ at the expense of this symmetry. In this gauge the special conformal connection $\mathfrak{F}_{AB}$ is 
not required for the covariance of $\nabla_A$ 
and may
be separated
\begin{align}
	\nabla_A = \cD_A - \mathfrak{F}_{AB} K^B~.
\end{align}
Here the degauged covariant derivative $\cD_A$ involves only the Lorentz connection and obeys the algebra \eqref{5.108}. The connection $\mathfrak{F}_{AB}$ was described in \eqref{5.107a} and \eqref{5.107b}. 
The constraints \eqref{5.107c} are now replaced with 
\begin{align}
	\psi_{---} = \cD_{--} G_{-} - \cD_+ \mathfrak{F}_{--,--} ~, \qquad \chi_+ = - \cD_{++} G_- + \cD_{--} \mathfrak{F}_{+,++}~.
\end{align}
These relations determine the curvature tensors $\j_{---}$ and $\c_+$ in terms of $\cD_A$ and $\mathfrak{F}_{AB}$.
Keeping \eqref{5.149} in mind, it is clear that this is a $(1,0)$ extension of \eqref{5.48}. In particular, for vanishing $\psi_{---}$  and $\chi_+$, the connections $\mathfrak{F}_{--,--}$  and $\mathfrak{F}_{+,++}$ are  non-local functions of the supergravity multiplet.
	
\section{Compactified Minkowski superspace} \label{Appendix5B}

Superconformal groups \eqref{5.123} do not act on Minkowski superspace,
since the special conformal and $S$-supersymmetry transformations are singular at some points. However, there exists a well defined action of \eqref{5.123} on a compactified $(p,q)$ Minkowski superspace $\overline{\mathbb M}^{(2|p,q)}$, for some $p,q$. In general, $\overline{\mathbb M}^{(2|p,q)}$ has the form
\bea
\overline{\mathbb M}^{(2|p,q)} = S^{1|p}_L \times S^{1|q}_R~,
\label{5.145}
\eea
where the bosonic body of $S^{1|n}$ is a circle $S^1$. The left superspace $S^{1|p}_L$ is a homogeneous space of the subgroup $G_L$
of \eqref{5.123}, and similarly in the right sector.

In a recent paper \cite{KT-M2021}, $\overline{\mathbb M}^{(2|p,q)}$ was realised
as a homogeneous space of the superconformal group
${\sOSp}_0 (p|2; {\mathbb R} ) \times  {\sOSp}_0 (q|2; {\mathbb R} )$.
Here we present a different construction for the case that $p$ is even, $p =2n$.
Specifically, we describe
$\overline{\mathbb M}^{(2|2n,q)}$
as a homogeneous space of the superconformal group
\bea
G=G_L\times G_R = \mathsf{SU} (1,1|n )
\times  {\sOSp}_0 (q|2; {\mathbb R} )~.
\eea

We begin by describing the action of $\mathsf{SU} (1,1|n )$ on $S^{1|2n}$.
The supergroup $\mathsf{SU} (1,1|n )$ is spanned by supermatrices of the form
\bea
g \in \sSL(2 | n;{\mathbb C} ) ~, \qquad
g^\dagger \,\O \,g = \O~,  \qquad
\O=
\left(
\begin{array}{cc|c}
	1&  0 ~&0\\
	0 &-1 ~&0\\
	\hline
	0 & 0& {\mathbbm 1}_{n}
\end{array}
\right) ~.
\eea
This supergroup naturally acts on the space of even supertwistors ${\mathbb C}^{2|n}$
\bea
X = \left(
\begin{array}{c}
	z \\
	w\\
	\hline
	\vf^i
\end{array}
\right) ~,\qquad i=1, \dots , n~,
\eea
where $z,w$ are complex bosonic variables, and $\vf^i$ are complex Grassmann variables.
We identify $S^{1|2n}$ with the space of null lines in ${\mathbb C}^{2|n}$.
By definition,
a null supertwistor $X$ is characterised by the conditions
\bea
X^\dagger \O X=0~, \qquad  \left(
\begin{array}{c}
	z \\
	w
\end{array}
\right) \neq 0~.
\eea
Two null supertwistors $X$ and $X'$ are said to be equivalent if
\bea
X' = {\mathfrak c} X~, \qquad {\mathfrak c} \in {\mathbb C}\setminus \{0\}~.
\label{5.158}
\eea
Any equivalence class in the set of null supertwistors is called a null line. Given a null supertwistor $X$ both bosonic components $z$ and $w$ are non-zero.
Making use of the equivalence relation \eqref{5.158} allows us to choose, for each null line,
a representative
\bea
X = \left(
\begin{array}{c}
	z \\
	1\\
	\hline
	\vf^i
\end{array}
\right) ~,\qquad |z|^2 = 1 -\vf^\dagger \vf~,
\eea
which
is uniquely defined for the null line under consideration. It is seen that the quotient space is $S^{1|2n}$.

In order to make contact to ordinary Minkowski superspace, it is useful to switch to a different parametrisation of $\sSU(1,1|n)$ and the associated supertwistor space.
Let us introduce the supermatrix
\bea
\S= \frac{1 }{ \sqrt{2} }
\left(
\begin{array}{cr|c}
	{1}  ~ & - {1} ~& 0\\
	1 ~&    1 ~& 0  \\
	\hline
	0 & 0 ~ & \sqrt{2} \,{\mathbbm 1}_{n}
\end{array}
\right)~, \qquad \S^\dagger \S= {\mathbbm 1}_{2+n}~,
\eea
and associate with it the following similarity transformation:
\bea
g ~& \to & ~ \hat{ g} = \S \, g\, \S^{-1} ~, \quad g \in \sSU(1,1|n)~;
\qquad
X ~ \to  ~ \hat{X} = \S \, T~, \quad X \in {\mathbb C}^{2|n}~.
\label{sim2}
\eea
The supertwistor metric $\O$ turns into
\bea
\hat{ \O} =  \S \, \O\, \S^{-1}
=
\left(
\begin{array}{cc|c}
	0&  {1} ~&0\\
	{1} &0 ~&0\\
	\hline
	0 & 0& {\mathbbm 1}_{n}
\end{array}
\right) ~.
\eea
In the new frame, it is not guaranteed that both bosonic components $\hat z$ and $\hat w$ of  a null supertwistor $\hat X$ are non-zero. However, at least one of $\hat z$ and $\hat w$ is non-vanishing, and we can introduce an open subset of $S^{1|p}$ which is parametrised by null supertwistors of the form
\bea
\hat{X} = \left(
\begin{array}{c}
	1 \\
	- \ri {\bm x}^{++} \\
	\hline
	\sqrt{2}\q^{+i}
\end{array}
\right) ~, \qquad {\bm x}^{++} - \bar{\bm x}^{++} = 2\ri \bar \q^+_i \q^{+i} ~,
\quad \bar \q^+_i := \overline{\q^{+ i}}~.
\eea
The constraint on ${\bm x}^{++}$ is solved by
\bea
{\bm x}^{++} = {x}^{++} + \ri \bar \q^+_i \q^{+i} ~, \qquad
\overline{ x^{++}} = x^{++}~.
\eea
The variables ${\bm x}^{++}$ and $\q^{+i} $ parametrise a chiral subspace of $\mathbb{M}^{(2|2n,q)}$. To deduce the superconformal transformations of this subspace it is necessary to act on $\hat{X}$ with a generic group element $\hat{g} \in \sSU(1,1|n)$:
\begin{align}
	\hat{g} = \re^{\L_L} ~, \qquad \L_L = 
	\left(
	\begin{array}{cc|c}
		-\hf(\s + K) - \frac{\ri n }{n-2} \chi&  {\ri b_{++}} ~&\sqrt{2} \eta_{+j}\\
		{- \ri a^{++}} & \hf(\s + K) - \frac{\ri n }{n-2} \chi ~&\sqrt{2} \bar{\epsilon}^+_{j}\\
		\hline
		\sqrt{2} \epsilon^{+i} & \sqrt{2} \bar{\eta}^{i}_+& \l^{i}{}_j- \frac{2 \ri n }{n-2} \chi \d^{i}_j
	\end{array}
	\right)~,
\end{align}
where all scalar and vector parameters are real and 
\begin{align}
	\l^{\dagger} = - \l ~, \qquad {\rm tr} \ \l = 0~.
\end{align}
Taking the parameters to be small, one may show that the most general infinitesimal superconformal transformations on this subspace are:
\begin{subequations}
	\label{5.167}
	\begin{align}
		\d \bm{x}^{++} &= (\s + K) \bm{x}^{++} + a^{++} + 2 \ri \bar{\epsilon}_i^+ \q^{+i} - \bm{x}^{++} b_{++} \bm{x}^{++} - 2\bm{x}^{++} \eta_{+i} \q^{+i} ~, \\
		\d \q^{+i} &= \hf(\s+K) \q^{+i} -\frac{\ri n \chi}{n-2} \q^{+i} + \epsilon^{+i} + \l^{i}{}_j \q^{+j} - \q^{+i} b_{++} \bm{x}^{++} \non \\
		&\phantom{=}- \ri \bar{\eta}_{+}^i \bm{x}^{++}
		-2 \q^{+i} \eta_{+j} \q^{+j}  ~.
	\end{align}
\end{subequations}
The constant bosonic parameters in \eqref{5.167} correspond to dilatations $(\s)$, Lorentz transformations $(K)$, spacetime translations $(a^{++})$, special conformal transformations $(b_{++})$, chiral transformations $(\chi)$ and $\sSU(n)$ rotations $(\l^{i}{}_{j})$. The constant fermionic parameters correspond to $Q$-supersymmetry $(\epsilon^{+i})$ and $S$-supersymmetry $(\eta_{+ j})$ transformations.

Next, we consider the action of $\mathsf{OSp}_0 (q|2;\mathbb{R})$ on $S^{1|q}$, see \cite{KT-M2021} for more details.
This supergroup is spanned by supermatrices of the form
\bea
h \in \sSL(2 | q;{\mathbb R} ) ~, \qquad
h^{\rm sT} \, \mathbb{J} \,h = \mathbb{J} ~,  \qquad
\mathbb{J}=
\left(
\begin{array}{cc|c}
	0&  1 ~&0\\
	-1 & 0 ~&0\\
	\hline
	0 & 0& {\ri \mathbbm 1}_{n}
\end{array}
\right) ~,
\eea
and naturally acts on the space of even supertwistors ${\mathbb R}^{2|q}$
\bea
Y = \left(
\begin{array}{c}
	a \\
	b\\
	\hline
	\s^{\UI}
\end{array}
\right) ~,\qquad \UI=\underline{1}, \dots , \underline{q}~,
\eea
where $a,b$ denote real bosonic variables which are not both zero, and are $\s^\UI$ real Grassmann variables. Two supertwistors $Y$ and $Y'$ are equivalent if $Y' = \g Y$, where $\g \in \mathbb{R} \setminus \{ 0 \}$. Assuming that $a \neq 0$, we can choose the representative
\bea
Y = \left(
\begin{array}{c}
	1 \\
	-x^{--}\\
	\hline
	\ri \q^{-\UI}
\end{array}
\right) ~,
\eea
where $(x^{--},\q^{-\UI})$ constitute inhomogeneous coordinates for $S^{1|q}$. To deduce the superconformal transformations of this subspace it is necessary to act on $Y$ with a generic group element $h \in {\sOSp}_0 (q|2; {\mathbb R} )$
\begin{align}
	h = \re^{\L_R} ~, \qquad \L_R = 
	\left(
	\begin{array}{cc|c}
		-\hf(\s - K) &  {- b_{--}} ~& - \eta_-^{\UJ}\\
		{- a^{--}} & \hf(\s - K) &\sqrt{2} {\epsilon}^{- \UJ}\\
		\hline
		\ri \epsilon^{- \UI} & \ri {\eta}^{\UI}_-& \r^{\UI \UJ}
	\end{array}
	\right)~.
\end{align}
Here all parameters are real and $\r^{\UI \UJ} = - \r^{\UJ \UI}$. Taking the parameters to be small, it may be shown that the most general infinitesimal superconformal transformations on this subspace are:
\begin{subequations}
	\label{5.172}
	\begin{align}
		\d x^{--} &= (\s - K) x^{--} + a^{--} - x^{--} b_{--} x^{--} + \ri \epsilon^{- \UI} \q^{- \UI} + \ri x^{--} \eta_-^{\UI} \q^{- \UI}  ~, \\
		\d \q^{- \UI} &= \hf (\s - K) \q^{- \UI} - \q^{- \UI} b_{--} x^{--} + \epsilon^{- \UI} + \r^{\UI \UJ} \q^{- \UJ} - \eta_-^{\UI} x^{--} - \ri \q^{- \UI} \eta_-^{ \UJ} \q^{- \UJ}~.
	\end{align}
\end{subequations}
The constant bosonic parameters in \eqref{5.172} correspond to dilatations $(\s)$, Lorentz transformations $(K)$, spacetime translations $(a^{--})$, special conformal transformations $(b_{--})$ and $\sSO(q)$ rotations $(\r^{\UI \UJ})$. The constant fermionic parameters correspond to $Q$-supersymmetry $(\epsilon^{-\UI})$ and $S$-supersymmetry $(\eta_-^{\UI})$ transformations. We emphasise that the parameters $\s$ and $K$ in \eqref{5.167} and \eqref{5.172} are the same.

\chapter{Discussion} \label{Chapter6}

Over the course of this thesis, we have employed the principles of superconformal symmetry to derive a plethora of new results for several distinct topics of study. They are as follows: (i) geometric symmetries of curved backgrounds and higher symmetries of several matter multiplets in four dimensions; (ii) (super)conformal higher-spin gauge theories in four dimensions; and (iii) superspace formulations for conformal $(p,q)$ supergravity in two dimensions. Since we have already discussed the key results of this thesis at the end of the appropriate chapters, here we will comment on their broader context within the literature and describe possible future avenues of research.

As we have repeatedly demonstrated, the formalism of conformal (super)space proves to be a powerful tool to probe the structure of (super)conformal field theories. This was first exemplified through the study of (conformal) isometries of curved backgrounds, where many calculations were significantly simplified by employing this framework.\footnote{In particular, the proof that the (super)conformal algebra of a given curved background is finite dimensional and that its dimension does not exceed that of Minkowski (super)space was made trivial by working within conformal (super)space.} A natural extension of the analysis we have presented is to describe such symmetries in $2 \leq d \leq 6$ dimensions,\footnote{As mentioned in chapter \ref{Chapter3},  progress in this area has been made in the $d=3,~\cN=1$ \cite{LO}, $d=5,~\cN=1$ \cite{KNT-M} and $d=6,~\cN=(1,0)$ \cite{KLRTM} cases.} where conformal superspace formulations exist. In the $d=2$ and $d=3$ cases the conformal superspace geometries described in chapter \ref{Chapter5} of the present work and \cite{BKNT-M1}, respectively, would be powerful tools to study this problem.

Further, as discussed in chapter \ref{Chapter3}, (conformal) isometries generate off-shell symmetries of every field theory propagating on a given background, and the corresponding equations of motion. As a natural generalisation, one can consider higher symmetries of a general kinetic operator as higher-rank extensions of the transformations induced by (conformal) isometries. We recall that the set of all higher symmetries of a given kinetic operator forms what is known as a (super)conformal higher-spin algebra, whose gauge theory is (super)conformal higher-spin theory. Hence, it would be interesting to extend the study of higher symmetries presented in this thesis to diverse dimensions and supersymmetries as it would lead to new realisations for such superalgebras.

The higher symmetries of a kinetic operator are in one-to-one correspondence with the rigid symmetries of the corresponding action, see e.g. \cite{BB}. Although it was not explicitly discussed in the main body, the latter are dual to, and thus may be used to compute conserved (super)current multiplets, see e.g. \cite{BJM,BBvD,BJM2,KPR22} for more details. Rather than pursuing this `constructive' approach, one may instead postulate the appropriate conservation equations and check their validity by employing the principles of (super)conformal symmetry, which was the approach advocated in this thesis. In general, each conserved (super)current $\mathfrak{J}^{\cA}$ is associated with a gauge (super)field $\U_{\cA}$, whose gauge freedom and (super)conformal properties follow from requring that the Noether coupling
\bea
\label{6.1}
\cS_{\text{N.C.}} = \int \rd \mu \,  \U_{\cA} \mathfrak{J}^{\cA} ~,
\eea 
is locally (super)conformal and gauge-invariant, where $\rd \mu$ is the appropriate measure. While this thesis was devoted to the four-dimensional story, it would be interesting to extend this analysis to diverse dimensions.

Further, though it was not emphasised, our attention throughout this thesis has been restricted to conserved currents and their dual gauge (super)fields of `minimal depth.' Simply put, this means that, at the component level, the gauge transformations of the latter are first-order in derivatives. Instead, one may consider conformal fields of depth-$t$. Such fields are characterised by having gauge transformations which are order-$t$ in derivatives. In conformally flat backgrounds, these fields were studied in \cite{KP} and their $\cN=1$ superconformal extension followed shortly afterwards \cite{KPR2}. It remains an open problem to determine superconformal extensions of these models for general $\cN$.

The free (S)CHS models described in this thesis possess another remarkable property, namely $\sU(1)$ duality. The latter follows from the fact that the Bianchi identities and free equations of motion for (S)CHS fields take the same functional form, and thus may be rotated into each other by an appropriate $\sU(1)$ transformation. One may then study nonlinear self-dual models for (S)CHS fields, though there is significant freedom in defining such a theory.\footnote{It should be emphasised that such theories possess many remarkable properties, including the self-duality of the corresponding Lagrangian under Legendre transformations.} This freedom is drastically reduced by requiring that the resuling theory is also (super)conformal, which was the approach taken in section \ref{Chapter4.3}. An important class of duality-invariant models not discussed in this section are those describing the propagation of SCHS gauge multiplets described by the complex superfields $H_{\a(m) \ad(n)}$, $m \neq n$ and the $\cN=2$ superconformal gravitino, whose gauge-invariant action was given in appendix \ref{Appendix4A2}. Analyses of such models are currently in progress \cite{WIP}. Additionally, $\sU(1)$ duality-invariant models for the higher-depth (super)fields mentioned in the previous paragraph are yet to be described in the literature. It would be interesting to pursue this line of study in the future.

Fundamental to (S)CHS theory is the identification of possible couplings between gauge (super)fields and matter multiplets. In addition to such analyses being important in their own right, they also may be used to define a theory of an infinite tower of interacting (S)CHS multiplets as an induced action, see e.g. \cite{Segal, Tseytlin, BJM,KMT,KP23,KR21}. At the cubic level, such interactions are described by constructing all possible conserved (super)currents $\mathfrak{J}^{\mathcal{A}}$ bilinear in some matter multiplet $\S$ and then appending the corresponding Noether couplings \eqref{6.1} to the free action of the latter, which we denote $\cS_{\text{Free}}[\S]$:
\bea
\label{6.2}
\cS_{\text{Cubic}}[\S,\U] = \cS_{\text{Free}}[\S] + \sum_{\cA} \int \rd \mu \,  \U_{\cA} \mathfrak{J}^{\cA}~.
\eea
We emphasise that $\sum_{\cA}$ denotes a summation over all possible Noether couplings. As was shown in section \ref{Chapter4.4}, for the cubic action $\cS_{\text{Cubic}}[\S,\U]$ to be gauge-invariant off-shell, it is necessary for the matter multiplet $\S$ to transform\footnote{Such transformations are uniquely determined by requiring consistency with all off-shell constraints and superconformal invariance.} under the gauge transformations of $\U_{\mathcal{A}}$ and to couple to an infinite tail of `auxiliary' gauge fields. The latter are auxiliary in the sense that they transform algebraically under gauge transformations, and hence may be gauged away. In this thesis we performed such an analysis for both a complex scalar and chiral multiplet, but the structure of such couplings in the case of an $\cN=2$ hypermultiplet remains an open problem. This is partially a consequence of the off-shell hypermultiplet possessing an infinite number of auxiliary fields, hence the framework of harmonic \cite{GIKOS,GIOS} or projective \cite{KLR,LR1,LR2} superspace must be employed. It would be interesting to complete this analysis.

As we have demonstrated throughout this thesis, symmetry principles have proven to be a powerful lens by which to study diverse aspects of (super)conformal field theories. Further, the four-dimensional conformal (super)space geometries, which we have made extensive use of, have proven to be a natural framework to capture such symmetries. In the final chapter of the main body, namely chapter \ref{Chapter5}, we proposed new conformal $(p,q)$ superspace geometries as a starting point for probing the structure of two-dimensional conformal supergravity and superconformal field theories. An important application of these geometries, which should be investigated in the future, is the determination of the various versions of extended Poincar\'e supergravity which may be obtained by coupling conformal supergravity to compensating multiplets. A natural approach to this problem is the universal one of \cite{KakuTownsend}. Further, it would be interesting to study couplings of two-dimensional $(p,q)$ supersymmetric nonlinear $\s$-models, see e.g. \cite{Hull:1985jv, GHR, Hull2, Lindstrom:2005zr, Hull:2018jkr,Lindstrom:2022lld}, to (conformal) supergravity.

\end{subappendices}

\end{document}